%% file: ms.tex
\pdfoutput=1
\documentclass[12pt]{article}

\usepackage[T1]{fontenc}
\usepackage[utf8]{inputenc}
\usepackage[english]{babel}

\usepackage{amsmath}
\usepackage{amssymb}
\usepackage{amsthm}
\usepackage{mathtools}
\mathtoolsset{showonlyrefs,showmanualtags}

\allowdisplaybreaks

\usepackage[natbibapa,nodoi]{apacite}
\usepackage{authblk}

\usepackage{bbm}
\newcommand{\bbone}{\mathbbm{1}}
\usepackage{booktabs}
\usepackage[format=plain,font=small,labelfont=bf,textfont=up]{caption}
\usepackage{multirow}
\usepackage{enumitem}
\setlist[enumerate]{itemsep=0.1ex}
\setlist[itemize]{itemsep=0.1ex}
\newcommand\expcommand\newcommand
\usepackage[letterpaper,margin=1in,bottom=1.4in]{geometry}
\usepackage{graphicx}
\usepackage[hidelinks]{hyperref}
\usepackage{subcaption}

\usepackage[nohints,tight]{minitoc}

\usepackage[linesnumbered,ruled,vlined]{algorithm2e}

\newcommand{\figurepath}{./figures}

\newcommand\mainref\ref
\newcommand\suppref\ref

\input{\texpath/stddef.tex}

\input{\texpath/commands.tex}

\makeatletter
\AddToHook{env/tabular/begin}{\let\input\@@input}
\makeatother

\title{\Large\textbf{Balancing Covariates in Randomized Experiments\\with the Gram--Schmidt Walk Design}}

\author[1]{Christopher Harshaw}
\author[2]{Fredrik Sävje}
\author[2]{Daniel A. Spielman}
\author[3]{Peng Zhang}
\affil[1]{Massachusetts Institute of Technology}
\affil[2]{Yale University}
\affil[3]{Rutgers University}

\date{}

\begin{document}

\makeatletter%
\begin{NoHyper}\gdef\@thefnmark{}\@footnotetext{\hspace{-1em}%
We thank Edo Airoldi, P.\ M.\ Aronow, Chen Chen, Nicholas Christakis, Peng Ding, Xavier D'Haultfœuille, Maximilian Kasy, Rad Niazadeh, David Pollard, Cyrus Samii, Jasjeet Sekhon and Johan Ugander for helpful comments and discussions.
We thank Akshay Ramachandran for allowing us to include his proof of Lemma~\suppref{lemma:technical_inequality} in the supplement, which is shorter than our original proof.
This work was supported in part by NSF Grant CCF-1562041, ONR Awards N00014-16-2374 and N00014-20-1-2335, a Simons Investigator Award to Daniel Spielman, and an NSF Graduate Research Fellowship (DGE1122492) awarded to Christopher Harshaw.
Computing infrastructure was supplied by the Yale Center for Research Computing.%
}\end{NoHyper}%
\makeatother%

\maketitle

\begin{abstract}
The design of experiments involves a compromise between covariate balance and robustness.
This paper provides a formalization of this trade-off and describes an experimental design that allows experimenters to navigate it.
The design is specified by a robustness parameter that bounds the worst-case mean squared error of an estimator of the average treatment effect.
Subject to the experimenter's desired level of robustness, the design aims to simultaneously balance all linear functions of potentially many covariates.
Less robustness allows for more balance.
We show that the mean squared error of the estimator is bounded in finite samples by the minimum of the loss function of an implicit ridge regression of the potential outcomes on the covariates.
Asymptotically, the design perfectly balances all linear functions of a growing number of covariates with a diminishing reduction in robustness, effectively allowing experimenters to escape the compromise between balance and robustness in large samples.
Finally, we describe conditions that ensure asymptotic normality and provide a conservative variance estimator, which facilitate the construction of asymptotically valid confidence intervals.

\vspace{0.2in}
\end{abstract}
\vspace{0.2in}

\newpage
\setcounter{parttocdepth}{1}
\doparttoc
\faketableofcontents
\mtcaddpart[Main]
\parttoc

\newpage

\section{Introduction}\label{sec:intro-gsw}

Randomized experiments are considered the most reliable way to estimate causal effects.
Properly implemented randomization ensures that treatment effect estimators are unbiased.
However, randomization does not ensure that estimators capture the true effect for any specific assignment of treatments.
In an effort to make the estimators more precise, experimenters sometimes restrict the randomization to achieve covariate balance between treatment groups.
A concern with this approach is that unobserved characteristics, including potential outcomes, may not be similar between the groups even if the observed characteristics are.

An idea that goes back to at least \citet{Efron1971Forcing} is that the design of experiments involves a compromise between covariate balance and robustness.
Randomization does not balance observed covariates to the same degree as a non-random assignment that specifically targets covariate balance, but randomization provides protection against imbalances on unobserved characteristics.
Experimenters must weigh the robustness granted by randomness against possible gains in precision granted by balancing prognostically important covariates.

The first contribution of this paper is a new formalization of the trade-off between covariate balance and robustness.
The formalization clarifies some ideas previously discussed by other authors and provides several new insights.
We describe quantitative measures of both covariate balance and robustness, and we motivate the measures by showing that they characterize the precision of the Horvitz--Thompson estimator of the average treatment effect.
There is a fundamental tension between the two measures, as an experimenter cannot simultaneously achieve maximal robustness and fully balance covariates.

The second contribution of the paper is the development of the Gram--Schmidt Walk design, which allows experimenters to navigate the trade-off between balance and robustness.
The design is specified by a parameter that bounds the worst-case mean squared error of the estimator.
The design aims to simultaneously balance all linear functions of the covariates specified by the experiment subject to meeting the worst-case guarantee.
We describe several characterizations of the behavior of the design in finite samples.
The main results are tight bounds on the mean squared error and on the tails of the sampling distribution of the treatment effect estimator.

We next investigate the asymptotic behavior of the estimator under the design.
Under mild assumptions on the potential outcomes and the covariates, we show that the estimator is root-$n$ consistent and that its limiting variance is the same as when all linear functions of the covariates are perfectly balanced.
This means that the Gram--Schmidt Walk design allows experimenters to escape the balance--robustness trade-off in large samples.
The limiting variance of the estimator under the Gram--Schmidt Walk design is less than or equal to the limiting variance of other commonly used designs, such as rerandomization.

The final contribution of the paper is to describe methods for inference.
We provide a central limit theorem for the Horvitz--Thompson estimator under the Gram--Schmidt Walk design, and provide a consistent, conservative estimator of the variance of the point estimator.
Together, these results allow experimenters to construct conservative confidence intervals that are asymptotically valid.

A key discovery facilitating results in this paper is a translation of the experimental design problem to a new type of problem in algorithmic discrepancy.
A central problem of discrepancy theory is to partition a collection of vectors into two sets so that the sum of the vectors in each set is similar \citep{Spencer1985Six}.
This problem directly corresponds to finding a treatment assignment that maximizes covariate balance.
However, algorithms for discrepancy minimization aim to produce a single partition, corresponding to a single assignment.
Experimenters generally seek a distribution of assignments when they assign treatments, so as to achieve robustness from randomization.
We argue that the experimental design problem is best interpreted as a \emph{distributional} discrepancy problem.
To tackle this problem, we take advantage of the Gram--Schmidt Walk algorithm of \citet{Bansal2019Gram}.
This is a randomized algorithm, but the original authors used the randomization simply as a means to solve a non-distributional discrepancy problem.
We leverage and deliberately amplify the randomized aspect to solve the distributional discrepancy problem.
We also tighten and extend the analysis of the algorithm compared to \citet{Bansal2019Gram} to be relevant for the experimental design problem.
While we find the connection between these two fields insightful and important, an understanding of discrepancy theory is not required to understand the results in this paper.

\section{Related Work}\label{sec:related-works-gsw}

The debate about the virtues of randomization goes back to the dawn of statistics.
\citet{Student1938Comparison} argued that randomization often is harmful because random assignments can only make treatment groups less comparable than what they would be under the most balanced assignment.
This idea has more recently been discussed and extended by \citet{Bertsimas2015Power}, \citet{Kasy2016Why}, \citet{Deaton2018Understanding}, and \citet{Kallus2018Optimal}.
On the other hand, \citet{Fisher1925Statistical,Fisher1926Arrangement} argued that randomization is desirable because it provides a certain level of robustness, in the form of unbiasedness, and facilitates well-motivated confidence intervals and testing.
The first of Fisher's points was extended by \citet{Wu1981Robustness} to worst-case mean squared error, which is a more general robustness concept than unbiasedness \citep[see also][]{Kallus2018Optimal,Bai2023Why,Basse2022Minimax,Nordin2022Properties}.
Fisher's second point is discussed and extended by \citet{Johansson2021Optimal}.

A compromise between these two viewpoints is possible.
While \citet{Wu1981Robustness} demonstrates that there is no room to seek balance when robustness is our only objective, we might still be willing to accept a less robust design if it provides balance along dimensions we believe are important.
This is the trade-off between balance and robustness mentioned in the introduction.
The idea can be traced back to \citet{Efron1971Forcing}, whose concept of ``accidental bias'' is closely related to our concept of robustness.
This work has been extended by \citet{Kapelner2021Harmonizing} and a related idea based on a decision theoretical framework has been explored by \citet{Banerjee2020Theory}.

It is rare that experimenters assign treatment deterministically as suggested by \citet{Student1938Comparison}, but they do not necessarily assign treatments fully at random.
Many designs fall in between the two extremes.
Examples include the matched pair design \citep{Greevy2004Optimal,Imai2009Essential,Bruhn2009Pursuit}, various stratified designs \citep{Fisher1935Design,Higgins2016Improving,Cytrynbaum2021Designing}, and rerandomization \citep{LockMorgan2012Rerandomization,Li2018Asymptotic}.
Existing analyses of these designs do not consider a formal balance--robustness trade-off nor provide ways to navigate one.

To the best of our knowledge, there are only two prior designs that explicitly account for some version of the balance--robustness trade-off.
\citet{Krieger2019Nearly} construct an algorithm that makes local changes to an assignment that is generated fully at random, aiming to produce a new assignment that is more balanced.
They show that if there are few covariates, then few changes are needed to reach a highly balanced assignment, so the final assignment vector is similar to the one that was generated at random.
\citet{Kapelner2022Optimal} investigate how to optimally select the acceptance criterion for rerandomization given a desired level of robustness.

\section{Preliminaries}\label{sec:preliminaries-gsw}

There are $n$ units in the experiment, indexed by integers $[n] = \setb{1, \dotsc, n}$.
The experimenter randomly assigns a treatment $\zi \in \setb{\pm 1}$ to each unit, and the assignments are collected in the random vector $\zv = \paren{\ze{1}, \dotsc, \ze{n}}$.
We use $\treated = \setb{i \in [n] : \zi = 1}$ and $\controls = \setb{i \in [n] : \zi = -1}$ to denote the random partition of the units into treatment and control groups.
The \emph{design} of the experiment is the distribution over the assignment vectors $\zv \in \setb{\pm 1}^n$.

Each unit has two potential outcomes: $\poai$, which is observed if $\zi = 1$, and $\pobi$, which is observed if $\zi = -1$.
We assume these potential outcomes are well-defined throughout the paper, meaning that we rule out interference and other hidden versions of treatment.
The observed outcome $\ooi$ for each unit is the random variable taking the value $\poai$ when $\zi = 1$ and $\pobi$ when $\zi = -1$.
It will prove convenient to collect the outcome variables into vectors:
\begin{equation}
\poav = \paren{\poae{1}, \poae{2}, \dotsc, \poae{n}},
\qquad
\pobv = \paren{\pobe{1}, \pobe{2}, \dotsc, \pobe{n}},
\qquad
\oov = \paren{\ooe{1}, \ooe{2}, \dotsc, \ooe{n}}.
\end{equation}

Each unit has a vector of $d$ covariates: $\xvi \in \Reals^d$.
The largest covariate norm among the units is denoted $\maxnormX = \max_{i \in [n]} \norm{\xvi}$.
The covariates are known to the experimenter prior to treatment assignment, so the experimental design may depend on them.
The only randomness in the experiment comes from the assignment of treatment.
The potential outcomes and covariates of the units are non-random and fixed, and we impose no assumptions on them at this point other than their existence.

The causal quantity of interest is the \emph{average treatment effect}: $\ate = n^{-1} \sum_{i=1}^n \paren{\poai - \pobi}$.
The average treatment effect cannot be directly observed, so it must be estimated.
In this paper, we restrict our attention to the Horvitz--Thompson estimator
\begin{equation}
\htest = \frac{1}{n} \sum_{i \in \treated} \frac{\ooi}{\Pr{\zi = 1}}
	- \frac{1}{n} \sum_{i \in \controls} \frac{\ooi}{\Pr{\zi = -1}}.
\end{equation}
This estimator is unbiased under designs that satisfy the positivity condition that the assignment probabilities are bounded away from zero and one for all units \citep{Aronow2013Class}.
The aim of the experimenter when designing the experiment is to improve the precision of the estimator.
To make the task concrete, we will primarily focus on the estimator's mean squared error, $\E{\paren{\ate - \htest}^2}$, as our measure of precision.

For expositional purposes, we restrict our attention throughout the paper to \emph{symmetric designs} for which each unit is equally likely to receive either treatment: $\Pr{\zi = 1} = 1/2$.
The extension of our results to settings with $\Pr{\zi = 1} \in (0,1)$ is straightforward but notionally cumbersome, and it is therefore discussed in Section~\suppref{sec:non-uniform-probs-supp} in the supplement.

The error of the Horvitz--Thompson estimator for a particular assignment can be shown to depend on the potential outcomes only through their sum: $\pomv = \poav + \pobv$.
For short, we refer to $\pomv$ as the \emph{potential outcome vector}.
This insight allows us to derive the mean square error of the estimator under an arbitrary design.

\expcommand{\mseexpression}{%
	For all symmetric experimental designs, the mean squared error of the Horvitz--Thompson estimator is
	\begin{equation}
		\E{\paren{\htest - \ate}^2} = n^{-2} \pomv^\tran \Cov{\zv} \pomv.
	\end{equation}
}

\begin{lemma}\label{lemma:mse-expression}
	\mseexpression
\end{lemma}

The lemma demonstrates that the mean squared error is a quadratic form in the covariate matrix of the treatment assignment vector, $\Cov{\zv}$, evaluated at the (unknown) potential outcome vector $\pomv$.
Properties of the design that affect the mean squared error are therefore completely captured by $\Cov{\zv}$.
This is a central insight motivating our work in this paper, informing both our interpretation of the experimental design problem as well as the proposed design.
This insight has been used previously to inform investigations of experimental designs, and the characterization of the precision of the estimator in Lemma~\ref{lemma:mse-expression} is similar to those given by \citet{Efron1971Forcing} and \citet{Kapelner2021Harmonizing}.

\section{The Balance--Robustness Trade-off}\label{sec:new-perspective}

\subsection{A Measure of Robustness}\label{sec:robustness}

Researchers use experiments because they provide credible causal inferences without need for strong assumptions.
For example, under mild moment conditions on the potential outcomes and independent treatment assignment, the Horvitz--Thompson estimator is unbiased and converges to the average treatment effect at a root-$n$ rate no matter what the potential outcomes might be.
Experiments are in this sense robust.
An important insight for what is to come is that all experiments are not equally robust.
We use a worst-case concept to quantify robustness:
an experiment is said to be robust if the estimator is sufficiently precise for all possible potential outcomes under its design.
Building on the work of \citet{Efron1971Forcing} and \citet{Kapelner2021Harmonizing}, we show that the operator norm of $\Cov{\zv}$ characterizes the worst-case performance of the design.

\expcommand{\lemmarobustness}{%
For all symmetric experimental designs, the worst-case mean squared error over the set of all potential outcomes with bounded magnitude is
\[
	\max_{\pomv \in \POspace{M}}
	\E{(\ate - \htest)^2}
	= \frac{M}{n} \norm{\Cov{\zv}},
	\qquadwhere
	\POspace{M} = \setb[\big]{\pomv \in \Reals^n : n^{-1} \norm{\pomv}^2 \leq M}.
\]
}

\begin{lemma}\label{lemma:robustness}
\lemmarobustness
\end{lemma}

Lemma~\ref{lemma:robustness} shows that the operator norm $\norm{\Cov{\zv}}$ captures how robust a design is.
The norm increases as the correlation between the assignments becomes stronger, so designs with greater correlation are less robust.
An implication is that designs with no correlation are most robust, as captured by the following proposition.

\expcommand{\corobernoulliminmax}{%
All symmetric experimental designs satisfy the inequality $\norm{\Cov{\zv}} \geq 1$, and equality holds for the Bernoulli design.
Thus, the Bernoulli design is min-max optimal for potential outcomes with bounded average magnitude, $\POspace{M}$, for any $M$.
}

\begin{proposition}\label{coro:bernoulli-min-max}
\corobernoulliminmax
\end{proposition}

The Bernoulli design assigns treatments independently between units, so $\Cov{\zv} = \unitM$ and $\norm{\Cov{\zv}} = 1$.
The operator norm cannot be made smaller than one because the diagonal entries of $\Cov{\zv}$ are always one for symmetric designs, and the operator norm is at least the maximum entry.
Thus, Proposition~\ref{coro:bernoulli-min-max} shows that an experimenter who seeks to maximize robustness, when formalized in this way, should use the Bernoulli design.

The operator norm $\norm{\Cov{\zv}}$ can be seen as a unitless measure of robustness, in the sense that it measures the multiplicative increase in the worst-case mean squared error compared to the min-max design.
For example, if some design has $\norm{\Cov{\zv}} = 2$, then its worst-case mean squared error is twice as large as the worst-case mean squared error under the min-max design.
The largest possible value of $\norm{\Cov{\zv}}$ is $n$, achieved by a minimally random design that assigns some $\zv' \in \setb{\pm 1}^n$ with probability $1/2$, and otherwise its negation $-\zv'$.

\subsection{A Measure of Covariate Balance}\label{sec:covariate-balance}

A robust design ensures that the estimator is reasonably precise no matter what the potential outcomes might be.
It is possible to make the estimator more precise if the experimenter has prior knowledge about the units and uses this knowledge when designing the experiment.
The experimenter would then forgo some robustness to improve precision for certain potential outcomes.
If the prior knowledge is in the form of pre-treatment covariates that are known to be predictive of the potential outcomes, then precision is improved by using a design that ensures balance between the treatment groups with respect to those covariates.

We collect the units' covariate vectors $\xv{1}, \dotsc, \xv{n}$ as rows of an $n$-by-$d$ matrix $\xM$.
It will prove convenient to use the maximum row norm $\maxnormX = \max_{i \in [n]} \norm{\xvi}$ as a measure the magnitude of the covariates $\xM$.
To introduce and illustrate our notion of covariate balance, assume for the moment that the covariates are perfectly linearly predictive of the outcomes, so there exists a function $\lfx$ such that $\pomv = \xM \lfx$.
This will not be necessary for any of our results, but it will be helpful to illustrate our concept of covariate balance in this subsection.
Using Lemma~\ref{lemma:mse-expression}, we can write the mean square error as
\begin{equation}
n^2 \E{\paren{\htest - \ate}^2}
= \lfx^\tran \xM^\tran \Cov{\zv} \xM \lfx
= \lfx^\tran \Cov{\xM^\tran \zv} \lfx.
\end{equation}

To make the estimator more precise in this setting, we should pick a design that makes $\lfx^\tran \Cov{\xM^\tran \zv} \lfx$ small.
However, even if we somehow knew that the covariates were perfectly predictive, we would generally not know the function $\lfx$; we must consider a set of possible functions.
As above, we can use an operator norm bound for this purpose.
For all linear functions $\lfx \in \Reals^\xdim$, we have $\lfx^\tran \Cov{\xM^\tran \zv} \lfx \leq \norm{\Cov{\xM^\tran \zv}} \times \norm{\lfx}^2$.

Holding the magnitude $\norm{\lfx}$ fixed, the operator norm $\norm{\Cov{\xM^\tran \zv}}$ provides a guarantee on the mean square error when the covariates are perfectly predictive.
If the operator norm is small, we know that the mean square error is small no matter how the potential outcomes are related to the covariates.
The bound is sharp, so conversely, if the operator norm is large, then we know that there exists a function $\lfx$ for which the mean square error is large compared to the magnitude $\norm{\lfx}$.
Importantly, $\norm{\Cov{\xM^\tran \zv}}$ does not depend on the potential outcomes, so experimenters can target it when designing their experiments.
For these reasons, we will use the operator norm $\norm{\Cov{\xM^\tran \zv}}$ as our measure of covariate balance.
The dependence on $\norm{\lfx}$ in the bound is inescapable because the mean square error depends on the magnitude of the outcomes, and $\norm{\lfx}$ captures the relative scaling of the covariates and potential outcomes.
Holding the relative scaling fixed, $\norm{\lfx}$ can be seen as a type of complexity measure of the function $\lfx$.

In Section~\suppref{sec:computational-hardness} of the supplement, we show  $\norm{\Cov{\xM^\tran \zv}}$ cannot be made smaller than $\maxnormX^2$ without imposing additional restrictions or assumptions.
A practically relevant upper bound on $\norm{\Cov{\xM^\tran \zv}}$ is given by the Bernoulli design, which achieves $\norm{\Cov{\xM^\tran \zv}} = \norm{\xM^\tran \xM}$.
Under the assumptions we use for our large sample analysis, $\maxnormX^2 = \bigO{d \logf{n}}$ and $\norm{\xM^\tran \xM} = \bigO{n}$, so the relevant interval for $\norm{\Cov{\xM^\tran \zv}}$ is $[d \logf{n}, n]$ up to constant factors.

\subsection{The Trade-off}\label{sec:trade-off}

A design that is maximally robust requires uncorrelated treatment assignments, but a design that achieves maximal covariate balance typically requires highly correlated assignments.
It is therefore not possible to construct a design that achieves both maximum robustness and maximum covariate balance, in the sense that it is not possible to make the operator norms $\norm{\Cov{\zv}}$ and $\norm{\Cov{\xM^\tran \zv}}$ small simultaneously.

\expcommand{\proptradeoff}{%
If the largest singular value of the covariate matrix is larger than the maximum norm of the covariate vectors, $\sigma_{\max}(\xM) > \maxnormX = \max_{i \in [n]} \norm{\xvi}$, then there does not exist a design that simultaneously minimizes $\norm{\Cov{\zv}}$ and $\norm{\Cov{\xM^\tran \zv}}$.
}

\begin{proposition}\label{prop:trade-off}
\proptradeoff
\end{proposition}

The proposition captures a tension between $\norm{\Cov{\zv}}$ and $\norm{\Cov{\xM^\tran \zv}}$.
We refer to this tension as the \emph{balance--robustness trade-off}.
The condition of the proposition ensures that there is not a single covariate vector that dominates the balance properties of the experiment.
We always have $\sigma_{\max}(\xM) \geq \maxnormX$, so the condition rules out the edge case $\sigma_{\max}(\xM) = \maxnormX$.
As we will discuss in Section~\ref{sec:asymptotic-analysis}, the typical rate for $\sigma_{\max}(\xM)$ is $\sqrt{n}$, while the typical rate for $\maxnormX$ is $\sqrt{d \logf{n}}$, meaning that $\sigma_{\max}(\xM)$ generally is much larger than $\maxnormX$.

Because they cannot achieve both balance and robustness, experimenters must navigate the balance--robustness trade-off when they design their experiments.
To better understand what is at stake, consider when the covariates are only somewhat predictive, so covariate balance would be useful, but they are not perfectly predictive, so the analysis in the previous subsection does not apply.
As above, let $\lfx \in \Reals^d$ be some linear function.
For the purpose of the current discussion, it is not important exactly what this function is---whatever it might be, it is unknown when the experiment is designed.
Decompose the potential outcome vector into the linear function evaluated at the covariates, $\xM \lfx$, and a residual term, $\rpomv = \pomv - \xM \lfx$, so that $\pomv = \xM \lfx + \rpomv$.
If the covariates were perfectly predictive, then there exists a function $\lfx$ such that $\rpomv = \zerovec$.
However, we here consider when the covariates are only partially predictive, in which case $\rpomv \neq \zerovec$ no matter the choice of $\lfx$.

For an arbitrary function $\lfx \in \Reals^d$, the mean square error decomposes as
\begin{equation}
	n^2 \E{\paren{\ate - \htest}^2}
	= \lfx^\tran \Cov{\xM^\tran \zv} \lfx + \rpomv^\tran \Cov{\zv} \rpomv + 2 \lfx^\tran \Cov{\xM^\tran \zv, \zv} \rpomv.
\end{equation}
The first term of this expression, $\lfx^\tran \Cov{\xM^\tran \zv} \lfx$, corresponds to the part of the potential outcomes that can be explained by the function $\lfx$.
Following the logic of the previous section, making $\norm{\Cov{\xM^\tran \zv}}$ small makes the first term small for any function $\lfx$ of fixed magnitude.
The second term, $\rpomv^\tran \Cov{\zv} \rpomv$, corresponds to what cannot be explained by the function $\lfx$.
If the covariates are completely unpredictive, then we cannot do better than $\lfx = \zerovec$, so $\pomv = \rpomv$, and the second term corresponds exactly to what was studied in Section~\ref{sec:robustness}.
Hence, making $\norm{\Cov{\zv}}$ small, makes the second term small.

The third term of the decomposition is a cross term: $2 \lfx^\tran \Cov{\xM^\tran \zv, \zv} \rpomv$.
It is possible to characterize this cross term by considering properties of the matrix $\Cov{\xM^\tran \zv, \zv}$.
Indeed, we do this for the design described in this paper, the Gram--Schmidt Walk design.
However, for the purpose of illustrating the key tension in the balance--robustness trade-off, such a detailed characterization would be a distraction.
Instead, to understand the trade-off, note that the cross term is bounded by the first two terms:
\begin{equation}
	2 \lfx^\tran \Cov{\xM^\tran \zv, \zv} \rpomv
	\leq
	\lfx^\tran \Cov{\xM^\tran \zv} \lfx + \rpomv^\tran \Cov{\zv} \rpomv.
\end{equation}
Therefore, controlling the first two terms indirectly controls the cross-term; if the first two terms are small, so is the cross term.

When the covariates are only partially predictive of the potential outcomes, the decomposition tells us that the mean square error is determined by both $\norm{\Cov{\zv}}$ and $\norm{\Cov{\xM^\tran \zv}}$.
Ideally, we would want to both operator norms small, but this is not possible.
Because of the balance-robustness trade-off, we can generally make one of them small only by accepting that the other becomes larger.
Experimenters therefore have to choose between a design that balances the covariates well or a design that is highly robust, or something in-between.

\subsection{Balance--Robustness Design Guarantees}\label{sec:bal-rob-guarantees}

While it is not possible to make $\norm{\Cov{\zv}}$ and $\norm{\Cov{\xM^\tran \zv}}$ simultaneously small, it is possible to make them simultaneously large.
That is, there are designs that provide both poor covariate balance and poor robustness.
We seek to avoid such designs.

\begin{definition}
An experimental design is said to provide a balance--robustness guarantee of $(\zsc, \xsc)$ if it ensures that $\norm{\Cov{\zv}} \leq \zsc$ and $\norm{\Cov{\xM^\tran \zv}} \leq \xsc$.
\end{definition}

A design that provides a guarantee that both $\zsc$ and $\xsc$ are reasonably small navigates the balance--robustness trade-off well.
While it is not possible to attain the minimums of $\zsc$ and $\xsc$ simultaneously, we can consider the minimal pairs $(\zsc, \xsc)$.
That is, a design that provides maximum covariate balance for a given level of robustness, or maximum robustness for a given level of covariate balance.
The set of all such designs constitutes a Pareto frontier of the balance--robustness trade-off.
We argue that experimenters should, if possible, use designs that are on or close to this Pareto frontier.

Designs that provide a balance--robustness guarantee also implicitly yields a guarantee on the mean square error of the treatment effect estimator.
Therefore, a design that better navigates the balance--robustness trade-off, in the sense of being closer to the Pareto frontier, provides a sharper guarantee on the mean square error.

\expcommand{\mseimplications}{%
For any symmetric experimental design with balance--robustness guarantee $(\zsc, \xsc)$, the mean squared error of the Horvitz--Thompson estimator is bounded as
\[
	n \E[\big]{\paren{\htest - \ate}^2}
	\leq
	\min_{\lfx \in \Reals^d}
	\bracket[\bigg]{
		\frac{\zsc}{n} \norm{\pomv - \xM \lfx}^2
		+ \frac{\xsc}{n} \norm{\lfx}^2
		+ \frac{2 \sqrt{\zsc \xsc}}{n} \norm{\pomv - \xM \lfx} \, \norm{\lfx}
	}.
\]
}

\begin{theorem}\label{thm:utility-of-new-discrepancy-formulation}
	\mseimplications
\end{theorem}

The theorem is a generalization of the characterizations of the mean square error in Sections~\ref{sec:robustness}~and~\ref{sec:covariate-balance} to settings with partially predictive covariates.
The first term captures how well a linear function $\lfx$ predicts the potential outcomes using the covariates.
This term can be made small if the potential outcome vector $\pomv$ is close to the span of the covariates.
The second term captures the magnitude of the function $\lfx$.
This term can be made small by using a function of small magnitude, typically meaning that the function does not predict the potential outcomes well.
The third term is the cross term discussed in the previous subsection.
The balance--robustness guarantee $(\zsc, \xsc)$ determines the trade-off between the terms, assigning more focus to either finding a function that predicts the outcomes well or one that is of small magnitude.
If the covariates are predictive, in the sense that there exists a function $\lfx \in \Reals^d$ such that the norm of $\rpomv = \pomv - \xM \lfx$ is small, then making $\xsc$ small will be more beneficial than making $\zsc$ small.
However, if no such function exists, so that the minimum of $\norm{\rpomv}$ is approximately the same as $\norm{\pomv}$, then making $\xsc$ small could cause harm by making $\zsc$ large.
The magnitude of the cross term is bounded by the geometric mean of the two leading terms, so if either of those terms are small, so will the cross term be.

The bound in Theorem~\ref{thm:utility-of-new-discrepancy-formulation} is tight, in the sense that it holds with equality for some potential outcomes and covariates, but there are situations in which the bound is quite loose.
It is not the purpose of the theorem to give an exact characterization of the mean square error.
Lemma~\ref{lemma:mse-expression} gives an exact characterization, but it depends on the full covariance matrix of the assignment, so it is considerably more complex than Theorem~\ref{thm:utility-of-new-discrepancy-formulation}.
The purpose of the theorem is instead to show that the balance--robustness trade-off, as formalized by the operator norms $\norm{\Cov{\zv}}$ and $\norm{\Cov{\xM^\tran \zv}}$, is widely applicable and relevant.
Crucially, the operator norms and the balance--robustness guarantee do not depend on the potential outcomes, so they can be used during the design stage of the experiment, before observing any potential outcomes.
A sharper characterization of the mean square error would need to consider more intricate aspects of the design and how they interact with the potential outcomes, making the characterization less useful for the purpose of designing experiments, because experimenters generally do not have access to such information at the design stage.

\section{The Gram--Schmidt Walk Design}\label{sec:gsw-design}

The Gram--Schmidt Walk design is constructed to navigate the balance--robustness trade-off.
It is parameterized by $\vto \in [0,1]$, which controls its balance--robustness guarantee.

A central aspect of the design is the construction of an \emph{augmented covariate vector} $\gscv{i} \in \Reals^{n+d}$ for each unit.
This is a scaled concatenation of the unit's raw covariate vector and a unit-unique indicator variable:\begin{equation}
	\gscv{i} = \begin{bmatrix}
		\sqrt{\vto} \basisv{i}
		\\[0.25em]
		\maxnormX^{-1} \sqrt{1 - \vto} \xvi
	\end{bmatrix},
\end{equation}
where $\basisv{i} = \paren{0, \dotsc, 0, 1, 0,  \dotsc, 0}$ is the $i$th standard basis vector of dimension $n$ and $\maxnormX = \max_{i \in [n]} \norm{\xvi}$ is the maximum covariate norm.
We collect the augmented vectors $\gscv{1}, \dotsc, \gscv{n}$ as columns of an $(n+d)$-by-$n$ matrix $\gsM$.

The design uses the augmented covariate vectors as input to a slight modification of the Gram--Schmidt Walk algorithm of \citet{Bansal2019Gram}.
This algorithm produces a random assignment vector $\zv \in \setb{ \pm 1}^n$ with the property that the (random) difference between the within-group sums of the augmented vectors concentrates around zero with high probability.
That is, $\gsM \zv = \sum_{i \in \treated} \gscvi - \sum_{i \in \controls} \gscvi \approx \zerovec$.
By balancing the augmented covariate vectors, the Gram--Schmidt Walk design balances both the original raw covariate vectors and the unit-unique basis vectors $\basisv{i}$.

The parameter $\vto$ determines to what extent the augmented covariate vectors resemble either the raw covariate vectors or the orthogonal basis vectors, and thus to what extent each of these sets of vectors are balanced.
The basis vectors are best balanced by assigning treatment fully at random, so the Gram--Schmidt Walk design induces less correlation between treatments when augmented covariate vectors mostly resemble the basis vectors.
This is the way the design navigates the balance--robustness trade-off.
When $\vto = 1$, the augmented covariate vectors are exactly the orthogonal basis vectors.
In that case, the Gram--Schmidt Walk design recovers the Bernoulli design.

The algorithm for sampling from the Gram--Schmidt Walk is described in Algorithm~\ref{alg:gswdesign-alg}.
It builds on a relaxation of the assignments from the integral values $\braces{\pm 1}$ to the interval $\bracket{-1,1}$.
We refer to assignments in the interior of this interval as \emph{fractional}.
The algorithm constructs the assignments by iteratively updating a vector of fractional assignments $\gsvt$ until it takes values in $\braces{\pm 1}$.
The initial fractional assignments are zero: $\gsv{1} = \zerovec$.
This means that the augmented covariate vectors start out perfectly balanced, because $\gsM \gsv{1} = \gsM \zerovec = \zerovec$.
However, the initial assignments are not acceptable, because they are not in $\braces{\pm 1}^n$.
As the algorithm updates the fractional assignments, the fundamental tension is between maintaining good balance, as measured by $\gsM \gsvt$, and making the assignments integral.
The algorithm navigates this tension by updating the assignments in a direction that does not increase
the imbalances too much, while ensuring that the update is large enough to be a sizable step towards integrality.

\begin{algorithm}[t]\caption{The Gram--Schmidt Walk}\label{alg:gswdesign-alg}
	Initialize an index $t \gets 1$.\\
	Initialize a vector of \emph{fractional assignments} $\gsv{1} \gets \paren{0, 0, \dotsc, 0}$.\\
	Select a \emph{pivot} unit $p$ uniformly at random from $[n]$.\\
	\While{ $\gsvt \notin \braces{\pm 1}^n$}
	{
		Create the set $\alive \gets \braces{ i \in [n] : \abs{\gsve{t}{i}} < 1 }$. \\
		If $p \notin \alive$, select a new pivot $p$ from $\alive$ uniformly at random. \\
		Compute a \emph{step direction} as \label{alg:step-direction}
		\begin{equation*}
			\begin{aligned}
				\gsuvt \gets \
				& \argmin_{ \gsuv{} \in \Reals^n}
				& & \norm{ \gsM \gsunv }^2 \\
				& \text{subject to}
				& & \gsun{p} = 1, \text{ and } \gsuni = 0 \text{ for all } i \notin \alive \\
			\end{aligned}
		\end{equation*} \\
		Set $\gsdnp \gets \abs{\max\Delta}$ and $\gsdnm \gets \abs{\min\Delta}$ where $\Delta = \braces{ \gsdn \in \Reals : \gsvt + \gsdn \gsuvt \in \bracket{-1, 1}^n}$. \\
		Select a \emph{step size} at random according to
		\begin{equation*}
			\gsdt \gets
			\begin{cases}
				\phantom{-}\gsdnp & \text{ with probability } \gsdnm / \paren{\gsdnp + \gsdnm},
				\\
				- \gsdnm & \text{ with probability } \gsdnp / \paren{\gsdnp + \gsdnm}.
			\end{cases}
		\end{equation*} \\
		Update the fractional assignments: $\gsv{t+1} \gets \gsvt + \gsdt \gsuvt$. \\
		Increment the index: $t \gets t + 1$. \\
	}
	\Return{assignment vector $\gsvt \in \braces{\pm 1}^n$.}
\end{algorithm}

A general implementation of the Gram--Schmidt Walk algorithm that explicitly constructs and solves the system of linear equations from scratch at each iteration would run in $\bigO{n^4 + n^3 d}$ time.
However, the structure of the augmented covariates allows us to construct a customized implementation that maintains a Cholesky factorization between iterations, improving the run time to $\bigO{n^2 d}$.
Section~\suppref{sec:gsw-implementation} in the supplement describes this implementation and proves its computational properties.

There are similarities between the Gram--Schmidt Walk design and the Cube Method of \citet{DevilleEfficient2004}, which is used in survey sampling.
Both methods build on the idea that assignment vectors can be represented as vertices of a hypercube and that an assignment can be obtained through a random walk inside that hypercube.
Indeed, many discrepancy minimization algorithms are based on such geometric interpretations.
The Cube Method can be seen as a randomized version of an algorithm by \citet{Beck1981Integer} for discrepancy minimization, followed by a rounding procedure.
To the best of our knowledge, this connection has gone unnoticed by both the survey sampling and theoretical computer science communities.
Unlike the Cube Method, which has two distinct phases, the iterations of the Gram--Schmidt Walk design all take a similar form.
The two-phase structure prevents the Cube Method from achieving balance--robustness guarantees comparable to those of the Gram--Schmidt Walk design, as the first phase does not consider how its updates affect the second phase.

\section{Finite-Sample Properties}\label{sec:finite-analysis}

\subsection{Martingale and Unbiasedness}

A central property of the Gram--Schmidt Walk design is that the sequence of the fractional assignment vectors forms a martingale.
This implies that the expectation of the assignments sampled from the design is zero, $\E{\zv} = \gsv{1} = \zerovec$, which in turn ensures unbiasedness of the Horvitz--Thompson estimator for the average treatment effect.
These insights are formalized in the following lemma and corollary.

\expcommand{\martingale}{%
	The sequence of fractional assignments $\gsv{1}, \gsv{2}, \dotsc$ forms a martingale.
}

\begin{lemma}\label{lemma:martingale}
	\martingale
\end{lemma}

\begin{corollary}\label{coro:marginal-prob-and-unbiased}
	Under the Gram--Schmidt Walk design, $\Pr{\zi = 1} = 1/2$ for all $i \in [n]$.
	Thus, the Horvitz--Thompson estimator is unbiased under the design.
\end{corollary}

The relation $\E{\zv} = \gsv{1}$ holds for any initial fractional assignments, which provides control over the first moment of the assignment vector.
We use this fact to extend the design to non-uniform assignment probabilities in Section~\suppref{sec:non-uniform-probs-supp} in the supplement.

\subsection{Navigating the Trade-off}\label{sec:solving-discrepancy}

The Gram--Schmidt Walk design is able to navigate the balance--robustness trade-off because it balances the augmented covariate vectors well, as described in the following theorem.
The proof, which is provided in the supplement, interprets the algorithm as implicitly constructing a random basis for the column space of $\gsM$, which reveals the connection between the Gram--Schmidt Walk and its namesake, the Gram--Schmidt orthogonalization procedure.

\begin{theorem}\label{thm:cov-bound}
	Under the Gram--Schmidt Walk design, the covariance matrix of the vector of imbalances for the augmented covariates $\gsM \zv$ is bounded as $\Cov{\gsM \zv} \preceq \PgsM$, where $\PgsM = \gsM \paren[\big]{ \gsM^\tran \gsM }^{-1} \gsM^\tran$ is the orthogonal projection onto the subspace spanned by the columns of $\gsM$.
\end{theorem}

The covariance matrix $\Cov{\gsM \zv}$ in Theorem~\ref{thm:cov-bound} captures how balanced the augmented covariates are.
The theorem states that the augmented covariates are balanced because the projection matrix $\PgsM$ in the upper bound is small by construction: it has at most $n$ eigenvalues that are one and $\xdim$ eigenvalues that are zero.
With this result in hand, we are ready to investigate what balance--robustness guarantee the design provides.

\expcommand{\theoremgswsatisfiesdiscrepancy}{%
The Gram--Schmidt Walk design with parameter $\vto \in [0, 1]$ provides the balance--robustness guarantee
\[
	\zsc = \frac{1}{\vto}
	\qquadand
	\xsc = \frac{\maxnormX^{2}}{1 - \vto}.
\]
}

\begin{theorem}\label{theorem:gsw-satisfies-discrepancy}
	\theoremgswsatisfiesdiscrepancy
\end{theorem}

When $\vto = 1$, the Gram--Schmidt Walk design places all emphasis on robustness and the min-max optimal robustness guarantee of $\zsc = 1$ is obtained.
When $\vto = 0$, all emphasis is instead placed on covariate balance and the balance guarantee $\xsc = \maxnormX^2$ is obtained.
As we noted in Section~\ref{sec:covariate-balance}, $\xsc$ cannot be made smaller than $\maxnormX^2$ unless restrictions are imposed on the covariates.
Intermediate values of the design parameter, $\vto \in (0,1)$, interpolate between these two extremes.
In this way, the design navigates the balance--robustness trade-off, and it lets experimenters select a guarantee that is appropriate for their applications.

The balance--robustness guarantee $(\zsc, \xsc)$ in Theorem~\ref{theorem:gsw-satisfies-discrepancy} can be loose relative to the actual performance of the Gram--Schmidt Walk design, especially for values of $\vto$ near $0$ and $1$.
For example, the Gram--Schmidt Walk design with $\vto = 1$ is exactly the Bernoulli design.
In this case, we know that $\norm{\Cov{\xM^\tran \zv}} = \norm{\xM^\tran \xM}$, but Theorem~\ref{theorem:gsw-satisfies-discrepancy} gives the vacuous bound $\norm{\Cov{\xM^\tran \zv}} \leq \infty$.
Similarly, when $\vto = 0$, we know that $\norm{\Cov{\zv}} \leq n$, but Theorem~\ref{theorem:gsw-satisfies-discrepancy} again gives the vacuous bound of $\norm{\Cov{\zv}} \leq \infty$.
This is largely a consequence of Theorem~\ref{theorem:gsw-satisfies-discrepancy} giving a balance--robustness guarantee for arbitrary covariates.
We would need to consider specific covariates to provide a sharper guarantee, and that would lead to a more complex bound.
The purpose of Theorem~\ref{theorem:gsw-satisfies-discrepancy} is to provide a finite-sample guarantee that is easy to understand and work with in practice.
The Gram--Schmidt Walk design can sometimes perform considerably better than this guarantee.

\subsection{Mean Squared Error}\label{sec:finite-mse-analysis}

Theorems~\ref{thm:utility-of-new-discrepancy-formulation}~and~\ref{theorem:gsw-satisfies-discrepancy} together provide a bound on the mean square error of the Horvitz--Thompson estimator under the Gram--Schmidt Walk design.
We use our understanding of the design to sharpen this bound, as described by the following theorem.

\expcommand{\htmsebound}{%
The mean squared error of the Horvitz--Thompson estimator under the Gram--Schmidt Walk design is at most the minimum of the loss function of an implicit ridge regression of the sum of the potential outcome vectors $\pomv = \poav + \pobv$ on the covariates:
\[
	n \E[\big]{\paren{\htest - \ate}^2}
	\leq
	\ridgeloss
	=
	\min_{\lfx \in \Reals^d} \bracket[\Bigg]{
		\frac{1}{\vto n} \norm[\big]{ \pomv - \xM\lfx}^2
		+ \frac{\maxnormX^2}{\paren{1 - \vto} n} \norm[\big]{\lfx}^2
	}.
\]
}

\begin{theorem}\label{thm:ht-mse-bound}
	\htmsebound
\end{theorem}

The bound in Theorem~\ref{thm:ht-mse-bound} is the same as the bound in Theorem~\ref{thm:utility-of-new-discrepancy-formulation}, but with the cross-term removed.
Theorem~\ref{thm:cov-bound} allows us to address the cross term directly, rather than use the cruder bound we used in Section~\ref{sec:new-perspective}.
A cosmetic difference between Theorems~\ref{thm:utility-of-new-discrepancy-formulation}~and~\ref{thm:ht-mse-bound} is that we here have written the bound in terms of the design parameter $\vto$ rather than the balance--robustness guarantee $(\zsc, \xsc)$.

The right-hand side of the bound in Theorem~\ref{thm:ht-mse-bound} is the scaled minimum loss of a ridge regression of the potential outcomes on the covariates.
The design parameter $\vto$ determines the regularization penalty of the regression, giving more weight to either functions $\lfx$ that explain the potential outcomes well or functions of small magnitude.
This is a manifestation of the balance--robustness trade-off.

While the mean square error is bounded by the loss of a ridge regression, no regression is ever run.
The estimator is the ordinary, unadjusted Horvitz--Thompson estimator.
Indeed, the regression can never be run, because it involves all potential outcomes, and we only observe half of them.
Theorem~\ref{thm:ht-mse-bound} instead highlights that the design assigns treatments in a way that makes Horvitz--Thompson estimator behaves as if such a regression had been run.

We can use Theorem~\ref{thm:ht-mse-bound} to characterize when it is beneficial to deviate from the mini-max design and set $\vto < 1$.
We already know from the balance--robustness trade-off that $\vto = 1$ is optimal when the covariates are completely unpredictive of the outcomes, but the trade-off by itself does not tell us how predictive the covariates must be to make it useful to seek covariate balance.
In supplement Section~\suppref{sec:when-balancing-improves}, we show that it is almost always beneficial to seek at least some covariate balance by setting $\vto < 1$ when using the Gram--Schmidt Walk design.
One exception is small experiments with nearly unpredictive covariates.

\subsection{Tail Behavior}\label{sec:tail-behavior}

Our characterization of the mean square error in the previous section gives only a limited view of the behavior of the Gram--Schmidt Walk design.
To paint a more complete picture, we provide finite-sample valid tail bounds on the discrepancy of the augmented covariates, $\gsM \zv$, and the Horvitz--Thompson estimator.

\citet{Bansal2019Gram} used the martingale inequality of \citet{Freedman1975Tail} to show that the Gram--Schmidt Walk algorithm ensures that $\gsM \zv$ is a subgaussian random vector with variance parameter $\sigma^2 \leq 40$.
However, tail bounds based on $\sigma^2 = 40$ will generally be too loose to be informative and useful in a statistical context.
The following theorem strengthens the analysis to variance parameter $\sigma^2 = 1$, which is tight.
To achieve this result, we develop a new proof technique for establishing martingale concentration, which might be of independent interest.
The proof technique is described in the supplement.

\begin{theorem}\label{thm:sub-gaussian}
	Under the Gram--Schmidt Walk design, the vector $\gsM \zv$ is subgaussian with variance parameter $\sigma^2 = 1$. That is, $\E[\big]{ \expf[\big]{ \iprod{\gsM \zv, \arbvec} } } \leq \expf[\big]{ \norm{\arbvec}^2 / 2}$ for all $\arbvec \in \Reals^{n + \xdim}$.
\end{theorem}

The proof appears in supplement Section~\suppref{sec:subguassian-proof-supp}, and is based on
a bound on the conditional expectation of an exponential quantity during a pivot phase.
\citet{Bansal2019Gram} bound this quantity using a lossy Taylor series approximations.
In contrast, we analyze it directly.

Theorem~\ref{thm:sub-gaussian} demonstrates that linear functions of the augmented covariates are well concentrated.
Because the augmented covariates contain the raw covariates, this implies concentration of the imbalance of any linear function of the covariates.
If we in Theorem~\ref{thm:sub-gaussian} set $\arbvec = n^{-1} \gsM \paren{\gsM^\tran \gsM}^{-1} \pomv$, we get $\iprod{\gsM \zv, \arbvec} = \htest - \ate$.
This allows us to use the theorem to also derive a finite-sample tail bound for the Horvitz--Thompson estimator itself.

\expcommand{\httailbound}{%
	Under the Gram--Schmidt Walk design, the tails of the sampling distribution of the Horvitz--Thompson estimator are bounded in finite samples such that, for all $\teerr > 0$,
	\begin{equation}
	\Pr[\big]{\abs{\htest - \ate} \geq \teerr} \leq 2 \expf[\bigg]{\frac{- \teerr^2 n}{2 \ridgeloss}}
	\quadwhere
	\ridgeloss
	=
	\min_{\lfx \in \Reals^d} \bracket[\Bigg]{
		\frac{1}{\vto n} \norm[\big]{ \pomv - \xM\lfx}^2
		+ \frac{\maxnormX^2}{\paren{1 - \vto} n} \norm[\big]{\lfx}^2
	}.
	\end{equation}
}

\begin{corollary}\label{coro:ht-tail-bound}
	\httailbound
\end{corollary}

\section{Large-Sample Properties}\label{sec:asymptotic-analysis}

\subsection{Asymptotic Regime and Assumptions}

Following convention in the design-based causal inference literature, we consider a sequence of finite populations of growing size in our large sample analysis.
All aspects of the experiment, including the potential outcomes and the design parameter, are thus indexed by $n$.
However, we leave the indexing implicit for notational clarity.
Our analysis focuses on the limiting behavior of the estimator and design under conditions on the sequence of experiments.

\begin{assumption}[Outcome regularity]\label{ass:outcome-reg}
	The fifth moments of the potential outcomes are asymptotically bounded: $n^{-1} \pnorm{5}{\poav}^5 = \bigO{1}$ and $n^{-1} \pnorm{5}{\pobv}^5 = \bigO{1}$.
\end{assumption}

\begin{assumption}[Covariate regularity]\label{ass:covariate-reg}
	The singular values of the covariate matrix are asymptotically bounded as $\sigma_{\min}(\xM) = \bigOmega{n^{1/2}}$ and $\sigma_{\max}(\xM) = \bigO{n^{1/2}}$.
\end{assumption}

\begin{assumption}[No extreme outliers]\label{ass:no-outliers}
	The maximum squared norm of the covariate vectors grows at most at the rate $\maxnormX^2 = \max_{i \in [n]} \norm{\xvi}^2 = \bigO{d \logf{n}}$.
\end{assumption}

\begin{assumption}[Covariate dimensions]\label{ass:cov-dim}
	The number of covariates grows at most at the rate $d = \bigO{n^{1/10 - \varepsilon}}$ for some $\varepsilon > 0$.
\end{assumption}

Outcome regularity (Assumption~\ref{ass:outcome-reg}) ensures that there are no extreme outliers among the potential outcomes.
The lower bound on the smallest singular value in Assumption~\ref{ass:covariate-reg} ensures that the moment matrix of the covariates, $n^{-1} \xM^\tran \xM$, is invertible.
Together with the upper bound on the largest singular value, this ensures that the matrix is well-conditioned.
No extreme outliers (Assumption~\ref{ass:no-outliers}) ensures that the magnitude of the largest covariate vector $\xvi$ is not too large, and Assumption~\ref{ass:cov-dim} ensures that there are not too many covariates relative to the number of units.
The last three assumptions concern the covariates, which are observed at the design stage.
Experimenters can therefore calculate and inspect the quantities in the assumptions before committing to a design.
They can also transform the covariates, for example by deleting columns, so as to better satisfy the assumptions.
If the covariates were to be drawn at random from some large population, Assumptions~\ref{ass:covariate-reg}~and~\ref{ass:no-outliers} would be satisfied with a probability approaching one if, for example, the population distribution of the covariate vector was sub-Gaussian and the population moment matrix was invertible.
In the supplement, we show that some of these assumptions can be made more general or otherwise relaxed, at the cost of added complexity of the theorems and proofs.
For example, if Assumption~\ref{ass:outcome-reg} is strengthened to uniformly bounded outcomes, then Assumption~\ref{ass:cov-dim} can be weakened to $d = \bigO{n^{1/6 - \varepsilon}}$ for some $\varepsilon > 0$.

\subsection{Consistency}\label{sec:consistency}

The Horvitz--Thompson estimator is root-$n$ consistent under most designs, including the Bernoulli design, and we want to ensure that the Gram--Schmidt Walk design does not lead to a slower rate of convergence.

\expcommand{\consistency}{%
Suppose that outcome regularity holds (Assumption~\mainref{ass:outcome-reg}) and that the design parameter is asymptotically bounded away from zero, $\vto = \bigOmega{1}$.
Then, the Horvitz--Thompson estimator under the Gram--Schmidt Walk design is root-$n$ consistent for the average treatment effect: $\htest - \ate = \bigOp{n^{-1/2}}$.
}

\begin{theorem}\label{thm:consistency}
\consistency
\end{theorem}

The theorem shows that the Gram--Schmidt Walk design achieves root-$n$ consistency as long as experimenters do not let the design parameter approach zero, assigning at least some weight to robustness in the design trade-off.
The theorem uses bounded fifth outcome moments as stipulated by Assumption~\mainref{ass:outcome-reg}, but the proof, which appears in the supplement, makes clear that this can be relaxed to bounded second moments.
This is the same condition required for root-$n$ consistency under conventional designs.
Indeed, if the second moments are not bounded, and no other assumptions are imposed on the outcomes, then there exists no design that is root-$n$ consistent.

\subsection{Limiting Variance}\label{sec:asymptotic-variance}

The limiting variance of the estimator under the Gram--Schmidt Walk design depends on the sequence of the design parameter $\vto$.
All else equal, it is easier to achieve a certain level of covariate balance when the sample is larger.
By letting $\vto$ approach one as the sample grows, it is possible to approach a setting with both maximal covariate balance and maximal robustness, effectively escaping the balance--robustness trade-off in large samples.
The following theorem formalizes this insight.

\expcommand{\asympvariance}{%
Suppose outcome and covariate regularity holds (Assumptions~\mainref{ass:outcome-reg}~and~\mainref{ass:covariate-reg}).
Further suppose that the design parameter approaches one at a sufficiently slow rate, so that $1 - \vto = \littleO{1}$ and $1 - \vto = \littleOmega{\maxnormX^2 / n}$.
Then, a tight asymptotic upper bound on the normalized variance of the Horvitz--Thompson estimator under the Gram--Schmidt Walk design is
\begin{equation}
\limsup_{n \to \infty} \bracket[\Big]{ n \Var[\big]{\htest} - \limvar } \leq 0,
\end{equation}
where $\limvar = n^{-1} \min_{\lfx} \norm{ \pomv - \xM \lfx }^2$ is the mean square residuals from a best least squares linear approximation of the potential outcomes using the covariates.
}

\begin{theorem}\label{thm:asymp-variance}
\asympvariance
\end{theorem}

The theorem describes the precision of the estimator in large samples when the design parameter approaches one.
It is an upper bound because $\limvar$ does not fully characterize the behavior of the design in the orthogonal complement of the covariate space.
While we have not found any sequences of potential outcomes for which the design performs better than the bound, we have not shown that none exist.
However, the bound is instance tight, in the sense that there always exist sequences of potential outcomes such that it holds with equality, no matter what the covariates might be.
We conjecture that the bound characterizes the asymptotic variance for most potential outcomes, in the sense that the bound holds with equality for all potential outcomes under mild regularity conditions.
The motivation for this conjecture is that most eigenvalues of $\Cov{\zv}$ will approach one when $\vto \to 1$.

Note that $\limvar = n^{-1} \min_{\lfx} \norm{ \pomv - \xM \lfx }^2$ would be attainable as the variance in finite samples if we somehow had access to all potential outcomes, so we could calculate the best linear approximation, and then use the residuals from this regression as outcomes in the experiment.
This procedure is of course infeasible, because we never observe all potential outcomes.
Nevertheless, $\limvar$ marks the lowest variance achievable by balancing linear functions.
Theorem~\ref{thm:asymp-variance} shows that we can attain this lower limit asymptotically using the Gram--Schmidt Walk design.
Another way to achieve $\limvar$ as the limiting variance is to do covariate adjustment in the estimation stage, as described by \citet{Lin2013Agnostic}.
However, such ex post covariate adjustment is not well-understood in finite samples, and introduces the risk of specification searching, so called p-hacking.
The Gram--Schmidt Walk design achieves $\limvar$ as the limiting variance by design, using the unadjusted Horvitz--Thompson estimator in the estimation stage.

It is important to let the design parameter approach one, $\vto \to 1$, but never set it exactly to one.
If we were to set $\vto = 1$, we would get the Bernoulli design.
The normalized variance would then be $\norm{ \pomv }^2 / n$, which can be considerably larger than $\limvar$.
Theorem~\ref{thm:asymp-variance} also requires that $\vto$ approaches one at a sufficiently slow rate, so that $n (1 - \vto) / \maxnormX^2 \to \infty$.
Given that $\maxnormX^2$ typically will be of considerably lower order than $n$, this rate condition is quite forgiving.
That is, the theorem describes the asymptotic behavior of the estimator for a wide range of sequences of $\vto$, and experimenters have substantial latitude in selecting the design parameter.
For example, if Assumption~\ref{ass:no-outliers} holds, so $\maxnormX^2 = \bigO{d \logf{n}}$, and $d = \littleO{n^\alpha}$ for some $0 < \alpha < 1$, then setting $\vto  = 1 - C n^{\alpha - 1} \logf{n}$, for some constant $C \in \Reals^+$, ensures that the rate condition holds.
Note, however, that Theorem~\ref{thm:asymp-variance} does not require Assumption~\ref{ass:no-outliers} to hold.
Note also that $\maxnormX^2$ and $n$ are known by the experimenter at the design stage, so they can select $\vto$ to ensure that the rate condition holds.

In Section~\suppref{sec:supp-limvar} in the online supplement, we analyze the limiting variance when $\vto$ is fixed asymptotically, relaxing the condition that the design parameter approaches one.
\citet{Chatterjee2023Central} provide an improved analysis of the limiting variance under the Gram--Schmidt Walk design when $\vto$ is fixed asymptotically under slightly different assumptions than the ones we use.

\subsection{Asymptotic Normality}\label{sec:clt}

The finite-sample tail bounds for the Horvitz--Thompson estimator described in Section~\ref{sec:tail-behavior} will often be loose in large samples.
The following theorem describes when the distribution of the estimator approaches a normal distribution as the sample grows.

\begin{theorem}\label{thm:clt}
Suppose that Assumptions~\ref{ass:outcome-reg},~\ref{ass:covariate-reg},~\ref{ass:no-outliers},~and~\ref{ass:cov-dim} hold.
Further suppose that the limiting distribution of the estimator is non-degenerate, in the sense that $n \Var{\htest} = \bigOmega{1}$.
Then, if the design parameter is asymptotically bounded away from zero, $\vto = \bigOmega{1}$, the limiting distribution of the Horvitz--Thompson estimator under the Gram--Schmidt Walk design is the standard normal distribution:
\begin{equation}
	\frac{\htest - \ate}{\sqrt{\Var{\htest}}} \darrow \func{\mathcal{N}}{0, 1}.
\end{equation}
\end{theorem}

We require that $n \Var{\htest} = \bigOmega{1}$ to avoid situations in which the estimator converges faster than the parametric rate.
There are sequences of experiments that satisfy our conditions for which $n \Var{\htest} \to 0$, but they are all knife-edge cases that are of little practical relevance, so non-degeneracy can be seen as a regularity condition.
The non-degeneracy condition has been used previously in the design-based causal inference literature; examples include Condition 6 in \citet{Aronow2017Estimating} and Assumption 5 in \citet{Leung2022Causal}.

To the best of our knowledge, the technique we use to prove Theorem~\ref{thm:clt} has not previously been used in the design-based causal inference literature.
Central limit theorems build on the insight that an appropriately scaled sum of sufficiently many weakly dependent random variables tend to be close to a normal distribution.
The conventional proof strategy in this setting is to analyze the terms of a linear estimator, which in our case would be $2 \ooi \zi / n$.
Instead of following this convention, we reinterpret the estimator as being the sum of the updates of the assignments (to all units) in each iteration of the Gram--Schmidt Walk algorithm.
That is, using the notation from Section~\ref{sec:gsw-design}, we interpret the estimator to be the sum of terms of the form $\gsdt \gsuvt^\tran \pomv / n$ over the iterations $t \in [T]$.
This allows us to form a martingale difference sequence for $\htest - \ate$, to which we can apply the
martingale central limit theorem by \citet{McLeish1974Dependent}.
The proof appears in the supplement.

\section{Inference}\label{sec:inference}

\subsection{Variance Bound and Estimator}\label{sec:var-est}

The first step towards constructing our confidence intervals is to estimate the variance of the estimator under the Gram–Schmidt Walk design.
However, the variance depends on joint features of the two potential outcomes, which are inherently unobservable, so it is not directly estimable.
This is a common problem in design-based causal inference.
We follow the conventional solution of estimating an upper bound for the variance, which acts as a conservative estimator.
The bound we use is based on the following decomposition of the limiting variance.

\expcommand{\rewritelimvar}{
	The limiting variance of the Horvitz--Thompson estimator under the Gram--Schmidt Walk design can be written
	\begin{equation}
		n \limvar
		= \min_{\lfx} \norm{ \pomv - \xM \lfx }^2
		= 2\min_{\lfx} \norm{ \poav - \xM \lfx }^2
		+ 2\min_{\lfx} \norm{ \pobv - \xM \lfx }^2
		- \min_{\lfx} \norm{ \boldsymbol{\tau} - \xM \lfx }^2,
	\end{equation}
	where $\boldsymbol{\tau} = \poav - \pobv$ is the vector of all individual treatment effects.
}

\begin{proposition}\label{prop:rewrite-limvar}
	\rewritelimvar
\end{proposition}

\expcommand{\vbisvb}{
A tight upper bound on the limiting variance of the Horvitz--Thompson estimator under the Gram--Schmidt Walk design is
\begin{equation}
	\vb = \frac{2}{n} \min_{\lfx} \norm{ \poav - \xM \lfx }^2
	+ \frac{2}{n} \min_{\lfx} \norm{ \pobv - \xM \lfx }^2
	\geq \limvar.
\end{equation}
}

\begin{corollary}\label{corollary:vb-is-vb}
\vbisvb
\end{corollary}

The fact that $\vb$ is an upper bound follows from $\min_{\lfx} \norm{ \boldsymbol{\tau} - \xM \lfx }^2 \geq 0$.
The fact that it is tight follows from that $\min_{\lfx} \norm{ \boldsymbol{\tau} - \xM \lfx }^2 = 0$ when $\boldsymbol{\tau} = \zerovec$.
If a constant is included among the covariates, the bound is tight whenever the treatment effects are constant, because then $\boldsymbol{\tau} = \tau \onevec$ for some $\tau \in \Reals$.
This mirrors the behavior of the Neyman variance bound \citep{Neyman1923Application}.
However, unlike the Neyman bound, the current bound is also tight whenever the covariates are perfectly predictive of the treatment effects.

To estimate $\vb$, we first estimate $\lfx_a = \paren{\xM^\tran \xM}^{-1} \xM^\tran \poav$, and then plug it into an estimate of $\norm{ \poav - \xM \lfx_a }^2 = \min_{\lfx} \norm{ \poav - \xM \lfx }^2$.
Our assumptions ensure that $\lfx_a$ exists.
Using Horvitz--Thompson-type estimators for both steps yields the estimator
\begin{equation}
	\evb = \frac{1}{n} \norm[\big]{ \diag{\onevec + \zv} \paren[\big]{\oov - \xM \widehat{\lfx}_a } }^2
	+ \frac{1}{n} \norm[\big]{ \diag{\onevec - \zv} \paren[\big]{\oov - \xM \widehat{\lfx}_b } }^2,
\end{equation}
where $\widehat{\lfx}_a = \paren{\xM^\tran \xM}^{-1} \xM^\tran \diag{\onevec + \zv} \oov$ and $\widehat{\lfx}_b = \paren{\xM^\tran \xM}^{-1} \xM^\tran \diag{\onevec - \zv} \oov$.
It is possible to use other estimators than Horvitz--Thompson-type estimators for these quantities, but we will not explore these alternative variance estimators in this paper.

Because the variance bound is a quadratic form, the estimator is not unbiased, despite being based on the Horvitz--Thompson estimation principle.
However, the estimator is consistent, as described in the following theorem.

\begin{theorem}\label{thn:var-est-cons}
Suppose outcome and covariate regularity holds (Assumptions~\mainref{ass:outcome-reg}~and~\mainref{ass:covariate-reg}).
Further suppose that the design parameter is bounded away from zero, $\vto = \bigOmega{1}$.
Then, the variance bound estimator converges to the variance bound at the rate $\evb - \vb = \bigOp[\big]{d n^{-1/2} \logf{n}}$.
\end{theorem}

When the number of covariates is bounded, $d = \bigO{1}$, the theorem states that the variance bound estimator is root-$n$ consistent, up to a logarithmic factor.
Under Assumption~\ref{ass:cov-dim}, stipulating that $d = \littleO{n^{1/10 - \epsilon}}$, the convergence rate is somewhat slower at $n^{-(2/5 + \epsilon)}$, again ignoring the logarithmic factor.

\subsection{Confidence Intervals}\label{sec:confidence-intervals}

Our confidence intervals are based on a normal approximation, motivated by Theorem~\ref{thm:clt}.
Let $\hat\sigma = \evb{\vphantom{\vb}}^{1/2}$ be the square root of the estimated variance bound, acting as a conservative estimator of the standard error of the treatment effect estimator.
Furthermore, let $z_\ciprob = \Phi^{-1}(1 - \ciprob)$ be the tails of the standard normal distribution, where $\Phi^{-1}$ is its quantile function.
A confidence interval at the $1 - \ciprob$ confidence level is then given by endpoints $\htest \pm n^{-1/2} z_\ciprob \hat\sigma$.

\begin{theorem}\label{prop:confidence-intervals}
Suppose that Assumptions~\ref{ass:outcome-reg},~\ref{ass:covariate-reg},~\ref{ass:no-outliers},~and~\ref{ass:cov-dim} hold.
Further suppose that the design parameter approaches one at a sufficiently slow rate, so that $1 - \vto = \littleO{1}$ and $1 - \vto = \littleOmega{\maxnormX^2 / n}$.
Then, the random interval centered at $\htest$ with radius $n^{-1/2} z_\ciprob \hat\sigma$ is an asymptotically valid $\paren{1 - \ciprob}$-confidence interval:
\begin{equation}
	\liminf_{n\to \infty} \Pr[\Big]{- z_\ciprob \hat\sigma \leq n^{1/2} \paren{\htest - \ate} \leq z_\ciprob \hat\sigma} \geq 1 - \ciprob.
\end{equation}
\end{theorem}

The confidence interval in Theorem~\ref{prop:confidence-intervals} uses several asymptotic approximations.
It is possible to modify the interval to improve its finite-sample validity, at the cost of additional conservativeness.
One important asymptotic approximation is that the interval is based on the limiting variance upper bound from Theorem~\ref{thm:asymp-variance}, which showed that experimenters can escape the balance--robustness trade-off asymptotically.
However, it is not possible to escape the trade-off in finite samples, so the limiting variance bound can be overly optimistic when the sample is small.
In Section~\suppref{sec:supp-alt-cis} in the supplement, we describe an alternative confidence interval with better finite-sample coverage, which is based on a more conservative variance estimator.
This confidence interval is also valid when parameter $\vto$ does not approach one asymptotically.
We discuss several other alternative confidence intervals in Section~\suppref{sec:supp-alt-cis}.

\section{Comparison with Other Designs}\label{sec:other-designs}

\subsection{Rerandomization}\label{sec:comp-rerand}

Rerandomization is a commonly used design approach to achieve covariate balance in experiments.
The design is a uniform distribution over a set of assignment vectors that satisfy some acceptance criterion based on a measure of covariate balance, and it is often implemented by rejection sampling.
Rerandomization implicitly navigates the balance--robustness trade-off through the strictness of its acceptance criterion.
In the version described by \citet{LockMorgan2012Rerandomization}, the acceptance criterion is based on the Mahalanobis distance between the means of covariates in the two treatment groups.

The properties of rerandomization are currently only well-understood in large samples, as described by \citet*{Li2018Asymptotic}.
Using Theorem~\ref{thm:asymp-variance}, we can compare the limiting variance of the Gram--Schmidt Walk design with the limiting variance of rerandomization.
In what follows, let $\complimvar = n^{-1} \min_{\beta} \norm{ \pomv - \onevec \beta }^2$ denote the limiting variance under complete randomization, let $\limvar = n^{-1} \min_{\lfx} \norm{ \pomv - \xM \lfx }^2$ denote the upper bound on the limiting variance of the Gram--Schmidt Walk design from Theorem~\ref{thm:asymp-variance}, and let $\rerandlimvar$ denote the limiting variance of rerandomization, as described by \citet{Li2018Asymptotic}.

\expcommand{\domrerand}{%
Suppose that Condition~1 in \citet{Li2018Asymptotic} holds, that the second moment of the potential outcomes is asymptotically bounded, $n^{-1} \norm{\pomv}^2 = \bigO{1}$, and that a constant is included among the covariates, so that the first column of $\xM$ is $\onevec$.
Then, the limiting variance of the difference-in-means estimator under rerandomization, as described by \citet{Li2018Asymptotic}, is greater or equal to the limiting variance of the Horvitz--Thompson estimator under the Gram--Schmidt Walk design when $\vto$ satisfies the rate condition in Theorem~\mainref{thm:asymp-variance}:
$
\limvar \leq \rerandlimvar \leq \complimvar.
$
Equality holds, $\limvar = \rerandlimvar$, only when $\rerandlimvar = \complimvar$.
}

\begin{proposition}\label{prop:dominates-rerandomization}
\domrerand
\end{proposition}

The proposition shows that the variance under the Gram--Schmidt Walk design always dominates the variance under rerandomization in large samples when a constant is included among the covariates.
The only setting in which rerandomization and the Gram--Schmidt Walk design have the same limiting variance is when the covariates (excluding the constant) are completely uninformative of the potential outcomes.
In that case, the limiting variance under both designs is equal to the limiting variance under complete randomization.

Proposition~\ref{prop:dominates-rerandomization} requires that a constant be included among the covariates to push the Gram--Schmidt Walk design towards treatment groups of equal sizes.
Because the acceptance criterion of rerandomization is stated in terms of demeaned covariates and because it produces treatment groups of equal sizes by construction, its behavior is unchanged by the inclusion of the constant column.
We discuss this further in Section~\suppref{sec:treatment-group-size-analysis} in the supplement.
The proposition also requires that the second moment of the potential outcomes is bounded in addition to Condition~1 in \citet{Li2018Asymptotic}.
This is because \citet{Li2018Asymptotic} consider central moments in their analysis, while we consider raw moments.

The central insight underlying Proposition~\ref{prop:dominates-rerandomization} is that the limiting variance under rerandomization is a convex combination of the limiting variance under complete randomization and the limiting variance under Gram--Schmidt Walk design in Theorem~\ref{thm:asymp-variance}.
In particular, we show in the supplement that $\rerandlimvar = v_{K, a} \complimvar + \paren{1 - v_{K, a}} \limvar$, where $v_{K, a} \in (0, 1)$ and $\complimvar \geq \limvar$.
The coefficient $v_{K, a}$ is defined in Proposition 2 in \citet{Li2018Asymptotic}.
It is the variance of a truncated random variable, which is shown to be the same as the ratio of the cumulative distribution functions of two chi-squared random variables: $v_{K, a} = \Pr{\chi^2_{K+2} \leq a} / \Pr{\chi^2_{K} \leq a}$, where $\chi^2_{K}$ denotes a chi-squared random variable with degrees of freedom $K$.
In the notation of \citet{Li2018Asymptotic}, $K$ is the number of covariates, excluding the constant, and $a$ is the balance acceptance threshold for the rerandomization procedure.
As noted by \citet{Li2018Asymptotic}, it is an open question how to select $a$, but the authors suggest setting $a$ so that $\Pr{\chi^2_{K} \leq a} = 0.001$.
Following this suggestion, we would have $v_{K, a} = 0.03$ when $K = 5$, meaning that rerandomization would be almost as asymptotically efficient as the Gram--Schmidt Walk design in that setting.
However, we would have $v_{K, a} = 0.31$ when $K = 25$, which is a sizeable difference if the covariates are informative of the potential outcomes.
That is, rerandomization yields less than $70\%$ of the variance improvement over complete randomization compared to the Gram--Schmidt Walk design in this setting.
Larger $K$ makes this difference even more pronounced.

For rerandomization to achieve a limiting variance that is comparable to the Gram--Schmidt Walk design, experimenters must set the acceptance threshold $a$ to be close to zero.
\citet{Wang2022Rerandomization} provide a formal investigation along these lines.
The authors show that when the acceptance criterion $a$ approaches zero, meaning that experimenters reject an increasing share of drawn assignments as the sample grows, rerandomization achieves the same limiting variance as the Gram--Schmidt Walk design.
However, setting the acceptance threshold close to zero will make the probability of accepting an assignment very small, often making the procedure infeasible to use in practice because computational resources are limited.
For example, when $K = 25$, one needs to set $a = 0.27$ to achieve $99\%$ of the improvement in asymptotic variance of the Gram--Schmidt Walk design, in the sense of $v_{K, a} = 0.01$.
The probability of accepting an assignment is then less than $10^{-20}$.
The run time of the Gram--Schmidt Walk design is unaffected by the choice of design parameter $\vto$.

\subsection{Matched pair design}\label{sec:comp-matched-pair}

The matched pair design is another common experimental design to achieve covariate balance \citep{Greevy2004Optimal}.
Units are here matched into pairs to minimize some objective function, which typically is the sum of Euclidean or Mahalanobis distances between the covariate vectors of paired units.
After the pairs have been constructed, exactly one unit in each pair is assigned active treatment and the other unit control, independently between pairs.

The matched pair design achieves covariate balance by introducing dependence between paired units.
This works well if paired units are nearly identical with respect to their covariates.
The concern is that such nearly identical pairs are rare, even when matching on only a moderate number of covariates.
Many, if not most,
pairs will often consist of units that are quite different from each other,
and covariate balance will then not improve much
despite considerable restrictions to randomization (and therefore robustness).
For this reason, the matched pair design sacrifices a lot of robustness to achieve relatively little covariate balance, according to the operator norm measures.
The following proposition formalizes this by considering
randomly chosen covariate vectors to reflect typical problem instances.
The argument implies there exist non-random vectors for which the same lower bound holds.

\expcommand{\comparematching}{%
	Suppose $n$ is an even integer and $\xv{1}, \dotsc, \xv{n}$ are drawn independently and uniformly from the $d$-dimensional unit ball with $d \geq 2$.
	For all matched pair designs,
	\[
	\norm[\big]{ \Cov{\zv} } = 2 \quad a.s.,
	\qquadand
	\E[\Big]{ \maxnormX^{-2} \norm[\big]{ \Cov{\xM^\tran \zv} }} \geq \frac{n^{1-2/d}}{8 d},
	\]
	where the covariances are taken with respect to the experimental design and the expectation is taken with respect to the random covariate vectors.
}

\begin{proposition}\label{prop:compare-w-matching}
	\comparematching
\end{proposition}

This shows that there is a limit on the amount of covariate balance that can be achieved by a matched pair design.
When there are $d \geq 3$ covariates, the lower bound on the imbalance grows with $n$.
If $d = 10$, the best possible balance guarantee that the matched pair design can provide is $\xsc \geq \maxnormX^{2} n^{4/5} / 80$.
This is better than the guarantee provided by the Bernoulli design, for which $\xsc$ will be of order $n$, but it is not much better.
The Gram--Schmidt Walk design provides the guarantee $\xsc = \maxnormX^{2} / (1 - \vto)$, regardless of $n$.
The matched pair design has a fixed robustness guarantee of $\zsc = 2$, independent of the covariates and sample size.
When $\vto = 1/2$, the Gram--Schmidt Walk design provides the guarantee $\zsc = 2$ and $\xsc = 2 \maxnormX^{2}$.
So, the Gram--Schmidt Walk design with $\vto = 1/2$ provides better guarantees than the matched pair design on both robustness and covariate balance in large samples whenever $d \geq 3$.

Note that we are here comparing a lower bound for the matched pair design with an upper bound for the Gram--Schmidt Walk design.
Even in situations where the lower bound for the matched pair design is lower than the upper bound for the Gram--Schmidt Walk design, the matched pair design will not necessarily provide better covariate balance.

The matched pair design differs from both the Gram--Schmidt Walk design and rerandomization by its targeting of all smooth functions of the covariates.
This could be helpful if the potential outcomes are explained well by non-linear functions of the covariates.
However, this comes at the cost of not providing much balance, neither on linear nor non-linear functions.
It is possible to emulate this behavior with the Gram--Schmidt Walk design by including an increasing number of transformations of the covariates as the sample grows.
This has the benefit of being more targeted than trying to balance all aspects of the covariates all at once, but it also requires experimenters to make considered choices about which transformations to target.
The same approach is not feasible with rerandomization, as acceptable assignments will be exceedingly rare when there are many things to balance, so it would take an insurmountable amount of time to find them using rejection sampling.

\section{Additional Results and Extensions}\label{sec:extensions}

We describe several extensions to the Gram--Schmidt Walk design and our analysis in the supplement.
In Section~\suppref{sec:non-uniform-probs-supp}, we relax the requirement of a symmetric design.
With this relaxation, the experimenter provides a vector $\pzv = \paren{\pz{1}, \dotsc, \pz{n}} \in \paren{0, 1}^n$ that specifies a desired assignment probability for each unit, and the design assigns the treatment accordingly.
In Section~\suppref{sec:ext-other-matrix-funs}, we derive finite-sample results for the Gram--Schmidt Walk design for matrix functions other than the operator norm, such as the trace norm and the Frobenius norm.
In Section~\suppref{sec:different-moment-defs}, we consider the balance--robustness trade-off for other types of moment conditions than the bounded average magnitude condition, $\POspace{M}$.

In Section~\suppref{sec:supp-balanced-gsw}, we provide additional results on the sizes of the treatment groups under the Gram--Schmidt Walk design.
We show that if $\vto < 1$ and at least a small constant is included among the covariates, the Gram--Schmidt Walk design will provide more balance on the group sizes than the Bernoulli design.
However, the design will not ensure exact balance, in the sense of treatment groups that contain exactly $n / 2$ units.
To achieve this, we describe a modification of the design that guarantees that treatment group sizes are exactly balanced.
This modification breaks certain orthogonality properties of the updates of the algorithm, so our analysis does not apply to this modified design.

The supplement contains in-depth numerical illustrations of the behavior of the Gram--Schmidt Walk design and other commonly used designs.
These simulations corroborate the theoretical results, showing that designs that balance the covariates well yield lower mean square error when the covariates are predictive, but are less robust.
The Gram--Schmidt Walk design is shown to navigate the balance--robustness trade-off well, providing more balance for a given level of robustness than other designs.
The simulations also show that the confidence intervals cover the true average treatment effect at the nominal rate for moderate and large samples.

\section{Concluding Remarks}

Randomized experiments are useful because they provide robustness, but experimenters are often tempted to balance covariates with the aim of improving precision.
The motivating idea of this paper is that a compromise between balance and robustness is at the heart of the experimental design problem.
At one extreme, we can resolve this trade-off cautiously by assigning treatments independently at random.
This yields a design that is maximally robust.
At the other extreme, we can make all assignments perfectly dependent.
This yields a design that performs exceptionally well for some potential outcomes, but it will perform exceptionally poorly for other outcomes.
Most experimenters would not prefer either of these extremes.
Instead, they prefer intermediate designs that introduce weak dependencies between the assignments to achieve some balance at the cost of some robustness.
The purpose of the Gram--Schmidt Walk design is to provide control over and efficiently navigate the trade-off between covariate balance and robustness.

The question of which covariates should be balanced and how to trade off balance and robustness can only fully be answered by an experimenter's preferences and substantive knowledge about the study at hand.
In general terms, experimenters should prioritize balance over robustness, by setting the design parameter $\vto$ to a lower value, when they have access to high-quality covariates that are known to be predictive of the potential outcomes.
Experimenters should also ensure that that the set of covariates they balance is as linearly predictive as possible, by adding transformations and removing irrelevant covariates.
We discuss practical considerations and heuristics related to the design in Section~\suppref{sec:practical-recommendations} of the online supplement.

One of the chief short-comings of the Gram--Schmidt Walk design is that it solely focuses on linear functions.
Experimenters can address this short-coming by balancing non-linear transformation of the raw covariates, but this requires an active choice which transformations to target.
It is possible to extend the design to automatically balance non-linear functions using kernel methods, but such an extension is beyond the scope of the current paper.
Another extension that is beyond the scope of the current paper is an online version of Gram--Schmidt Walk design, where the experimenter must assign treatments to units in sequence without knowing the characteristics of future units.

\nocite{*}

\input{gsw-design-refs.bbl}
\clearpage

\mtcaddpart[Supplement]

\renewcommand\thesection{S\arabic{section}}
\renewcommand\thetable{S\arabic{table}}
\renewcommand\thefigure{S\arabic{figure}}
\renewcommand\theequation{S\arabic{equation}}

\setcounter{section}{0}
\setcounter{table}{0}
\setcounter{figure}{0}
\setcounter{equation}{0}
\setcounter{theorem}{0}

\begin{center}
{\Large \textbf{Supplement to ``Balancing covariates in randomized\\experiments with the Gram--Schmidt Walk Design''}}
\end{center}

\setcounter{parttocdepth}{2}
\parttoc

\clearpage

\input{\texpath/practical}

\input{\texpath/supp-discrepancy-connection}

\input{\texpath/supp-gsw-analysis}

\input{\texpath/supp-clt-proof}

\input{\texpath/supp-balancing-analysis}

\input{\texpath/supp-implementation.tex}

\input{\texpath/supp-other-designs}

\input{\texpath/supp-extensions}

\input{\texpath/supp-additional-proofs}

\input{\texpath/supp-simulations.tex}

\end{document}

%% file: tex/stddef.tex
\DeclarePairedDelimiter\paren\lparen\rparen
\DeclarePairedDelimiter\bracket\lbrack\rbrack
\DeclarePairedDelimiter\braces\lbrace\rbrace

\DeclarePairedDelimiter\abs\lvert\rvert

\providecommand{\bbone}{\mathbf{1}}

\DeclarePairedDelimiterXPP\indicator[1]{\bbone}{\lbrack}{\rbrack}{}{#1}

\DeclarePairedDelimiterXPP\expf[1]{\exp}{\lparen}{\rparen}{}{#1}
\DeclarePairedDelimiterXPP\logf[1]{\log}{\lparen}{\rparen}{}{#1}
\DeclarePairedDelimiterXPP\maxf[1]{\max}{\lparen}{\rparen}{}{#1}
\DeclarePairedDelimiterXPP\minf[1]{\min}{\lparen}{\rparen}{}{#1}

\DeclarePairedDelimiterXPP\func[2]{#1}{\lparen}{\rparen}{}{#2}

\newcommand{\set}[1]{#1}

\DeclarePairedDelimiter\setb\lbrace\rbrace

\newcommand{\Reals}{\mathbb{R}}

\newcommand{\Naturals}{\mathbb{N}}

\DeclareMathOperator*{\argmax}{arg\,max}
\DeclareMathOperator*{\argmin}{arg\,min}

\newcommand{\mat}[1]{\boldsymbol{#1}}
\renewcommand{\vec}[1]{\boldsymbol{#1}}

\newcommand{\arbvec}{\vec{v}}
\newcommand{\onevec}{\vec{1}}
\newcommand{\zerovec}{\vec{0}}

\newcommand{\unitM}{\mat{I}}

\newcommand{\eigvec}[1]{\vec{\eta}_{#1}}
\newcommand{\eigval}[1]{\lambda_{#1}}

\newcommand{\eigvali}{\eigval{i}}
\newcommand{\eigvalmin}{\eigval{\min}}
\newcommand{\eigvalmax}{\eigval{\max}}

\DeclarePairedDelimiterXPP{\eigvecf}[2]{\vec{\eta}_{#1}}{\lparen}{\rparen}{}{#2}
\DeclarePairedDelimiterXPP{\eigvalf}[2]{\lambda_{#1}}{\lparen}{\rparen}{}{#2}
\DeclarePairedDelimiterXPP{\eigvecfi}[1]{\eigvec{i}}{\lparen}{\rparen}{}{#1}
\DeclarePairedDelimiterXPP{\eigvalfi}[1]{\eigval{i}}{\lparen}{\rparen}{}{#1}
\DeclarePairedDelimiterXPP{\eigvalfmin}[1]{\eigval{\min}}{\lparen}{\rparen}{}{#1}
\DeclarePairedDelimiterXPP{\eigvalfmax}[1]{\eigval{\max}}{\lparen}{\rparen}{}{#1}

\newcommand{\basisv}[1]{\vec{e}_{#1}}

\newcommand{\pinv}{\dagger}

\makeatletter
\newcommand*{\tran}{{\mathpalette\@tran{}}}
\newcommand*{\@tran}[2]{\raisebox{\depth}{$\m@th#1\intercal$}}
\makeatother

\DeclarePairedDelimiter\iprod\langle\rangle

\DeclarePairedDelimiter\norm\lVert\rVert
\DeclarePairedDelimiterXPP\tnorm[1]{}{\lVert}{\rVert_{1}}{}{#1}
\DeclarePairedDelimiterXPP\enorm[1]{}{\lVert}{\rVert_{2}}{}{#1}
\DeclarePairedDelimiterXPP\inorm[1]{}{\lVert}{\rVert_{\infty}}{}{#1}
\DeclarePairedDelimiterXPP\pnorm[2]{}{\lVert}{\rVert_{#1}}{}{#2}

\DeclarePairedDelimiterXPP\detf[1]{\det}{\lparen}{\rparen}{}{#1}

\DeclareMathOperator{\trsym}{tr}
\DeclarePairedDelimiterXPP\tr[1]{\trsym}{\lparen}{\rparen}{}{#1}

\DeclareMathOperator{\diagsym}{diag}
\DeclarePairedDelimiterXPP\diag[1]{\diagsym}{\lparen}{\rparen}{}{#1}

\DeclareMathOperator{\vectorizesym}{vec}
\DeclarePairedDelimiterXPP\vectorize[1]{\vectorizesym}{\lparen}{\rparen}{}{#1}

\providecommand\given{}
\newcommand\givensymbol[1]{\nonscript\:#1\vert\allowbreak\nonscript\:\mathopen{}}

\let\Prsym\Pr
\let\Pr\relax
\DeclarePairedDelimiterXPP\Pr[1]{\Prsym}{\lparen}{\rparen}{}{%
	\renewcommand\given{\givensymbol{\delimsize}}%
	#1}

\DeclarePairedDelimiterXPP\Prsub[2]{\Prsym_{#1}}{\lparen}{\rparen}{}{%
	\renewcommand\given{\givensymbol{\delimsize}}%
	#2}

\DeclareMathOperator{\Esym}{\mathbb{E}}
\DeclarePairedDelimiterXPP\E[1]{\Esym}{\lbrack}{\rbrack}{}{%
	\renewcommand\given{\givensymbol{\delimsize}}%
	#1}

\DeclarePairedDelimiterXPP\Esub[2]{\Esym_{#1}}{\lbrack}{\rbrack}{}{%
	\renewcommand\given{\givensymbol{\delimsize}}%
	#2}

\DeclareMathOperator{\Varsym}{Var}
\DeclarePairedDelimiterXPP\Var[1]{\Varsym}{\lparen}{\rparen}{}{%
	\renewcommand\given{\givensymbol{\delimsize}}%
	#1}

\DeclarePairedDelimiterXPP\Varsub[2]{\Varsym_{#1}}{\lparen}{\rparen}{}{%
	\renewcommand\given{\givensymbol{\delimsize}}%
	#2}

\DeclarePairedDelimiterXPP\EstVar[1]{\widehat{\Varsym}}{\lparen}{\rparen}{}{%
	\renewcommand\given{\givensymbol{\delimsize}}%
	#1}

\DeclarePairedDelimiterXPP\signf[1]{\textrm{sign}}{\lparen}{\rparen}{}{%
	\renewcommand\given{\givensymbol{\delimsize}}%
	#1}

\DeclareMathOperator{\Covsym}{Cov}
\DeclarePairedDelimiterXPP\Cov[1]{\Covsym}{\lparen}{\rparen}{}{%
	\renewcommand\given{\givensymbol{\delimsize}}%
	#1}

\DeclarePairedDelimiterXPP\Covsub[2]{\Covsym_{#1}}{\lparen}{\rparen}{}{%
	\renewcommand\given{\givensymbol{\delimsize}}%
	#2}

\DeclareMathOperator{\Corrsym}{Corr}
\DeclarePairedDelimiterXPP\Corr[1]{\Corrsym}{\lparen}{\rparen}{}{%
	\renewcommand\given{\givensymbol{\delimsize}}%
	#1}

\DeclareMathOperator{\Volsym}{Vol}
\DeclarePairedDelimiterXPP\Vol[1]{\Volsym}{\lparen}{\rparen}{}{%
	\renewcommand\given{\givensymbol{\delimsize}}%
	#1}

\makeatletter
\newcommand{\indep}{\protect\mathpalette{\protect\@indep}{\perp}}
\newcommand*{\@indep}[2]{\mathrel{\rlap{$#1#2$}\mkern3mu{#1#2}}}
\makeatother

\newcommand{\darrow}{\overset{d}{\longrightarrow}}

\newcommand{\bigOsym}{\mathcal{O}}
\DeclarePairedDelimiterXPP\bigO[1]{\bigOsym}{\lparen}{\rparen}{}{#1}

\newcommand{\littleOsym}{o}
\DeclarePairedDelimiterXPP\littleO[1]{\littleOsym}{\lparen}{\rparen}{}{#1}

\newcommand{\bigOpsym}{\bigOsym_p}
\DeclarePairedDelimiterXPP\bigOp[1]{\bigOpsym}{\lparen}{\rparen}{}{#1}

\newcommand{\littleOpsym}{\littleOsym_p}
\DeclarePairedDelimiterXPP\littleOp[1]{\littleOpsym}{\lparen}{\rparen}{}{#1}

\newcommand{\bigOmegasym}{\Omega}
\DeclarePairedDelimiterXPP\bigOmega[1]{\bigOmegasym}{\lparen}{\rparen}{}{#1}

\newcommand{\littleOmegasym}{\omega}
\DeclarePairedDelimiterXPP\littleOmega[1]{\littleOmegasym}{\lparen}{\rparen}{}{#1}

\newcommand{\bigThetasym}{\Theta}
\DeclarePairedDelimiterXPP\bigTheta[1]{\bigThetasym}{\lparen}{\rparen}{}{#1}

\newcommand{\sumin}{\sum_{i=1}^n}

\newcommand{\quadtext}[1]{\quad\text{#1}\quad}
\newcommand{\qquadtext}[1]{\qquad\text{#1}\qquad}

\newcommand{\quadand}{\quadtext{and}}
\newcommand{\qquadand}{\qquadtext{and}}

\newcommand{\quadwhere}{\quadtext{where}}
\newcommand{\qquadwhere}{\qquadtext{where}}

\theoremstyle{plain}
\newtheorem{theorem}{Theorem}[section]

\newtheorem{corollary}[theorem]{Corollary}
\newtheorem{lemma}[theorem]{Lemma}
\newtheorem{proposition}[theorem]{Proposition}

\newtheorem{appcorollary}[theorem]{Corollary}

\newtheorem{applemma}[theorem]{Lemma}

\newtheorem{appproposition}[theorem]{Proposition}

\newtheorem{apptheorem}[theorem]{Theorem}

\newtheorem{innerrefcorollary}{Corollary}
\newenvironment{refcorollary}[1]
{\renewcommand\theinnerrefcorollary{#1}\innerrefcorollary}
{\endinnerrefcorollary}

\newtheorem{innerreflemma}{Lemma}
\newenvironment{reflemma}[1]
{\renewcommand\theinnerreflemma{#1}\innerreflemma}
{\endinnerreflemma}

\newtheorem{innerrefproposition}{Proposition}
\newenvironment{refproposition}[1]
{\renewcommand\theinnerrefproposition{#1}\innerrefproposition}
{\endinnerrefproposition}

\newtheorem{innerreftheorem}{Theorem}
\newenvironment{reftheorem}[1]
{\renewcommand\theinnerreftheorem{#1}\innerreftheorem}
{\endinnerreftheorem}

\theoremstyle{definition}
\newtheorem{assumption}[theorem]{Assumption}

\newtheorem{definition}[theorem]{Definition}
\newtheorem{problem}[theorem]{Problem}

\theoremstyle{remark}

%% file: tex/commands.tex
\newcommand{\zv}{\vec{z}}
\newcommand{\ze}[1]{z_{#1}}
\newcommand{\zi}{\ze{i}}

\newcommand{\trgroup}{\set{Z}}
\newcommand{\treated}{\trgroup^+}
\newcommand{\controls}{\trgroup^-}

\newcommand{\ooe}[1]{y_{#1}}
\newcommand{\ooi}{\ooe{i}}

\newcommand{\oov}{\vec{y}}

\newcommand{\poae}[1]{a_{#1}}
\newcommand{\poai}{\poae{i}}
\newcommand{\poaj}{\poae{j}}
\newcommand{\poav}{\vec{a}}

\newcommand{\pobe}[1]{b_{#1}}
\newcommand{\pobi}{\pobe{i}}
\newcommand{\pobj}{\pobe{j}}
\newcommand{\pobv}{\vec{b}}

\newcommand{\pomv}{\vec{\mu}}
\newcommand{\pom}[1]{\mu({#1})}
\newcommand{\poim}{\pom{i}}

\DeclareMathOperator{\POsym}{PO}
\DeclarePairedDelimiterXPP\POspace[1]{\POsym}{\lparen}{\rparen}{}{#1}

\newcommand{\atesym}{\tau}
\newcommand{\ate}{\atesym}

\newcommand{\htest}{\widehat{\ate}}

\newcommand{\xv}[1]{\vec{x}_{#1}}
\newcommand{\xvi}{\xv{i}}
\newcommand{\xvj}{\xv{j}}

\newcommand{\xM}{\mat{X}}
\newcommand{\xdim}{d}

\newcommand{\maxnormX}{\xi}

\newcommand{\lfx}{\vec{\beta}} % linear function of X
\newcommand{\arblfx}{\vec{\theta}} % arbitrary linear function of X

\newcommand{\fpomv}{\widehat{\pomv}}
\newcommand{\rpomv}{\vec{\varepsilon}}
\newcommand{\lsfx}{\lfx_{\normalfont\textsc{ls}}}

\newcommand{\zsc}{\gamma_z}
\newcommand{\xsc}{\gamma_x}

\newcommand{\gswdesign}{{\normalfont\textsc{GSW-Design}}\xspace}

\newcommand{\gsM}{\mat{B}} % gram schmidt matrix (B)
\newcommand{\gscv}[1]{\vec{b}_{#1}} % gram schmidt column vector (b_1 ... b_n)
\newcommand{\gscvi}{\gscv{i}}
\newcommand{\vto}{\phi} % trade off parameter

\newcommand{\alive}{\mathcal{A}}

\newcommand{\gsvsym}{z}
\newcommand{\gsv}[1]{\vec{\gsvsym}_{#1}}
\newcommand{\gsvt}{\gsv{t}}
\newcommand{\gsve}[2]{\gsv{#1}(#2)}

\newcommand{\gsusym}{u}
\newcommand{\gsuv}[1]{\vec{\gsusym}_{#1}}
\newcommand{\gsuvt}{\gsuv{t}}
\newcommand{\gsuve}[2]{\gsusym_{#1}(#2)}

\newcommand{\gsunv}{\vec{\gsusym}}
\newcommand{\gsun}[1]{\gsusym({#1})}
\newcommand{\gsuni}{\gsun{i}}

\newcommand{\gsdsym}{\delta}
\newcommand{\gsd}[1]{\gsdsym_{#1}}
\newcommand{\gsdp}[1]{\gsdsym^+_{#1}}
\newcommand{\gsdm}[1]{\gsdsym^-_{#1}}

\newcommand{\gsdn}{\gsdsym}
\newcommand{\gsdnp}{\gsdsym^+}
\newcommand{\gsdnm}{\gsdsym^-}

\newcommand{\gsdt}{\gsd{t}}
\newcommand{\gsdtp}{\gsdp{t}}
\newcommand{\gsdtm}{\gsdm{t}}

\newcommand{\pzsym}{\pi}
\newcommand{\pzv}{\vec{\pzsym}}
\newcommand{\pz}[1]{\pzsym_{#1}}
\newcommand{\pzi}{\pz{i}}

\newcommand{\pphasesym}{S}
\newcommand{\pphase}[1]{\pphasesym_{#1}}
\newcommand{\pphasei}{\pphase{i}}
\newcommand{\pspace}[1]{V_{#1}}
\newcommand{\pspacei}{\pspace{i}}

\newcommand{\projBsym}{\mat{P}}
\newcommand{\PgsM}{\projBsym} % Projection gram schmidt matrix (column space)

\newcommand{\phaseprojsym}{\projBsym}
\newcommand{\phaseproj}[1]{\phaseprojsym_{#1}}
\newcommand{\phaseproji}{\phaseproj{i}}

\newcommand{\projX}{\mat{\Pi}}

\newcommand{\ridgeloss}{L}
\newcommand{\rub}{LB}
\newcommand{\erub}{\widehat{LB}}

\newcommand{\maxeigxG}{\lambda_{G}}

\newcommand{\teerr}{\gamma}
\newcommand{\ciprob}{\alpha}

\newcommand{\gsdpt}{\gsdp{t}}
\newcommand{\gsdmt}{\gsdm{t}}

\newcommand{\reorder}{\boldsymbol{\sigma}} % re-ordering
\newcommand{\invreorder}{\boldsymbol{\sigma}^{-1}} % inverse ordering
\newcommand{\reorderi}[1]{\reorder(#1)} % re-ordering index
\newcommand{\invreorderi}[1]{\invreorder(#1)} % re-ordering index

\newcommand{\obv}[1]{\vec{w}_{#1}} % orthogonal basis vectors

\DeclarePairedDelimiterXPP\spanv[1]{\mathrm{span}}{\lbrace}{\rbrace}{}{#1}

\newcommand{\gsuvi}[2]{\gsuv{#1}(#2)}

\newcommand{\grad}{\nabla}

\newcommand{\lfxr}{\lfx^*}
\newcommand{\rM}{\mat{R}}

\newcommand{\minloss}[1]{\mathcal{L}(#1)}

\newcommand{\rlMatv}[1]{\rlMat(#1)}

\newcommand{\aM}{\mat{A}} % A matrix
\newcommand{\lM}{\mat{L}} % L matrix
\newcommand{\bv}{\vec{b}} % b vector
\newcommand{\vv}{\vec{v}} % rank 1 vv^T
\newcommand{\at}[1]{\vec{a}_{t}^{(#1)}} % a vector used in faster implementation

\newcommand{\rlMat}{\mat{Q}}

\newcommand{\gpomv}{\widetilde{\vec{\mu}}}

\newcommand{\event}{\mathcal{E}}
\newcommand{\notevent}{\overline{\event}}

\newcommand{\limvar}{V_{{\normalfont\textsc{gsw}}}}

\newcommand{\complimvar}{V_{{\normalfont\textsc{co}}}}
\newcommand{\rerandlimvar}{V_{{\normalfont\textsc{re}}}}

\newcommand{\xMx}[1]{\mat{X}_{(#1)}}

\newcommand{\vb}{{\normalfont\textrm{VB}}}
\newcommand{\evb}{\widehat{\vb}}

\newcommand{\afn}{\alpha(n)}

%% file: tex/practical.tex
\section{Practical Considerations and Recommendations}\label{sec:practical-recommendations}

The question of which covariates should be balanced and how to trade off balance and robustness can only be answered by an experimenter's preferences and substantive knowledge about the study at hand.
In general terms, experimenters should prioritize balance over robustness, by setting the design parameter $\vto$ to a lower value, when they have access to high-quality covariates that are known to be predictive of the potential outcomes.
This could, for example, be baseline measures of the outcome.
However, experimenters should keep in mind that the exchange rate between balance and robustness becomes worse as $\vto$ decreases.
For example, Theorem~\mainref{thm:ht-mse-bound} tells us that going from $\vto = 0.1$ to $\vto = 0.01$ yields an improvement in the bound on the mean squared error of about $10\%$ in the best-case scenario where the covariates are perfectly predictive of the potential outcomes, but the mean squared error is ten times as large in the worse-case scenario where the covariates are not predictive at all.
Furthermore, to achieve the limiting variance described in Section~\mainref{sec:asymptotic-variance}, the design parameter must approach one, $\vto \to 1$.
This suggests that experimenters should not set the parameter too low.

As a rough heuristic, we believe it rarely is motivated to set $\vto$ lower than $1/2$.
This ensures that the worst-case performance of the Gram--Schmidt Walk design is never worse than the worst-case performance under the matched-pair design, but it still gives the design enough room to balance the covariates.
Another heuristic is to set $\vto$ no lower than $\min_{\lfx} \norm{ \pomv - \xM \lfx }^2 / \norm{ \pomv }^2$, which is the fraction of $\pomv$ that cannot be explained by the covariates.
The ratio $\min_{\lfx} \norm{ \pomv - \xM \lfx }^2 / \norm{ \pomv }^2$ cannot be directly observed or estimated in the design stage, but experimenters might be comfortable reasoning about the quantity given its similarity with the coefficient of determination.
While these heuristics can be helpful, the parameter should ultimately be decided by the experimenter's knowledge of the context at hand.

Experimenters should consider transforming the covariates to make them as linearly predictive as possible relative to their dimensions.
This could include removing or downweighting covariates which are suspected not to be predictive and including higher-order terms and interactions if they are believed to be prognostically important.
Experimenters should avoid covariates that are on very different scales, unless the scaling reflects the predictiveness of the covariates, because the design will put disproportionally more effort towards balancing covariates on large scales.
It may therefore be useful to normalize or decorrelate the covariates prior to the design stage, although that is not necessary.

By the same token, experimenters should generally avoid including covariates not believed to be predictive because the design will attempt to balance these uninformative covariates at the expense of important covariates.
Experimenters do not always know which covariates are prognostically important, and they may therefore want to seek balance on many covariates.
The Gram--Schmidt Walk design can accommodate this, but the balance on each covariate will naturally be worse than when only a small set of covariates is targeted.
The design is especially adept at balancing many covariates in large samples, as demonstrated in our asymptotic analysis, so experimenters can be more liberal in their inclusion of many covariates if the sample is large.

%% file: tex/supp-discrepancy-connection.tex
\section{Connection to Discrepancy Theory}

We now highlight the way in which the Gram--Schmidt Walk design is solving a \emph{distributional discrepancy} problem.
The problem statement is given below:

\begin{problem} \label{problem:distributional-discrepancy}
	Given covariate vectors $\xv{1}, \xv{2}, \dotsc \xv{n} \in \Reals^d$ arranged as rows of a matrix $\xM$, what are minimal pairs $(\zsc, \xsc)$
	for which there exists a distribution on $\zv \sim \setb{\pm 1}^n$ satisfying $\Pr{\zi = 1} = 1/2$ for all $i \in [n]$ and
	\[
	\norm{\Cov{\zv}} \leq \zsc
	\quadand
	\norm{\Cov{ \xM^\tran \zv }} \leq \xsc
	\enspace?
	\]
\end{problem}
 Problem~\ref{problem:distributional-discrepancy} is a \emph{distributional discrepancy} problem, in contrast to typical discrepancy problems which construct a single assignment.
Given a set of vectors $\xv{1}, \xv{2}, \dots \xv{n} \in \Reals^d$, the \emph{discrepancy vector} of an assignment $\zv \in \setb{\pm 1}^n$ is the difference of within-group sums:
\[
\xM^\tran \zv = \sumin \zi \xvi = \sum_{i \in \treated} \xvi - \sum_{i \in \controls} \xvi \enspace.
\]
The \emph{discrepancy} of an assignment is a measurement of the magnitude of the corresponding discrepancy vector, typically with the squared Euclidean norm or the infinity norm.
The squared Euclidean norm may be expressed in the following variational way:
\[
\norm{\xM^\tran \zv}^2
= \max_{\substack{\arblfx \in \Reals^d \\ \norm{\arblfx} = 1}} \iprod{\arblfx, \xM^\tran \zv}^2
=\max_{\substack{\arblfx \in \Reals^d \\ \norm{\arblfx} = 1}} \paren[\Big]{\sum_{i \in \treated} \iprod{\arblfx, \xvi} - \sum_{i \in \controls} \iprod{\arblfx, \xvi} }^2
\]
The operator norm $\norm{\Cov{\xM^\tran \zv}}$ may be expressed in a similar variational manner.
In particular, the operator norm is then the maximum of the mean squared inner product over linear functions of the covariate difference:
\begin{equation}
	\norm{\Cov{\xM^\tran \zv}}
	= \max_{\norm{\lfx} = 1} \E[\Big]{ \paren[\Big]{ \sum_{i \in \treated} \iprod{\lfx, \xvi} - \sum_{i \in \controls} \iprod{\lfx, \xvi} }^2 }
	\enspace.
\end{equation}
Hence, the norm captures the maximum mean squared imbalance between the covariate vectors in the two groups, as measured by linear functions.

In this sense, $\norm{\Cov{\xM^\tran \zv}}$ may be understood as the distributional extension of the squared Euclidean discrepancy.
Of course, the key aspect of Problem~\ref{problem:distributional-discrepancy} is the trade-off between this and the dependence of assignments, captured by $\norm{\Cov{\zv}}$.

We emphasize here that Problem~\ref{problem:distributional-discrepancy} is not subsumed or solved by previously considered discrepancy problems in the literature.
Indeed, the goal of most discrepancy problems is to produce a \emph{single} assignment vector $\zv$ which minimizes a norm of the discrepancy vector.
A naive application of discrepancy minimization to Problem~\ref{problem:distributional-discrepancy} is to choose $\zv^*$ to be the assignment which minimizes the squared Euclidean norm $\norm{\xM^\tran \zv}^2$ and construct a distribution by choosing either $\zv^*$ or $-\zv^*$ with equal probability.
This naive experimental design may result in substantial covariate balance, as in this case $\norm{\Cov{\xM^\tran \zv}} = \norm{ \xM^\tran \zv}^2$; however, this design affords virtually no robustness, as it yields $\norm{\Cov{\zv}} = n$.
Thus, Problem~\ref{problem:distributional-discrepancy} is a new discrepancy problem which requires new insights and algorithmic considerations.

The only other distributional discrepancy problem that we are aware of is the subgaussian discrepancy problem introduced by \cite{Dadush2019Towards}, which led to the development of the Gram--Schmidt Walk algorithm of \cite{Bansal2019Gram}.
This discrepancy problem is similar to Problem~\ref{problem:distributional-discrepancy} above but differs in the sense that it bounds all the higher moments and only considers the covariate balance, i.e. $\xsc^2$.

%% file: tex/supp-gsw-analysis.tex
\section{Analysis of the Gram--Schmidt Walk Algorithm}\label{sec:supp-gsw-algo-analysis}

In this section, we  present our analysis of the Gram--Schmidt Walk algorithm of \citet{Bansal2019Gram}.
We begin by restating the algorithm and introducing notation that will be used in the proofs.
Next, we describe a formal connection to the Gram--Schmidt orthogonalization process which is also used in our proofs.
We then provide proofs of the martingale property (Lemma~\mainref{lemma:martingale}), covariance bound (Theorem~\mainref{thm:cov-bound}) and the subgaussian concentration (Theorem~\mainref{thm:sub-gaussian}) of the Gram--Schmidt Walk algorithm.

\input{\texpath/supp-restate-algorithm.tex}

\input{\texpath/supp-gsw-and-gs-ortho.tex}

\input{\texpath/supp-martingale-assignments}

\input{\texpath/supp-cov-bound.tex}

\input{\texpath/supp-subgaussian.tex}

%% file: tex/supp-restate-algorithm.tex
\subsection{Gram--Schmidt Walk algorithm}\label{sec:gsw-algorithm}
In this section, we restate the Gram--Schmidt Walk algorithm using more detailed notation.
This more detailed notation contains explicit references to the iteration index and will be used in the proofs in this supplement.
Algorithm~\ref{alg:gs_walk} below is the Gram--Schmidt Walk algorithm of \citet{Bansal2019Gram}.
The algorithm presented in Section~\mainref{sec:gsw-design} sets the initial point $\zv_1 = \vec{0}$, but we allow for an arbitrary initial point in the present analysis.

\begin{algorithm}
	\SetAlgoLined
	\caption{Gram--Schmidt Walk}\label{alg:gs_walk}
	\SetKwInOut{Input}{Input}
	\SetKwInOut{Output}{Output}
	\Input{Vectors $\gscv{1}, \gscv{2}, \dots \gscv{n} \in \Reals^m$ arranged as columns in the matrix $\gsM$ and an initial point $\gsv{1} \in [-1, 1]^n$}
	\Output{$\zv \in \{ \pm 1 \}^n$ }
	Set the pivot ordering, a permutation $\pi : [n] \to [n]$ of units uniformly at random. \\
	Set iteration index $t \gets 1$ and alive set $\alive_{1} \gets [n]$.\\
	Set the first pivot $p_0 \gets \pi(n)$  \label{line:select_pivot1}\\
	\While{$\alive_t \neq \emptyset$ \label{line:begin_while}} {
		\uIf{$p_{t-1} \notin \alive_t$}{
			Set the pivot with largest pivot ordering, $p_t \gets \argmax_{i \in \alive_t} \pi(i)$. \label{line:select_pivot2}}
		\Else{$p_t \gets p_{t-1}$}
		Compute the step direction \label{line:condition_here}
		\begin{equation}
		\gsuvt \gets \argmin_{\gsunv \in \set{U}} \norm{ \gsM \gsunv },
		\end{equation}
		where $\set{U}$ is the set of all $\gsunv \in \Reals^n$ such that $\gsunv(p_{t}) = 1$ and $\gsunv(i) = 0$ for all $i \notin \alive_t$. \\
		Set $\gsdpt \gets \abs{\max\Delta}$ and $\gsdmt \gets \abs{\min\Delta}$ where $\Delta = \braces{ \gsdn \in \Reals : \gsvt + \gsdn \gsuvt \in \bracket{-1, 1}^n}$.\\
		Set the step size $\gsdt$ at random according to
		\begin{equation}
		\gsdt \gets
		\begin{cases}
		\phantom{-}\gsdpt & \text{ with probability } \gsdmt / \paren{\gsdpt + \gsdmt},
		\\
		- \gsdmt & \text{ with probability } \gsdpt / \paren{\gsdpt + \gsdmt}.
		\end{cases}
		\end{equation}\\
		Update the fractional assignment $\gsv{t+1} \gets \gsv{t} + \delta_{t} \gsuv{t}$ \label{line:assingment_update}\\
		Update set of alive units $\alive_{t+1} \gets  \setb{i \in [n] : \abs*{\gsve{t}{i}} < 1} $\\
		Increment the iteration index $t \gets t + 1$
	}
	\Return $\zv \gets$ the final iterate $\gsv{T+1}$
\end{algorithm}

The Gram--Schmidt Walk algorithm considered here differs from \citet{Bansal2019Gram} in that the pivots are selected in a randomized, rather than deterministic, way.
In the main paper, the pivot $p_t$ is selected uniformly at random from alive units $\alive_t$ at a given iteration.
In Algorithm~\ref{alg:gs_walk} presented in this section, the pivot ordering $\pi$ is chosen (uniformly at random from all permutations) at the beginning of the algorithm and the pivot $p_t$ is selected to be the alive unit with largest pivot ordering, i.e. $p_t \gets \argmax_{i \in \alive_t} \pi(i)$.
These two ways of randomly selecting pivots yield identical distributions of all variables in the algorithm.
We choose to work with this second pivot selection procedure in the appendix because it simplifies proofs via a conditioning argument, i.e. conditioning on the pivot ordering $\pi$.
In fact, the majority of the proofs appearing in Sections~\ref{sec:cov-bound-supp} and \ref{sec:subguassian-proof-supp} derive the main results conditioned on an arbitrary pivot ordering, and then proceed by marginalizing over all pivot orderings.
The randomized pivot ordering is critical only in the proof of the Central Limit Theorem, presented in Section~\ref{sec:supp-clt-proof}.

We remark on some of the differences between the notation in Algorithm~\ref{alg:gs_walk} here and the pseudo-code presented in Section~\mainref{sec:gsw-design} of the main body.
First, the Gram--Schmidt Walk algorithm takes as input arbitrary vectors $\gscv{1}, \gscv{2}, \dots \gscv{n} \in \Reals^m$.
For purposes of analysis, we often assume that the $\ell_2$ norms of these input vectors is at most $1$.
By construction, this assumption is satisfied for the input vectors (i.e. augmented covariate vectors) considered in the main body of the paper.
Additionally, the notation presented here contains more reference to iteration indices.
In particular, the notation of the pivot unit $p_t$, the alive set $\alive_t$, and the choice of update steps $\gsdpt$, $\gsdmt$ all feature the iteration index in the subscript.
We also use the notation that $\gsuve{t}{i}$ denotes the $i$th coordinate of the vector $\gsunv$ at time $t$.

We denote the (random) number of iterations by $T$.
We now introduce a notational convention which improves the clarity of some further analysis.
Because the number of iterations $T$ is always at most $n$,
we may suppose that the algorithm runs for exactly $n$ iterations and that for iterations $t > T$, we set the update direction $\gsuvt = \vec{0}$ and the step size $\gsdt = 0$.
The same vector $\zv$ is returned and the output distribution of the algorithm is unchanged.
We remark that this convention is used sparingly throughout the analysis and does not change the algorithm.

The concept of pivot phases was central to the analysis in \citet{Bansal2019Gram} and it remains a central part of the analysis presented here as well.
For each unit $i \in [n]$, we define the \emph{pivot phase} $\pphasei$ to be the set of iterations for which unit $i$ is the pivot, i.e.
\begin{equation}
\pphasei = \setb{ t : p_t = i}.
\end{equation}
During a particular run of the algorithm, the pivot phase $\pphasei$ may be empty if unit $i$ is not chosen as a pivot unit during that run.

During the course of the algorithm, a unit $i \in [n]$ is said to be \emph{alive} if $\abs{\gsve{t}{i}} < 1$ and \emph{frozen} otherwise.
This is the convention is used by \citet{Bansal2019Gram} and it reflects that fact that once a unit is frozen, its fractional assignment becomes integral and it is no longer updated.
The set $\alive_t$ is referred to as the \emph{alive set} because it contains all alive units at the beginning of iteration $t$.
We refer to the vectors $\gscv{1}, \gscv{2}, \dots \gscv{n}$ as the input vectors.
We may slightly abuse our terminology and call an input vector $\gscv{i}$ alive or frozen when we mean that the corresponding unit $i$ is alive or frozen.

We say that a unit $i$ is \emph{decided by the algorithm} when it is either selected as the pivot (Lines~\ref{line:select_pivot1} or \ref{line:select_pivot2}) or frozen without being chosen as the pivot (Line~\ref{line:assingment_update}).
Throughout the proofs below, we often condition on the previous random decisions made by the algorithm.
We use $\Delta_i$ to denote all the random decisions made by the algorithm up to and including when unit $i$ was decided by the algorithm.
There is, however, some care to be taken in this definition to distinguish between units which are chosen as pivots and those which are not.
If $i$ is chosen as a pivot at the beginning of iteration $t$, then $\Delta_i$ includes all previous choices of step sizes $\gsd{1} \dots \gsd{t-1}$.
If $i$ is frozen at the end of iteration $t$ without being chosen as the pivot, then $\Delta_i$ includes all choices of step sizes $\gsd{1} \dots \gsd{t}$.
Other types of conditioning will be presented throughout the proofs as the needs arise.

%% file: tex/supp-gsw-and-gs-ortho.tex
\subsection{Connection to Gram--Schmidt orthogonalization} \label{sec:gram-schmidt-orthogonalization}

A key aspect in our analysis of the Gram--Schmdit Walk algorithm is a Gram--Schmidt orthogonalization applied to a random re-ordering of the input vectors.
We use the randomized Gram--Schmidt orthogonalization to obtain the tight bounds on the covariance matrix and the subgaussian constant in Theorems~\mainref{thm:cov-bound} and \mainref{thm:sub-gaussian}, respectively.
In this section, we describe this connection in detail, providing additional notation and several technical lemmas which will be used in the proofs of Theorems~\mainref{thm:cov-bound} and \mainref{thm:sub-gaussian}.

Before continuing, we make three remarks regarding the randomized Gram--Schmidt orthogonalization.
First, we emphasize that this re-ordering and orthogonalization is only for the purposes of analysis and is not executed by the algorithm.
Second, we remark that although \citet{Bansal2019Gram} discuss how the Gram--Schmidt Walk algorithm was inspired by Gram--Schmidt orthogonalization, an explicit connection is not made in that paper.
This is one of the technical differences in our analysis which allow us to obtain tighter bounds.
Third, we remark that this re-ordering and orthogonalization will be carried out conditioned on an arbitrary pivot ordering $\pi$.
Without loss of generality, we assume that $\pi$ is the identity permutation, i.e. $\pi(i) = i$ for all $i \in [n]$.

We begin this discussion by first describing the randomized re-ordering of the input vectors and then defining the Gram--Schmidt Orthogonalization processes applied to this re-ordering.
Let us introduce the notation of the re-ordering.
The inputs vectors $\gscv{1}, \gscv{2}, \dots \gscv{n} \in \Reals^m$ will be re-ordered as
\begin{equation*} \label{eq:reoder_seq}
\gscv{\reorderi{1}}, \gscv{\reorderi{2}}, \dots \gscv{\reorderi{n}}
\enspace,
\end{equation*}
where $\reorder$ is a bijection mapping positions in the re-ordering to the units.
Formally, $\reorder: [n] \rightarrow [n]$ and to avoid confusion in this notation, we reserve the symbol $r$ for a position in the re-ordering and the symbol $i$ for a unit.
In this way, we write $\reorderi{r} = i$ to mean that the $r$th position in the re-ordering is occupied by unit $i$.
We may also refer to the position of a specific unit in the re-ordering using the inverse function $\invreorder$.
That is, $\invreorderi{i} = r$ means that the unit $i$ is assigned to position $r$ in the re-ordering.

The re-ordering we consider is random and it is defined by the random choices made in the algorithm.
Recall that a unit $i$ is decided by the algorithm when it is either selected as the pivot (Lines~\ref{line:select_pivot1} or \ref{line:select_pivot2}) or frozen without being chosen as the pivot (Line~\ref{line:assingment_update}).
The ordering of the units $\reorderi{1}, \reorderi{2}, \dots \reorderi{n}$ will be the \emph{reverse order} in which they are decided, breaking ties arbitrarily.
In this way, as the algorithm decides units at each iteration, the randomized re-ordering is determined in reverse order.
For example, the first unit to be decided is the first pivot unit $p_1$ so that $\reorderi{n} = p_1 = n$.
If a single unit $j \neq p_{1}$ is frozen in the first iteration, then this is the next unit decided by the algorithm, in which case it is second to last in the re-ordering, i.e. $\reorderi{n-1} = j$.
On the other hand, if only the pivot $p_{1}$ is frozen in the first iteration, the next unit decided by the algorithm is the next pivot, which is $p_{2}$.
In this case, $\reorderi{n-1} = p_2$.

Next, we introduce the Gram--Schmidt orthogonalization process on this randomized re-ordering of the input vectors.
The Gram--Schmidt orthogonalization process is a method to construct a sequence of orthonormal vectors which form a basis for the span of a given set of vectors.
For our problem at hand, we denote this sequence of orthonormal basis vectors by
\begin{equation}
\obv{\reorderi{1}}, \obv{\reorderi{2}}, \dots \obv{\reorderi{n}}.
\end{equation}
They are recursively defined by the Gram--Schmidt orthogonalization process
\begin{equation}
\obv{\reorderi{1}}
= \frac{\gscv{\reorderi{1}}}{\norm{\gscv{\reorderi{1}}}}
\quad \text{and} \quad
\obv{\reorderi{r}}
= \frac{\gscv{\reorderi{r}} - \mat{A}_r \gscv{\reorderi{r}}}{\norm*{\gscv{\reorderi{r}} - \mat{A}_r \gscv{\reorderi{r}}}}
\quad \text{for $r =2, \dots n$},
\end{equation}
where $\mat{A}_r = \sum_{s < r} \obv{\reorderi{s}} \obv{\reorderi{s}}^\tran$ is the projection onto the span of the first $r-1$ input vectors $\gscv{\reorderi{1}} \dots \gscv{\reorderi{r-1}}$.
Because the random re-ordering of the input vectors is determined by the random choices of $\gsd{1} \dots \gsd{n}$ in the algorithm, the random sequence $\obv{\reorderi{1}} \dots \obv{\reorderi{n}}$ is also determined by the random choices made by the algorithm.
Regardless of the randomization, this sequence of vectors forms an orthonormal basis for the span of the input vectors.
Moreover, while the vector $\obv{\reorderi{r}}$ depends on the set of vectors
  $\setb{\gscv{\reorderi{1}}, \ldots, \gscv{\reorderi{r-1}}}$, it does not depend on their order.
For further reading on the Gram--Schmidt orthogonalization process, we refer readers to Chapter~4 of \citet{Strang09Introduction}.

The main benefit of using this Gram--Schmidt orthogonalization process is that we can cleanly analyze the behavior of the algorithm within pivot phases.
In particular, it provides a way to partition the span of the input vectors into orthogonal subspaces $\pspace{1}, \pspace{2}, \dots \pspace{n}$ corresponding to each of the $n$ units.
These subspaces are defined by the algorithm's random choices within the corresponding unit's pivot phase.
We begin by defining the subspaces for units that are chosen as pivots.
Let $i$ be a unit which is chosen as pivot and assume it has position $r = \invreorderi{i}$ in the reordering so that the $k+1$ vectors which are decided during this pivot phase appear in the ordering as $\gscv{\reorderi{r-k}}, \gscv{\reorderi{r-k+1}}, \dots \gscv{\reorderi{r}}$.
The subspace $\pspacei \subset \Reals^m$ is
  defined to be the span of the vectors
	$\gscv{\reorderi{r-k}}, \gscv{\reorderi{r-k+1}}, \dots \gscv{\reorderi{r}}$ after they have been projected orthogonal to $\gscv{\reorderi{1}}, \gscv{\reorderi{2}}, \dots, \gscv{\reorderi{r-k-1}}$.
As the set $\setb{ \reorderi{1}, \ldots, \reorderi{r-k-1} }$ is determined at this time, the projection is well-defined.
The vectors
\begin{equation}
\obv{\reorderi{r-k}}, \obv{\reorderi{r-k+1}}, \dots, \obv{\reorderi{r}}
\end{equation} form an orthonormal basis for the subspace $\pspacei$ and the projection matrix onto this subspace is
\begin{equation}
\phaseproji
= \sum_{s=0}^k \obv{\reorderi{r-s}} \obv{\reorderi{r-s}}^\tran .
\end{equation}
If a unit $i$ is never chosen as a pivot unit, then $\pspace{i}$ is the zero subspace and so the projection matrix $P_i$ is the zero matrix.

The following lemma follows directly from the definition of the subspaces but may also be verified by orthonormality of the vector sequence produced by Gram--Schmidt orthogonalization.

\begin{applemma} \label{lem:orth_proj_mat}
	The subspaces are $\pspace{1}, \pspace{2}, \dots \pspace{n}$ are orthogonal and their union is $\spanv{\gscv{1}, \gscv{2}, \dots \gscv{n}}$.
	Equivalently, the corresponding projection matrices $\phaseproj{1} \dots \phaseproj{n}$ satisfy
	\begin{equation}
		\sum_{i=1}^n \phaseproji = \PgsM,
	\end{equation}
	where $\PgsM$ is the projection matrix onto $\spanv{\gscv{1}, \gscv{2}, \dots \gscv{n}}$.
\end{applemma}

Next, we will show that the fractional balance update $\gsM \gsuvt$ is contained in the subspace corresponding to the current pivot, $\pspace{p_t}$.
We will show a stronger property, but in order to make these statements precise, we need additional notation which connects an iteration $t$ with the re-ordered positions of the units that have already been decided during in the current pivot phase.
We define $\ell_t$ and $g_t$ to be the least and greatest re-ordering positions that were decided during the current pivot phase before Line~\ref{line:condition_here} at iteration $t$.
The first unit to be decided in any pivot phase is the pivot unit.
Thus the greatest re-ordering position of any unit which was decided during the current pivot phase is $g_t = \invreorderi{p_t}$.
Note that when we arrive at Line~\ref{line:condition_here}, $\alive_t \setminus p_t$ is the set of units which have not yet been decided.
Thus, these are the units which will appear earliest in the re-ordering (although their ordering is not yet determined) and so we have that $\ell_t = \abs{ \alive_t \setminus p_t} + 1 = \abs{\alive_t}$.
In the first iteration of a pivot phase, we have $\ell_t = g_t$ because only the pivot has been decided before Line~\ref{line:condition_here} at this iteration.

Using this notation, at Line~\ref{line:condition_here} of iteration $t$, the input vectors whose units have been decided during the current pivot phase are
\begin{equation}
	\gscv{\reorderi{\ell_t}}, \gscv{\reorderi{\ell_t + 1}}, \dots \gscv{ \reorderi{g_t}}.
\end{equation}
The next lemma demonstrates that $\gsM \gsuvt$ is the projection of the pivot onto the subspace spanned by $\obv{\reorderi{\ell_t}}, \obv{\reorderi{\ell_t + 1}}, \dots \obv{\reorderi{g_t}}$.

\begin{applemma} \label{lemma:writing_in_basis}
At each iteration $t$, we can write $\gsM \gsuvt$ in the orthonormal basis
$\obv{\reorderi{1}} \dots \obv{\reorderi{n}}$ as
\begin{equation}
\gsM \gsuvt = \sum_{r = \ell_t}^{g_t} \iprod*{ \obv{\reorderi{r}}, \gscv{p_t}} \obv{\reorderi{r}}.
\end{equation}
\end{applemma}
\begin{proof}
Recall that the step direction $\gsuvt$ is determined by a least squares problem.
That is, the undecided coordinates of the step direction, $\gsuve{t}{\alive_t \setminus p_t}$, are the minimizers of the least squares program
\begin{equation}
\gsuve{t}{\alive_t \setminus p_t}
= \argmin_{u_i : i \in \alive_t \setminus p_t} \norm[\Big]{\gscv{p_t} + \sum_{i \in \alive_t \setminus p_t} u_i \gscv{i}}^2.
\end{equation}
Because the step direction is the minimizer, it must satisfy the normal equations
	\begin{equation}
		\gsM \gsuvt = \gscv{p_t} - \mat{A}_t \gscv{p_t},
	\end{equation}
where $\mat{A}_t$ is the projection matrix onto the span of the alive vectors which are not the pivot.
That is, $\gscv{i}$ for $i$ in $\alive_t \setminus p_t = \setb{\reorderi{1}, \ldots, \reorderi{\ell_t - 1}}$.
By the construction of the re-ordering and the Gram--Schmidt orthogonalization, we have that $\mat{A}_t = \sum_{s < \ell_t} \obv{\reorderi{s}} \obv{\reorderi{s}}^\tran$.
Writing the fractional balance update $\gsM \gsuvt$ in the orthonormal basis, we have that
\begin{align}
\gsM \gsuvt
&= \sum_{r=1}^n \iprod{ \obv{\reorderi{r}} , \gsM \gsuvt} \obv{\reorderi{r}}
	&\text{(orthonormal basis)}\\
&= \sum_{r=1}^n \iprod{ \obv{\reorderi{r}} , \gscv{p_t} - \mat{A}_t \gscv{p_t}} \obv{\reorderi{r}}
	&\text{(normal equations)}\\
&= \sum_{r=1}^n \bracket[\Big]{ \iprod{ \obv{\reorderi{r}} , \gscv{p_t} } - \iprod{ \obv{\reorderi{r}} , \mat{A}_t \gscv{p_t}} } \obv{\reorderi{r}}
	&\text{(linearity)}\\
&= \sum_{r=1}^n \bracket[\Big]{ \iprod{ \obv{\reorderi{r}} , \gscv{p_t} } - \iprod{ \mat{A}_t \obv{\reorderi{r}} , \gscv{p_t}} } \obv{\reorderi{r}}.
	&\text{(projection matrix, $\mat{A}_t^\tran = \mat{A}_t$)}
\end{align}
We now examine each term in this sum.
If $r < \ell_t$ then $\mat{A}_t \obv{\reorderi{r}} = \obv{\reorderi{r}}$ because $\obv{\reorderi{r}}$ is a vector in the subspace associated with the projection $\mat{A}_t$.
Thus, the two terms in the bracket are the same, so the terms corresponding to $r < \ell_t$ are zero and do not contribute to the sum.
If $r \geq \ell_t$, then by the construction of the re-ordering and Gram--Schmidt orthogonalization, $\obv{\reorderi{r}}$ is orthogonal to the subspace corresponding to $\mat{A}_t$ and so $\mat{A}_t \obv{\reorderi{r}} = 0$.
This means that for $\ell_t \leq r \leq g_t$, the second term in the brackets is zero, and only the first term in brackets contributes to the sum.
On the other hand, if $r > g_t$, then by the re-ordering and Gram--Schmidt orthogonalization, $\obv{\reorderi{r}}$ is orthogonal to $\gscv{\reorderi{g_t}} = \gscv{p_t}$.
In this case, both terms in the brackets are zero and the terms corresponding to $r > g_t$ contribute nothing to the sum.
Thus, we have shown that
\begin{equation}
\gsM \gsuvt = \sum_{r = \ell_t}^{g_t} \iprod*{ \obv{\reorderi{r}}, \gscv{p_t}} \obv{\reorderi{r}}. \tag*{\qedhere}
\end{equation}
\end{proof}

%% file: tex/supp-martingale-assignments.tex
\subsection{Martingale Property (Lemma~\mainref{lemma:martingale})}

In this section, we prove that the sequence of fractional assignments in the Gram--Schmidt Walk design forms a martingale.

\begin{reflemma}{\mainref{lemma:martingale}}
	\martingale
\end{reflemma}
\begin{proof}
	Recall that the fractional assignments are updated as $\gsv{t + 1} = \gsvt + \gsdt \gsuvt$.
	Consider the conditional expectation of the assignments updated at iteration $t$:
	\[
	\E{ \gsv{t + 1} \given \gsv{1}, \dotsc, \gsvt} = \gsvt + \E{ \gsdt \gsuvt \given \gsv{1}, \dotsc,
		\gsvt}.
	\]
	By the law of iterated expectations,
	\[
	\E{ \gsdt \gsuvt \given \gsv{1}, \dotsc, \gsvt}
	=
	\E[\big]{ \E{ \gsdt \given \gsdtp, \gsdtm} \gsuvt \given \gsv{1}, \dotsc, \gsvt},
	\]
	because $\gsdt$ is conditionally independent of $\paren{\gsv{1}, \dotsc, \gsvt, \gsuvt}$ given $\paren{\gsdtp, \gsdtm}$.
	The step size $\gsdt$ takes the values $\gsdtp$ and $\gsdtm$ with probabilities inversely proportional to their magnitudes, so
	\[
	\E{ \gsdt \given \gsdtp, \gsdtm}
	= \gsdtp \paren[\bigg]{\frac{\gsdtm}{\gsdtp + \gsdtm}} - \gsdtm \paren[\bigg]{\frac{\gsdtp}{\gsdtp +
			\gsdtm}}
	= 0.
	\]
	It follows that the expected update is zero: $\E{ \gsdt \gsuvt \given \gsv{1}, \dotsc, \gsvt} = \zerovec$.
\end{proof}

%% file: tex/supp-cov-bound.tex
\subsection{Covariance bound (Theorem~\mainref{thm:cov-bound})} \label{sec:cov-bound-supp}

This section contains a proof of an extended version of the covariance bound in Theorem~\mainref{thm:cov-bound}.
As stated in the previous section, all analysis will be carried out conditioned on an arbitrary pivot ordering $\pi$ and without loss of generality, we suppose that $\pi(i) = i$ for all $i \in [n]$.
We begin by deriving a form of the covariance matrix of the assignment vector in terms of the update quantities in the algorithm.

\begin{applemma} \label{lemma:cov_x}
	The covariance matrix of the assignment vector is given by
	\begin{equation}
	\Cov{\zv} = \E[\Big]{ \sum_{t=1}^{T} \gsdt^2 \gsuv{t} \gsuv{t}^{\tran} } .
	\end{equation}
\end{applemma}
\begin{proof}
First, observe that
\begin{equation}
\Cov{\zv}
= \E{\zv \zv^{\tran}} - \E{\zv} \E{\zv}^{\tran}
= \E{\zv \zv^{\tran}} - \gsv{1} \gsv{1}^{\tran}
\end{equation}
where the second equality uses $\E{\zv} = \gsv{1}$, which in a consequence of the martingale property (Lemma~\mainref{lemma:martingale}).
By the update rule $\gsv{t+1} \leftarrow \gsvt + \gsdt \gsuv{t}$,
\begin{equation}
\gsv{t+1} \gsv{t+1}^{\tran}
= \paren*{\gsvt + \gsdt \gsuv{t}}\paren*{\gsvt + \gsdt \gsuv{t}}^{\tran}
= \gsvt \gsvt^{\tran}
+ \gsdt \paren*{\gsuv{t} \gsvt^{\tran} + \gsvt \gsuv{t}^{\tran}}
+ \gsdt^{2} \gsuv{t} \gsuv{t}^{\tran} .
\end{equation}
Iteratively applying this over all iterations $t\in\braces{1, 2, \dots}$ and using that the returned vector is
$\zv = \gsv{T+1}$, we have that
\begin{equation}
\zv \zv^{\tran}
= \gsv{T+1} \gsv{T+1}^{\tran}
= \gsv{1} \gsv{1}^{\tran}
+ \sum_{t=1}^{T} \gsdt \paren*{\gsuv{t} \gsvt^{\tran} + \gsvt \gsuv{t}^{\tran}}
+ \sum_{t=1}^{T} \gsdt^{2} \gsuv{t} \gsuv{t}^{\tran} .
\end{equation}
Substituting this expression of $\zv \zv^{\tran}$ into $\E{\zv \zv^{\tran}}$ in the earlier covariance calculation, we obtain that
\begin{equation}\label{eq:second_cov_calc}
\Cov{\zv}
=  \E*{\sum_{t=1}^{T} \gsdt^{2} \gsuv{t} \gsuv{t}^{\tran}}
+ \E*{\sum_{t=1}^{T} \gsdt \paren*{\gsuv{t} \gsvt^{\tran} + \gsvt \gsuv{t}^{\tran}} }
\end{equation}

We will now show that the last term is zero because the step size $\gsdt$ is zero in
expectation.
By linearity of expectation and using the convention that the algorithm runs for $n$ iterations with $\gsdt = 0$ and $\gsuvt = \vec{0}$ for $t > T$,
\begin{equation}
\E*{\sum_{t=1}^{T} \gsdt \paren*{\gsuv{t} \gsvt^{\tran} + \gsvt \gsuv{t}^{\tran}} }
= \sum_{t=1}^{n} \E*{ \gsdt \paren*{\gsuv{t} \gsvt^{\tran} + \gsvt \gsuv{t}^{\tran}}}
\end{equation}
For a fixed iteration $t$, consider the individual term $\E{\gsdt \paren*{\gsuv{t} \gsvt^{\tran} + \gsvt \gsuv{t}^{\tran}}}$ in the sum above.
Observe that if we condition on all previous random decisions made by the algorithm before step size $\gsdt$ is chosen (i.e. choices of step sizes $\gsd{1} \dots \gsd{t-1}$), then the step direction $\gsuvt$ and fractional assignment $\gsvt$ are both determined, so that $ \gsuvt \gsvt^\tran + \gsvt \gsuvt^\tran$ is a deterministic quantity.
In this way, $\gsdt$ is conditionally independent of $ \gsuvt \gsvt^\tran + \gsvt \gsuvt^\tran$ conditioned on all previous random decisions made by the algorithm.
Using the fact that the expected step size $\gsdt$ is zero, we have that
\begin{equation}
\E{\gsdt \paren*{\gsuv{t} \gsvt^{\tran} + \gsvt \gsuv{t}^{\tran}} \given \gsd{1} \dots \gsd{t-1}}
=  \paren*{\gsuv{t} \gsvt^{\tran} + \gsvt \gsuv{t}^{\tran}} \cdot \E{\gsdt \given \gsd{1} \dots \gsd{t-1}}
= 0
\end{equation}
for all iterations $t$. By the law of total expectation,
$\E{\gsdt \paren*{\gsuv{t} \gsvt^{\tran} + \gsvt \gsuv{t}^{\tran}}} = 0$ and so
that the second term in \eqref{eq:second_cov_calc} is zero.
\end{proof}

Next, we prove a lemma stating that the expected sum of the squared step sizes in the remainder of a pivot phase is not too large in expectation.
To do this, we introduce notation that connects a position in the re-ordering to the subsequent iterations in a pivot phase.
For each position $r$ in the re-ordering, we define
\begin{equation}
L_r = \setb{ t : \ell_t \leq r \leq g_t} .
\end{equation}
The set $L_r$ allows us to discuss what happens in the remaining iterations of a pivot phase after the unit in position $r$ has been decided.
For example, if a unit $i$ is chosen as the pivot and assigned to position $r$, then $L_r$ is the entire pivot phase $S_i$.
If a non-pivot unit $i$ is frozen and assigned to position $r$, then $L_r$ are the remaining iterations in the pivot phase.
Note that $L_r$ may be empty if a non-pivot unit is frozen along with pivot at the last iteration of the pivot phase.
We are now ready to state a lemma on the expected sum of the squared step sizes throughout the remainder of a pivot phase.

\begin{applemma} \label{lem:bound_delta_squared}
	For each $r \in [n]$, conditional on the random decisions made up until unit $\reorderi{r}$ is decided,
    the expected sum of squared step sizes in the remainder of its pivot phase is at most one.
	That is, for each unit $i \in [n]$ with re-ordering position $r = \invreorderi{i}$,
	\begin{equation}
	\E[\Bigg]{ \sum_{t \in L_{r}} \delta_{t}^{2} \given \Delta_{\reorderi{r}}} \leq 1 .
	\end{equation}
\end{applemma}
\begin{proof}
	Because only one pivot phase is being considered, we drop the iteration subscripts here and write the pivot as $p$.
	Recall that $\Delta_{\reorderi{r}}$ denotes all the random decisions made by the algorithm up to and including when unit $i$ was decided by the algorithm.
	If $L_r$ is empty, then the statement is trivially true.
	Otherwise, $L_r$ is a (random) contiguous set of iterations $t_0, t_0+1, \dots t_0 + k$, where $t_0+k$ is the last iteration in the pivot phase.
	Because the pivot phase terminates when the pivot $p$ is frozen, $\abs{\gsve{t_0+ k}{p}} = 1$.
	It follows that
	\begin{align*}
	1 - \gsve{t_0}{p}^{2}
	&= \gsve{t_0+k}{p}^{2} - \gsve{t_0}{p}^{2}
	&\text{($\abs{\gsve{t_0+k}{p}} = 1$)}\\
	&= \sum_{s=0}^{k - 1} \bracket*{\gsve{t_0+s+1}{p}^{2} - \gsve{t_0+s}{p}^{2}}
	&\text{(telescoping sum)}\\
	&= \sum_{s=0}^{k - 1} \bracket*{ \paren*{\gsve{t_0+s}{p} + \gsd{t_0+s} \gsuvi{t_0 + s}{p} }^{2} - \gsve{t_0+s}{p}^{2}}
	&\text{(update rule)}\\
	&= \sum_{s=0}^{k - 1}  \bracket*{ \gsd{t_0+s}^{2} \gsuvi{t_0 + s}{p}^{2} +2  \gsd{t_0+s} \gsuvi{t_0 + s}{p} \gsve{t_0+s}{p}}
	&\text{(cancelling terms)}\\
	\end{align*}
	Taking conditional expectations of both sides and using linearity of expectation, we have that
	\begin{equation} \label{eq:equality-in-squared-delta}
	1 - \gsve{t_0}{p}^{2}
	= \E[\Bigg]{\sum_{t \in L_{r}} \delta_{t}^{2} \given \Delta_{\reorderi{r}}}
	+ 2 \E[\Bigg]{\sum_{t \in L_{r}} \delta_{t} \gsuvi{t}{p} \gsve{t}{p} \given \Delta_{\reorderi{r}}},
	\end{equation}
	because the left hand side is a deterministic quantity under this conditioning.
	We now seek to show that the second term on the right hand side is zero.
	To this end, observe that we may extend the sum from iterations $t \in L_{r}$ to all remaining iterations because $\gsuvi{t}{p} = 0$ for iterations $t$ after the current pivot phase, i.e.,
	\[
	\E[\Bigg]{\sum_{t \in L_{r}} \delta_{t} \gsuvi{t}{p} \gsve{t}{p} \given \Delta_{\reorderi{r}}}
	= \E[\Bigg]{\sum_{t \geq t_0} \delta_{t} \gsuvi{t}{p} \gsve{t}{p} \given \Delta_{\reorderi{r}}}
	= \sum_{t \geq t_0} \E[\Big]{\delta_{t} \gsuvi{t}{p} \gsve{t}{p} \given \Delta_{\reorderi{r}}}.
	\]
	We now show that each term $ \E{\delta_{t} \gsuvi{t}{p} \gsve{t}{p} \given \Delta_{\reorderi{r}}}$ is zero for each $t$.
	Suppose that we further condition on all previous random decisions made by the algorithm before step size $\gsdt$ is chosen.
	In this case, the quantity $\gsuvi{t}{p} \gsve{t}{p}$ is completely determined and so $\delta_{t}$ is independent of $\gsuvi{t}{p} \gsve{t}{p}$.
	Moreover, the step size has mean zero, as shown in the proof of Lemma~\mainref{lemma:martingale}.
	Thus, for $t \geq t_0$,
	\[
	\E{\delta_{t} \gsuvi{t}{p}  \gsve{t}{p} \given \gsd{1} \dots \gsd{t-1}}
	= \gsuvi{t}{p} \gsve{t}{p}\cdot \E{\delta_{t} \given \gsd{1} \dots \gsd{t-1}}
	= 0
	\]
	By the law of total expectation, it follows that the term
	$\E{\delta_{t} \gsve{t}{p} \given \Delta_{\reorderi{r}}}$ is zero for $t \geq t_0$.
	Thus, the second term in \eqref{eq:equality-in-squared-delta} is zero and so we have that
	\begin{equation}
	\E[\Bigg]{\sum_{t \in L_{r}} \delta_{t}^{2} \given \Delta_{\reorderi{r}}}
	= 1 - \gsve{t_0}{p}^{2}
	\leq 1 ,
	\end{equation}
	where the inequality follows from $\gsve{t_0}{p} \in \paren{-1,1}$.
\end{proof}

At this point, we are ready to prove the covariance bound.

\begin{reftheorem}{\mainref{thm:cov-bound}*}
If all input vectors $\gscv{1} \dots \gscv{n}$ have $\ell_2$ norm at most one, then
the covariance matrix of the vector of imbalances $\gsM \zv$ is bounded in the Loewner order by the orthogonal projection onto the subspace spanned by the columns of $\gsM$:
\begin{equation}
	\Cov{\gsM \zv} \preceq \PgsM = \gsM \paren[\big]{ \gsM^\tran \gsM }^{\pinv} \gsM^\tran,
\end{equation}
where we recall that $\mat{A}^\pinv$ denotes the pseudoinverse of the matrix $\mat{A}$.
\end{reftheorem}

\begin{proof}
	Condition on an arbitrary pivot ordering $\pi$ and without loss of generality, suppose that $\pi(i) = i$ for all $i \in [n]$.
To prove the matrix inequality in the statement of the theorem, we seek to show that
\begin{equation}
{ \arbvec^{\tran} \Cov{ \gsM \zv} \arbvec \leq \arbvec^{\tran} \PgsM \arbvec}
\quad \quad \text{ for all } \arbvec \in \Reals^m
\end{equation}
Using Lemma~\ref{lemma:cov_x} for the form of $\Cov{\zv}$ and linearity of expectation, we have that
\begin{equation}
\arbvec^\tran \Cov{ \gsM \zv} \arbvec
= \arbvec^\tran \gsM \Cov{\zv} \gsM^\tran \arbvec
= \arbvec^\tran \gsM
\E*{ \sum_{t=1}^{T} \gsdt^{2} \gsuv{t} \gsuv{t}^{\tran} }  \gsM^{\tran} \arbvec
= \E*{ \sum_{t=1}^{T} \gsdt^2 \iprod{\gsM \gsuv{t} , \arbvec }^2  }
\enspace.
\end{equation}
Thus, we seek to show that for all $\arbvec \in \Reals^m$,
\begin{equation}
\E*{ \sum_{t=1}^{T} \gsdt^{2} \iprod{\gsM \gsuv{t} , \arbvec }^{2}  }
\leq
\arbvec^{\tran} \PgsM \arbvec .
\end{equation}
Next, we compute an upper bound on the quadratic forms in the sum. For each iteration $t$,
\begin{align*}
\iprod{\gsM \gsuv{t} , \arbvec}^2
&= \iprod*{ \sum_{r=\ell_{t}}^{g_t} \iprod*{ \obv{\reorderi{i}}, \gscv{p_t}} \obv{\reorderi{i}}, \arbvec }^2
	&\text{(Lemma~\ref{lemma:writing_in_basis})}\\
&= \paren*{ \sum_{r=\ell_{t}}^{g_t} \iprod*{\obv{\reorderi{r}} , \gscv{p_{t}}} \iprod{\obv{\reorderi{r}} , \arbvec}}^2
	&\text{(linearity)} \\
&\leq \paren*{\sum_{r=\ell_{t}}^{g_t} \iprod{\obv{\reorderi{r}},  \gscv{p_{t}}}^2} \paren*{\sum_{r=\ell_{t}}^{g_t} \iprod{\obv{\reorderi{r}},  \arbvec}^2}
	&\text{(Cauchy--Schwarz)} \\
&\leq \norm*{\gscv{p_{t}}}^2 \cdot \paren*{ \sum_{r=\ell_{t}}^{g_t} \iprod{\obv{\reorderi{r}},  \arbvec}^2}
	&\text{($\obv{\reorderi{r}}$ are orthonormal)} \\
&\leq \paren*{ \sum_{r=\ell_{t}}^{g_t} \iprod{\obv{\reorderi{r}},  \arbvec}^2}.
	&\text{(by assumption, $\norm*{\gscv{p_{t}}}^2 \leq 1$)} \\
\end{align*}
Using this upper bound, we obtain an upper bound for the expected quantity of interest,
\begin{align*}
\E*{ \sum_{t=1}^{T} \gsdt^{2} \iprod{\gsM \gsuv{t} , \arbvec }^{2}  }
&\leq  \E*{ \sum_{t=1}^{T} \gsdt^{2} \paren*{ \sum_{r=\ell_{t}}^{g_t} \iprod{\obv{\reorderi{r}}, \arbvec}^{2}}  }
	&\text{(from above)} \\
&= \E*{\sum_{r=1}^{n} \iprod{\obv{\reorderi{r}}, \arbvec}^{2} \sum_{t \in L_{r}} \gsdt^{2}  }
	&\text{(rearranging terms)} \\
&= \sum_{r=1}^{n} \E*{\iprod{\obv{\reorderi{r}}, \arbvec}^{2} \sum_{t \in L_{r}} \gsdt^{2}}
	&\text{(linearity of expectation)}
\end{align*}
We examine each of the terms in this sum.
Fix a position $r$ in the random re-ordering.
Suppose that we further condition on $\Delta_{\reorderi{r}}$, which contains all random decisions made by the algorithm up to and including when unit $\reorderi{r}$ was decided by the algorithm.
Under this conditioning, the vector $\obv{\reorderi{r}}$ is completely
determined and so the quantity $\iprod{\obv{\reorderi{r}},\arbvec}^{2}$ is also completely determined.
In this way, the random term $\sum_{t \in L_{r}} \gsdt^{2}$ is conditionally independent of
$\iprod{\obv{\reorderi{r}},\arbvec}^{2}$ given $\Delta_{\reorderi{r}}$. Thus, we have that
\begin{equation}
\E*{\iprod{\obv{\reorderi{r}},\arbvec}^{2} \sum_{t \in L_{r}} \gsdt^{2} \middle\vert \Delta_{\reorderi{r}}}
= \iprod{\obv{\reorderi{r}},\arbvec}^{2} \cdot \E*{\sum_{t \in L_{r}} \gsdt^{2} \middle\vert \Delta_{\reorderi{r}}}
\leq \iprod{\obv{\reorderi{r}},\arbvec}^{2},
\end{equation}
where the equality is due to conditional independence and the inequality follows from
Lemma~\ref{lem:bound_delta_squared}.
Using iterated expectation, it follows that
\begin{equation}
\E*{\iprod{\obv{\reorderi{r}}, \arbvec}^{2} \sum_{t \in L_{r}} \gsdt^{2}}
\leq \E[\Big]{\iprod{\obv{\reorderi{r}}, \arbvec}^{2}} .
\end{equation}
Substituting this bound and using linearity of expectation yields
\[
\E*{\sum_{t=1}^{T} \gsdt^{2} \iprod{\gsM \gsuv{t} , \arbvec }^{2}}
\leq \sum_{r=1}^{n} \E[\Big]{\iprod{\obv{\reorderi{r}}, \arbvec}^{2}}
= \arbvec^{\tran}  \E*{\sum_{r=1}^{n} \obv{\reorderi{r}} \obv{\reorderi{r}}^{\tran}  } \arbvec
= \arbvec^{\tran} \PgsM \arbvec \enspace,
\]
where the last equality follows from the fact that the vectors $\obv{\reorderi{1}}, \obv{\reorderi{2}}, \dots, \obv{\reorderi{n}}$ form an orthonormal basis for the span of input vectors, thus $ \sum_{r=1}^{n} \obv{\reorderi{r}} \obv{\reorderi{r}}^{\tran} = \PgsM$ holds deterministically, regardless of the randomized re-ordering.

Thus, we have that conditioned on any pivot ordering $\pi$, $\Cov{\gsM \zv} \preceq \PgsM$.
The desired result is obtained by marginalizing over all pivot orderings.
\end{proof}

%% file: tex/supp-subgaussian.tex
\subsection{Subgaussian bound (Theorem~\mainref{thm:sub-gaussian})}\label{sec:subguassian-proof-supp}

In this section, we prove an extended version of the subgaussian concentration inequality of Theorem~\mainref{thm:sub-gaussian}.
As stated in the previous sections, all analysis will be carried out conditioned on an arbitrary pivot ordering $\pi$ and without loss of generality, we suppose that $\pi(i) = i$ for all $i \in [n]$.
We begin by presenting the main technical inequality (Lemma~\ref{lemma:technical_inequality}) which is stated in terms of operator monotonicity and proved using basic calculus.
Next, we present Lemma~\ref{lemma:sub-gaussian-pivot}, which analyzes the behavior of the Gram--Schmidt Walk algorithm in one pivot phase using a backwards induction style argument.
Finally, we prove the subgaussian concentration inequality by showing how we may repeatedly apply Lemma~\ref{lemma:sub-gaussian-pivot}.

The main technical inequality is stated in terms of operator monotonicity, which we briefly describe here.
Let $\mathcal{D}$ be a set of $n$-by-$n$ symmetric matrices.
A real-valued matrix function $f: \mathcal{D} \rightarrow \Reals$ is said to be \emph{operator monotone increasing} if
\begin{equation}
\mat{A}, \mat{B} \in \mathcal{D} \text{ with } \mat{A} \preceq \mat{B}
\Rightarrow
f(\mat{A}) \leq f(\mat{B}).
\end{equation}
Intuitively, a real-valued matrix function $f$ is monotone increasing if ``larger'' matrices (as determined by the Loewner order) are assigned larger values.
We say that $f$ is \emph{operator monotone decreasing} if $ \mat{A} \preceq \mat{B}$ implies instead that $f(\mat{A}) \geq f(\mat{B})$.
Although there is a well developed theory of operator monotonicity, we use only very basic facts here which are mostly self contained.
For more information on operator monotonicity, we refer readers to Chapter 5 of \citet{Bhatia97Matrix}.

\begin{applemma} \label{lemma:technical_inequality}
	For all $x \in \bracket{-1,1}$ the function
	\begin{equation}
	f_x \begin{pmatrix} \alpha \ & \  \eta \\ \eta \ & \ \beta \end{pmatrix}
	= \exp\paren[\bigg]{ -\frac{1}{2}\alpha \beta }
	\bracket*{ \frac{1+x}{2} \exp\paren{\paren{1-x}\eta } + \frac{1-x}{2} \exp\paren{-\paren{1+x}\eta } }
	\end{equation}
	is operator monotone decreasing over the set of $2$-by-$2$ positive semidefinite matrices.
\end{applemma}
\begin{proof}
Operator monotonicity of a function $g: \mathcal{D} \rightarrow \Reals$ is preserved under composition with any monotone increasing $h: \Reals \rightarrow \Reals$.
Using this and observing that $f_x$ takes positive values for $x \in \bracket{-1, 1}$, we have that $f_x$ is operator monotone decreasing if and only if $\log f_x$ is operator monotone decreasing.
Moreover, a differentiable function $g: \mathcal{D} \rightarrow \Reals$ is operator monotone decreasing if and only if $-\grad g (\mat{A})$ is positive semidefinite for all $\mat{A} \in \mathcal{D}$.
The function $f_x$ under consideration is differentiable and thus, to prove the lemma, it suffices to show that
\begin{equation}
- \grad \log f_x  \begin{pmatrix} \alpha \ & \  \eta \\ \eta \ & \ \beta \end{pmatrix}
\end{equation}
is positive semidefinite when the $2$-by-$2$ input matrix is positive semidefinite, i.e., $\alpha, \beta \geq 0$ and $\alpha \beta \geq \eta^2$.

We begin by defining the shorthand
	\begin{equation}
		\psi_x(\eta)
		= \log \bracket*{ \frac{1+x}{2} \exp\paren{{\paren{1-x}\eta }} + \frac{1-x}{2} \exp\paren{{-\paren{1+x}\eta }} }
	\end{equation}
for the $\log$ of the bracketed term in the definition of $f_x$.
Using this, we may write the function $\log f_x$ as
\begin{equation}
\log f_x \begin{pmatrix} \alpha \ & \  \eta \\ \eta \ & \ \beta \end{pmatrix}
= \psi_x(\eta) - \frac{1}{2} \alpha \beta.
\end{equation}
From the above expression, it is clear that $\partial_\alpha \log f_x = - \beta / 2$, $\partial_\beta \log f_x = - \alpha /2$, and $\partial_\eta \log f_x = \partial_\eta \psi_x$.
Thus, the matrix gradient may be computed:
\begin{equation}
-2 \grad \log f_x =
\begin{pmatrix}
\beta \ & \ -\partial_\eta \psi_x(\eta) \\
-\partial_\eta \psi_x(\eta) & \alpha
\end{pmatrix}.
\end{equation}
Recall that when computing the matrix gradient, we scale the off diagonals by $1/2$, as they appear twice in the trace inner product.
We seek to show that the matrix above is positive semidefinite when the input matrix is positive semidefinite.
Because the matrix above is $2$-by-$2$, proving that it is positive semidefinite is equivalent to showing the three inequalities $\alpha, \beta \geq 0$ and $\alpha \beta \geq \paren{\partial_\eta \psi_x(\eta)}^2$.
Because the input matrix is positive semidefinite, we already have that $\alpha, \beta \geq 0$.
To show the final inequality, we show in the next part of the proof that $\eta^2 \geq \paren{\partial_\eta \psi_x(\eta)}^2$.
Because the input matrix already satisfies $\alpha \beta \geq \eta^2$, this will imply the final inequality.

So for the final part of the proof, we focus on showing the inequality
	\begin{equation}
		\paren{\partial_\eta \psi_x(\eta)}^2 \leq \eta^2 \quad \text{for all } x \in \bracket{-1,1}.
	\end{equation}
To this end, we use an enveloping argument to show that $\abs{\partial_\eta \psi_x(\eta)} \leq \abs{\eta}$ for all $x \in \bracket{-1,1}$.
We begin by computing the first and second derivatives of $\psi_x(\eta)$.
First, we rewrite the function $\psi_x(\eta)$ as
\begin{align}
\psi_x(\eta)
&= \log \bracket*{ \frac{1+x}{2} \exp{\paren{1-x}\eta } + \frac{1-x}{2} \exp{-\paren{1+x}\eta } } \\
&= \log \bracket*{\frac{1}{2} \paren*{ e^{\eta - x\eta} + x e^{\eta - x \eta} + e^{-\eta - x\eta} - x e^{-\eta - x \eta}} } \\
&= \log \bracket*{\frac{e^{-x \eta}}{2} \paren{ e^\eta + xe^{\eta} + e^{-\eta} - x e^{-\eta}} } \\
&= \log \bracket*{ \frac{1}{2} \paren{ e^\eta + xe^{\eta} + e^{-\eta} - x e^{-\eta}}} -x \eta \\
&= \log \bracket*{\cosh(\eta) + x \sinh(\eta)} - x \eta .
\end{align}
Next, we compute the derivative $\partial_\eta \psi_x(\eta)$ by using chain rule and derivatives of $\log$ and hyperbolic trigonometric functions:
\begin{equation}
\partial_\eta \psi_x(\eta)
= \frac{\sinh(\eta) + x \cosh(\eta)}{\cosh(\eta) + x \sinh(\eta)} - x .
\end{equation}
Finally, we compute the second derivative of $\psi_x(\eta)$ using the above result, the quotient rule, and derivatives for the hyperbolic functions:
\begin{equation}
\partial^2_\eta \psi_x(\eta)
= 1 -\paren*{\frac{\sinh(\eta) + x\cosh(\eta)}{ \cosh(\eta) + x \sinh(\eta)} }^2
= 1 - \paren{\partial_\eta \psi_x(\eta) + x}^2 .
\end{equation}
We now establish the basis of our enveloping argument.
That is, we show that the second derivative of $\psi_x(\eta)$ is bounded above and below by
\begin{equation}
0 \leq \partial^2_\eta \psi_x(\eta) \leq 1 \quad \text{for all} \quad \eta \in \Reals \quad \text{and} \quad x \in \bracket{-1,1}.
\end{equation}
The upper bound is immediate from the earlier expression, as $\partial^2_\eta \psi_x(\eta)  = 1 - \paren{\partial_\eta \psi_x(\eta) + x}^2  \leq 1$.
The lower bound is a consequence of $x \in \bracket{-1,1}$.
To see this, observe that
\begin{align}
\partial^2_\eta \psi_x(\eta)
&= 1 -\paren*{\frac{\sinh(\eta) + x\cosh(\eta)}{ \cosh(\eta) + x \sinh(\eta)} }^2 \geq 0 \\
&\Leftrightarrow \paren{\cosh(\eta) + x \sinh(\eta)}^2 \geq \paren{\sinh(\eta) + x\cosh(\eta)}^2 \\
&\Leftrightarrow \cosh^2(\eta) + x^2 \sinh^2(\eta) \geq \sinh^2(\eta) + x^2 \cosh^2(\eta) \\
&\Leftrightarrow \cosh^2(\eta) - \sinh^2(\eta) \geq x^2 \paren{ \cosh^2(\eta) - \sinh^2(\eta) } \\
&\Leftrightarrow 1 \geq x^2
\end{align}
Now, we make our enveloping argument.
First, we observe that $\partial_\eta \psi_x(0) = 0$.
Next, for $\eta > 0$, we can bound the value of $\partial_\eta \psi_x(\eta)$ from above and below by
\begin{align}
\partial_\eta \psi_x(\eta)
&= \partial_\eta \psi_x(0) + \int_{y=0}^\eta \partial^2_\eta \psi_x(y) dy
\leq 0 + \int_{y=0}^\eta 1 dy
= \eta \\
\partial_\eta \psi_x(\eta)
&= \partial_\eta \psi_x(0) + \int_{y=0}^\eta \partial^2_\eta \psi_x(y) dy
\geq 0 + \int_{y=0}^\eta 0 dy
= 0 .
\end{align}
Written together, these inequalities state that $0 \leq \partial_\eta \psi_x(\eta) \leq \eta$ for values $\eta \geq 0$.
A similar enveloping argument shows that $-\eta \leq \partial_\eta \psi_x(\eta) \leq 0$ for values $\eta \leq 0$.
Putting these two together, we have that $\abs{\partial_\eta \psi_x(\eta)} \leq \abs{\eta}$ for all $\eta \in \Reals$ and $x \in \bracket{-1,1}$, as desired.
\end{proof}

\begin{applemma}\label{lemma:sub-gaussian-pivot}
	Let $p$ be a unit that is chosen as the pivot and let $\Delta_p$ denote all random decisions made by the algorithm up until the beginning of pivot phase $p$.
	If $\norm{\gscv{p}} \leq 1$,
	then for all $\arbvec \in \Reals^m$,
	\begin{equation}
	\E[\Bigg]{\expf[\Bigg]{
	\sum_{t \in \pphase{p}} \gsdt \iprod{\gsM \gsuvt , \arbvec} - \frac{1}{2} \norm{\phaseproj{p} \gscv{p}}^2 \cdot \norm{\phaseproj{p} \arbvec}^2
	} \given \Delta_p } \leq 1,
	\end{equation}
	where $\pphase{p}$ is the set of iterations for which $p$ is the pivot.
\end{applemma}
\begin{proof}
Let $t_p$ be the iteration at which $p$ is first chosen to be the pivot.
This iteration $t_p$ is a deterministic quantity conditioned on $\Delta_p$.

We begin by describing a convention which we adopt for the purposes of this analysis.
Recall that the number of iterations in a pivot phase is generally a random quantity; however, the number of iterations in a pivot phase is at most $n$.
In fact, because $t_p-1$ iterations have already occurred, the number of iterations in the pivot phase $S_p$ is at most $n - t_p + 1$.
For the purposes of this proof, we adopt a convention which deterministically fixes the number of iterations within the pivot phase to be $n - t_p + 1$.
We adopt this convention because fixing the number of iterations in a pivot phase to be a deterministic quantity simplifies our backwards induction style argument.
Once the pivot is frozen at iteration $t$, all remaining iterations of the pivot phase $s  > t$ have step size zero, i.e. $\gsd{s} = 0$.
In this way, the fractional assignment is not updated in the remainder of the pivot phase after the pivot is frozen and thus this convention does not change the behavior of the algorithm.
We emphasize again that this convention is for purposes of the current analysis and does not change the algorithm itself.

Using this convention and writing the iterations in the pivot phase as $\pphase{p} = \setb{t_p \dots n}$, we seek to show that
\begin{equation} \label{eq:pivot_subgaussian_desired_inequality}
\E[\Bigg]{\expf[\Bigg]{
		\sum_{t=t_p}^n \gsdt \iprod{\gsM \gsuvt , \arbvec} - \frac{1}{2} \norm{\phaseproj{p} \gscv{p}}^2 \cdot \norm{\phaseproj{p} \arbvec}^2
	} \given \Delta_p } \leq 1.
\end{equation}
All expectations in the remainder of the proof are conditioned on $\Delta_p$ and so we drop this notation.

We now rewrite the terms in the exponent by using the sequence of orthonormal basis vectors produced by the Gram--Schmidt orthogonalization process, as described in Section~\ref{sec:gram-schmidt-orthogonalization}.
Suppose that the pivot unit has position $r = \invreorderi{p}$ in the reordering so that the $k+1$ vectors which are decided during this pivot phase appear in the ordering as
\begin{equation}
 \gscv{\reorderi{r-k}}, \gscv{\reorderi{r-k+1}}, \dots \gscv{\reorderi{r}},
\end{equation}
where the pivot vector is the last in this re-ordering, i.e., $\reorderi{r} = p$, and so $\gscv{\reorderi{r}} = \gscv{p}$.
The corresponding basis vectors produced by the Gram--Schmidt orthogonalization are
\begin{equation}
\obv{\reorderi{r-k}}, \obv{\reorderi{r-k+1}}, \dots \obv{\reorderi{r}}.
\end{equation}

We now define a way to partition these reordering positions according to the iterations when they were decided.
For each iteration $t=t_p, \dots n$ in this pivot phase, we define $Q_t$ to be the reordering positions of the units that are frozen during the fractional assignment update in Line~\ref{line:assingment_update} during iteration $t$.
By our convention, it may happen that $\gsdt = 0$ and in this case, $Q_t = \emptyset$.
We also define $Q_{t_p-1} = \setb{g_p} = \setb{\invreorderi{p}}$, which is the re-ordering index of the pivot.
We remark that this reordering position is deterministic given the conditioning $\Delta_p$ and the subscript $t_p-1$ is chosen for notational convenience.
Note that the reordering positions are determined in the order $Q_{t_p-1}, Q_{t_p}, \dots Q_n$ and this forms a partition of the reordering positions decided in this pivot phase.

Lemma~\ref{lemma:writing_in_basis} shows that for each iteration $t$,
\begin{equation}
\gsM \gsuvt = \sum_{s =t_p-1 }^{t-1} \sum_{r \in Q_s} \iprod{\obv{\reorderi{r}} , \gscv{p}} \obv{\reorderi{r}}
\quad \text{and so} \quad
\iprod{\gsM \gsuvt, \arbvec}
= \sum_{s = t_p-1 }^{t-1} \sum_{r \in Q_s} \iprod{\obv{\reorderi{r}} , \gscv{p}} \iprod{ \obv{\reorderi{r}} , \arbvec}.
\end{equation}

Recall that the projection matrix $\phaseproj{p}$ is defined as
\begin{equation}
\phaseproj{p}
= \sum_{s=t_p-1}^n \sum_{r \in Q_s} \obv{\reorderi{r}} \obv{\reorderi{r}}^\tran
\end{equation}
and thus we have that
\begin{equation}
\norm{\phaseproj{p} \gscv{p}}^2 = \sum_{s=t_p-1}^n \sum_{r \in Q_s} \iprod{ \obv{\reorderi{r}} , \gscv{p}}^2
\quad \text{and} \quad
\norm{\phaseproj{p} \arbvec}^2 = \sum_{s=t_p-1}^n \sum_{r \in Q_s} \iprod{ \obv{\reorderi{r}} , \arbvec}^2
\end{equation}
For notational convenience, for each reordering position $r$, let $\alpha_r =  \iprod{\obv{\reorderi{r}} , \gscv{p}}$ and $\beta_r =  \iprod{\obv{\reorderi{r}} , \arbvec}$.

Substituting these terms into \eqref{eq:pivot_subgaussian_desired_inequality}, we have that the desired inequality may be written as
\begin{equation}
\E[\Bigg]{\expf[\Bigg]{
		\sum_{t=t_p}^n \gsdt \sum_{s = t_p-1}^{t-1} \sum_{r \in Q_s} \alpha_r \beta_r
		- \frac{1}{2} \paren[\Big]{\sum_{s=t_p-1}^n \sum_{r \in Q_s} \alpha_r^2}
		\cdot \paren[\Big]{\sum_{s=t_p-1}^n \sum_{r \in Q_s} \beta_r^2}
	}}
\leq 1.
\end{equation}

We will prove this inequality using a backwards induction style argument.
We use the main technical inequality of Lemma~\ref{lemma:technical_inequality} to show that, conditioned on the first $n-1$ iterations, the expectation above is maximized when $\alpha_r = \beta_r = 0$ for all $r \in Q_n$.
In some sense, this is identifying the worst-case values that $\setb{ (\alpha_r , \beta_r) : r \in Q_n}$ may take.
We then continue backwards and show that given the values of $\setb{ (\alpha_r , \beta_r) : r \in Q_t}$ for $t < R$, the values of $\setb{ (\alpha_r,  \beta_r) : r \in \cup_{s=R}^n Q_s}$ which maximize the expectation are $\alpha_r = \beta_r = 0$.

We now proceed more formally. For each $R = 0, 1, \dots n$, we define the quantity
\begin{equation}
g(R) =
\E[\Bigg]{\expf[\Bigg]{ \paren[\Big]{
		\sum_{t=t_p}^n \gsdt \sum_{s = t_p-1}^{\min\setb{R,t-1}} \sum_{r \in Q_s} \alpha_r \beta_r }
		- \frac{1}{2} \paren[\Big]{\sum_{s=t_p-1}^R \sum_{r \in Q_s} \alpha_r^2}
		\cdot \paren[\Big]{\sum_{s=t_p-1}^R \sum_{r \in Q_s} \beta_r^2}
	}}
\end{equation}
Note that $g(R)$ is similar to the expectation we are interested in bounding, except that $\alpha_r = \beta_r = 0$ for all $r \in \cup_{s > R} Q_s$.
Note that $g(n)$ is exactly the expectation that we seek to upper bound by 1.
We prove this upper bound by establishing the following chain of inequalities
\begin{equation}
g(n) \leq g(n-1) \leq \dots \leq g(t_p) \leq  1.
\end{equation}
We prove this chain of inequalities in three steps.
The first step is to establish that $g(n) \leq g(n-1)$.
This inequality is the simplest one to establish because it follows directly from the definition of $g(R)$.
In particular, observe that the term
$
\sum_{t=t_p}^n \gsdt \sum_{s = t_p-1}^{\min\setb{R,t-1}} \sum_{r \in Q_s} \alpha_r \beta_r
$
is the same for $R = n$ and $R = n-1$, while the term
$
\frac{1}{2} \paren[\Big]{\sum_{s=t_p-1}^R \sum_{r \in Q_s} \alpha_r^2}
\cdot \paren[\Big]{\sum_{s=t_p-1}^R \sum_{r \in Q_s} \beta_r^2}
$
is larger for $R = n$ than for $R = n-1$.
Thus, $g(n) \leq g(n-1)$.

We now show the second chunk of inequalities: $g(R) \leq g(R-1)$ for $t_p < R \leq n-1$.
Before continuing, we show how to use the main technical inequality (Lemma~\ref{lemma:technical_inequality}) to prove that for all $R$ in this range,
\begin{align}\label{eq:application_of_technical_inequality}
&\E[\Bigg]{
	\expf[\Bigg]{ \paren[\Big]{
		\sum_{t=R+1}^n \gsdt \sum_{s = t_p-1}^{R} \sum_{r \in Q_s} \alpha_r \beta_r }
		- \frac{1}{2} \paren[\Big]{\sum_{s=t_p-1}^{R} \sum_{r \in Q_s} \alpha_r^2}
		\cdot \paren[\Big]{\sum_{s=t_p-1}^R \sum_{r \in Q_s} \beta_r^2}
	} \given \Delta_R} \\
\quad \quad &\leq
\E[\Bigg]{
	\expf[\Bigg]{ \paren[\Big]{
		\sum_{t=R+1}^n \gsdt \sum_{s = t_p-1}^{R-1} \sum_{r \in Q_s} \alpha_r \beta_r }
		- \frac{1}{2} \paren[\Big]{\sum_{s=t_p-1}^{R-1} \sum_{r \in Q_s} \alpha_r^2}
		\cdot \paren[\Big]{\sum_{s=t_p-1}^{R-1} \sum_{r \in Q_s} \beta_r^2}
	} \given \Delta_R} ,
\end{align}
where $\Delta_R$ denotes the step sizes, $\gsd{t_p}, \gsd{t_p+1}, \dots \gsd{R}$, in addition to the previous randomness in the algorithm denoted by $\Delta_p$.
Under this conditioning, the values of $\setb{ (\alpha_r, \beta_r) : r \in  \cup_{s=t_p-1}^R Q_s}$ are decided and the only random quantity in the expression above is $\sum_{t=R+1}^n \gsdt$.
We claim that this random variable is precisely
\begin{equation}
\sum_{t = R+1}^n \gsdt
=
\left\{
\begin{array}{lr}
1  - \gsve{R+1}{p} &\text{with probability } {(1 + \gsve{R+1}{p})} / {2}\\
-(1  + \gsve{R+1}{p} )&\text{with probability } {(1 - \gsve{R+1}{p})} / {2}
\end{array}
\right.
\end{equation}
To see this, observe that because the step direction satisfies $\gsuve{t}{p} = 1$ in the pivot phase $p$ and the update procedure is $\gsv{t+1} \gets \gsvt + \gsdt \gsuvt$,
\begin{equation}
\gsve{n}{p}
= \sum_{t = R+1}^n \gsdt \gsuve{t}{p} + \gsve{R+1}{p}
= \sum_{t = R+1}^n \gsdt + \gsve{R+1}{p}
\quad \text{and thus} \quad
\sum_{t = R+1}^n \gsdt = \gsve{n}{p}  - \gsve{R+1}{p}.
\end{equation}
Because $\gsve{n}{p}$ takes values $\pm 1$, we have that the sum $\sum_{t = R+1}^n \gsdt$ only takes two values.
Moreover, because all step sizes have mean zero, we have that $\E{\sum_{t = R+1}^n \gsdt} = 0$.
This determines the probabilities of each of the two values.

Because we know exactly the distribution of the random sum $\sum_{t = R+1}^n \gsdt$, we may derive the expectation in the left hand side of \eqref{eq:application_of_technical_inequality} exactly as
\begin{align} \label{eq:exact-form-expectation}
&\frac{1 + \gsve{R+1}{p}}{2} \expf[\Bigg]{\paren{1  - \gsve{R+1}{p}} \sum_{s = t_p-1}^{R} \sum_{r \in Q_s} \alpha_r \beta_r
	- \frac{1}{2} \paren[\Big]{\sum_{s=t_p-1}^{R} \sum_{r \in Q_s} \alpha_r^2}
	\cdot \paren[\Big]{\sum_{s=t_p-1}^R \sum_{r \in Q_s} \beta_r^2}} \\
&+ \frac{1 - \gsve{R+1}{p}}{2} \expf[\Bigg]{-\paren{1  + \gsve{R+1}{p}} \sum_{s = t_p-1}^{R} \sum_{r \in Q_s} \alpha_r \beta_r
	- \frac{1}{2} \paren[\Big]{\sum_{s=t_p-1}^{R} \sum_{r \in Q_s} \alpha_r^2}
	\cdot \paren[\Big]{\sum_{s=t_p-1}^R \sum_{r \in Q_s} \beta_r^2}}
\end{align}
We now demonstrate how this expectation may be recognized as the matrix function appearing in Lemma~\ref{lemma:technical_inequality}.
Let $\mat{A}$ and $\mat{A}_R$ be the 2-by-2 matrices given by
\begin{equation}
\mat{A} = \sum_{s=t_p-1}^{R-1} \sum_{r \in Q_s}
\begin{pmatrix}
 \alpha_r^2 & \alpha_r \beta_r \\
\alpha_r \beta_r & \beta_r^2
\end{pmatrix},
\quad
\mat{A}_R =
\sum_{r \in Q_R}
\begin{pmatrix}
\alpha_r^2 & \alpha_r \beta_r \\
\alpha_r \beta_r & \beta_r^2
\end{pmatrix}.
\end{equation}
These matrices are the sum of $2$-by-$2$ positive semidefinite matrices and so they are themselves positive semidefinite.
Recall that the matrix function in Lemma~\ref{lemma:technical_inequality} is defined for $x \in \bracket{-1,1}$ as
\begin{align}
f_x \begin{pmatrix} \alpha \ & \  \eta \\ \eta \ & \ \beta \end{pmatrix}
&= e^{-\frac{1}{2}\alpha \beta}
\bracket*{ \frac{1+x}{2} \exp\paren{\paren{1-x}\eta } + \frac{1-x}{2} \exp\paren{-\paren{1+x}\eta } } \\
&= \frac{1+x}{2} \expf[\Big]{\paren{1-x}\eta -\frac{1}{2}\alpha \beta} + \frac{1-x}{2} \expf[\Big]{-\paren{1+x}\eta -\frac{1}{2}\alpha \beta }.
\end{align}
Observe that the expectation in \eqref{eq:exact-form-expectation} is equal to $f_{\gsve{R}{p}} \paren{\mat{A} + \mat{A}_R} $.
By Lemma~\ref{lemma:technical_inequality}, the function is operator monotone decreasing over positive semidefinite matrices so that
\begin{equation}
f_{\gsve{R}{p}} \paren{\mat{A} + \mat{A}_R}
\leq
f_{\gsve{R}{p}} \paren{\mat{A}}.
\end{equation}
The proof of inequality \eqref{eq:application_of_technical_inequality} is completed by observing that $f_{\gsve{R}{p}} \paren{\mat{A}}$ is equal to the expectation on the right hand side of \eqref{eq:application_of_technical_inequality}.

Now we are ready to show that $g(R) \leq g(R-1)$ for $t_p < R \leq n-1$.
For notational convenience, we define
\begin{equation}
X_R = \expf[\Bigg]{\sum_{t=t_p}^{R} \gsdt \sum_{s=t_p-1}^{t-1} \alpha_r \beta_r}.
\end{equation}
By rearranging terms, applying iterated expectations, and using the inequality \eqref{eq:application_of_technical_inequality}, we have that
\begin{align}
&g(R) \\
&=
\E[\Bigg]{\expf[\Bigg]{
		\sum_{t=t_p}^n \gsdt \sum_{s=t_p-1}^{\min\setb{R,t-1}} \sum_{r \in Q_s} \alpha_r \beta_r
		- \frac{1}{2} \paren[\Big]{\sum_{s=t_p-1}^R \sum_{r \in Q_s} \alpha_r^2}
		\cdot \paren[\Big]{\sum_{s=t_p-1}^R \sum_{r \in Q_s} \beta_r^2}
}} \\
&= \E[\Bigg]{
	X_R
	\cdot
	\expf[\Bigg]{
		\sum_{t=R+1}^n \gsdt \sum_{s = t_p-1}^{R} \sum_{r \in Q_s} \alpha_r \beta_r
		- \frac{1}{2} \paren[\Big]{\sum_{s=t_p-1}^{R} \sum_{r \in Q_s} \alpha_r^2}
		\cdot \paren[\Big]{\sum_{s=t_p-1}^R \sum_{r \in Q_s} \beta_r^2}
}} \\
&= \E[\Bigg]{
	X_R
	\cdot
	\E[\Bigg]{
	\expf[\Bigg]{
		\sum_{t=R+1}^n \gsdt \sum_{s = t_p-1}^{R} \sum_{r \in Q_s} \alpha_r \beta_r
		- \frac{1}{2} \paren[\Big]{\sum_{s=t_p-1}^{R} \sum_{r \in Q_s} \alpha_r^2}
		\cdot \paren[\Big]{\sum_{s=t_p-1}^R \sum_{r \in Q_s} \beta_r^2}
	} \given \Delta_R}
} \\
&\leq
	\E[\Bigg]{
	X_R
	\cdot
	\E[\Bigg]{
		\expf[\Bigg]{
			\sum_{t=R+1}^n \gsdt \sum_{s = t_p-1}^{R-1} \sum_{r \in Q_s} \alpha_r \beta_r
			- \frac{1}{2} \paren[\Big]{\sum_{s=t_p-1}^{R-1} \sum_{r \in Q_s} \alpha_r^2}
			\cdot \paren[\Big]{\sum_{s=t_p-1}^{R-1} \sum_{r \in Q_s} \beta_r^2}
		} \given \Delta_R}
} \\
&=
\E[\Bigg]{\expf[\Bigg]{
		\sum_{t=t_p}^n \gsdt \sum_{s = t_p-1}^{\min\setb{R-1,t-1}} \sum_{r \in Q_s} \alpha_r \beta_r
		- \frac{1}{2} \paren[\Big]{\sum_{s=t_p-1}^{R-1} \sum_{r \in Q_s} \alpha_r^2}
		\cdot \paren[\Big]{\sum_{s=t_p-1}^{R-1} \sum_{r \in Q_s} \beta_r^2}
}} \\
&= g(R-1)
\end{align}
This establishes the chain of inequalities
\begin{equation}
g(n) \leq g(n-1) \leq \dots \leq g(t_p).
\end{equation}
Establishing that $g(t_p) \leq 1$ may be done via a similar application of the operator monotonicity result of Lemma~\ref{lemma:technical_inequality}.
In particular,
\begin{align}
g(t_p)
&= \E[\Bigg]{\expf[\Bigg]{
		\paren[\Big]{\sum_{t=t_p}^n \gsdt} \iprod{\obv{p} , \gscv{p}} \iprod{\obv{p} , \arbvec}
		- \frac{1}{2} \iprod{\obv{p} , \gscv{p}}^2 \iprod{\obv{p} , \arbvec}^2
	}} \\
&= f_{\gsv{t_p}(p)} \paren[\Bigg]{
	\begin{bmatrix}
	\iprod{\obv{p} , \gscv{p}}^2 & \iprod{\obv{p} , \gscv{p}} \\
	\iprod{\obv{p} , \gscv{p}} & \iprod{\obv{p} , \arbvec}^2
	\end{bmatrix}} \\
&\leq f_{\gsv{t_p}(p)} \paren{\mat{0}} = 1. \qedhere
\end{align}
\end{proof}

We now present the proof of the subgaussian concentration result.

\begin{reftheorem}{\mainref{thm:sub-gaussian}*}
If the input vectors $\gscv{1} \dots \gscv{n}$ all have $\ell_2$ norm at most $1$, then the Gram--Schmidt Walk algorithm returns an assignment vector $\zv$ so that the vector of imbalances $\gsM \zv$ is subgaussian with variance parameter $\sigma^2 = 1$:
\begin{equation}
\E[\Big]{ \expf[\Big]{ \iprod{\gsM \zv, \arbvec} - \iprod{ \E{\gsM \zv}, \arbvec} } } \leq \expf[\big]{ \norm{\arbvec}^2 / 2}
\qquadtext{for all}
\arbvec \in \Reals^{n + \xdim}.
\end{equation}
\end{reftheorem}

\begin{proof}
	Condition on an arbitrary pivot ordering $\pi$ and without loss of generality, suppose that $\pi(i) = i$ for all $i \in [n]$.
We prove the stronger inequality
\begin{equation} \label{eq:stronger_subgaussian_inequality}
\E[\Big]{ \expf[\Big]{ \iprod{\gsM \zv, \arbvec} - \iprod{ \E{\gsM \zv}, \arbvec}  - \frac{1}{2} \sum_{i=1}^n \norm{\phaseproji \gscv{i}}^2 \norm{\phaseproji \arbvec}^2}}
\leq 1
\quad
\text{for all}
\quad
\arbvec \in \Reals^m.
\end{equation}
To see that inequality \eqref{eq:stronger_subgaussian_inequality} is stronger, we use the contractive property of projection matrices and the assumption that all input vectors have $\ell_2$ norm at most $1$ to show
\begin{equation}
\sum_{i=1}^n \norm{\phaseproji \gscv{i}}^2 \norm{\phaseproji \arbvec}^2
\leq \sum_{i=1}^n \norm{\gscv{i}}^2 \norm{\phaseproji \arbvec}^2
\leq \sum_{i=1}^n \norm{\phaseproji \arbvec}^2
= \norm{\PgsM \arbvec}^2
\leq \norm{\arbvec}^2.
\end{equation}
Using this, we have that \eqref{eq:stronger_subgaussian_inequality} implies that
\begin{align*}
& \E[\Big]{ \expf[\Big]{ \iprod{\gsM \zv, \arbvec} - \iprod{ \E{\gsM \zv}, \arbvec} } } \cdot \expf[\big]{ \norm{- \arbvec}^2 / 2} \\
&\leq \E[\Big]{ \expf[\Big]{ \iprod{\gsM \zv, \arbvec} - \iprod{ \E{\gsM \zv}, \arbvec}  - \frac{1}{2} \sum_{i=1}^n \norm{\phaseproji \gscv{i}}^2 \norm{\phaseproji \arbvec}^2}} \\
&\leq 1
	\enspace,
\end{align*}
and rearranging terms yields the desired result.
Thus, it remains for us to prove  \eqref{eq:stronger_subgaussian_inequality}.
At this point, we drop the ``for all $\arbvec \in \Reals^m$'' qualifier and assume that an arbitrary $\arbvec \in \Reals^m$ is given.
We re-write the quantity $\iprod{\gsM \zv, \arbvec} - \iprod{ \E{\gsM \zv}, \arbvec}$ in terms of the fractional updates in the algorithm:
\begin{equation}
\iprod{\gsM \zv, \arbvec}
=  \iprod[\Big]{ \gsM \paren[\Big]{\sum_{t=1}^T \gsdt \gsuvt + \gsv{1}} , \arbvec}
= \sum_{t=1}^T \gsdt \iprod{ \gsM \gsuvt , \arbvec} + \iprod{\gsM \gsv{1}, \arbvec}
= \sum_{i=1}^n \sum_{t \in \pphasei} \gsdt \iprod{ \gsM \gsuvt , \arbvec} + \iprod{\gsM \gsv{1}, \arbvec}.
\end{equation}
Note that by the martingale property of the fractional updates (Lemma~\mainref{lemma:martingale}), $\E{\zv} = \gsv{1}$.
Thus,
\begin{equation}
\iprod{ \E{\gsM \zv}, \arbvec}
= \iprod{ \gsM \E{\zv}, \arbvec}
= \iprod{ \gsM \gsv{1}, \arbvec}
\end{equation}
and so the difference is given by
\begin{equation}
\iprod{\gsM \zv, \arbvec} - \iprod{ \E{\gsM \zv}, \arbvec}
= \sum_{i=1}^n \sum_{t \in \pphasei} \gsdt \iprod{ \gsM \gsuvt , \arbvec}.
\end{equation}
Using this expression for the difference, we may write the desired inequality, which features a sum over units in the exponent, as follows:
\begin{equation}
\E[\Bigg]{ \expf[\Bigg]{ \sum_{i=1}^n \paren[\Big]{ \sum_{t \in \pphasei} \gsdt \iprod{ \gsM \gsuvt , \arbvec}  - \frac{1}{2} \norm{\phaseproji \gscv{i}}^2 \norm{\phaseproji \arbvec}^2}} }
\leq 1.
\end{equation}
A unit $i \in [n]$ which is not chosen as the pivot does not contribute to this sum because the corresponding pivot phase $\pphasei$ is empty and the projection matrix $\phaseproji$ is the zero.
Thus, we may write the sum over units which are chosen as the pivot.
We denote the sequence of pivot units as $p_1, p_2, \dots p_k$ where the subscripts denote the order in which the pivots are chosen by the algorithm.
We seek to show that
\begin{equation}
\E[\Bigg]{ \expf[\Bigg]{ \sum_{j=1}^k \paren[\Big]{ \sum_{t \in \pphase{p_j}} \gsdt \iprod{ \gsM \gsuvt , \arbvec}  - \frac{1}{2} \norm{\phaseproj{p_j} \gscv{p_j}}^2 \norm{\phaseproj{p_j} \arbvec}^2}} }
\leq 1.
\end{equation}
To this end, we define the sequence of random variables $X_1, X_2, \dots X_k$ by
\begin{equation}
X_j = \sum_{t \in \pphase{p_j}} \gsdt \iprod{ \gsM \gsuvt , \arbvec}  - \frac{1}{2} \norm{\phaseproj{p_j} \gscv{p_j}}^2 \norm{\phaseproj{p_j} \arbvec}^2,
\end{equation}
where each $X_j$ corresponds to the $j$th pivot that was chosen by the algorithm.\footnote{
	This highlights that the subgaussian bound will be loose when $\norm{\phaseproj{p_j} \gscv{p_j}}^2 \leq 1$ is a loose inequality.
}
We show that $\E{\expf{\sum_{j=1}^k X_j }} \leq 1$ by proving the chain of inequalities
\begin{equation}
\E[\Big]{\expf[\Big]{\sum_{j=1}^k X_j }}
\leq
\E[\Big]{\expf[\Big]{\sum_{j=1}^{k-1} X_j }}
\leq \dots \leq \E{\expf{X_1 }}
\leq \E{\expf{0}} = 1.
\end{equation}
Consider some $1 \leq \ell \leq k$.
Let $\Delta_\ell$ be all random decisions made by the algorithm up until the beginning of pivot phase $\ell$.
Then observe that
\begin{align}
\E[\Big]{\expf[\Big]{\sum_{j=1}^\ell X_j }}
&= \E[\Big]{\expf[\Big]{\sum_{j=1}^{\ell-1} X_j } \cdot \expf{X_\ell}}
	&\text{(property of exponential)}\\
&= \E[\Big]{\expf[\Big]{\sum_{j=1}^{\ell-1} X_j } \cdot \E{\expf{X_\ell} \given \Delta_\ell} }
	&\text{(iterated expectations)}\\
&\leq \E[\Big]{\expf[\Big]{\sum_{j=1}^{\ell-1} X_j }},
	&\text{(by Lemma~\ref{lemma:sub-gaussian-pivot})}
\end{align}
which completes the induction.

This establishes that conditioned on any pivot ordering $\pi$, the subgaussian bound holds.
The desired result is obtained by marginalizing over all pivot orderings.
\end{proof}

%% file: tex/supp-clt-proof.tex
\section{Asymptotic Analysis and Inference} \label{sec:supp-large-sample-inference}

In this section, we provide proofs of our asymptotic analyses.
The section is organized into the following parts:

\begin{enumerate}
	\item In Section~\ref{sec:supp-asymptotic-framework}, we present the asymptotic framework and the general assumptions we make on the asymptotic sequence.
	\item In Section~\ref{sec:supp-outcome-regularity-implications}, we present implications of the asymptotic regularity conditions which will be used in our analyses.
	\item In Section~\ref{sec:rates}, we establish the rate of convergence of the Horvitz--Thompson estimator under the Gram--Schmidt Walk Design (Theorem~\mainref{thm:consistency}).
	\item In Section~\ref{sec:supp-limvar}, we bound the limiting variance of the Horvitz--Thompson estimator under the Gram--Schmidt Walk Design (Theorem~\mainref{thm:asymp-variance}).
	\item In Section~\ref{sec:supp-clt-proof}, we present a proof of the Central Limit Theorem (Theorem~\mainref{thm:clt}).
	\item In Section~\ref{sec:supp-var-est}, we present analysis of the variance estimator (Theorem~\mainref{thn:var-est-cons}).
	\item In Section~\ref{sec:supp-valid-interval-proof}, we prove asymptotic validity of the confidence intervals (Theorem~\mainref{prop:confidence-intervals}).
	\item In Section~\ref{sec:supp-ridge-estimator}, we present an estimator of the variance upper bound.
\end{enumerate}

The assumptions and results contained in the supplement are sometimes more general than those presented in the main paper.
For this reason, we use an asterisk to denote theorems in the supplement which are more general versions of theorems presented in the main body, e.g. Theorem~\mainref{thm:clt}* is the more general version of Theorem~\mainref{thm:clt} presented in the main body.

\subsection{Asymptotic Assumptions} \label{sec:supp-asymptotic-framework}

Our large sample results are stated in terms of a finite population asymptotic regime, where we consider a sequence of experiments.
As discussed in the main body of the paper, this asymptotic regime is the convention in the design-based literature.
Formally, the asymptotic regime is defined as follows: for each positive integer $n \in \Naturals$, we consider an experiment consisting of $n$ units with potential outcomes $a_i^{(n)}$ and $b_i^{(n)}$ and covariates $\xvi^{(n)}$.
Aside from the regularity assumptions below, we do not assume that the units, their outcomes, nor their covariates are related in any special way in the sequence.
Each unit $i \in [n]$ receives a treatment assignment $\zi^{(n)} \in \setb{\pm 1}$ and the joint distribution of treatment assignments is given by the Gram--Schmidt Walk design with parameter $\vto^{(n)} \in [0,1]$.
This induces a sequence of estimands $\setb{\ate^{(n)}}_{n=1}^\infty$ which are real values, and a sequence of Horvitz--Thompson estimators $\setb{ \htest^{(n)} }_{n=1}^\infty$, which are random variables.
All asymptotic statements, such as limiting theorems, are with respect to this asymptotic sequence.
As is common in the literature, we will often drop the superscript $n$ notation for clarity.

We will now collect the asymptotic assumptions made in the main paper.
We will also generalize the asymptotic assumptions, so that certain trade-offs are made clear.
For the sake of transparency in our proof techniques, we unpack all constants implicit in the asymptotic order notation used in the main paper.

The first assumption is on the design parameter.
Namely, we place an assumption that the design parameter $\vto$ is bounded away from zero by a constant in the asymptotic sequence.
This assumption is made in the statement of theorems, but we explicitly make it here for simplicity.

\begin{assumption}[Design Assumption] \label{supp-assumption:design}
	There exists a constant $c > 0$ such that the design parameter is bounded away from zero by $1 / \vto \leq c$.
\end{assumption}

Next, we place assumptions on the sequence of potential outcomes.
While the main paper deals with $p = 5$, we consider general outcome regularity conditions here.
As $p$ increases, the regularity conditions become more restrictive.
For example, $p = \infty$ bounds each of the outcomes uniformly.
The general assumption below subsumes Assumption~\mainref{ass:outcome-reg} in the main paper, where $p = 5$.

\begin{assumption}[Outcome Regularity] \label{supp-assumption:outcome-regularity}
	There exists a constants $c_1 > 0$ and $p \geq 5$ such that
	\[
	\paren[\Big]{\frac{1}{n} \sumin \abs{\poai}^p}^{1/p} \leq c_1
	\quadand
	\paren[\Big]{\frac{1}{n} \sumin \abs{\pobi}^p}^{1/p} \leq c_1
	\enspace.
	\]
\end{assumption}

We will frequently use the following consequence of Assumption~\ref{supp-assumption:outcome-regularity}, which follows by the Power Mean Inequality, which is also known as the Generalized Mean Inequality: for all $q \leq p$, and in particular for all $q \leq 5$,
\[
	\paren[\Big]{\frac{1}{n} \sumin \abs{\poai}^q}^{1/q} \leq c_1
	\quadand
	\paren[\Big]{\frac{1}{n} \sumin \abs{\pobi}^q}^{1/q} \leq c_1
	\enspace.
	\]

Next, we provide regularity conditions on the covariates.
The following is a restatement of Assumptions~\mainref{ass:covariate-reg} and \mainref{ass:no-outliers} in the main paper, with the constants made explicit for transparency.

\begin{assumption}[Covariate Regularity] \label{supp-assumption:whole-covariate-regularity}
	There exists constants $c_2$, $c_3$, and $c_3'$ such that for all $n$:
	\begin{enumerate}
		\item The maximum row norm of the covariate matrix, $\maxnormX = \max_{i \in [n]} \norm{\xvi}$, is bounded as $\maxnormX^2 \leq c_2 d \log(n)$.
		\item The smallest singular value of the covariate matrix is bounded as $\sigma_{\min}(\xM) \geq c_3 \sqrt{n}$.
		\item The largest singular value of the covariate matrix is bounded as $\sigma_{\max}(\xM) \leq c_3' \sqrt{n}$.
	\end{enumerate}
\end{assumption}

Next, we place assumptions on the dimension of the covariates.
The following assumptions are more general than Assumption~\mainref{ass:cov-dim} which appeared in the main body, because there will be a trade-off between regularity conditions on the covariates and the outcomes.
In particular, Assumption~\mainref{ass:cov-dim} stipulates that $d = \bigO{n^{1/10 - \epsilon}}$ for some $\epsilon > 0$; on the other hand, the following assumption allows for general trade-offs between the regularity placed on the potential outcomes (in terms of $p$) and the conditions placed on the covariates (growth of $d$).

\begin{assumption}[Covariate Dimension] \label{supp-assumption:covariate-dimension}
	The growth of the dimension of the covariates is bounded as
	\[
	d = \littleO[\Big]{\frac{n^{(1/6) \cdot (1 - 2/p)}}{\log(n)^{2}}}
	\enspace.
	\]
\end{assumption}

The more restrictive assumptions we place on the outcomes (i.e. larger $p$), the less restrictive the assumptions become on the covariates (i.e. larger dimension).
For example, if $p = \infty$ so that all outcomes are uniformly bounded then Assumption~\ref{supp-assumption:covariate-dimension} stipulates that $d = \littleO{n^{1/6} / \log(n)^{2}}$.
On the other hand, if $p = 5$ so that only the fifth moment of the outcomes are bounded, then Assumption~\ref{supp-assumption:covariate-dimension} stipulates that $d = \littleO{n^{1/10} / \log(n)^{2}}$.

Finally, we introduce the assumption of non-super-efficiency, which states that the variance of the Horvitz--Thompson estimator under the Gram--Schmidt Walk Design cannot be smaller than the parametric rate.
As described in the main body, this is best viewed as a regularity condition on the outcomes themselves.

\begin{assumption}[Not Super efficient] \label{supp-assumption:not-superefficient}
	There exists a constant $c_5$ such that $n \cdot \Var{\htest} \geq c_5$ for all $n$.
\end{assumption}

Next, we recall the definition of the incoherence of a matrix
  from \citet{candes2009exact}.
We remark that it is equal to the largest leverage score (see \cite{hoaglin1978hat, Drineas2012Fast}).

\begin{definition}[Incoherence]
	Let $\mat{A}$ be an $n$-by-$d$ matrix ($n \geq d$) with singular value decomposition $\mat{A} = \mat{U} \mat{\Sigma} \mat{V}^\tran$.
	For each $i \in [n]$, let $\mat{U}_{(i)}$ denote the $i$th row of the matrix $\mat{U}$.
	The \emph{incoherence} of the matrix $\mat{A}$, denoted $\Xi(\mat{A})$, is defined as
	\[
	\Xi(\mat{A}) = \max_{i \in [n]} \norm{\mat{U}_{(i)}}_2^2
	\enspace.
	\]
\end{definition}

From a different perspective, the incoherence of a matrix is a measure of the similarity between the column span of $\mat{A}$ and the standard basis vectors in that space.
More precisely, let $\mat{H}$ be the orthogonal projection  onto the span of the columns of $\mat{A}$.
Then, the incoherence is $\Xi(\mat{A}) = \max_{k \in [m]} \norm{ \mat{H} \vec{e}_k }^2$, where $\vec{e}_k$ is the $k$th standard basis vector in $\Reals^m$.
From the above definition, it follows that the incoherence of a matrix is at most $1$ and is invariant to scaling the matrix $\mat{A}$.

\begin{corollary} \label{corollary:entire-incoherence}
	Assumption~\ref{supp-assumption:whole-covariate-regularity} implies that the incoherence of the covariate matrix is bounded as
	\[
	\Xi(\xM) \leq \frac{c_2}{c_3^2} \cdot \frac{d}{n} \log(n)
	\enspace.
	\]
\end{corollary}
\begin{proof}
	Let $\mat{H} = \xM (\xM^\tran \xM)^{-1} \xM^\tran$.
	Recall that the incoherence is defined as $\Xi(\xM) = \max_{i \in [n]} \norm{\mat{H} \vec{e}_i}_2^2$.
	Observe that the incoherence may be upper bounded as
	\begin{align*}
		\Xi(\xM)
		&= \max_{i \in [n]} \norm{\mat{H} \vec{e}_i}_2^2 \\
		&\leq \max_{i \in [n]} \vec{e}_i^\tran \xM (\xM \xM)^{-1} \xM^\tran \vec{e}_i \\
		&\leq \max_{i \in [n]} \xvi^\tran (\xM^\tran \xM)^{-1} \xvi \\
		&\leq \max_{i \in [n]}  \norm{ (\xM^\tran \xM)^{-1} } \cdot \norm{\xvi}_2^2 \\
		&\leq \frac{\max_{i \in [n]} \norm{\xvi}_2^2 }{ \sigma_{\min}(\xM)^2 }
		\intertext{
			By Assumption~\ref{supp-assumption:whole-covariate-regularity}, we have that $\max_{i \in [n]} \norm{\xvi}_2^2 \leq c_2 d \log(n)$ and that $\sigma_{\min}(\xM_s) \geq c_3 \sqrt{n}$.
			Putting these together yields the desired result: }
		&\leq \frac{c_2}{c_3^2} \cdot \frac{d}{n} \log(n)
		\enspace.
		\qedhere
	\end{align*}
\end{proof}

\subsection{Implications of Outcome Regularity} \label{sec:supp-outcome-regularity-implications}

In this section, we present several implications of Assumption~\ref{supp-assumption:outcome-regularity}, which place regularity conditions on the potential outcomes.
First, we show that bounds on the $p$th moments of the treatment and control outcomes, $\poav$ and $\pobv$ respectively, translate to bounds on the $p$th moments of the outcome vector $\pomv$.

\begin{lemma} \label{lemma:moment-on-mu}
	Under Assumption~\ref{supp-assumption:outcome-regularity}, we have that
	\[
	\paren[\Big]{ \frac{1}{n} \sumin \abs{\poim}^p }^{1/p} \leq 2 c_1
	\enspace.
	\]
\end{lemma}
\begin{proof}
	This follows from Minkowski's (also known as the triangle) inequality.
\end{proof}

The following lemma, which follows from H{\"o}lder's inequality, relates different vector norms.

\begin{lemma} \label{lemma:p-norm-inequality}
	For all $x_1, \dots x_n \geq 0$, subsets $S \subset [n]$, and $r,p$ such that $r \leq p$, we have that
	\[
	\sum_{i \in S} x_i^r \leq |S|^{1 - r/p} \paren[\Big]{ \sum_{i \in S} x_i^p }^{r/p}
	\enspace.
	\]
\end{lemma}

Finally, we present a helpful lemma which yields the bounds on the outcomes that we use in the asymptotic proofs.

\begin{lemma} \label{lemma:outcome-bounds}
	Let $S \subseteq[n]$ be a subset of units.
	Under Assumption~\ref{supp-assumption:outcome-regularity}, we have that for all lower moments $q \leq p$,
	\[
	\paren[\Big]{ \sum_{i \in S} \abs{ \poim }^q }^{1/q}
	\leq 2 c_1 \cdot \abs{S}^{1/q - 1/p} \cdot n^{1/p}
	\enspace.
	\]
\end{lemma}
\begin{proof}
	Using Lemma~\ref{lemma:p-norm-inequality} together with Lemma~\ref{lemma:moment-on-mu} we have that
	\begin{align*}
		\sum_{i \in S} \abs{ \poim }^q
		&\leq \abs{S}^{1 - q/p} \paren[\Big]{ \sum_{i \in S} \abs{\poim}^{p} }^{q/p} \\
		&= \abs{S}^{1 - q/p} \cdot n^{q/p} \cdot \bracket[\Bigg]{ \paren[\Big]{ \frac{1}{n} \sum_{i \in S}  \abs{\poim}^{p} }^{1/p} }^q \\
		&\leq \abs{S}^{1 - q/p} \cdot n^{q/p} \cdot (2 c_1)^q
		\enspace.
	\end{align*}
	Raising both sides to the $1/q$ yields
	\[
	\paren[\Big]{ \sum_{i \in S} \abs{ \poim }^q }^{1/q}
	\leq (2c_1) \cdot \abs{S}^{1/q - 1/p} \cdot n^{1/p}
	\enspace.
	\qedhere
	\]
\end{proof}

\subsection{Rate of Convergence (Theorem~\mainref{thm:consistency})} \label{sec:rates}

In this section, we establish that the Horvitz--Thompson estimator under the Gram Schmidt Walk design is $\sqrt{n}$-consistent.
We prove Theorem~\mainref{thm:consistency}, restating it in terms of the more general assumptions in the supplementary.

\begin{reftheorem}{\mainref{thm:consistency}*}
	Suppose Assumptions~\ref{supp-assumption:design} and \ref{supp-assumption:outcome-regularity} hold.
	Then, the Horvitz--Thompson estimator under the Gram--Schmidt Walk design is root-$n$ consistent for the average treatment effect: $\htest - \ate = \bigOp{n^{-1/2}}$.
\end{reftheorem}
\begin{proof}
	Using Theorem~\mainref{thm:ht-mse-bound}, we may bound the normalized variance of the Horvitz--Thompson estimator under the Gram--Schmidt Walk design as
	\begin{align*}
		n \cdot \Var{\htest}
		&\leq \min_{\lfx \in \Reals^d} \bracket[\Big]{  \frac{1}{\vto \cdot n} \norm{\pomv - \xM \lfx}^2 + \frac{1}{n(1-\vto)} \norm{\lfx}^2 } \\
		&\leq \frac{1}{\vto} \cdot \frac{1}{n} \sumin \poim^2
		\tag{$\lfx = \zerovec$}
		\\
		&\leq c \cdot 4 c_1^2
		\enspace,
	\end{align*}
	where the last inequality follows by Assumption~\ref{supp-assumption:design} and Lemma~\ref{lemma:moment-on-mu}, which uses Assumption~\ref{supp-assumption:outcome-regularity}.
	The result follows by applying Chebyshev's inequality.
\end{proof}

\subsection{Asymptotic Variance (Theorem~\mainref{thm:asymp-variance})} \label{sec:supp-limvar}

In this section, we establish the limiting variance of the Horvitz--Thompson estimator under the Gram--Schmidt Walk design in two settings.
First, where $\vto$ is chosen to slowly approach one and second where $\vto$ remains a constant.
Recall that
\[
\limvar = n^{-1} \min_{\lfx} \norm{ \pomv - \xM \lfx }^2
\]
is the mean square residuals from a best least squares linear approximation of the potential outcomes using the covariates.

\begin{reftheorem}{\mainref{thm:asymp-variance}*}
	Under Assumptions~\ref{supp-assumption:design}, \ref{supp-assumption:outcome-regularity}, and \ref{supp-assumption:whole-covariate-regularity}, and further supposing that the design parameter approaches one at a sufficiently slow rate, so that $1 - \vto = \littleOmega{\maxnormX^2 / n}$, then
	a tight asymptotic upper bound on the normalized variance of the Horvitz--Thompson estimator under the Gram--Schmidt Walk design is
	\[
	\limsup_{n \to \infty} \bracket[\Big]{ n \Var[\big]{\htest} - \limvar } \leq 0
	\enspace.
	\]
\end{reftheorem}

\begin{proof}
	Let $\lsfx = \argmin_{\lfx} \norm{ \pomv - \xM \lfx }^2 = \paren{\xM^\tran \xM}^{-1} \xM^\tran \pomv$ be the linear function that best approximates the potential outcomes using the covariates.
	The covariate regularity (Assumption~\ref{supp-assumption:whole-covariate-regularity}) ensures that this function exists and is unique.
	Theorem~\mainref{thm:ht-mse-bound} together with unbiasedness of $\htest$ allow us to write
	\begin{equation}
		n \Var[\big]{\htest}
		= n \E[\big]{\paren{\htest - \ate}^2}
		\leq
		\min_{\lfx \in \Reals^d} \bracket[\Bigg]{
			\frac{1}{\vto n} \norm[\big]{ \pomv - \xM\lfx}^2
			+ \frac{\maxnormX^2}{\paren{1 - \vto} n} \norm[\big]{\lfx}^2
		}.
	\end{equation}
	Because the right-hand side is the minimum over $\lfx \in \Reals^d$, which includes $\lsfx \in \Reals^d$, we have
	\begin{equation}
		n \Var[\big]{\htest}
		\leq
		\frac{1}{\vto n} \norm{ \rpomv }^2
		+ \frac{\maxnormX^2}{\paren{1 - \vto} n} \norm[\big]{\lsfx}^2,
	\end{equation}
	where $\rpomv = \pomv - \xM\lsfx$.
	Note that $\limvar = n^{-1} \norm{ \rpomv }^2 = n^{-1} \min_{\lfx} \norm{ \pomv - \xM \lfx }^2$.
	We have $\vto > 0$, so we can write
	\begin{equation}
		\frac{1}{\vto n} \norm{ \rpomv }^2 = \frac{1}{n} \norm{ \rpomv }^2 + \frac{1 - \vto}{\vto n} \norm{ \rpomv }^2.
	\end{equation}
	It follows that
	\begin{equation}
		n \Var[\big]{\htest} - \frac{1}{n} \norm{ \rpomv }^2
		\leq \frac{1 - \vto}{\vto n} \norm{ \rpomv }^2 + \frac{\maxnormX^2}{\paren{1 - \vto} n} \norm[\big]{\lsfx}^2.
	\end{equation}

	Starting with the first term on the right-hand side, note that
	\begin{equation}
		\norm{ \rpomv }^2 = \min_{\lfx} \norm{ \pomv - \xM \lfx }^2 \leq \norm{ \pomv }^2.
	\end{equation}
	Furthermore, when $1 - \vto = \littleO{1}$, then $\vto \geq 1/2$ for sufficiently large $n$.
	Taken together, we have that for sufficiently large $n$,
	\begin{equation}
		\frac{1 - \vto}{\vto n} \norm{ \rpomv }^2 \leq 2 \paren{1 - \vto} n^{-1} \norm{ \pomv }^2 = \littleO{1},
	\end{equation}
	which follows from $1 - \vto = \littleO{1}$ and $n^{-1} \norm{\pomv}^2 = \bigO{1}$, the latter of which follows from Assumption~\ref{supp-assumption:outcome-regularity}.

	Next, consider the second term:
	\begin{equation}
		\frac{\maxnormX^2}{\paren{1 - \vto} n} \norm[\big]{\lsfx}^2.
	\end{equation}
	Write
	\begin{equation}
		\norm[\big]{\lsfx}^2
		= \pomv^\tran \xM \paren{\xM^\tran \xM}^{-2} \xM^\tran \pomv
		\leq \norm{\xM \paren{\xM^\tran \xM}^{-2} \xM^\tran} \times \norm{\pomv}^2,
	\end{equation}
	where the inequality is an operator norm bound.
	The operator norm is given by
	\begin{equation}
		\norm{\xM \paren{\xM^\tran \xM}^{-2} \xM^\tran}
		= \frac{1}{\sigma_{\min}(\xM)^2},
	\end{equation}
	where $\sigma_{\min}(\xM)$ is the smallest singular value of $\xM$.
	We can now write
	\begin{equation}
		\norm[\big]{\lsfx}^2
		\leq \frac{\norm{\pomv}^2}{ \sigma_{\min}(\xM)^2}.
	\end{equation}
	By Assumption~\ref{supp-assumption:whole-covariate-regularity}, $\sigma_{\min}(\xM) = \bigOmega{n^{1/2}}$, and by Assumption~\ref{supp-assumption:outcome-regularity}, $n^{-1} \norm{\pomv}^2 = \bigO{1}$, so it follows that $\norm[\big]{\lsfx}^2 = \bigO{1}$.
	Therefore,
	\begin{equation}
		\frac{\maxnormX^2}{\paren{1 - \vto} n} \norm[\big]{\lsfx}^2
		= \littleO{1},
	\end{equation}
	because $1 - \vto = \littleOmega[\big]{\maxnormX^2 / n}$.
	Both terms are bounded from below by zero, so we have
	\begin{equation}
		\limsup_{n \to \infty} \bracket[\bigg]{ n \Var[\big]{\htest} - n^{-1} \min_{\lfx} \norm{ \pomv - \xM \lfx }^2 }
		\leq \lim_{n \to \infty} \bracket[\bigg]{\frac{1 - \vto}{\vto n} \norm{ \rpomv }^2 + \frac{\maxnormX^2}{\paren{1 - \vto} n} \norm[\big]{\lsfx}^2}
		= 0.
	\end{equation}

	To show that the bound is tight, recall from Lemma~\mainref{lemma:mse-expression} that
	\begin{equation}
		n \E{\paren{\htest - \ate}^2}
		= n \Var{\htest}
		= n^{-1} \pomv^\tran \Cov{\zv} \pomv.
	\end{equation}
	Every diagonal entry of $\Cov{\zv}$ is 1, so
	for a vector $\pomv$ chosen uniformly at random from $\{\pm 1\}^{n}$,
	$\E{\pomv^\tran \Cov{\zv} \pomv} = n$.
	So, there is a vector $\pomv \in \{\pm 1\}^{n}$ for which
	$\pomv^\tran \Cov{\zv} \pomv \geq n$.
	This vector satisfies $n^{-1} \norm{\pomv}^2 = 1$ and can be written as the sum of vectors
  $\poav$ and $\pobv$ that satisfy Assumption~\ref{supp-assumption:whole-covariate-regularity}.
	Then,
	\begin{equation}
		n \Var{\htest}
		= n^{-1} \pomv^\tran \Cov{\zv} \pomv
		\geq n^{-1} \norm{\pomv}^2.
	\end{equation}
	Note that $\norm{\pomv}^2 \geq \min_{\lfx} \norm{ \pomv - \xM \lfx }^2$ for any $\pomv \in \Reals^n$.
	Thus, this sequence of potential outcomes $\pomv \in \Reals^n$ satisfies
	\[
	\liminf_{n \to \infty} \bracket[\bigg]{ n \Var[\big]{\htest} - n^{-1} \min_{\lfx} \norm{ \pomv - \xM \lfx }^2 }
	\geq 0.
	\]
	Therefore, there exists at least one sequence of potential outcomes $\pomv \in \Reals^n$ with asymptotically bounded second moment such that
	\begin{equation}
		\lim_{n \to \infty} \bracket[\bigg]{ n \Var[\big]{\htest} - n^{-1} \min_{\lfx} \norm{ \pomv - \xM \lfx }^2 } = 0 \enspace.
		\tag*{\qedhere}
	\end{equation}
\end{proof}

The following result derives the asymptotic mean squared error for a fixed design parameter $\vto < 1$.

\begin{proposition} \label{coro:asymp-mse-fixed-vto}
	Under
	Assumptions~\ref{supp-assumption:design}, \ref{supp-assumption:outcome-regularity}, and \ref{supp-assumption:whole-covariate-regularity} and further supposing that the design parameter is asymptotically constant, then an asymptotic upper bound on the normalized variance of the Horvitz--Thompson estimator under the Gram--Schmidt Walk design is
	\[
	\limsup_{n \to \infty} \bracket[\Big]{ n \Var{\htest} - \frac{1}{\vto} \limvar  }
	\leq 0
	\enspace.
	\]
\end{proposition}

\begin{proof}
	Let $\lsfx = \argmin_{\lfx} \norm[\big]{ \pomv - \xM\lfx}^2 = \paren{ \xM^\tran \xM }^{-1} \xM^\tran \pomv$.
	Using the mean squared error bound of Theorem~\mainref{thm:ht-mse-bound}, we have that
	\[
	n \Var{\htest}
	= n \E{\paren{\ate - \htest}^2}
	\leq \frac{1}{\vto n} \norm[\big]{ \pomv - \xM\lsfx}^2 + \frac{\maxnormX^2}{\paren{1 - \vto} n} \norm[\big]{\lsfx}^2 \enspace.
	\]
	Using the definition of $\limvar = n^{-1} \norm{\pomv - \xM\lsfx}^2$ and rearranging terms yields
	\begin{equation}
		\bracket[\bigg]{ n \Var{\htest} - \frac{1}{\vto} \limvar }
		\leq \frac{\maxnormX^2}{\paren{1 - \vto} n} \norm[\big]{\lsfx}^2.
	\end{equation}
	Observe that by the operator norm inequality, together with Assumptions~\ref{supp-assumption:outcome-regularity} and \ref{supp-assumption:whole-covariate-regularity} we have that
	\[
	\norm{\lsfx}^2
	\leq \norm{ \paren{\xM^\tran \xM }^{-1} \xM^\tran } \cdot \norm{\pomv}^2
	= \frac{\norm{\pomv}^2}{\sigma_{\min}(\xM)^2}
	\leq \frac{(2c_1)^2 n^{1+2/p}}{c_3^2 n}
	= (2 c_1 / c_3)^2 \cdot n^{2/p}
	\enspace.
	\]
	Note that for $p > 2$, we have that $\norm{\lsfx}^2 = \littleO{n}$.
	The result is obtained by observing that for fixed $\vto > 0$, we have that
	\begin{equation}
		\limsup_{n \to \infty} \bracket[\Big]{ n \Var{\htest} - \frac{1}{\vto} \limvar  }
		\leq
		\limsup_{n \to \infty} \frac{\maxnormX^2}{\paren{1 - \vto} n} \norm[\big]{\lsfx}^2
		= 0.
		\tag*{\qedhere}
	\end{equation}
\end{proof}

\subsection{Central Limit Theorem (Theorem~\mainref{thm:clt})} \label{sec:supp-clt-proof}

In this section, we prove the central limit theorem for the Horvitz--Thompson estimator under the Gram--Schmidt Walk Design, which appears as Theorem~\mainref{thm:clt} in Section~\mainref{sec:clt}.

To prove a central limit theorem for the Horvitz--Thompson estimator under the Gram--Schmidt Walk Design, we take advantage of the martingale property of the construction of the assignment vector.
Formally, we use the following martingale CLT due to \citet{McLeish1974Dependent}.

\begin{apptheorem}[Theorem~2.3 of \citealp{McLeish1974Dependent}] \label{thm:martingale-clt-result}
	For each $n \in \Naturals$, let $\setb{ X_t^{(n)} }_{t=1}^n$ be a martingale difference sequence, i.e. $\E{ X_t^{(n)} \mid X_1^{(n)}, \dots X_{t-1}^{(n)} } = 0$, satisfying the following conditions:
	\begin{itemize}
		\item There exists a constant $C > 0$ such that $\sqrt{ \E[\Big]{ \paren[\big]{\max_{t \leq n} \abs{X_{t}^{(n)} } }^2 } } \leq C$ for all $n \in \Naturals$.
		\item $\max_{t \leq n} \abs{X_{t}^{(n)} }  \xrightarrow{p} 0$.
		\item $\sum_{t=1}^n \paren{X_{t}^{(n)}}^2 \xrightarrow{p} 1$.
	\end{itemize}
	Let $S^{(n)} = \sum_{t=1}^n X_t^{(n)}$.
	Under these conditions, $S^{(n)} \xrightarrow{d} \mathcal{N}(0,1)$.
\end{apptheorem}

Central limit theorems concern the sum of a large number of random variables which are small in some sense.
Typical applications of central limit theorems in the design-based literature view the terms in the sum as corresponding to the experimental units, e.g. the errors of individual effect estimators.
In contrast, we sum terms corresponding to the errors of the effect estimator incurred at iterations of the Gram--Schmidt Walk algorithm.
In particular, we apply Theorem~\ref{thm:martingale-clt-result} to the following martingale sequence:
\begin{equation}\label{eqn:Xt}
X_t = \frac{1}{\sigma} \gsdt \iprod{ \gsuvt , \pomv } \enspace,
\end{equation}
where $\sigma^2 = n^2 \cdot \Var{ \htest }$ is the $n^2$-normalized variance of the estimator, $\pomv$ is the potential outcome vector, and $\gsdt$ and $\gsuvt$ are the step size and direction, respectively.
Observe that
\[
S
= \sum_{t=1}^n X_t
= \frac{1}{\sigma} \sum_{t=1}^n \gsdt \iprod{ \gsuvt , \pomv }
= \frac{1}{\sigma} \iprod[\Big]{ \sum_{t=1}^n \gsdt \gsuvt , \pomv }
= \frac{\iprod{\zv , \pomv} / n}{\sqrt{\Var{\htest}}}
= \frac{\htest - \ate}{\sqrt{\Var{\htest}}} \enspace.
\]
Thus, a central limit theorem for $S$ implies a central limit theorem for the Horvtiz--Thompson estimator under the Gram--Schmidt Walk design.
The remainder of the section is devoted to showing that, under the asymptotic assumptions stated in Section~\ref{sec:supp-asymptotic-framework}, the conditions of Theorem~\ref{thm:martingale-clt-result} are satisfied.

\subsubsection{Covariate Regularity Under Randomized Pivots} \label{sec:supp-covariate-regularity-implications}

In this section, we set up the notation to begin reasoning about the regularity of the covariates under the randomized pivot rule.
We emphasize that it is in this part of the analysis where the randomized pivot rule plays a crucial role.
At a high level, the rest of the analysis proceeds as follows:
\begin{itemize}
	\item We define two events $\event_1$ and $\event_2$, which yield the combined event $\event = \event_1 \cap \event_2$.
	\item We will show that the combined event $\event$ occurs with high probability as $n$ increases.
	\item We will show that the combined event $\event$ implies a certain regularity in the Gram--Schmidt Walk updates, which will allow us to reason about the CLT conditions.
\end{itemize}

Recall that the randomized pivot ordering is a uniformly chosen permutation $\pi : [n] \to [n]$, which is chosen at the beginning of the algorithm before any fractional updates are made.
This ordering is used to choose the next pivot when a new one is needed, provided that the corresponding coordinate is not frozen.
While $\pi$ will not in general be the order of the pivots, we prove that under our assumptions it will almost definitely be the order of all but the last few pivots:
we will probably have $p_t = \pi(n-t+1)$ for iterations $t$ between $1$ and $n - \afn$, where
  $\afn$ is a slowly growing function of $n$ to be defined shortly.

For integers $s$ in the range $1 \leq s \leq n-1$, we define the following random quantities which depend (only) on the randomized pivot ordering.
Define the matrix $\xM_s$ to be the $(n-s) \times d$ row-submatrix of $\xM$ consisting of rows $\xv{\pi(1)} \dots \xv{\pi(n-s)}$, i.e. the first $n-s$ covariate vectors in the pivot ordering.
Let $P_s$ be the (random) $n \times n$ matrix that projects onto the coordinates $\pi(1) \dots \pi(n-s)$.
More concretely, $P_s$ is the $n \times n$ diagonal matrix where the diagonal entry $P_s(i,i) = 1$ if and only if $\pi(i) \leq n-s$ and $P_s(i,i) = 0$ otherwise.
Define the vector $\vec{w}_s \in \Reals^n$ to be the vector whose first $n-s$ entries in the pivot ordering are given as
\[
\vec{w}_s( \pi(1) : \pi(n-s) )
= - (1-\vto) \paren[\Big]{ \vto \unitM + (1-\vto) \maxnormX^{-2} \xM_s \xM_s^\tran }^{-1}
\paren{ \maxnormX^{-1} \xM_s }
\paren{\maxnormX^{-1} \xv{n-s+1}}
\enspace,
\]
and the remaining entries are given as $\vec{w}_s(\pi(n-s+1)) = 1$ and $\vec{w}_s(\pi(i)) = 0$ for all $i > n -s +1$.
We emphasize here that the row submatrices $\xM_s$, the projection matrices $P_s$, and the vectors $\vec{w}_s$ are determined at the beginning of the algorithm once the randomized pivot order is determined, and they do not depend on the fractional updates made by the algorithm.
Thus, conditioning on the pivot ordering $\pi$ completely determines these quantities.

We now define the two events which are central to our proof of the Central Limit Theorem.
In what follows, we define the quantity $\afn = n^{(1/2) \frac{1 - 2/p}{1 - 1/p}} / \log(n)$ which grows with $n$ in a way that depends on the moment assumptions on the outcomes.
We also define a new constant $c_6 = \sqrt{ 2 \sqrt{2} \cdot c_2^2 / {c_3^3}}$, which will be useful in our proofs.

\begin{align*}
	\event_1 &= \setb[\Bigg]{ \sigma_{\min}(\xM_s) \geq \frac{c_3}{\sqrt{2}} \sqrt{n-s} \text{ for all } s \leq n - \afn  }  \\
	\event_2 &= \setb[\Bigg]{
		\abs[\Big]{ \sum_{s \leq t}  \gsd{s} \vec{w}_s(\pi(i)) }
		\leq  \frac{\gamma \cdot c_6 \cdot d \log (n)^{3/2} }{\sqrt{n-t}}
		\text{ for all $t \leq n - \afn$ and $i \leq n - t$} } \enspace,
\end{align*}
where we set $\gamma = 4$.
For clarity in our proofs, we use $\gamma$ throughout the analysis, rather than writing an explicit constant.
Throughout this section, we use the notation $\notevent$ to denote the complement of an event $\event$.

The first event $\event_1$ depends only on the randomized pivot ordering.
In particular, it states that all of the row submatrices $\xM_1, \dots \xM_s$ for $s \leq n - \afn$ have large singular values.
Recall that these are the matrices of size at least $\afn$ in the prefix determined by the randomized pivot ordering.
The second event depends on both the randomized pivot ordering (as it involves $\vec{w}_s$) as well as the fractional updates in the Gram--Schmidt Walk algorithm (as it involves the step sizes $\delta_1 \dots \delta_t$ for $t \leq n - \afn$).
Informally speaking, this event ensures that for units which appear early in the randomized pivot ordering, the signed sum of the corresponding coordinates of $\vec{w}_s$ does not grow large in any of the early iterations.
In this context, early iterations means $t \leq n - \afn$ and units which are early in the randomized pivot order means those units assigned to the first $n-t$ positions, i.e. $\pi(1) \dots \pi(n-t)$.
We also consider the combined event $\event = \event_1 \cap \event_2$.

We now present two propositions which motivate the definition of the first event $\event_1$.
Taken together, they show that under $\event_1$, the incoherence of the row submatrices $\xM_s$ for $1 \leq s \leq n - \afn$ is bounded and thus all coordinates of the vectors $P_s \vec{w}_s$ are bounded.

\begin{appproposition} \label{prop:small-incoherence}
	Under Assumptions~\ref{supp-assumption:whole-covariate-regularity} and conditioned on event $\event_1$, the following holds with probability 1:
	\[
	\Xi(\xM_s) \leq \frac{2 c_2}{c_3^2}  \cdot \frac{d}{n-s} \log(n)
	\quad \text{for all } 1 \leq s \leq n - \afn
	\enspace.
	\]
\end{appproposition}
\begin{proof}
	Let $\mat{H}_s = \xM_s (\xM_s^\tran \xM_s)^{-1} \xM_s^\tran$.
	Recall that the incoherence is defined as $\Xi(\xM_s) = \max_{i \in [n]} \norm{\mat{H}_s \vec{e}_i}_2^2$.
	As in the proof of Corollary~\ref{corollary:entire-incoherence},
	the incoherence may be upper bounded as
\begin{equation*}
		\Xi(\xM_s)
		 = \max_{i \in [n]} \norm{\mat{H}_s \vec{e}_i}_2^2 \\
		\leq \frac{\max_{i \in [n]} \norm{\xvi}_2^2 }{ \sigma_{\min}(\xM_s)^2 }
\end{equation*}
Conditioned on event $\event_1$, $\sigma_{\min}(\xM_s) \geq \frac{c_3}{\sqrt{2}} \sqrt{n-s}$. By Assumption~\ref{supp-assumption:whole-covariate-regularity}, we have that $\max_{i \in [n]} \norm{\xvi}_2^2 \leq c_2 d \log(n)$.
Putting these together yields the desired result:
\begin{equation*}
	\Xi(\xM_s)
		\leq \frac{2 c_2}{c_3^2} \cdot \frac{d}{n-s} \log(n)
		\enspace.
		\qedhere
\end{equation*}
\end{proof}

\begin{appproposition} \label{prop:wt-small-norm}
	Under Assumption~\ref{supp-assumption:whole-covariate-regularity} and conditioned on event $\event_1$, the following holds with probability 1:
	\[
	\norm{P_s \vec{w}_s}_{\infty} \leq c_6 \cdot \frac{ d \log(n) }{n-s}
	\quad \text{for all } 1 \leq s \leq n - \afn
	\enspace.
	\]
\end{appproposition}
\begin{proof}
	Observe that by construction,
	\[
	\norm{ P_s \vec{w}_s }_\infty
	= \norm{ \vec{w}_s( \pi(1) : \pi(n-s) ) }_\infty
	\enspace,
	\]
	where $\vec{w}_s( \pi(1) : \pi(n-s) )$ is the $n-s$ dimensional vector obtained via the first $n-s$ coordinates according to the pivot ordering.
	Using the definition of $\vec{w}_s$, we have that
	\begin{align*}
		\vec{w}_s( \pi(1) : \pi(n-s) )
		&= - (1-\vto) \paren[\Big]{ \vto \unitM + (1-\vto) \maxnormX^{-2} \xM_s \xM_s^\tran }^{-1}
		\paren{ \maxnormX^{-1} \xM_s }
		\paren{\maxnormX^{-1} \xv{n-s+1}} \\
		&= - \paren[\Big]{ \frac{\vto}{1-\vto} \maxnormX^2 \unitM + \xM_s \xM_s^\tran }^{-1} \xM_s \xv{n-s+1} \\
		&= - \paren[\Big]{ \beta \unitM + \xM_s \xM_s^\tran }^{-1} \xM_s \xv{n-s+1}
		\enspace,
		\intertext{where $\beta = \frac{\vto}{1-\vto} \maxnormX^2$ is used for notational convenience.
			Let the Singular Value Decomposition of $\xM_s$ be given as $\xM_s = \mat{U} \mat{\Sigma} \mat{V}^\tran$.
			Substituting the singular value decomposition into the above, we obtain:}
		&= - \mat{U} \paren[\Big]{ \beta \unitM + \mat{\Sigma}^2 }^{-1} \mat{\Sigma} \mat{V}^\tran \xv{n-s+1} \\
		&= - \mat{U} \vec{y} \enspace,
	\end{align*}
	where the vector $\vec{y}$ is defined as $\vec{y} = \paren[\Big]{ \beta \unitM + \mat{\Sigma}^2 }^{-1} \mat{\Sigma} \mat{V}^\tran \xv{n-s+1}$.
	Using the operator norm, we can bound the $\ell_2$ norm of $\vec{y}$ as
	\[
	\norm{\vec{y}}_2
	= \norm[\Big]{\paren[\Big]{ \beta \unitM + \mat{\Sigma}^2 }^{-1} \mat{\Sigma} \mat{V}^\tran \xv{n-s+1}}
	\leq \norm[\Big]{\paren[\Big]{\beta \unitM + \mat{\Sigma}^2 }^{-1} \mat{\Sigma} } \norm{\mat{V}^\tran \xv{n-s+1}}_2
	\enspace.
	\]
	The singular values of the matrix $\paren{\beta \unitM + \mat{\Sigma}^2 }^{-1} \mat{\Sigma}$ are of the form
	\[
	\frac{\sigma}{\beta + \sigma^2} \enspace,
	\]
	where $\sigma$ is a singular value of $\xM_s$.
	Because $\beta$ is positive, we have that $\frac{\sigma}{\beta + \sigma^2} \leq \frac{1}{\sigma}$.
	Thus, the operator norm of $\paren{\beta \unitM + \mat{\Sigma}^2 }^{-1} \mat{\Sigma}$ is at most $1 / \sigma_{\min}$, where $\sigma_{\min} \triangleq \sigma_{\min}(\xM_s)$.
	Note that $\mat{V}$ is an orthogonal matrix, so that $\norm{\mat{V}^\tran \xv{n-s+1}}_2 = \norm{\xv{n-s+1}}_2 \leq \maxnormX$.
	Thus, we have that $\norm{\vec{y}}_2 \leq \maxnormX / \sigma_{\min}$.

	Recall that $\vec{w}_s( \pi(1) : \pi(n-s) ) = - \mat{U} \vec{y}$.
	In particular, the $r$th entry of $P_s \vec{w}_s$ is the inner product of the $r$th row of $\mat{U}$ and the vector $\vec{y}$.
	Using Cauchy--Schwarz together with the definition of incoherence, we have that
	\begin{align*}
		\norm{ \vec{w}_s( \pi(1) : \pi(n-s) ) }_\infty
		&= \max_{k \in [n-s]} \abs[\Big]{ \langle \mat{U}_{(k)} , \vec{y} \rangle } \\
		&\leq \max_{k \in [n-s]} \norm{ \mat{U}^T \vec{e}_{k} }_2 \cdot \norm{ \vec{y} }_2 \\
		&\leq \sqrt{ \Xi( \xM_s  ) } \cdot \frac{\maxnormX}{\sigma_{\min}}
		\intertext{Now, we use the fact we are conditioning on $\event_1$.
			Given $\event_1$, we have that $\sigma_{\min}(\xM_s) \geq \frac{c_3}{\sqrt{2}} \sqrt{n-s}$ and using Proposition~\ref{prop:small-incoherence}, we have that $\Xi( \xM_s ) \leq \frac{2 c_2}{c_3^2}  \cdot \frac{d}{n-s} \log(n)$.
			Taking these together yields the bound,}
		&\leq \sqrt{ \frac{2 c_2}{c_3^2}  \cdot \frac{d}{n-s} \log(n) } \cdot \sqrt{\frac{ \sqrt{2} c_2 d \log(n)}{c_3 (n-s)}} \\
		&= \sqrt{\frac{2 \sqrt{2} c_2^2}{c_3^3}} \cdot \frac{d \log(n)}{n-s} \\
		&\triangleq c_6 \cdot \frac{d \log(n)}{n-s}
		\qedhere
	\end{align*}
\end{proof}

Before continuing, we present the following simple proposition which keeps track of the growth of $d$ relative to the term $\afn$ guaranteed by the asymptotic Assumption~\ref{supp-assumption:covariate-dimension}.

\begin{proposition}\label{prop:helpful-asymptotics}
	Recall that $\afn = n^{(1/2) \frac{1 - 2/p}{1 - 1/p}} / \log(n)$.
	Assumption~\ref{supp-assumption:covariate-dimension} states that $d = \littleO{ n^{(1/6)(1 - 2/p)} \log(n)^{-2} }$, which implies the following:
	\begin{enumerate}
		\item [a.] $d = \littleO{ \afn / \log(n) }$
		\item [b.] $d = \littleO[\Big]{ \frac{\afn^{1/2}}{\log(n)^{3/2}}}$
		\item [c.] $d = \littleO[\Big]{ \frac{n^{1/2 - 1/p} \cdot \afn^{1/p}}{\log(n)} }$
	\end{enumerate}
\end{proposition}
\begin{proof}
	To prove relation $a$, observe that for $p \geq 5$,
	\[
	\frac{n^{(1/6)(1 - 2/p)} \log(n)^{-2}}{n^{(1/2) \frac{1 - 2/p}{1 - 1/p}} \log(n)^{-2}}
	= n^{(1-2/p)(1/6 - \frac{1}{2-2/p})  }
	\leq n^{-11/40 }
	\to 0
	\enspace.
	\]
	To prove relation $b$, observe that for $p \geq 5$,
	\[
	\frac{n^{(1/6)(1 - 2/p)} \log(n)^{-2}}{n^{(1/4) \frac{1 - 2/p}{1 - 1/p}} \log(n)^{-1/2} \log(n)^{-3/2}}
	=  n^{(1-2/p)(1/6 - \frac{1}{4-4/p})  } \cdot \frac{1}{\log(n)^{1/2}}
	\leq n^{-7/80  } \cdot \frac{1}{\log(n)^{1/2}}
	\to 0
	\enspace.
	\]
	To establish the relation $c$, we show the equivalent statement that $d^2 = \littleO[\Big]{ \frac{n^{1 - 2/p} \cdot \afn^{2/p}}{\log(n)^2} }$.
	To this end, observe that
	\[
	\frac{n^{(1/3)(1 - 2/p)} \log(n)^{-4}}{n^{1 - 2/p} \cdot \afn^{2/p} \log(n)^{-2}}
	= \paren[\Big]{\frac{1}{n^{2/3}}}^{1 - 2/p} \cdot \paren[\Big]{ \frac{1}{\afn} }^{2/p} \cdot \frac{1}{\log(n)^{2}}
	\to 0
	\enspace.
	\qedhere
	\]
\end{proof}

\subsubsection{The Probability of \texorpdfstring{$\event$}{E}}

In this section, we will show that both events $\event_1$ and $\event_2$ have high probability under the random pivot order and the randomized fractional updates in the Gram--Schmidt Walk algorithm.
Using a union bound, we will show then that the combined event $\event = \event_1 \cap \event_2$ has high probability.

First, we show that $\event_1$ happens with high probability.
To this end, we appeal to the matrix Chernoff bound from \citet{TroppUser2012}, which we restate here for ease of reference.

\begin{appproposition}[Corollary 5.2 of \citet{TroppUser2012}]\label{prop:tropp}
		Let $\mat{A}_1 \dots \mat{A}_n$ be independent, random, self-adjoint $d$-dimensional matrices that satisfy $\mat{A}_i \succeq 0$ and $\eigvalmax(\mat{A}_i) \leq R$ almost surely.
		Then, for $0 \leq \delta \leq 1$,
		\[
		\Pr[\Bigg]{ \eigvalmin \paren[\Big]{ \sum_{i=1}^n \mat{A}_i } < \delta \cdot \mu } \leq d \paren[\Big]{\frac{e^{-\delta}}{(1-\delta)^{(1 - \delta)}}}^{\mu / R}
		\enspace,
		\]
		where $\mu = \eigvalmin(\sum_{i=1}^n \E{ \mat{A}_i})$.
\end{appproposition}

\begin{proposition}\label{prop:event1-no-assumptions}
	For any $1 \leq k \leq n-1$,
	\[
	\Pr[\Bigg]{ \exists \ s \leq n-k : \sigma_{\min}(\xM_s) < \sqrt{\frac{n-s}{2n}} \sigma_{\min}(\xM)}
	\leq  d n^2 \expf[\Bigg]{ - \frac{1}{10} \frac{ (n-k) \sigma_{\min}( \xM)^2 }{n \maxnormX^2} }
	\enspace.
	\]
\end{proposition}
\begin{proof}
	Recall that for each $1 \leq s \leq n-1$, the random matrix $\xM_s$ is distributed uniformly over row submatrices of $\xM$ with exactly $n-s$ rows.
	We will bound the ``failure'' probability that the minimum singular value of a single submatrix $\xM_s$ falls below $\sqrt{\frac{n-s}{2n}} \sigma_{\min}(\xM)$ and then apply a union bound over all $s \leq n-k$.

	Fix a value of $s \leq n-k$.
	In order to bound the tail of the smallest singular value of $\xM_s$, we study a different random matrix whose smallest singular value is more amenable to the matrix Chernoff bound.
	At a high level, the idea is that sampling a subset of $n-s$ rows uniformly at random is the same as sampling rows independently with probability $(n-s)/n$ and then conditioning on the event that exactly $n-s$ rows were chosen.
	Let $a_1 \dots a_n$ be independent Bernoulli random variables which take $1$ with probability $p \triangleq (n-s)/n$ and $0$ otherwise.
	Define the random matrix $\widetilde{\xM}_s$ to be the row submatrix of $\xM$ which contains row $\xvi$ if $a_i = 1$.
	Observe that $\mat{M}_s = \widetilde{\xM}_s^\tran \widetilde{\xM}_s$ is a $d$-by-$d$ random matrix given by
	\[
	\mat{M}_s = \sum_{i=1}^n a_i  \xvi \xvi^\tran  \enspace.
	\]
	The two matrices are related as $\lambda_{\min}(\mat{M}_s) = \sigma_{\min}(\widetilde{\xM}_s)^2$.
	We seek to apply Tropp's matrix Chernoff bound, Proposition~\ref{prop:tropp}, to the random matrix $\mat{M}_s$.
	Observe that by setting $\mat{A}_i = a_i \xvi \xvi^\tran$ we have that $\mat{M}_s = \sum_{i=1}^n \mat{A}_i$.
	The random matrix $\mat{A}_i$ is always positive semidefinite and satisfies
	\[
	\eigvalmax( \mat{A}_i )
	\leq \norm{\xvi}^2
	\leq \maxnormX^2
	\enspace.
	\]
	So that the random matrices $\mat{A}_i$ satisfy Tropp's condition with $R = \maxnormX^2$.
	Further observe that $\E{a_i} = p$ so that $\E{ \mat{A}_i } = p \cdot \xvi \xvi^\tran$ and thus
	\[
	\mu
	= \lambda_{\min} \paren[\Big]{ \sum_{i=1}^n \E{\mat{A}_i} }
	= \lambda_{\min} \paren[\Big]{ \sum_{i=1}^n p \cdot \xvi \xvi^\tran } \\
	= p \cdot \sigma_{\min}( \xM )^2 \\
	= \frac{n-s}{n} \sigma_{\min}( \xM )^2
	\enspace.
	\]
	By setting $\delta = 1/2$ and applying Proposition~\ref{prop:tropp}, we obtain
	\begin{align*}
	\Pr[\Bigg]{ \sigma_{\min}( \widetilde{\mat{X}}_s  ) < \sqrt{\frac{1}{2} \frac{n-s}{n} } \sigma_{\min}( \xM )   }
	&= \Pr[\Bigg]{ \lambda_{\min}( \mat{M}_s ) < \frac{1}{2} \frac{n-s}{n} \sigma_{\min}( \xM )^2 } \\
	&\leq d \cdot \paren[\Big]{ \sqrt{\frac{2}{e}} }^{ \frac{n-s}{n} \cdot \frac{\sigma_{\min}(\xM)^2}{\maxnormX^2}} \\
	&= d \expf[\Bigg]{ \log\paren[\Big]{ \sqrt{\frac{2}{e}} } \cdot \frac{n-s}{n} \cdot \frac{\sigma_{\min}(\xM)^2}{\maxnormX^2} } \\
	&\leq d \expf[\Bigg]{ - \frac{1}{10} \frac{n-s}{n} \cdot \frac{\sigma_{\min}(\xM)^2}{\maxnormX^2} }
	\end{align*}

	Next, we relate tails on the singular values of $\widetilde{\xM}_s$ and $\xM_s$.
	Recall that $\xM_s$ has the same distribution as $\widetilde{\xM}_s$ conditioned on $\sum_{i=1}^n a_i = n-s$.
	Observe that $\Pr{ \sum_{i=1}^n a_i = n-s } \geq 1/n$ because $n-s$ is the mode of the random variable $\sum_{i=1}^n a_i$ which takes at most $n$ values.
	This fact together with Bayes' Rule gives
	\begin{align*}
		& \Pr[\Bigg]{ \sigma_{\min}( \xM_s  ) < \sqrt{ \frac{n-s}{2n} } \sigma_{\min}( \xM )   } \\
		&= \Pr[\Bigg]{ \sigma_{\min}( \widetilde{\mat{X}}_s  ) < \sqrt{\frac{n-s}{2n} } \sigma_{\min}( \xM ) \mid \sum_{i=1}^n a_i = n-s   } \\
		&= \frac{ \Pr[\Bigg]{ \sigma_{\min}( \widetilde{\mat{X}}_s  ) < \sqrt{\frac{n-s}{2n} } \sigma_{\min}( \xM ) \text{ and } \sum_{i=1}^n a_i = n-s   } }{\Pr{ \sum_{i=1}^n a_i = n-s }} \\
		&\leq \frac{\Pr[\Bigg]{ \sigma_{\min}( \widetilde{\mat{X}}_s  ) < \sqrt{ \frac{n-s}{2n} } \sigma_{\min}( \xM ) }}{\Pr{ \sum_{i=1}^n a_i = n-s }} \\
		&\leq n \cdot \Pr[\Bigg]{ \sigma_{\min}( \widetilde{\mat{X}}_s  ) < \sqrt{ \frac{n-s}{2n} } \sigma_{\min}( \xM ) } \\
		&\leq n d \expf[\Bigg]{ - \frac{1}{10} \frac{n-s}{n} \cdot \frac{\sigma_{\min}(\xM)^2}{\maxnormX^2} }
		\enspace.
	\end{align*}
	The desired result follows by taking a union bound over all $s \leq n-k$, which contributes at most a factor $n$.
\end{proof}

\begin{appproposition} \label{prop:first-event-high-prob}
	Under Assumptions~\ref{supp-assumption:whole-covariate-regularity} and \ref{supp-assumption:covariate-dimension} the event $\event_1$ occurs with exponentially high probability:
	\[
	\Pr{\notevent_1} \leq \expf[\Big]{ - \bigOmega{ n^{1/4} } }
	\enspace.
	\]
\end{appproposition}
\begin{proof}
	Recall that the first event $\event_1$ is defined as
	\[
	\event_1 = \setb[\Bigg]{ \sigma_{\min}(\xM_s) \geq \frac{c_3}{\sqrt{2}} \sqrt{n-s} \text{ for all } s \leq n - \afn  }
	\enspace.
	\]
	We upper bound the probability that $\event_1$ fails by using Assumption~\ref{supp-assumption:whole-covariate-regularity} and then applying Proposition~\ref{prop:event1-no-assumptions} with $k = n - \afn$ to obtain that
	\begin{align*}
	\Pr{\notevent_1}
	&= \Pr[\Bigg]{ \exists \ s \leq n-k : \sigma_{\min}(\xM_s) < \frac{c_3}{\sqrt{2}} \sqrt{n-s} } \\
	&\leq \Pr[\Bigg]{ \exists \ s \leq n-k : \sigma_{\min}(\xM_s) < \sqrt{\frac{n-s}{2n}} \sigma_{\min}(\xM)} \\
	&\leq d n^2 \expf[\Bigg]{ - \frac{1}{10} \frac{ \afn \sigma_{\min}( \xM)^2 }{n \maxnormX^2} } \\
	&\leq d n^2 \expf[\Big]{ - \frac{c_3^2}{10 \cdot c_2} \frac{\afn}{d \log(n)} }
	\enspace,
	\end{align*}
	where the final inequality used Assumption~\ref{supp-assumption:whole-covariate-regularity} to bound the smallest singular value $\sigma_{\min}( \xM)$ and the largest norm of any covariate $\maxnormX$.
	Using Assumption~\ref{supp-assumption:covariate-dimension}, we can bound the non-constant term in the exponential as
	\[
	\frac{\afn}{d \log(n)}
	\geq \frac{n^{ (1/2) \cdot (\frac{1 - 2/p}{1 - 1/p}) }}{\littleO{ n^{(1/6) \cdot (1-2/p) } \log(n)^{-2} } \log(n)^2}
	\geq \bigOmega{ n^{1/4}}
	\enspace.
	\]
	The polynomial factor $dn^2 \leq \bigO{n^{1/6} \cdot n^2}$ can be pulled inside the $\bigOmega{n^{1/4}}$ factor in the exponential.
\end{proof}

Next, we show that conditioned on $\event_1$, the event $\event_2$ happens with high probability.
The proof is based on using the Azuma-Hoeffding bound together with Proposition~\ref{prop:wt-small-norm}.

\begin{appproposition} \label{prop:second-event-high-prob}
	For $n$ large enough that $\afn \geq 2$,
	under Assumption~\ref{supp-assumption:whole-covariate-regularity} and conditioned on $\event_1$, the event $\event_2$ holds with high probability:
	\[
	\Pr{ \event_2 \mid \event_1 } \geq 1 - 1 / n^{2}
	\enspace.
	\]
\end{appproposition}
\begin{proof}
	Fix any pivot order $\pi$ that is in the event $\event_1$.
	We proceed by bounding the probability of $\notevent_2$ conditioned on this pivot order.
	Recall that the step sizes $\gsd{1}, \dots \gsd{n}$ form a bounded martingale sequence.
	In particular, $\E{ \gsdt \mid \gsd{1} \dots \gsd{t-1} } = 0$ and $\abs{ \gsdt } \leq 2$ with probability 1.
	Thus, we may bound the tails of sums of the form $\abs[\Big]{ \sum_{s \leq t}  \gsd{s} c_s }$ with the Azuma-Hoeffding inequality.
	Let $m = n - \afn$.
	Using the union bound together with the Azuma--Hoeffding inequality, we get that
	\begin{align*}
		\Pr{\notevent_2 \mid \pi}
		&= \Pr[\Bigg]{ \exists \ t \leq m \text{ and } i < t \text{ such that } \abs[\Big]{ \sum_{s \leq t}  \gsd{s} \vec{w}_s(\pi(i)) }  > \frac{\gamma \cdot c_6 \cdot d \log (n)^{3/2}}{\sqrt{n-t}} } \\
		&\leq \sum_{t=1}^m \sum_{i=1}^{t-1} \Pr[\Bigg]{ \abs[\Big]{ \sum_{s \leq t}  \gsd{s} \vec{w}_s(\pi(i)) }  > \frac{\gamma \cdot c_6 \cdot d \log (n)^{3/2} }{\sqrt{n-t}}  } \\
		&\leq \sum_{t=1}^m \sum_{i=1}^{t-1} 2 \expf[\Bigg]{ - \frac{2 \paren[\Big]{ \frac{\gamma \cdot c_6 \cdot d \log (n)^{3/2}}{\sqrt{n-t}}}^2 }{ 4 \sum_{s=1}^t \vec{w}_s(\pi(i))^2 } } \\
		&\leq \sum_{t=1}^m \sum_{i=1}^{t-1} 2 \expf[\Bigg]{ -  \frac{\gamma^2 \cdot c_6^2 d^2 \log (n)^3}{2(n-t)} \cdot \frac{1}{\sum_{s=1}^t \norm{P_s \vec{w}_s}_\infty^2 } }
		\enspace.
	\end{align*}
	Under Assumption~\ref{supp-assumption:whole-covariate-regularity} and conditioned on $\pi \in \event_1$, Proposition~\ref{prop:wt-small-norm} tells us that $\norm{P_s \vec{w}_s}_\infty^2 \leq c_6^2 \cdot \frac{\paren{d \log(n)}^2}{(n-s)^2}$.
	This implies that
	\[
	\sum_{s=1}^t \norm{P_s \vec{w}_s}_\infty^2
	\leq  \sum_{s=1}^{t} c_6^2 \cdot \frac{\paren{d \log(n)}^2}{(n - s)^2} \\
	\leq  c_6^2 \cdot \frac{2 \paren{d \log(n)}^2}{n-t}
	\]
	where the last inequality follows from the fact that $t \leq n - \afn \leq n - 2$ and thus
	\[
	\sum_{s=1}^t \frac{1}{(n-s)^2}
	\leq \int_{x=1}^{t+1} \frac{1}{(n-x)^2}  dx
	= \frac{1}{n-(t+1)} - \frac{1}{n-1}
	\leq \frac{2}{n-t}
	\enspace.
	\]
	Plugging this into the tail bound above, we have that
	\begin{align*}
		\Pr{\notevent_2 \mid \pi}
		&\leq \sum_{t=1}^m \sum_{i=1}^{t-1} 2 \expf[\Bigg]{ - \frac{\gamma^2 \cdot c_6^2 \cdot  d^2 \log (n)^3}{2(n-t)} \cdot \frac{ n-t}{c_6^2 \cdot 2 \paren{d \log (n)}^2 }} \\
		&= \sum_{t=1}^m 2 (t-1) \expf[\big]{ - \gamma^2 \log(n)  / 4 } \\
		& \leq m^2 \expf[\big]{ - \log(n^{\gamma^2 / 4} )} \\
		&\leq n^2 \cdot n^{-\gamma^2 / 4} \\
		&= 1 / n^{\gamma^2 / 4 - 2}
		\enspace.
	\end{align*}
	To finish, we recall that $\gamma = 4$.
\end{proof}

By combining these two together, we get that the combined event $\event = \event_1 \cap \event_2$ happens with high probability.

\begin{appproposition}  \label{prop:event-is-high-prob}
	Under Assumption~\ref{supp-assumption:whole-covariate-regularity},
	for $n$ large enough that $\afn \geq 2$, the combined event $\event$ occurs with high probability:
	\[
	\Pr{\notevent} \leq \bigO[\Big]{ \frac{1}{n^{2}} }
	\enspace.
	\]
\end{appproposition}
\begin{proof}
	Compute
	\begin{multline*}
		\Pr{\notevent}
			= \Pr{ \notevent_1 \cup \notevent_2 }
			= \Pr{ \notevent_1} +  \Pr{\notevent_2 \cap \event_1} \\
			= \Pr{ \notevent_1 } + \Pr{\notevent_2 \mid \event_1} \Pr{\event_1}
			\leq \Pr{ \notevent_1 } + \Pr{\notevent_2 \mid \event_1}
			\enspace.
	\end{multline*}
	Propositions~\ref{prop:first-event-high-prob} and~\ref{prop:second-event-high-prob}
	  tell us that this sum is at most
	\[
	\frac{1}{n^{2}} + \expf{ - \bigOmega{n^{1/4}} }
	= \bigO[\Big]{ \frac{1}{n^{2}} }
	\enspace.
	\qedhere
	\]
\end{proof}

\subsubsection{Regularity in Gram--Schmidt Walk Updates}

Let $\widetilde{P}_t$ be the (random) matrix which projects a vector onto the coordinates $i \in \alive_t \setminus p_t$.
More concretely, $\widetilde{P}_t$ is an $n$-by-$n$ diagonal matrix whose entry $P_t(i,i) = 1$ if and only if $i \in \alive_t \setminus p_t$ and $0$ otherwise.
The following lemma shows that when all entries of the current fractional assignment are sufficiently small and the coordinates of the update direction are not too large, then the pivot unit is the only frozen unit at an iteration.
Additionally, the squared step size will be close to 1.

\begin{applemma} \label{lemma:gsw-alg-one-frozen-conditions}
	Let $t$ be an iteration of the Gram--Schmidt Walk and let $q \in [0,1]$.
	If $\abs{ \gsve{t}{i}} \leq q$ for all $i \in \alive_t$ and $\norm{\widetilde{P}_t \gsuvt}_\infty < \frac{1-q}{1+q}$, then with probability 1:
	\begin{itemize}
		\item [a.] The pivot is the only coordinate frozen at iteration $t$, i.e. $\alive_{t+1} = \alive_t \setminus p_t$.
		\item [b.] The squared step size is bounded from 1 as $\abs{1 - \gsdt^2} \leq 3 q$.
	\end{itemize}
\end{applemma}
\begin{proof}
	Recall that the random step size $\gsdt$ takes values $\gsdpt = \abs{\max\Delta}$ and $\gsdmt = -\abs{\min\Delta}$, where $\Delta = \setb{ \delta : \gsv{t} + \delta \gsuvt \in \bracket{-1,1}^n }$.
	We will first show that the conditions above imply that the single tight constraint in both directions is the $p_t$ coordinate, which will establish part $a$.

	Consider an alive coordinate which is not the pivot, $i \in \alive_t \setminus p_t$.
	The largest positive value of $\delta$ such that $\gsve{t}{i} + \delta \gsuve{t}{i} \in [-1,1]$ satisfies
	\[
	\gsve{t}{i} + \delta \gsuve{t}{i} = \signf{\gsuve{t}{i}}
	\Rightarrow
	\delta = \frac{\signf{\gsuve{t}{i}} - \gsve{t}{i} }{\gsuve{t}{i}}
	\enspace.
	\]
	Using the hypotheses above on the fractional assignment vector $\gsvt$ and the step direction coordinates, we have that the magnitude of this $\delta$ is at least
	\begin{align*}
		\abs{\delta}
		&= \frac{ \abs[\big]{ \signf{\gsuve{t}{i}} - \gsve{t}{i} } }{ \abs{\gsuve{t}{i}} } \\
		&\geq \frac{ \abs[\Big]{ \abs{\signf{\gsuve{t}{i}}} - \abs{\gsve{t}{i}} } }{ \abs{\gsuve{t}{i}} }
			&\text{(reverse triangle inequality)} \\
		&= \frac{ 1 - \abs{\gsve{t}{i}} }{ \abs{\gsuve{t}{i}} }
			&\text{$\abs{\gsve{t}{i}} \leq 1$ and $\abs{\signf{\gsuve{t}{i}}}=1$ } \\
		&\geq \frac{1-q}{ \frac{1-q}{1+q} }
			&\text{(hypotheses)} \\
		&= 1 + q \enspace.
			&\text{(cancelling terms)}
	\end{align*}
	Likewise, one can show that the the most negative value $\delta$ for which $\gsve{t}{i} + \delta \gsuve{t}{i} \in [-1,1]$ has magnitude strictly greater than $1 + q$.

	Let us now consider the pivot unit $p_t$.
	By construction, we have that $\gsuve{t}{p_t} = 1$.
	Thus, the largest positive and negative values of $\delta$ for which $\gsve{t}{p_t} + \delta \gsuve{t}{p_t} \in [-1,1]$ will be $1 + \gsve{t}{p_t}$ and $-1 - \gsve{t}{p_t}$, respectively.
	By assumption, $\abs{\gsve{t}{p_t}} \leq q$ so that both of these values have magnitude at most $1 + q$.

	This shows that the only tight constraint in the step size will be from the pivot $p_t$.
	Thus, with probability 1, the pivot will get frozen so that $\alive_{t+1} = \alive_t \setminus p_t$.

	To show part $b$, recall that
	the random step size $\gsdt$ will take one of the two values: $1 + \gsve{t}{p_t}$ or $1 - \gsve{t}{p_t}$.
	This implies that the square $\gsdt^2$ takes one of the two values: $\paren{1 + \gsve{t}{p_t}}^2$ or $\paren{1 - \gsve{t}{p_t}}^2$.
	The difference of these from 1 may be bounded as
	\[
	\abs{ 1 - \paren{1 \pm \gsve{t}{p_t}}^2 }
	= \abs{ \gsve{t}{p_t}^2 \pm 2 \gsve{t}{p_t} }
	\leq 3 \abs{\gsve{t}{p_t}}
	\leq 3 q
	\enspace.
	\qedhere
	\]
\end{proof}

Before continuing, let us establish a fixed $n^*$ which is ``sufficiently large'' for the purposes of our asymptotic argument.
Define $n^*$ to be the smallest integer such that for all $n \geq n^*$,
\[
\max \setb[\Big]{ d \paren[\Big]{\frac{\log (n)^3}{\afn} }^{1/2} , d \frac{\log(n)}{\afn}  }
<
\min \setb[\Big]{ \frac{1}{2 \cdot \gamma \cdot c_6} , \frac{1}{3 c_6}  }
\enspace.
\]
Such an integer $n^*$ is guaranteed to exist because $d = \littleO{ \afn^{1/2} / \log(n)^{3/2} }$ and $d = \littleO{ \afn / \log(n) }$, which are shown in parts $a$ and $b$ of Proposition~\ref{prop:helpful-asymptotics} to be implied by Assumption~\ref{supp-assumption:covariate-dimension}.

We now show that, conditioned on $\event$, the pivots and the step directions $\gsuvt$ are deterministically determined by $\pi$ for the first $t \leq n - \afn$ iterations.
Recall that we define the vector $\vec{w}_s \in \Reals^n$ to be the vector whose first $n-s$ entries in the pivot ordering are given as
\[
\vec{w}_s( \pi(1) : \pi(n-s) )
= - (1-\vto) \paren[\Big]{ \vto \unitM + (1-\vto) \maxnormX^{-2} \xM_s \xM_s^\tran }^{-1}
\paren{ \maxnormX^{-1} \xM_s }
\paren{\maxnormX^{-1} \xv{n-s+1}}
\enspace,
\]
and the remaining entries are given as $\vec{w}_s(\pi(n-s+1)) = 1$ and $\vec{w}_s(\pi(i)) = 0$ for all $i > n -s +1$.
Moreover, recall that this vector is determined completely by the pivot ordering $\pi$.

\begin{appproposition} \label{prop:known-pivots-and-step-directions}
	Under Assumption~\ref{supp-assumption:whole-covariate-regularity} and conditioned on the event $\event$, we have that for sufficiently large $n \geq n^*$, at each iteration $t \leq n - \afn$ the following hold with probability 1:
	\begin{itemize}
		\item The pivot is $p_{t} = \pi(n-t+1)$ and non-frozen coordinates are $\alive_t \setminus p_t = \setb{ \pi(1), \dots \pi(n-t)}$.
		\item The step directions are $\gsuvt = \vec{w}_t$.
	\end{itemize}
\end{appproposition}
\begin{proof}

	We proceed by proving the claim by induction on the iterations of the algorithm.
	Consider the first iteration $t = 1$.
	By definition, the first pivot is $p_1 = \pi(n)$ and thus $\alive_1 \setminus p_1 = \setb{\pi(1) \dots \pi(n-1)}$.
	The step direction depends only on the values of $p_t$ and $\alive_t \setminus p_t$, and therefore we have that $\gsuv{1} = \vec{w}_1$.
	This establishes the base case.

	Suppose that the induction hypotheses hold for all iterations up to and including iteration $t$.
	We will show that they hold also for iteration $t+1$.
	Observe that for all $i \in \alive_t$, $\gsve{i}{t} = \sum_{s < t} \gsd{s} \gsuve{s}{i}$.
	By induction, the alive coordinates are $\alive_t = \setb{ \pi(1), \dots \pi(n-t+1)}$, and so $\widetilde{P}_t = P_t$.
	Thus, the event $\event$ guarantees that for all $i \in \alive_t$,
	\[
	\abs[\Big]{ \gsve{i}{t} } \leq \frac{\gamma \cdot c_6 \cdot d \log (n)^{3/2}}{\sqrt{n-t}}
	\leq \frac{\gamma \cdot c_6 \cdot d \log(n)^{3/2}}{\afn^{1/2}}
	\enspace,
	\]
	where the second inequality follows from $t \leq n - \afn$.

	By induction,the step direction is $\gsuvt = \vec{w}_t$ and thus Proposition~\ref{prop:wt-small-norm} guarantees that
	\[
	\norm{\widetilde{P}_t \gsuvt }_\infty
	= \norm{P_t\vec{w}_t}_\infty
	\leq \frac{c_6 \cdot d \log(n)}{n-t}
	\leq \frac{c_6 \cdot d \log(n)}{\afn}
	\enspace.
	\]
	Our goal is now to show that the conditions of Lemma~\ref{lemma:gsw-alg-one-frozen-conditions} are met with $q = 1/2$ for $n \geq n^*$.
	Indeed, the definition of $n^*$ together with the above bound on $\abs{\gsve{i}{t}}$ ensures that
	\[
	\abs[\Big]{ \gsve{i}{t} }
	\leq \gamma \cdot c_6 \cdot \paren[\Big]{d \cdot \paren[\Big]{\frac{\log(n)^{3}}{\afn}}^{1/2} }
	< \frac{\gamma \cdot c_6}{2 \gamma c_6}
	= 1/2
	= q
	\enspace.
	\]
	Likewise, the definition of $n^*$ together with the above bound on $\norm{\widetilde{P}_t \gsuvt }_\infty$ ensures that
	\[
	\norm{\widetilde{P}_t \gsuvt }_\infty
	\leq c_6 \cdot \paren[\Big]{ \frac{d \log(n)}{\afn} }
	< c_6 \frac{1}{3 c_6}
	= 1/3
	= \frac{1 - 1/2}{1 + 1/2}
	= \frac{1 - q}{1 + q}
	\enspace.
	\]
	Thus, we may invoke Lemma~\ref{lemma:gsw-alg-one-frozen-conditions} to show that only the pivot $p_t$ is frozen.
	By definition of the pivot order, $p_{t+1} = \pi(n-t)$ and $\alive_{t+1} = \setb{\pi(1), \dots \pi(n-(t+1))}$.
	The step direction depends only on the values of $p_t$ and $\alive_t \setminus p_t$, and therefore we have that $\gsuv{t+1} = \vec{w}_{t+1}$.
	This establishes the claim by induction.
\end{proof}

The next Proposition shows that conditioned on $\event$, the squared step sizes $\gsdt^2$ are very close to $1$ for the first $t \leq n - \afn$ iterations.

\begin{appproposition} \label{prop:small-squared-step-sizes}
	Under Assumption~\ref{supp-assumption:whole-covariate-regularity},
	and conditioned on the event $\event$, we have that for sufficiently large $n \geq n^*$, at each iteration $t \leq n - \afn$, the squared step sizes are concentrated around their conditional mean:
	\[
	\abs{\gsdt^2 - \E{\gsdt^2 \mid \event}} \leq \frac{4 \gamma c_6 \cdot d \log(n)^{3/2}}{\sqrt{n-t}}
	\quadtext{with probability 1, conditioned on $\event$. }
	\]
\end{appproposition}
\begin{proof}
	We begin by deriving a formula for $\E{\gsdt^2 \mid \event}$, the conditional expectation of the squared step size under $\event$.
	As in the proof of Proposition~\ref{prop:known-pivots-and-step-directions}, we can show that the conditions of Lemma~\ref{lemma:gsw-alg-one-frozen-conditions} are met
	with $q = \gamma \cdot c_6 \cdot d \log (n)^{3/2} / \sqrt{n-t}$ at each iteration $t \leq n - \afn$.
	Thus, the two values that $\gsdt$ could take are $(1 + \gsve{t}{p_t})$ and $(1 - \gsve{t}{p_t})$.
	In this case, we have that the conditional expectation of the squared step size is
	\begin{align}
		\E{\gsdt^2 \mid \gsd{1}, \dots \gsd{t-1}}
		&= \paren{\gsdpt}^2 \paren{ \frac{\gsdmt}{\gsdpt + \gsdmt} } + \paren{\gsdmt}^2 \paren{ \frac{\gsdpt}{\gsdpt + \gsdmt} } \\
		&= \abs{\gsdpt \gsdmt} \\
		&= (1 + \gsve{t}{p_t}) (1 - \gsve{t}{p_t}) \\
		&= 1 - \gsve{t}{p_t}^2
		\enspace.
	\end{align}
	By the law of total expectation, the conditional expectation can be computed as
	\[
	\E{\gsdt^2 \mid \event}
	= \E[\Big]{ \E{\gsdt^2 \mid \gsd{1}, \dots \gsd{t-1}} \mid \event }
	= 1 - \E{\gsve{t}{p_t}^2 \mid \event}
	\]
	and the distance to 1 can be bounded as
	\[
	\abs[\Big]{1 - \E{\gsdt^2 \mid \event} }
	= \E[\Big]{\gsve{t}{p_t}^2 \mid \event} \\
	\leq \E[\Big]{\abs{\gsve{t}{p_t}} \mid \event} \\
	\leq \frac{\gamma \cdot c_6 \cdot d \log (n)^{3/2}}{\sqrt{n-t}}
	\enspace.
	\]
	Now we are ready to bound the distance from $\gsdt^2$ to the conditional mean.
	To do this, observe that we can use the part $b$ of Lemma~\ref{lemma:gsw-alg-one-frozen-conditions} to bound $\abs{1 - \delta_t^2} \leq 3 q$.
	Recall that, as stated above, we have that $q = \gamma \cdot c_6 \cdot d \log (n)^{3/2} / \sqrt{n-t}$.
	Using this together with triangle inequality, we have that
	\[
	\abs[\Big]{\gsdt^2 - \E{\gsdt^2 \mid \event} }
	\leq \abs[\Big]{1 - \gsdt^2} + \abs[\Big]{1 -\E{\gsdt^2 \mid \event} }
	\leq \frac{4 \gamma c_6 \cdot d \log (n)^{3/2}}{\sqrt{n-t}}
	\enspace,
	\]
	which holds with probability 1, conditioned on the event $\event$.
\end{proof}

\subsubsection{Verifying CLT Conditions of McLeish (1974)}

We begin by providing claims which bound the individual terms $X_t$ in absolute value with probability 1.
These bounds will generally be quite loose, and primarily useful when we are conditioning on a low probability event or considering only a few terms.

\begin{appproposition} \label{prop:large-bound-on-ut}
	For every iteration $t \in [n]$, $ \norm{\widetilde{P}_t \gsuvt }_2^2 \leq \frac{1}{4}  \cdot \frac{1 - \vto}{\vto} $ with probability 1.
\end{appproposition}
\begin{proof}
	Recall that the step direction restricted to the $\alive_t \setminus p_t$ coordinates is given by
	\[
	\widetilde{P}_t \gsuvt
	= - (1-\vto) \paren[\Big]{ \vto \unitM + (1-\vto) \maxnormX^{-2} \xM_t \xM_t^\tran }^{-1}
	\paren{ \maxnormX^{-1} \xM_t }
	\paren{\maxnormX^{-1} \xv{p_t}}
	\enspace,
	\]
	where $\xM_t$ is the matrix whose columns are given by covariate vectors $\xvi$ for $i \in \alive_t \setminus p_t$.
	We may bound the $\ell_2$ norm of this part of the update vector by using the operator norm:
	\begin{align*}
		\norm{\widetilde{P}_t \gsuvt }_2^2
		&= \xv{p_t}^\tran \xM_t^\tran  \paren[\Big]{ \frac{\vto}{1-\vto} \maxnormX^{2} \unitM + (1-\vto) \xM_t \xM_t^\tran }^{-2} \xM_t \xv{p_t} \\
		&\leq \norm[\Big]{ \xM_t^\tran  \paren[\Big]{ \frac{\vto}{1-\vto} \maxnormX^{2} \unitM + (1-\vto) \xM_t \xM_t^\tran }^{-2} \xM_t } \norm{\xv{p_t}}_2^2 \\
		&\leq \norm[\Big]{ \xM_t^\tran  \paren[\Big]{ \frac{\vto}{1-\vto} \maxnormX^{2} \unitM + (1-\vto) \xM_t \xM_t^\tran }^{-2} \xM_t } \maxnormX^2
		\enspace.
	\end{align*}
	The eigenvalues of the above matrix are of the form
	\[
	\frac{\sigma}{\paren[\big]{  \frac{\vto}{1-\vto} \maxnormX^{2} +  \sigma }^2}
	\enspace,
	\]
	where $\sigma$ is a singular value of $\xM_t$.
	The function $f(x) = x / (\beta + x)^2$ achieves it largest value of $1/(4 \beta)$ at the point $x = \beta$.
	Therefore, all of the eigenvalues{\textemdash}and thus the operator norm{\textemdash} are bounded above by $1/4 \cdot \frac{1-\vto}{\vto} \maxnormX^{-2}$.
	The result follows by observing that the $\maxnormX^{2}$ terms cancel.
\end{proof}

For notational simplicity, we introduce the new constant $c_7 = \frac{8 (2 \cdot c_1)^2}{c_5}$.

\begin{appproposition} \label{prop:large-bound-on-Xt}
	Under Assumptions~\ref{supp-assumption:design}, \ref{supp-assumption:outcome-regularity}, \ref{supp-assumption:not-superefficient},
	we have that for every iteration $t \in [n]$, with probability 1,
	\[
	X_t^2  \leq c \cdot c_7 \bracket[\Big]{ \paren[\Big]{\frac{1}{n}}^{1-2/p} + \paren[\Big]{ \frac{n-t}{n} }^{1 - 2/p} }
	\enspace,
	\]
	where we recall that $X_t$ is defined on line \eqref{eqn:Xt} at the start of Section~\ref{sec:supp-clt-proof}.
\end{appproposition}
\begin{proof}
	Recall that the step direction can be broken up into the pivot and alive non-pivot coordinates, $\gsuvt = \vec{e}_{p_t} + \widetilde{P}_t \gsuvt$ where $\vec{e}_{p_t}$ is the standard basis vector that has $1$ in coordinate $p_t$ and $0$ in all other coordinates.
	Using this, we observe that
	\begin{align*}
		\sigma^2 \cdot X_t^2
		&=  \gsdt^2 \iprod{ \gsuvt , \pomv }^2 \\
		&\leq 4 \iprod{ \gsuvt , \pomv }^2
			&\text{($\abs{\gsdt} \leq 2$ with probability 1)}\\
		&= 4 \iprod{ \vec{e}_{p_t} + \widetilde{P}_t \gsuvt , \pomv }^2
		&\text{(definition of $\gsuvt$)}\\
		&= 4 \paren[\Big]{ \pom{p_t} + \iprod{\widetilde{P}_t \gsuvt , \widetilde{P}_t \pomv} }^2
		\\
		&\leq 8 \paren[\Big]{ \pom{p_t}^2 + \iprod{\widetilde{P}_t \gsuvt , \widetilde{P}_t \pomv}^2  }
		&\text{(AM-GM inequality)}\\
		&\leq 8 \paren[\Big]{ \pom{p_t}^2  + \norm{\widetilde{P}_t \gsuvt }_2^2 \norm{\widetilde{P}_t \pomv}_2^2  }
		&\text{(Cauchy--Schwarz)}\\
		&\leq 8 \paren[\Big]{ \pom{p_t}^2 +  \frac{1 - \vto}{\vto} \norm{\widetilde{P}_t \pomv}_2^2}
		&\text{(Proposition~\ref{prop:large-bound-on-ut})}\\
		&\leq 8 \paren[\Big]{ (2 c_1)^2 n^{2/p} +  \frac{1 - \vto}{\vto} (2 c_1)^2  \cdot (n-t)^{1 - 2/p} \cdot n^{2/p}}
		&\text{(Lemma~\ref{lemma:outcome-bounds}, with $q=2$)} \\
		&= 8 (2 c_1)^2 n^{2/p} \cdot \paren[\Big]{ 1 + \frac{1 - \vto}{\vto} \cdot (n-t)^{1 - 2/p} } \\
		&\leq 8 (2 c_1)^2 n^{2/p} \cdot \paren[\Big]{ 1 + c \cdot (n-t)^{1 - 2/p} }
			&\text{(Assumption~\ref{supp-assumption:design})}
	\end{align*}
	Bringing $\sigma^2$ to the other side, using $\sigma^2 = n^2 \Var{\htest}$, and Assumption~\ref{supp-assumption:not-superefficient} that $n \Var{\htest} \geq c_5$ we obtain,
	\begin{align*}
	X_t^2
	&\leq \frac{8 (2 \cdot c_1)^2}{c_5} n^{-(1 - 2/p)} \cdot \paren[\Big]{ 1 + c \cdot (n-t)^{1 - 2/p} }  \\
	&= c_7 \cdot  n^{-(1 - 2/p)} \cdot \paren[\Big]{ 1 + c \cdot (n-t)^{1 - 2/p} }  \\
	&\leq c \cdot c_7 \paren[\Big]{ \paren[\Big]{\frac{1}{n}}^{1-2/p} + \paren[\Big]{ \frac{n-t}{n} }^{1 - 2/p} }
	\enspace.
	\qedhere
	\end{align*}
\end{proof}

Finally, we show that conditioned on event $\event$, the quantities $X_t^2$ are small for all but the last $\afn$ iterations.

\begin{appproposition} \label{prop:mu_dot_ut}
	Under Assumptions~\ref{supp-assumption:outcome-regularity} and~\ref{supp-assumption:whole-covariate-regularity},
	and conditioned on event $\event$,
	we have that for every iteration $t \leq n - \afn$, with probability 1,
	\[
	\iprod{ \vec{w}_t , \pomv }^2
	\leq
	8 c_1^2 \cdot n^{2/p}  \paren[\Big]{ 1 + (c_6 \cdot d \log(n) )^2 (n-t)^{-2/p}} \enspace.
	\]
\end{appproposition}
\begin{proof}
	\begin{align*}
 	\lefteqn{	\iprod{\vec{w}_t , \pomv}^2 }\\
		&= \paren[\Big]{ \pom{p_t} + \iprod{P_t \vec{w}_t , P_t \pomv } }^2 \\
		&\leq 2 \paren[\Big]{ \pom{p_t}^2 + \iprod{P_t \vec{w}_t , P_t \pomv }^2 }
		&\text{(AM-GM)} \\
		&\leq 2 \paren[\Big]{ \pom{p_t}^2 + \norm{P_t \vec{w}_t}_\infty^2 \norm{P_t \pomv}_1^2 }
		&\text{(H{\"o}lder's inequality)} \\
		&\leq 2 \paren[\Big]{ \pom{p_t}^2 + \frac{(c_6 \cdot d \log(n))^2}{(n-t)^2} \norm{P_t \pomv}_1^2 }
		&\text{(Proposition~\ref{prop:wt-small-norm})}\\
		&\leq 2 \paren[\Big]{ (2 c_1)^2 n^{2/p} + \frac{(c_6 \cdot d \log(n))^2}{(n-t)^2} (2 c_1)^2 \cdot (n-t)^{2 - 2/p} \cdot n^{2/p} }
		&\text{(Lemma~\ref{lemma:outcome-bounds}, with $q=1$)}\\
		&= 8 c_1^2 \cdot n^{2/p}  \paren[\Big]{ 1 + (c_6 \cdot d \log(n) )^2 (n-t)^{-2/p}}
		\enspace.
	\end{align*}
\end{proof}

\begin{appproposition} \label{prop:Xt-tighter-bound}
	Under Assumptions~\ref{supp-assumption:outcome-regularity}, \ref{supp-assumption:whole-covariate-regularity}, and \ref{supp-assumption:not-superefficient}
	and conditioned on the event $\event$, we have that for sufficiently large $n \geq n^*$, at each iteration $t \leq n - \afn$ the following holds with probability 1:
	\[
	X_t^2 \leq
	c_6 \cdot c_7 \bracket[\Big]{ \paren[\Big]{\frac{1}{n}}^{1 - 2/p} + \frac{(d \log(n))^2}{n}  \paren[\Big]{ \frac{n}{n-t} }^{2/p} }
	\enspace.
	\]
\end{appproposition}
\begin{proof}
	Conditioned on the event $\event$, Proposition~\ref{prop:known-pivots-and-step-directions} shows that for sufficiently large $n \geq n^*$ and iterations $t \leq n - \afn$, we have that $\gsuvt = \vec{w}_t$ with probability 1.
	Thus, by Proposition~\ref{prop:mu_dot_ut},
	\begin{align*}
		\sigma^2 \cdot X_t^2
		&= \gsdt^2 \iprod{\gsuvt , \pomv}^2 \\
		&\leq 4 \iprod{\vec{w}_t, \pomv}^2
		\\
		&\leq  8 (2c_1)^2 \cdot n^{2/p}  \paren[\Big]{ 1 + (c_6 \cdot d \log(n) )^2 (n-t)^{-2/p}}
		\enspace.
	\end{align*}
	The rest of the proof is completed by bringing $\sigma^2$ to the other side.
	In particular, doing this and using Assumption~\ref{supp-assumption:not-superefficient}, which implies $\sigma^2 \geq c_5 / n$, yields
	\begin{align*}
		 X_t^2
		 &\leq \frac{8 (2c_1)^2 }{ c_5} n^{- (1 -2/p )} \paren[\Big]{ 1 + (c_6 \cdot d \log(n) )^2 (n-t)^{-2/p}} \\
		 &= c_7 \bracket[\Big]{ \paren[\Big]{\frac{1}{n}}^{1 - 2/p} + c_6^2 \frac{(d \log(n))^2}{n}  \paren[\Big]{ \frac{n}{n-t} }^{2/p} } \\
		 &\leq c_6 \cdot c_7 \bracket[\Big]{ \paren[\Big]{\frac{1}{n}}^{1 - 2/p} + \frac{(d \log(n))^2}{n}  \paren[\Big]{ \frac{n}{n-t} }^{2/p} }
		 \enspace.
		 \qedhere
	\end{align*}
\end{proof}

We are now ready to prove two lemmas which will establish that the conditions of the martingale CLT (Theorem~\ref{thm:martingale-clt-result}) are satisfied.

\begin{applemma} \label{lemma:first-clt-condition}
	Under Assumptions~\ref{supp-assumption:design},
	\ref{supp-assumption:outcome-regularity},
	\ref{supp-assumption:whole-covariate-regularity},
	\ref{supp-assumption:covariate-dimension},
	and \ref{supp-assumption:not-superefficient}, we have that
	\[
	\lim_{n \to \infty} \E[\Big]{ \paren[\big]{ \max_{t \leq n} \abs{X_t}  }^2 } = 0
	\enspace.
	\]
\end{applemma}
\begin{proof}
	We begin by decomposing the expectation as
	\[
	\E[\Big]{ \paren[\big]{ \max_{t \leq n} \abs{X_t}  }^2 }
	= \E[\Big]{\max_{t \leq n} X_t^2  \mid \event } \Pr{\event}
	+ \E[\Big]{\max_{t \leq n} X_t^2  \mid \notevent } \Pr{\notevent}
	\enspace.
	\]
	Our approach to the proof will be to show that both of the terms go to zero under the stated assumptions.

	Let's begin with the second term.
	By Proposition~\ref{prop:large-bound-on-Xt}, we can show that the maximum of $X_t^2$ is at most a constant with probability 1, i.e.
	\[
	\max_{t \leq n} X_t^2
	\leq \max_{t \leq n} c \cdot c_7 \bracket[\Big]{ \paren[\Big]{\frac{1}{n}}^{1-2/p} + \paren[\Big]{ \frac{n-t}{n} }^{1 - 2/p} }
	\leq 2 c \cdot c_7
	\enspace,
	\]
	where the maximum is achieved at $t = 1$.
	Likewise, Proposition~\ref{prop:event-is-high-prob} bounds the probability of $\notevent$ as $\Pr{\notevent} \leq \bigO{n^{-2}}$.
	Together, these two propositions ensure that the second maximum term is bounded as
	\[
	\E[\Big]{\max_{t \leq n} X_t^2  \mid \notevent } \Pr{\notevent}
	\leq 2 c \cdot  c_7 \cdot \bigO[\Big]{ \frac{1}{n^{2}} }
	\enspace,
	\]
	which goes to zero.

	Now we show that the first term goes to zero.
	To do this, we break the maximum into two separate sets: the earlier iterations iterations $t \leq n - \afn$ and the later iterations $t > n - \afn$.
	Conditioned on $\event$, Proposition~\ref{prop:Xt-tighter-bound} ensures that for all $t \leq n - \afn$,
	\begin{align*}
	\max_{t \leq n - \afn} X_t^2
	&\leq \max_{t \leq n - \afn} c_6 \cdot c_7 \bracket[\Big]{ \paren[\Big]{\frac{1}{n}}^{1 - 2/p} + \frac{(d \log(n))^2}{n}  \paren[\Big]{ \frac{n}{n-t} }^{2/p} } \\
	&= c_6 \cdot c_7 \bracket[\Big]{ \paren[\Big]{\frac{1}{n}}^{1 - 2/p} + \frac{(d \log(n))^2}{n}  \paren[\Big]{ \frac{n}{\afn} }^{2/p} } \\
	&= c_6 \cdot c_7 \bracket[\Big]{ \paren[\Big]{\frac{1}{n}}^{1 - 2/p} + \paren[\Big]{d \cdot \frac{\log(n)}{n^{(1/2 - 1/p)} \cdot \afn^{1/p}} }^2  }
	\enspace,
	\end{align*}
	because the upper bound is maximized at $t = n - \afn$.
	The first term will go to zero because $p \geq 5 > 2$.
	The second term will go to zero because part $c$ of Proposition~\ref{prop:helpful-asymptotics} shows that $d = \littleO{ n^{(1/2 - 1/p)} \cdot \afn^{1/p} / \log(n) }$ is implied by Assumption~\ref{supp-assumption:covariate-dimension}.
	Thus, this maximum goes to zero almost surely, conditioned on $\event$.

	Likewise, Proposition~\ref{prop:large-bound-on-Xt} ensures that for iterations $t > n - \afn$, it holds with probability 1 that
	\begin{align*}
	\max_{t > n - \afn} X_t^2
	&\leq \max_{t > n - \afn} c \cdot c_7 \bracket[\Big]{ \paren[\Big]{\frac{1}{n}}^{1-2/p} + \paren[\Big]{ \frac{n-t}{n} }^{1 - 2/p} } \\
	&= c \cdot c_7 \bracket[\Big]{ \paren[\Big]{\frac{1}{n}}^{1-2/p} + \paren[\Big]{ \frac{\afn}{n} }^{1 - 2/p} } , \\
	\end{align*}
	because this upper bound is also maximized at $t = n - \afn$.
	This term goes to zero under our conditions, as $p \geq 5 > 2$ and $\afn$
	is defined so that $\afn / n$ goes to zero.
	Thus, this establishes that the first expected maximum term goes to zero.
\end{proof}

\begin{applemma} \label{lemma:second-clt-condition}
	Under Assumptions~
	\ref{supp-assumption:design},
	\ref{supp-assumption:outcome-regularity},
	\ref{supp-assumption:whole-covariate-regularity},
	\ref{supp-assumption:covariate-dimension}, and
	\ref{supp-assumption:not-superefficient},
	we have that
	\[
	\sum_{t=1}^n X_t^2 \xrightarrow{p} 1
	\enspace.
	\]
\end{applemma}
\begin{proof}
	We seek to show that $\sum_{t=1}^n X_t^2 \xrightarrow{p} 1$, which is equivalent to showing that
	\[
	\abs[\Big]{\sum_{t=1}^n X_t^2 - 1} \xrightarrow{p} 0
	\enspace.
	\]
	Observe that
	\[
	\E[\Big]{ \sum_{t=1}^n X_t^2 }
	= \frac{1}{\sigma^2} \E[\Big]{ \sum_{t=1}^n\gsdt^2 \iprod{\gsuvt , \pomv}^2 }
	= \frac{\sigma^2}{\sigma^2}
	= 1
	\enspace,
	\]
	where the second equality follows by Lemmas~\mainref{lemma:mse-expression} and \suppref{lemma:cov_x}.
	Using this, we decompose $\abs{\sum_{t=1}^n X_t^2 - 1}$ into three terms.
	Let $m = n - \afn$.
	By the triangle inequality and the law of total expectation, we have that
	\begin{align*}
		&\abs[\Big]{\sum_{t=1}^n X_t^2 - 1} \\
		&= \abs[\Big]{\sum_{t=1}^n \paren{X_t^2 - \E{X_t^2}} } \\
		&\leq
		\abs[\Big]{\sum_{t=1}^m \paren{X_t^2 - \E{X_t^2}} }
		+ \abs[\Big]{\sum_{t=m+1}^n \paren{X_t^2 - \E{X_t^2}} } \\
		&\leq \Pr{\event} \abs[\Big]{\sum_{t=1}^m \paren{X_t^2 - \E{X_t^2 \mid \event}} }
		+ \Pr{\notevent} \abs[\Big]{\sum_{t=1}^m \paren{X_t^2 - \E{X_t^2 \mid \notevent}} }
		+ \abs[\Big]{\sum_{t=m+1}^n \paren{X_t^2 - \E{X_t^2}} }
	\end{align*}
	In the remainder of the proof, we show that each of these terms goes to zero in probability.
	We restrict ourselves to sufficiently large $n \geq n^*$.

	Let us begin with the third term.
	To do analyze this term, we will use Proposition~\ref{prop:large-bound-on-Xt}, which bounds $X_t^2$ with probability 1 and thus in turn upper bounds the deviation of $X_t^2$ from its mean.
	In particular, this yields that
	\begin{align*}
		\abs[\Big]{\sum_{t=n - \afn + 1}^n \paren{X_t^2 - \E{X_t^2}} }
		&\leq \sum_{t = n - \afn + 1}^n c \cdot c_7 \bracket[\Big]{ \paren[\Big]{\frac{1}{n}}^{1-2/p} + \paren[\Big]{ \frac{n-t}{n} }^{1 - 2/p} } \\
		&= c \cdot c_7 \bracket[\Big]{ \frac{\afn}{n^{1 - 2/p}} + \frac{1}{n^{1 - 2/p}} \sum_{k=0}^{\afn - 1} k^{1 - 2/p}  } \\
		&\leq c \cdot c_7 \bracket[\Big]{ \frac{\afn}{n^{1 - 2/p}} + \frac{\afn^{2 - 2/p} }{n^{1 - 2/p}}  } \\
		&= c \cdot c_7 \cdot \bracket[\Big]{ \frac{\afn}{n^{1 - 2/p}} + \paren[\Big]{\frac{ \afn}{n^{(1/2) \frac{1 - 2/p}{1 - 1/p} } } }^{2 - 2/p } } \\
	\end{align*}
	By definition, $\afn = \littleO{n^{(1/2) \cdot \frac{1 - 2/p}{1 - 1/p} } } = \littleO{n^{1/2}}$, which implies that both terms go to zero.

	We now turn our attention to the second term.
	To analyze this term, we will again use Proposition~\ref{prop:large-bound-on-Xt}, which bounds $X_t^2$ with probability 1 and thus in turn upper bounds the deviation of $X_t^2$ from its mean.
	This bound will be looser than before, as we are considering a larger number of variables.
	However, the low probability of event $\notevent$, guaranteed by Proposition~\ref{prop:event-is-high-prob}, will cause the term to go zero.
	More precisely,
	\begin{align*}
		\Pr{\notevent} \abs[\Big]{\sum_{t=1}^n \paren{X_t^2 - \E{X_t^2 \mid \notevent}} }
		&\leq \bigO[\Big]{\frac{1}{n^{2}} } \sum_{t=1}^{n - \afn} c \cdot c_7 \bracket[\Big]{ \paren[\Big]{\frac{1}{n}}^{1-2/p} + \paren[\Big]{ \frac{n-t}{n} }^{1 - 2/p} } \\
		&\leq \bigO[\Big]{\frac{1}{n^{2}} } \bracket[\Big]{ \frac{n - \afn}{n^{1 - 2/p}} +  \frac{1}{n^{1 - 2/p}} \sum_{t=1}^{n - \afn} (n-t)^{1 - 2/p} } \\
		&\leq \bigO[\Big]{\frac{1}{n^{2}} } \bracket[\Big]{ \frac{n - \afn}{n^{1 - 2/p}} +  \frac{n^{1 - 2/p} \paren{ n - \afn }}{n^{1 - 2/p}} } \\
		&\leq \bigO[\Big]{\frac{1}{n^{2}} } \bracket[\Big]{ n^{2/p} + n } \\
		&\leq \bigO[\Big]{\frac{n}{n^{2}} } &\text{($p \geq 2$)} \\
		&= \bigO[\Big]{\frac{1}{n }}
		\enspace,
	\end{align*}
	which goes to zero.

	We now turn our attention back to the first term.
	For the rest of the analysis, we will condition on the event $\event$.
	By Proposition~\ref{prop:known-pivots-and-step-directions}, this conditioning ensures that $\gsuvt = \vec{w}_t$ with probability 1 for iterations $t \leq m = n - \afn$.
	Thus, we can write the sum in the first term as
	\begin{align*}
		& \abs[\Big]{\sum_{t=1}^{m} \paren{X_t^2 - \E{X_t^2 \mid \event}} } \\
		&= \frac{1}{\sigma^2} \abs[\Big]{ \sum_{t=1}^m \gsdt^2 \iprod{\gsuvt, \pomv}^2 - \E{ \gsdt^2 \iprod{\gsuvt, \pomv}^2 \mid \event }  } \\
		&= \frac{1}{\sigma^2} \abs[\Big]{ \sum_{t=1}^m \paren[\big]{ \gsdt^2 - \E{\gsdt^2 \mid \event} } \iprod{\vec{w}_t , \pomv}^2  } \\
		& \leq \frac{1}{c_5 n} \abs[\Big]{ \sum_{t=1}^m \paren[\big]{ \gsdt^2 - \E{\gsdt^2 \mid \event} } \iprod{\vec{w}_t , \pomv}^2  } \\
		&\leq  \sum_{t=1}^m \paren[\Big]{ \frac{4 \gamma c_6 \cdot d \log(n)^{3/2}}{c_5 n \sqrt{n-t}}} \iprod{\vec{w}_t , \pomv}^2
		\\
		&\leq  \sum_{t=1}^m \paren[\Big]{ \frac{4 \gamma c_6 \cdot d \log(n)^{3/2}}{c_5 n \sqrt{n-t}}} 8 c_1^2 \cdot n^{2/p}  \paren[\Big]{ 1 + (c_6 \cdot d \log(n) )^2 (n-t)^{-2/p}}
		\\
		& = \frac{2^5 c_1^2 \gamma c_6 \cdot d \log(n)^{3/2}}{c_5} \cdot n^{2/p - 1}
		  \paren[\Big]{ \sum_{t=1}^m \frac{1}{\sqrt{n-t}} +  (c_6 \cdot d \log(n) )^2  \sum_{t=1}^m \frac{1}{(n-t)^{1/2 + 2/p}} }
		  \enspace,
		\end{align*}
	where the first inequality
	  follows from Assumption~\ref{supp-assumption:not-superefficient}, which implies $\sigma^2 \geq c_5 n$,
	the second follows from Proposition~\ref{prop:small-squared-step-sizes}
	and the third from Proposition~\ref{prop:mu_dot_ut}.

	We now break the above sum into two parts.
	To bound the left-hand part, we recall that
	\[
		\sum_{k=1}^{n} \frac{1}{\sqrt{k}} \leq 2 \sqrt{n},
	\]
	and so
	\begin{align*}
		\frac{2^5 c_1^2 \gamma c_6 \cdot d \log(n)^{3/2}}{c_5} \cdot n^{2/p - 1}
	   \sum_{t=1}^m \frac{1}{\sqrt{n-t}}
	   & \leq
	   \frac{2^5 c_1^2 \gamma c_6 \cdot d \log(n)^{3/2}}{c_5} \cdot n^{2/p - 1}
	   \sum_{k=1}^n \frac{1}{\sqrt{k}}
	   \\
	   & \leq
	   \frac{2^6 c_1^2 \gamma c_6 \cdot d \log(n)^{3/2}}{c_5} \cdot n^{2/p - 1/2}
	   \\
	   & = \littleO[\Big]{n^{(1/6) \cdot (1 - 2/p)} \log(n)^{-2}
	   \log(n)^{3/2} n^{2/p - 1/2}}
	   \\
	   & = \littleO[\Big]{n^{-1/3 + 5/3p} \log(n)^{-1/2}}  ,
	\end{align*}
 	   which goes to zero for $p \geq 5$.

To bound the right-hand part, observe that
\[
	\sum_{t=1}^m \frac{1}{(n-t)^{1/2 + 2/p}}
	=
	\sum_{k=\afn}^{n-1} \frac{1}{k^{1/2 + 2/p}}
	\leq
	\int_{x=\afn - 1}^{n-1} \frac{1}{x^{1/2 + 2/p}} dx
	\leq \frac{1}{1/2 - 2/p} n^{1/2 - 2/p}.
\]
So,
\begin{align*}
\lefteqn{
	\frac{2^5 c_1^2 \gamma c_6 \cdot d \log(n)^{3/2}}{c_5} \cdot n^{2/p - 1}
   (c_6 \cdot d \log(n) )^2  \sum_{t=1}^m \frac{1}{(n-t)^{1/2 + 2/p}}
}	\\
& =
\frac{2^5 c_1^2 \gamma c_6^3 \cdot d^3 \log(n)^{7/2}}{c_5} \cdot n^{2/p - 1}
 \sum_{t=1}^m \frac{1}{(n-t)^{1/2 + 2/p}}
 \\
 & \leq
 \frac{2^5 c_1^2 \gamma c_6^3 \cdot d^3 \log(n)^{7/2}}{c_5 (1/2 - 2/p)} \cdot n^{2/p - 1}
 n^{1/2 - 2/p}
  \\
  & =
 \frac{2^5 c_1^2 \gamma c_6^3 \cdot d^3 \log(n)^{7/2}}{c_5 (1/2 - 2/p)} \cdot \frac{1}{n^{1/2}}
  \\
  & = \littleO[\Big]{
	n^{(1/2) \cdot (1 - 2/p)} \log(n)^{-6} \log(n)^{7/2} \frac{1}{n^{1/2}}
   },
\end{align*}
which also goes to zero for $p \geq 5$.

	This ensures that all terms go to zero in probability, which establishes the lemma.
\end{proof}

These two lemmas establish our central limit theorem, which we reproduce here in terms of the assumptions in the supplement.

\begin{reftheorem}{\mainref{thm:clt}*}
	Under Assumptions~
	\ref{supp-assumption:design},
	\ref{supp-assumption:outcome-regularity},
	\ref{supp-assumption:whole-covariate-regularity},
	\ref{supp-assumption:covariate-dimension}, and
	\ref{supp-assumption:not-superefficient},
	the limiting distribution of the Horvitz--Thompson estimator under the Gram--Schmidt Walk Design is the standard normal distribution:
	\[
	\frac{\htest - \ate}{\sqrt{\Var{\htest}}} \xrightarrow{d} \mathcal{N}(0,1)
	\enspace.
	\]
\end{reftheorem}
\begin{proof}
	Define the random variables
	$
	X_t = \frac{1}{\sigma} \gsdt \iprod{ \gsuvt , \pomv }
	$,
	where $\sigma^2 = n^2 \cdot \Var{ \htest }$ is the $n^2$-normalized variance of the estimator, $\pomv$ is the potential outcome vector, and $\gsdt$ and $\gsuvt$ are the step size and direction, respectively.
	Observe that $\sum_{t=1}^n X_t = (\htest - \ate) / \sqrt{\Var{\htest}}$.

	Lemma~\ref{lemma:first-clt-condition} establishes that the variable $\max_{t \leq n} \abs{ X_{t} }$ converges in mean square to zero.
	This implies the first two conditions of Theorem~\ref{thm:martingale-clt-result}, namely that the mean square of $\max_{t \leq n} \abs{X_{t} }$ is bounded uniformly and that $\max_{t \leq n} \abs{ X_{t} }$ converges in probability to $0$.
	Next, Lemma~\ref{lemma:second-clt-condition} show that the second condition of Theorem~\ref{thm:martingale-clt-result} is satisfied, namely that $\sum_{t=1}^n X_t \xrightarrow{p} 1$.
	Thus, we apply Theorem~\ref{thm:martingale-clt-result} to obtain the desired result.
\end{proof}

\subsection{Conservative Variance Estimator (Theorem~\mainref{thn:var-est-cons})} \label{sec:supp-var-est}

In this section, we show that the variance estimator proposed in Section~\mainref{sec:var-est} is consistent for  the upper bound on the asymptotic variance.
Recall that the asymptotic variance is bounded by
\[
\limvar = \min_{ \lfx \in \Reals^d } \frac{1}{n}  \sumin \paren[\Big]{ (\poai + \pobi) - \iprod{\xvi , \lfx} }^2
\enspace.
\]
We proposed the upper bound on the asymptotic variance,
\[
\vb = \frac{2}{n} R_a + \frac{2}{n} R_b \enspace,
\]
where
\[
R_a = \min_{\lfx \in \Reals^d} \sumin \paren{ \poai -   \iprod{\xvi , \lfx} }^2
\quadand
R_b = \min_{\lfx \in \Reals^d} \sumin \paren{ \pobi -   \iprod{\xvi , \lfx} }^2
\enspace.
\]

We begin by proving Proposition~\mainref{prop:rewrite-limvar}, which describes a decomposition of the limiting variance which motivates this upper bound.

\begin{refproposition}{\mainref{prop:rewrite-limvar}}
	\rewritelimvar
\end{refproposition}

\begin{proof}
	Define the $n$-by-$n$ symmetric matrix $\mat{M}$ as
	\[
	\mat{M} = \paren[\Big]{ \unitM - \xM \paren{\xM^\tran \xM}^{-1} \xM^\tran }
	\enspace,
	\]
	and observe that the quadratic form in $\mat{M}$ is the norm of the residual of regression covariates.
	In particular, for all vectors $\vec{v} \in \Reals^n$ , we have that
	\[
	\vec{v}^\tran \mat{M} \vec{v}
	= \min_{\lfx} \norm{ \vec{v} - \xM \lfx }^2
	\enspace.
	\]
	Using this, and rearranging terms, we have that
	\begin{align*}
		\min_{\lfx} \norm{ \pomv - \xM \lfx }^2
		&= \pomv^\tran \mat{M} \pomv \\
		&= (\poav + \pobv)^\tran \mat{M} (\poav + \pobv) \\
		&= \poav^\tran \mat{M} \poav + \pobv^\tran \mat{M} \pobv + 2 \poav^\tran \mat{M} \pobv \\
		&= 2 \poav^\tran \mat{M} \poav + 2 \pobv^\tran \mat{M} \pobv - \paren[\Big]{\poav^\tran \mat{M} \poav + \pobv^\tran \mat{M} \pobv - 2 \poav^\tran \mat{M} \pobv} \\
		&= 2 \poav^\tran \mat{M} \poav + 2 \pobv^\tran \mat{M} \pobv - (\poav - \pobv)^\tran \mat{M} (\poav - \pobv) \\
		&= 2 \min_{\lfx} \norm{ \poav - \xM \lfx }^2 + 2 \min_{\lfx} \norm{ \pobv - \xM \lfx }^2 - \min_{\lfx} \norm{ \vec{\ate} - \xM \lfx }^2
		\enspace.
		 \qedhere
	\end{align*}
\end{proof}

Corollary~\mainref{corollary:vb-is-vb} follows directly from this proposition, establishing that $\vb \geq \limvar$.

\begin{refcorollary}{\mainref{corollary:vb-is-vb}}
	\vbisvb
\end{refcorollary}
\begin{proof}
	Use the decomposition of the limiting variance in Proposition~\mainref{prop:rewrite-limvar}, and apply the inequality
	\[
	\min_{\lfx} \norm{ \vec{\ate} - \xM \lfx }^2
	\geq 0
	\enspace.
	\]
	This inequality is tight whenever the vector of individual treatment effects is perfectly explained by a linear function of the covariates.
\end{proof}

Recall that our variance estimator uses the Horvitz--Thompson estimation approach.
We define the estimated potential outcomes $\widehat{\poav} \in \Reals^n$ and $\widehat{\pobv} \in \Reals^n$ as
\[
	\widehat{\poav} = \paren[\Big]{ \frac{\indicator{\ze{1} = 1}}{\Pr{\ze{1} = 1}} \ooe{1},  \dotsc , \frac{\indicator{\ze{n} = 1}}{\Pr{\ze{n} = 1}} \ooe{n} }
	\quadand
	\widehat{\pobv} = \paren[\Big]{ \frac{\indicator{\ze{1} = -1}}{\Pr{\ze{1} = -1}} \ooe{1},  \dotsc , \frac{\indicator{\ze{n} = -1}}{\Pr{\ze{n} = -1}} \ooe{n} }
	\enspace.
\]
Define the estimated minimizers to be
\begin{align*}
	\widehat{\lfx}_a
	&= \argmin_{\lfx \in \Reals^d} \sum_{i=1}^n \paren[\Big]{ \widehat{\poai} - \iprod{\lfx, \xvi} }^2
	= (\xM^\tran \xM)^{-1} \xM^\tran \widehat{\poav} \\
	\quadand
	\widehat{\lfx}_b
	&= \argmin_{\lfx \in \Reals^d} \sum_{i=1}^n \paren[\Big]{ \widehat{\pobi} - \iprod{\lfx, \xvi} }^2
	= (\xM^\tran \xM)^{-1} \xM^\tran \widehat{\pobv}
	\enspace.
\end{align*}
Define the estimated residuals as
\[
	\widehat{R}_a = \sumin \frac{\indicator{\zi = 1}}{\Pr{\zi = 1}} \paren[\Big]{ \ooi -   \iprod{\xvi , \widehat{\lfx}_a } }^2
	\quadand
	\widehat{R}_b = \sumin \frac{\indicator{\zi = -1}}{\Pr{\zi = -1}} \paren[\Big]{ \ooi -   \iprod{\xvi , \widehat{\lfx}_b } }^2
	\enspace.
\]
Finally, define the variance bound estimator to be
\[
	\evb = \frac{2}{n} \widehat{R}_a  + \frac{2}{n} \widehat{R}_b
	\enspace.
\]

In the rest of the section, we show that the variance estimator converges in probability to the variance upper bound.
Consequently, this will yield that the ratio of the variance estimator to the limiting variance converges in probability to a value larger than 1.
This is the central aspect of the variance estimator which enables asymptotically valid confidence intervals.

First, we demonstrate that the fourth moment of the residuals of the individual regressions are bounded, under the covariate and outcome regularity conditions.

\begin{lemma} \label{lemma:fourth-moment-residuals}
	Under Assumptions~\ref{supp-assumption:outcome-regularity} and \ref{supp-assumption:whole-covariate-regularity},
	the fourth moment of the individual residual of the individual regressions are bounded as
	\begin{align*}
	\frac{1}{n} \sum_{i=1}^n \paren[\Big]{ \poai - \iprod{ \xvi , \lfx_a } }^4
	&\leq 2^3 c_1^4 \paren[\Big]{ 1 + \frac{c_2^{2}}{c_3^4} \cdot d^{2} \log(n)^{2} } \\
	\frac{1}{n} \sum_{i=1}^n \paren[\Big]{ \pobi - \iprod{ \xvi , \lfx_b } }^4
	&\leq 2^3 c_1^4 \paren[\Big]{ 1 + \frac{c_2^{2}}{c_3^4} \cdot d^{2} \log(n)^{2} }
	\enspace.
	\end{align*}
\end{lemma}
\begin{proof}
	We prove the first inequality involving residuals from the regression on the treatment outcomes, as the second inequality is identical.
	As $p \geq 5$, the Power Mean Inequality and Assumption~\ref{supp-assumption:outcome-regularity} tells us that
	\[
		\paren[\Big]{ \frac{1}{n} \sumin \poai^4 }^{1/4}
		\leq \paren[\Big]{ \frac{1}{n} \sumin \poai^p }^{1/p}
		\leq c_1.
	\]
	The Power Mean Inequality also tells us that $(a+b)^4 \leq 2^3(a^4 + b^4)$.
	Combining this inequality with the previous yields
	\begin{align*}
		\frac{1}{n} \sum_{i=1}^n \paren[\Big]{ \poai - \iprod{ \xvi , \lfx_a } }^4
		&\leq 2^3 \frac{1}{n} \sum_{i=1}^n \poai^4
			+ 2^3 \frac{1}{n} \sum_{i=1}^n \iprod{ \xvi , \lfx_a }^4
			\\
		&\leq 2^3 c_1^4 + 2^3 \frac{1}{n} \sum_{i=1}^n \iprod{ \xvi , \lfx_a }^4
		\enspace.
	\end{align*}
	We now examine the sum.
	By definition of $\lfx_a$, we have that
	\begin{align*}
		\frac{1}{n} \sum_{i=1}^n \iprod{ \xvi , \lfx_a }^4
		&= \frac{1}{n} \sum_{i=1}^n \iprod{ \xM^\tran \vec{e}_i , (\xM^\tran \xM)^{-1} \xM^\tran \poav }^4 \\
		&= \frac{1}{n} \sum_{i=1}^n \iprod{ \xM (\xM^\tran \xM)^{-1} \xM^\tran \vec{e}_i , \poav }^4 \\
		&= \frac{1}{n} \sum_{i=1}^n \iprod{ \mat{H} \vec{e}_i , \poav }^4 \\
		&\leq \frac{1}{n} \sum_{i=1}^n \norm{\mat{H} \vec{e}_i}_q^4 \cdot \norm{\poav}_4^4 \enspace,
		\intertext{
			where $\mat{H} = \xM (\xM^\tran \xM)^{-1} \xM^\tran$ is the projection matrix and $q = 4/3$ is chosen so that $1/4 + 1/q = 1$ and the inequality is an application of H{\"o}lder's inequality.
			Next, by rearranging terms we obtain
		}
		&= \paren[\Big]{ \frac{1}{n} \sumin \abs{\poai}^4 } \sum_{i=1}^n \norm{\mat{H} \vec{e}_i}_q^4 \\
		&\leq c_1^4 \sum_{i=1}^n \norm{\mat{H} \vec{e}_i}_q^4 \\
		\intertext{Because  $q \leq 2$  we can apply Lemma~\ref{lemma:p-norm-inequality}
		to upper bound the $q$-norm in terms of the $2$-norm}
		&\leq c_1^4
			\sum_{i=1}^n \left( n^{1/q - 1/2} \norm{\mat{H} \vec{e}_i}_2 \right)^4  \\
		& = c_1^4 n^{4 \cdot (1/q - 1/2)} \sum_{i=1}^n \norm{\mat{H} \vec{e}_i}_2^4 \\
		&\leq c_1^4 n \cdot n \cdot \max_{i \in [n]} \norm{\mat{H} \vec{e}_i}_2^4 \\
		&= c_1^4 n^{2} \cdot \Xi(\xM)^{2} \\
		&\leq c_1^4 n^{2} \cdot  \paren[\Big]{ \frac{c_2}{c_3^2} \cdot \frac{d}{n} \log(n) }^{2} \\
		&= \frac{c_1^4  c_2^{2}}{ c_3^4 } \cdot d^{2} \log(n)^{2}
		\enspace,
	\end{align*}
	where the final inequality follows from Corollary~\ref{corollary:entire-incoherence}, which bounds the incoherence of the covariate matrix.
	Thus, we have the desired result, which is
	\[
	\frac{1}{n} \sum_{i=1}^n \paren[\Big]{ \poai - \iprod{ \xvi , \lfx_a } }^4
	\leq 2^3 c_1^4 \paren[\Big]{ 1 + \frac{c_2^{2}}{c_3^4} \cdot d^{2} \log(n)^{2} }
	\enspace.
	\qedhere
	\]
\end{proof}

The following lemma shows that the error of the predicted best linear fit to the true best linear fit does not grow too quickly with $n$.

\begin{lemma} \label{lemma:beta-estimation}
	Under Assumptions~\ref{supp-assumption:design},
	\ref{supp-assumption:outcome-regularity}, and \ref{supp-assumption:whole-covariate-regularity},
	we have that
	\[
	\E[\Big]{ \norm{ \xM \paren[\big]{ \lfx_a - \widehat{\lfx}_a} }^2 }  \leq \frac{c c_1^2 c_2}{c_3^2} d \log(n)
	\quadand
	\E[\Big]{ \norm{ \xM \paren[\big]{ \lfx_b - \widehat{\lfx}_b} }^2 }  \leq \frac{c c_1^2 c_2}{c_3^2} d \log(n)
	\enspace.
	\]
\end{lemma}
\begin{proof}
	Recall that $\lfx_a = (\xM^\tran \xM)^{-1} \xM^\tran \poav$ and $\widehat{\lfx}_a = (\xM^\tran \xM)^{-1} \xM^\tran \widehat{\poav}$.
	Thus, we can write the error as
	\[
	\lfx_a - \widehat{\lfx}_a
	= (\xM^\tran \xM)^{-1} \xM^\tran (\poav - \widehat{\poav})
	\enspace.
	\]
	Observe that these error terms are
	\begin{align*}
	\widehat{\poav} - \poav
	&=  \paren[\Big]{ \frac{\indicator{\ze{1} = 1}}{\Pr{\ze{1} = 1}} \ooe{1},  \dotsc , \frac{\indicator{\ze{n} = 1}}{\Pr{\ze{n} = 1}} \ooe{n} }
		- \paren[\Big]{ \poae{1},  \dotsc , \poae{n} } \\
	&=  \paren[\Bigg]{ \paren[\Big]{\frac{\indicator{\ze{1} = 1}}{\Pr{\ze{1} = 1}} - 1} \cdot \poae{1},  \dotsc , \paren[\Big]{\frac{\indicator{\ze{n} = 1}}{\Pr{\ze{n} = 1}} - 1} \poae{n} } \\
	&= \paren[\big]{ \ze{1} \cdot \poae{1},  \dotsc , \ze{n} \cdot \poae{n} }  \\
	&= \zv \circ \poav \enspace,
	\end{align*}
	where the second to last equality follows
	because the treatment probabilities are uniform for each unit, i.e. $\Pr{\zi = 1} = 1/2$ for each $i \in [n]$.
	In the last equality, we use $\vec{x} \circ \vec{y}$ to denote the Hadamard product between vectors $\vec{x}$ and $\vec{y}$.
	Thus, we have that
	\[
	\norm{ \xM \paren[\big]{ \lfx_a - \widehat{\lfx}_a} }^2
	= \norm{ \xM (\xM^\tran \xM)^{-1} \xM^\tran ( \poav \circ \zv ) }
	= \norm{ \mat{H} ( \poav \circ \zv ) }
	\enspace,
	\]
	where $\mat{H} = \xM (\xM^\tran \xM)^{-1} \xM^\tran$ is the projection matrix onto the column span of $\xM$, also known as the ``hat matrix''.
	Let $\mat{U}$ be the $n$-by-$d$ orthonormal matrix such that $\mat{H} = \mat{U} \mat{U}^\tran$.
	Because $\mat{U}$ is orthornormal, we have that for all $\vec{v} \in \Reals^n$,
	\[
	\norm{ \mat{H} \vec{v} }_2
	= \norm{ \mat{U} \mat{U}^\tran \vec{v} }_2
	= \norm{ \mat{U}^\tran \vec{v} }_2
	\enspace,
	\]
	and thus by setting $\vec{v} = \poav \circ \zv$, we have that
	\[
	\norm{ \xM \paren[\big]{ \lfx_a - \widehat{\lfx}_a} }^2
	= \norm{ \mat{H} ( \poav \circ \zv ) }^2
	= \norm{ \mat{U}^\tran  ( \poav \circ \zv ) }^2
	\enspace.
	\]
	Using this, we have that the expected norm will be
	\begin{align*}
		\E[\Big]{ \norm{ \xM \paren[\big]{ \lfx_a - \widehat{\lfx}_a} }^2 }
		&= \E[\Big]{\norm{\mat{U}^\tran  ( \poav \circ \zv ) }^2 } \\
		&= \E[\Big]{ \sum_{\ell=1}^d \iprod{\vec{e}_\ell , \mat{U}^\tran ( \poav \circ \zv ) }^2 } \\
		&=  \sum_{\ell=1}^d \E[\Big]{ \iprod{\mat{U} \vec{e}_\ell ,   \poav \circ \zv }^2 }
		\enspace,
	\end{align*}
	where $\vec{e}_\ell \in \Reals^d$ is a standard basis vector.
	Note that for all $\vec{v} \in \Reals^n$, we have that
	\begin{align*}
		\E[\Big]{ \iprod{\vec{v}, \poav \circ \zv  }^2 }
		&= \E[\Big]{ \iprod{\zv, \vec{v} \circ \poav  }^2 } \\
		&= ( \vec{v} \circ \poav )^\tran \Cov{ \zv } ( \vec{v} \circ \poav ) \\
		&\leq \frac{1}{\vto} \norm{ ( \vec{v} \circ \poav ) }_2^2
			&\text{(Theorem~\mainref{theorem:gsw-satisfies-discrepancy})}\\
		&\leq c \norm{ ( \vec{v} \circ \poav ) }_2^2
			&\text{(Assumption~\ref{supp-assumption:design})}
		\enspace,
	\end{align*}
	where the first inequality follows from the operator norm bound on $\Cov{\zv}$ guaranteed by Gram--Schmidt Walk design (Theorem~\mainref{theorem:gsw-satisfies-discrepancy}) and the second inequality uses Assumption~\ref{supp-assumption:design}, which bounds $\vto$ away from zero.
	Using this above with $\vec{v} = \mat{U} \vec{e}_\ell$ together with the coherence bound on the covariate matrix (Corollary~\ref{corollary:entire-incoherence}) and the assumption of bounded outcomes (Assumption~\ref{supp-assumption:outcome-regularity}), we have that
	\begin{align*}
		\E[\Big]{ \norm{ \xM \paren[\big]{ \lfx_a - \widehat{\lfx}_a} }^2 }
		&= \sum_{\ell=1}^d \E[\Big]{ \iprod{\mat{U} \vec{e}_\ell ,   \poav \circ \zv }^2 } \\
		&\leq c \sum_{\ell=1}^d \norm{ ( \mat{U} \vec{e}_\ell \circ \poav ) }_2^2 \\
		&= c \sum_{\ell=1}^d \sumin \iprod{ \vec{e}_i , \mat{U} \vec{e}_\ell \circ \poav }^2 \\
		&= c  \sum_{\ell=1}^d \sumin \mat{U}(i, \ell)^2 \poai^2 \\
		&= c \sumin \poai^2 \cdot \sum_{\ell=1}^d \mat{U}(i, \ell)^2 \\
		&= c \sumin \poai^2 \cdot \norm{ \mat{H} \vec{e}_i }^2 \\
		&\leq c \paren[\Big]{ \sumin \poai^2 } \cdot \paren[\Big]{ \frac{c_2}{c_3^2} \frac{d}{n} \log(n) }
			&\text{(Corollary~\ref{corollary:entire-incoherence})}\\
		&= \frac{c c_2}{c_3^2} \paren[\Big]{ \frac{1}{n} \sumin \poai^2 } d \log(n) \\
		&\leq \frac{c c_1^2 c_2}{c_3^2} d \log(n)
		\enspace,
			&\text{(Assumption~\ref{supp-assumption:outcome-regularity})}
	\end{align*}
	which yields the desired result.
\end{proof}

Before continuing, we define standard order in probability notation.
The notation below is commonly used in mathematical statistics and probability theory, but we restate it here for completeness.
Given a sequence of real-valued random variables $\setb{ X_n }_{n=1}^\infty$ and a sequence of real numbers $\setb{a_n}_{n=1}^\infty$, we write $X_n = \bigOp{a_n}$ if for any $\epsilon > 0$, there exists a finite $M > 0$ and $N > 0$ such that
\[
\Pr[\Big]{ \abs[\Big]{ \frac{ X_n }{ a_n } } > M } < \epsilon
\quad
\forall \ n \geq N
\enspace.
\]
Similarly, we write $X_n = \littleOp{a_n}$ if $X_n / a_n$ converges to zero in probability, i.e. for all $\epsilon > 0$,
\[
\lim_{n \to \infty} \Pr[\Big]{ \abs[\Big]{ \frac{ X_n }{ a_n } } \geq \epsilon } = 0
\enspace.
\]

The final technical lemma combines the previously derived lemmas to show that the estimated individual residuals converge to the true residuals in probability at a sufficiently fast rate.

\begin{lemma} \label{lemma:regression-terms-converge}
	Under Assumptions~\ref{supp-assumption:design}, \ref{supp-assumption:outcome-regularity}, and \ref{supp-assumption:whole-covariate-regularity},
	the difference between the individual true and estimated regression terms is bounded in probability as follows:
	\begin{align*}
	\frac{2}{n} \paren[\Big]{ R_a - \widehat{R}_a } &= \bigOp[\Big]{ \frac{d \log(n)}{\sqrt{n}}  }  \\
	\frac{2}{n} \paren[\Big]{ R_b - \widehat{R}_b } &= \bigOp[\Big]{ \frac{d \log(n)}{\sqrt{n}}  } \enspace.
	\end{align*}
\end{lemma}
\begin{proof}
	We only prove the first inequality corresponding to $R_a$ term, as the proof for $R_b$ is the same.
	Because $\indicator{\zi = 1} = 0$ for $\zi = -1$,
	\[ \widehat{R}_a =
	\sumin \frac{\indicator{\zi = 1}}{\Pr{\zi = 1}} \paren[\Big]{ \poai -   \iprod{\xvi , \widehat{\lfx}_a } }^2 \enspace.
	\]
	So, the difference between the true $R_a$ and the estimator $\widehat{R}_a$ can be written as
	\[
	R_a - \widehat{R}_a
	= \sumin \paren{ \poai -   \iprod{\xvi , \lfx_a} }^2
	- \sumin \frac{\indicator{\zi = 1}}{\Pr{\zi = 1}} \paren[\Big]{ \poai -   \iprod{\xvi , \widehat{\lfx}_a } }^2
	\enspace.
	\]
	By rearranging and expanding terms, we have that
	\begin{align*}
		\paren[\Big]{ \poai -   \iprod{\xvi , \widehat{\lfx}_a } }^2
		&= \paren[\Big]{ \paren[\big]{\poai - \iprod{\xvi , \lfx_a} } - \paren[\big]{ \iprod{\xvi , \widehat{\lfx}_a } - \iprod{\xvi , \lfx_a} }}^2 \\
		&= \paren[\Big]{ \paren[\big]{\poai - \iprod{\xvi , \lfx_a} } - \paren[\big]{ \iprod{\xvi , \widehat{\lfx}_a - \lfx_a} }}^2 \\
		&= \paren[\Big]{\poai - \iprod{\xvi , \lfx_a}}^2 + \iprod{\xvi , \widehat{\lfx}_a - \lfx_a}^2
		- 2 \paren[\Big]{\poai - \iprod{\xvi , \lfx_a}} \iprod{\xvi , \widehat{\lfx}_a - \lfx_a}
		\enspace.
	\end{align*}
	Plugging this into the above, we have that the difference between the true and estimated residual can be broken into two terms:
	\begin{align*}
		\frac{2}{n} \paren[\Big]{ R_a - \widehat{R}_a }
		&=
		\underbrace{\frac{2}{n} \sumin \paren[\Big]{ 1 - \frac{\indicator{\zi = 1}}{\Pr{\zi = 1}} } \paren{ \poai -   \iprod{\xvi , \lfx_a} }^2 }_{ \triangleq T_1 \text{ (Term 1)}}
		+
		\underbrace{\frac{2}{n} \sumin \frac{\indicator{\zi = 1}}{\Pr{\zi = 1}} \iprod{\xvi , \widehat{\lfx}_a - \lfx_a}^2}_{ \triangleq T_2 \text{ (Term 2)}} \\
		&\qquad - \underbrace{\frac{4}{n} \sumin \frac{\indicator{\zi = 1}}{\Pr{\zi = 1}} \paren[\Big]{\poai - \iprod{\xvi , \lfx_a}} \iprod{\xvi , \widehat{\lfx}_a - \lfx_a}}_{ \triangleq T_3 \text{ (Term 3)}}
	\end{align*}

	We will now bound each of these terms in probability, which will yield the desired result.
	Throughout the proof, we define $\tilde{\poai} = \poai - \iprod{\xvi, \lfx_a}$ to be the residual corresponding to unit $i \in [n]$ and $\tilde{\poav} = \paren{ \tilde{\poae{1}} \dotsc  \tilde{\poae{n}} }$.
	In an abuse of notation, we will write $\tilde{\poav}^2 = \paren{ \tilde{\poae{1}}^2 \dotsc  \tilde{\poae{n}}^2 }$.

 	\textbf{Term 1}:
 	Observe that because the treatment probabilities are uniform for each unit, i.e. $\Pr{\zi = 1} = 1/2$ for each $i \in [n]$, we have that $1 - \indicator{\zi = 1} / \Pr{\zi = 1} = - \zi$.
 	Thus, we can write term 1 as
 	\[
 	T_1
 	= \frac{2}{n} \sumin \paren[\Big]{ 1 - \frac{\indicator{\zi = 1}}{\Pr{\zi = 1}} } \paren{ \poai -   \iprod{\xvi , \lfx_a} }^2
 	= -\frac{2}{n} \sumin \zi \tilde{\poai}^2 \enspace.
 	\]
 	Because $\E{ \zi } = 0$, we have that $\E{ T_1 } = 0$.
 	Moreover, we may bound the variance of the first term as
 	\begin{align*}
 	\Var{ T_1 }
 	&= \Var[\Bigg]{ -\frac{2}{n} \sumin \zi \tilde{\poai}^2  } \\
 	&= \frac{4}{n^2} (\tilde{\poav}^2)^\tran \Cov{\zv} (\tilde{\poav}^2) \\
 	&\leq \frac{4}{n^2} \norm{\Cov{\zv}} \sumin \tilde{\poai}^4 \\
 	&\leq \frac{4}{n^2} \cdot \frac{1}{\vto} \sumin \tilde{\poai}^4
	& \text{(by Theorem~\mainref{theorem:gsw-satisfies-discrepancy}) }\\
 	&\leq \frac{4 c}{n} \cdot \paren[\Big]{ \frac{1}{n} \sumin \tilde{\poai}^4 }
	& \text{(by Assumption~\ref{supp-assumption:design})}\\
 	&\leq \frac{4 c}{n} \cdot 2^3 c_1^4 \paren[\Big]{ 1 + \frac{c_2^{2}}{c_3^4} \cdot d^{2} \log(n)^{2} }
 	\enspace,
 	\end{align*}
	where the last inequality follows from Lemma~\ref{lemma:fourth-moment-residuals}.
 	Thus, we have that $\Var{T_1} = \bigO{d^2 \log(n)^2/n}$,
 	and by Chebyshev's inequality,  $T_1 = \bigOp{ d \log(n) / \sqrt{n} }$.

 	\textbf{Term 2}:
 	To analyze the second term, we use Markov's inequality together with Lemma~\ref{lemma:beta-estimation}:
 	\begin{align*}
 		\E[\Big]{T_2}
 		&= \E[\Big]{ \frac{2}{n} \sumin \frac{\indicator{\zi = 1}}{\Pr{\zi = 1}} \iprod{\xvi , \widehat{\lfx}_a - \lfx_a}^2 } \\
 		&\leq \frac{4}{n} \E[\Big]{ \sumin  \iprod{\xvi , \widehat{\lfx}_a - \lfx_a}^2 }  \\
 		&= \frac{4}{n} \E[\Big]{ \norm{\xM \paren{\widehat{\lfx}_a - \lfx_a}}^2 } \\
 		&\leq 4 \frac{c c_1^2 c_2}{c_3^2} \frac{d \log(n)}{n}
 		\enspace.
 	\end{align*}
 	As $T_2$ is a non-negative random variable, Markov's inequality implies that $T_2 = \bigOp[\big]{\frac{d \log(n)}{n}}$.
 	Observe that the term $T_2$ is asymptotically order dominated by $T_1$.

	\textbf{Term 3}:
	Observe that by the Cauchy--Schwarz inequality we can bound the expectation of the absolute value of the third term as
	\begin{align*}
	\E[\Big]{\abs{T_3}}
	&= \E[\Bigg]{ \abs[\Bigg]{ \frac{4}{n} \sumin \frac{\indicator{\zi = 1}}{\Pr{\zi = 1}} \paren[\Big]{\poai - \iprod{\xvi , \lfx_a}} \iprod{\xvi , \widehat{\lfx}_a - \lfx_a}  } }  \\
	&\leq \frac{4}{n} \E[\Bigg]{ \paren[\Bigg]{ \sumin \frac{\indicator{\zi = 1}}{\Pr{\zi = 1}^2}\paren[\Big]{\poai - \iprod{\xvi , \lfx_a}}^2 }^{1/2} \cdot \paren[\Bigg]{ \sumin \iprod{\xvi , \widehat{\lfx}_a - \lfx_a}^2  }^{1/2} } \\
	&\leq \frac{8}{n} \E[\Bigg]{ \paren[\Bigg]{ \sumin \paren[\Big]{\poai - \iprod{\xvi , \lfx_a}}^2 }^{1/2} \cdot \paren[\Bigg]{ \sumin \iprod{\xvi , \widehat{\lfx}_a - \lfx_a}^2  }^{1/2} } \\
	&\leq  \frac{8}{n} \paren[\Big]{ \sumin \poai^2}^{1/2} \cdot \E[\Bigg]{ \paren[\Bigg]{ \sumin \iprod{\xvi , \widehat{\lfx}_a - \lfx_a}^2  }^{1/2} }
	\enspace,
	\intertext{where the final inequality uses the fact that the sum of squared residuals is at most the sum of squared outcomes, and that both of these terms are non-random and so may be pulled out of the expectation. Now, using Assumption~\ref{supp-assumption:outcome-regularity}, we have that we may bound the outcomes as }
	&= \frac{8}{n^{1/2}} \cdot  \paren[\Big]{ \frac{1}{n} \sumin \poai^2}^{1/2} \cdot \E[\Bigg]{ \paren[\Bigg]{ \sumin \iprod{\xvi , \widehat{\lfx}_a - \lfx_a}^2  }^{1/2} } \\
	&\leq 8 c_1 \cdot \frac{1}{n^{1/2}} \cdot  \E[\Bigg]{ \paren[\Bigg]{ \sumin \iprod{\xvi , \widehat{\lfx}_a - \lfx_a}^2  }^{1/2} } \\
	\intertext{Using Jensen's inequality and Lemma~\ref{lemma:beta-estimation}, we have that}
	&\leq 8 c_1 \cdot \frac{1}{n^{1/2}} \cdot \sqrt{ \E[\Big]{ \norm{ \xM \paren[\big]{ \lfx_a - \widehat{\lfx}_a} }^2 } }
		&\text{(Jensen)} \\
	&\leq 8 c_1 \cdot \frac{1}{n^{1/2}} \cdot \sqrt{ \frac{c c_1^2 c_2}{c_3^2} d \log(n) }
		&\text{(Lemma~\ref{lemma:beta-estimation})} \\
	&= \frac{8 c_1^2 \sqrt{c c_2} }{c_3} \cdot \sqrt{ \frac{d \log(n)}{n} }
	\enspace.
	\end{align*}
	Thus, by Markov's inequality, we have that $T_3$ is $\bigOp{ \sqrt{d \log(n) / n}}$.
	Observe that this is asymptotically order dominated by $T_1$, which is $\bigOp{ d \log(n) / \sqrt{n} }$.
	Thus, $T_1$ is the leading term among the three terms, which establishes the desired bound.
\end{proof}

The following theorem bounds the error between variance estimator and the variance upper bound in probability.
This theorem subsumes Theorem~\mainref{thn:var-est-cons} in the main paper, which assumes that $p = 5$ and that $d$ is growing sufficiently small in $n$.

\begin{reftheorem}{\mainref{thn:var-est-cons}*}
	Under Assumptions~\ref{supp-assumption:design}, \ref{supp-assumption:outcome-regularity}, and \ref{supp-assumption:whole-covariate-regularity},
	the difference between the variance estimator and the variance upper bound converges in probability at the following rate:
	\[
	\evb - \vb = \bigOp[\Big]{ \frac{d \log(n)}{\sqrt{n}}  }
	\enspace.
	\]
\end{reftheorem}
\begin{proof}
	Observe that the difference between the variance estimator and the variance bound may be written as
	\[
	\evb - \vb
	= \paren[\Big]{\frac{2}{n} \widehat{R}_a + \frac{2}{n} \widehat{R}_b} - \paren[\Big]{\frac{2}{n} R_a + \frac{2}{n} R_b }
	= \frac{2}{n} \paren[\Big]{ \widehat{R}_a - R_a } + \frac{2}{n} \paren[\Big]{ \widehat{R}_b - R_b }
	\enspace.
	\]
	The result follows by applying Lemma~\ref{lemma:regression-terms-converge}.
\end{proof}

\subsection{Asymptotically Valid Confidence Intervals (Theorem~\mainref{prop:confidence-intervals})} \label{sec:supp-valid-interval-proof}

In this section, we establish that the confidence intervals proposed in Section~\mainref{sec:confidence-intervals} are asymptotically valid.
Throughout the proof, we use $\Phi^{-1} : [0,1] \to \Reals$ to denote the quantile function of a standard normal distribution and define $\hat\sigma = \sqrt{\evb}$.

\begin{reftheorem}{\mainref{prop:confidence-intervals}*}
	Under Assumptions~\ref{supp-assumption:design}, \ref{supp-assumption:outcome-regularity}, \ref{supp-assumption:whole-covariate-regularity}, \ref{supp-assumption:covariate-dimension}, and \ref{supp-assumption:not-superefficient},
	 and further supposing that the design parameter approaches one at a sufficiently slow rate, so that $1 - \vto = \littleOmega{\maxnormX^2 / n}$,
	 the proposed confidence intervals asymptotically cover at least at the nominal rates:
	 \[
	 \liminf_{n\to \infty} \Pr[\Big]{ - \Phi^{-1}(1 - \ciprob / 2) \hat\sigma \leq n^{1/2} \paren{ \ate - \htest } \leq \Phi^{-1}(1 - \ciprob / 2) \hat\sigma } \geq 1 - \ciprob.
	 \]
\end{reftheorem}
\begin{proof}
	Define the random variable $Z = \frac{\ate - \htest}{\sqrt{\Var{\htest}}}$.
	Under these assumptions, Theorem~\mainref{thm:clt}* ensures that $Z \xrightarrow{d} \mathcal{N}(0,1)$.
	Define the random variable $Z' = \sqrt{n} \cdot \frac{\ate - \htest}{\sqrt{\evb}}$ and observe that
	\[
	Z'
	= \sqrt{n} \cdot \frac{\ate - \htest}{\sqrt{\evb}}
	= \frac{\ate - \htest}{\sqrt{\Var{\htest}}}
	\cdot \sqrt{ \frac{n \cdot \Var{\htest}}{\limvar} \cdot \frac{\limvar}{\vb} \cdot \frac{\vb}{\evb} }
	= Z \cdot \sqrt{ \frac{n \cdot \Var{\htest}}{\limvar} \cdot \frac{\limvar}{\vb} \cdot \frac{\vb}{\evb} }
	\enspace.
	\]
	Because $1 - \vto = \littleOmega{\maxnormX^2 / n}$, we have by Theorem~\mainref{thm:asymp-variance} that $\limsup_{n \to \infty} {n \Var{\htest}} / ({\limvar}) \leq 1$.
	By Corollary~\mainref{corollary:vb-is-vb}, $\limsup_{n \to \infty } {\limvar}/{\vb} \leq 1$.
	By Theorem~\mainref{thn:var-est-cons}, $ {\vb } / {\evb } \xrightarrow{p} 1$.
	Thus, by Slutsky's theorem, we have that $Z'$ is asymptotically stochastically dominated by  $\mathcal{N}(0,1)$.

	Now, we can evaluate the asymptotic probability of coverage.
	Namely,
	\begin{align*}
		\liminf_{n \to \infty} & \Pr[\Big]{ - \Phi^{-1}(1 - \ciprob / 2) \hat\sigma \leq n^{1/2} \paren{ \ate - \htest } \leq \Phi^{-1}(1 - \ciprob / 2) \hat\sigma }  \\
		&= \liminf_{n \to \infty} \Pr[\Big]{- \Phi^{-1}(1 - \ciprob / 2) \leq n^{1/2} \cdot  \frac{\ate - \htest}{\sqrt{\evb}} \leq \Phi^{-1}(1 - \ciprob / 2) }  \\
		&= \liminf_{n \to \infty} \Pr[\Big]{- \Phi^{-1}(1 - \ciprob / 2) \leq Z' \leq \Phi^{-1}(\ciprob / 2) }
		\enspace,
		\intertext{where the last equality follows by definition of $Z'$ and symmetry of the normal distribution.
		Let $\setb{F_n}_{n=1}^\infty$ be the sequence of CDF of the sequence of random variables $Z'$.
		Using the definition of the CDF and asymptotic stochastic dominance by standard normal,}
		&= \liminf_{n \to \infty} F_n \paren[\Big]{ \Phi^{-1}(1 - \ciprob / 2) } - F_n \paren[\Big]{ \Phi^{-1}(\ciprob / 2) } \\
		&\geq \Phi \paren[\Big]{ \Phi^{-1}(1 - \ciprob / 2) } - \Phi \paren[\Big]{  \Phi^{-1}(\ciprob / 2) } \\
		&= 1 - \ciprob / 2 - (\ciprob / 2) \\
		&= 1 - \ciprob \enspace. \qedhere
	\end{align*}
\end{proof}

\subsection{Estimator of the Ridge Regression Loss} \label{sec:supp-ridge-estimator}

In this section, we propose an estimator for the ridge regression loss, which mirrors the estimator for the least squares loss.
In particular, it is a consistent estimator for an upper bound on the ridge loss.
Under the asymptotic conditions that we propose in Section~\ref{sec:supp-asymptotic-framework}, the two estimators should converge in distribution; however, this estimator is more conservative as it is based on an upper bound of the finite sample variance (rather than an asymptotic variance) and thus may be more suitable for inference in small samples.

Because the estimator described here mirrors the variance estimator described in Section~\ref{sec:supp-var-est}, we provide sketches of its analysis, rather than detailed proofs.

Recall that the ridge regression loss is defined as
\[
\ridgeloss
= \frac{1}{n} \cdot \min_{\lfx \in \Reals^d} \bracket[\Bigg]{ \frac{1}{\vto} \norm{ \pomv - \xM \lfx }^2 + \frac{\maxnormX^2}{1 -\vto} \norm{\lfx}^2 }
\enspace.
\]
Observe that the ridge loss is defined with respect to the vector which sums the potential outcomes, $\pomv = \poav + \pobv$.
Similar to Corollary~\mainref{corollary:vb-is-vb}, one can decompose the ridge loss to obtain an upper bound.
In particular, define the ridge losses $R_a$ and $R_b$ on treatment and control outcomes as
\[
R_{a} = \min_{\lfx \in \Reals^d} \bracket[\Bigg]{ \frac{1}{\vto} \norm{ \poav - \xM \lfx }^2 + \frac{\maxnormX^2}{1 -\vto} \norm{\lfx}^2 }
\quadand
R_{b} = \min_{\lfx \in \Reals^d} \bracket[\Bigg]{ \frac{1}{\vto} \norm{ \pobv - \xM \lfx }^2 + \frac{\maxnormX^2}{1 -\vto} \norm{\lfx}^2 }
\enspace.
\]
We define the \emph{ridge loss upper bound} as
\[
\rub = \frac{2}{n} R_a + \frac{2}{n} R_b
\enspace.
\]
Using similar techniques used to prove Corollary~\mainref{corollary:vb-is-vb}, we can show that $\rub \geq \ridgeloss$ and, in particular,
\[
\rub - \ridgeloss
= \frac{1}{n}  \min_{\lfx \in \Reals^d} \bracket[\Bigg]{ \frac{1}{\vto} \norm{ \vec{\ate} - \xM \lfx }^2 + \frac{\maxnormX^2}{1 -\vto} \norm{\lfx}^2 }
\geq 0
\enspace.
\]
In order to construct an estimator for the ridge loss upper bound $\rub$, we use the Horvitz--Thompson estimation principle, which mirrors the estimator for the limiting variance discussed in Section~\ref{sec:supp-var-est}.
We define the estimated potential outcomes as $\widehat{\poav} \in \Reals^n$ and $\widehat{\pobv} \in \Reals^n$ as
\[
\widehat{\poav} = \paren[\Big]{ \frac{\indicator{\ze{1} = 1}}{\Pr{\ze{1} = 1}} \ooe{1} \dotsc \frac{\indicator{\ze{n} = 1}}{\Pr{\ze{n} = 1}} \ooe{n} }
\quadand
\widehat{\pobv} = \paren[\Big]{ \frac{\indicator{\ze{1} = -1}}{\Pr{\ze{1} = -1}} \ooe{1} \dotsc \frac{\indicator{\ze{n} = -1}}{\Pr{\ze{n} = -1}} \ooe{n} }
\enspace.
\]
Define the estimated regularized minimizers to be
\begin{align*}
	\widehat{\lfx}_a
	&= \argmin_{\lfx \in \Reals^d} \frac{1}{\vto} \sum_{i=1}^n \paren[\Big]{ \widehat{\poai} - \iprod{\lfx, \xvi} }^2 + \frac{\maxnormX^2}{1-\vto} \norm{\lfx}^2
	= \paren[\Big]{ \frac{\maxnormX^2 (1-\vto)}{\vto} \unitM + \xM^\tran \xM }^{-1} \xM^\tran \widehat{\poav} \\
	\quadand
	\widehat{\lfx}_b
	&= \argmin_{\lfx \in \Reals^d} \frac{1}{\vto} \sum_{i=1}^n \paren[\Big]{ \widehat{\pobi} - \iprod{\lfx, \xvi} }^2 + \frac{\maxnormX^2}{1-\vto} \norm{\lfx}^2
	= \paren[\Big]{ \frac{\maxnormX^2 (1-\vto)}{\vto} \unitM + \xM^\tran \xM }^{-1} \xM^\tran \widehat{\pobv}
	\enspace.
\end{align*}
Define the estimated outcome specific ridge losses as
\begin{align*}
&\widehat{R}_a = \sumin \frac{\indicator{\zi = 1}}{\Pr{\zi = 1}} \paren[\Big]{ \ooi -   \iprod{\xvi , \widehat{\lfx}_a } }^2 + \norm{\widehat{\lfx_a}}^2 \\
\quadand
&\widehat{R}_b = \sumin \frac{\indicator{\zi = -1}}{\Pr{\zi = -1}} \paren[\Big]{ \ooi -   \iprod{\xvi , \widehat{\lfx}_b } }^2 + \norm{\widehat{\lfx_b}}^2
\enspace.
\end{align*}
Finally, define the ridge regression loss bound estimator to be
\[
\erub = \frac{2}{n} \widehat{R}_a  + \frac{2}{n} \widehat{R}_b
\enspace.
\]

Under our asymptotic conditions, we will have that $\rub - \erub = \bigOp{ d \log(n) / \sqrt{n}}$, which mirrors the convergence rates of $\evb - \vb$, guaranteed by Theorem~\mainref{thn:var-est-cons}*.
There are several ingredients required to re-derive these results.

First, define the two matrices $\mat{M}$ and $\mat{M}_\vto$ used the variance estimator and the ridge loss estimator, respectively, as
\[
\mat{M} = \paren[\Big]{ \xM^\tran \xM }^{-1} \xM
\quadand
\mat{M}_\vto = \paren[\Big]{ \frac{\maxnormX^2 (1-\vto)}{\vto} \unitM + \xM^\tran \xM }^{-1} \xM
\enspace.
\]
Observe that the singular values of these two matrices are related in the following way: the $i$th singular values are given as
\[
\sigma_i (\mat{M}) = \frac{1}{\sigma_i(\xM)}
\quadand
\sigma_i (\mat{M}_\vto) = \frac{\sigma_i(\xM)}{ \frac{\maxnormX^2 (1-\vto) }{\vto} + \sigma_i(\xM)^2 }
\enspace,
\]
where $\sigma_i(\xM)$ is the $i$th singular value of $\xM$.
Our asymptotic assumptions stipulate that $\vto$ is bounded away from zero and $\maxnormX^2$ grows at a small rate.
Thus, the singular values of $\mat{M}$ and $\mat{M}_{\vto}$ are asymptotically very similar.
Similarly, the singular vectors of $\mat{M}$ and $\mat{M}_{\vto}$ are precisely the same, so that incoherence properties are preserved.
These are the two main properties one needs to use when translating the rate analysis of $\evb - \vb$ into a formal rate analysis on $\erub - \rub$.

%% file: tex/supp-balancing-analysis.tex
\section{Analysis of Covariate Balancing} \label{sec:balancing-analysis}

In this section, we investigate the covariate balancing properties of the Gram--Schmidt Walk design.
First, we obtain a more refined analysis of the covariate balance than what is guaranteed by Problem~\ref{problem:distributional-discrepancy}.
Then, we use an existing hardness result to show that improving the covariate balance by even a constant factor is computationally intractable.

\subsection{Refined bound on covariate balance}

We now present a more refined analysis of the covariate balancing properties of the Gram--Schmidt Walk design than the operator norm bound in Problem~\ref{problem:distributional-discrepancy}.
We begin by presenting a matrix bound on the covariance matrix of the discrepancy vector of covariates.

\expcommand{\xMcovbound}{%
	Under the Gram--Schmidt Walk design, the covariance matrix of $\xM^\tran \zv$ is bounded in the Loewner order by
	\[
	\Cov{\xM^\tran \zv}
	\preceq
	\paren[\Big]{ \vto \paren{\xM^\tran \xM}^\pinv + \paren{1 - \vto} \paren{ \maxnormX^{2} \projX}^\pinv }^\pinv,
	\]
	where $\projX$ is the orthogonal projection onto the rows of the covariate matrix $\xM$ and $\mat{A}^\pinv$ denotes the pseudo-inverse of $\mat{A}$.
}

\begin{proposition}\label{prop:xM-cov-bound}
	\xMcovbound
\end{proposition}

The proof of Proposition~\ref{prop:xM-cov-bound} appears at the end of this section.
The matrix in the upper bound is the weighted harmonic mean of two $\xdim$-by-$\xdim$ matrices: the Gram matrix $\xM^\tran \xM$ and the scaled projection matrix $\maxnormX^2 \projX$.
When $\vto = 1$, the bound is the Gram matrix, which is the value the covariate matrix takes when the assignments are pair-wise independent.
When $\vto = 0$, the bound is $\maxnormX^2 \projX$, which is a scaled version of the projection onto the span of the covariate vectors.
When the covariate vectors span the entire vector space, $\projX$ is the identity matrix; otherwise, we may interpret $\projX$ as being the identity matrix on the subspace containing the data.
Intermediate values interpolate between the two extremes.

The matrix bound in Proposition~\ref{prop:xM-cov-bound} yields a bound on the variance of the difference between the within-group sums of any linear functions of the covariate vectors.
In particular, applying the definition of the Loewner order and evaluating the quadratic form, we have that for any linear function $\arblfx \in \Reals^d$,
\[
\E[\bigg]{\paren[\Big]{ \sum_{i \in \treated} \iprod{\arblfx, \xvi} - \sum_{i \in \controls} \iprod{\arblfx, \xvi} }^2 }
\leq \arblfx^\tran \paren[\Big]{ \vto \paren{\xM^\tran \xM}^\pinv + \paren{1 - \vto} \paren{ \maxnormX^{2} \projX}^\pinv }^\pinv \arblfx  \enspace.
\]
When $\arblfx$ is a basis vector, then the inequality above bounds the discrepancy of a single covariate between the two groups.
The inequality may be hard to interpret for a general linear function, but experimenters may use the quadratic form on the right hand side to investigate an imbalance in the covariates before running the experiment.
In any case, we may use the operator norm bound on the quadratic form on the right hand side to obtain a worst-case bound over all linear functions:
\begin{equation}\label{eq:var-ub-worst-case}
	\E[\bigg]{\paren[\Big]{ \sum_{i \in \treated} \iprod{\arblfx, \xvi} - \sum_{i \in \controls} \iprod{\arblfx, \xvi} }^2 }
	\leq \frac{\norm{\arblfx}^2}{ \vto \maxeigxG^{-1} + \paren{1 - \vto} \maxnormX^{-2} } \enspace,
\end{equation}
where $\maxeigxG$ is the largest eigenvalue of the Gram matrix $\xM^\tran \xM$.
This bound mirrors the matrix bound in Proposition~\ref{prop:xM-cov-bound}, in that it is a weighted harmonic mean between $\maxeigxG$ and $\maxnormX^2$.
At the extremes, when $\vto$ is either one or zero, the bound is $\maxeigxG$ and $\maxnormX^{2}$, respectively.
Intermediate values of $\vto$ interpolate between the two end points.

The interpolation is monotone: the bound decreases with $\vto$.
This is because $\maxeigxG \geq \maxnormX^{2}$.
This indicates that the imbalance for the worst-case linear function tends to decrease as the parameter approaches zero.
Moreover, \eqref{eq:var-ub-worst-case} shows that the magnitude of $\maxeigxG$ relative to $\maxnormX$ determines the slope of the decrease.
The eigenvalue $\maxeigxG$ is typically considerably larger than the norm $\maxnormX$, so the imbalance tends to decrease quickly with $\vto$.
To see this, let $k \in [n]$ be such that $\norm{\xv{k}} = \maxnormX = \max_{i \in [n]} \norm{\xvi}$, and observe that
\[
\maxeigxG
= \max_{ \norm{\arblfx} \leq 1 } \sumin \iprod{\xvi, \arblfx}^2
\geq \max_{ \norm{\arblfx} \leq 1 } \iprod{\xv{k}, \arblfx}^2
= \norm{\xv{k}}^2 = \maxnormX^2.
\]
The gap introduced by the inequality is large as long as there is not a unit whose covariate vector has disproportionately large norm and is nearly orthogonal to the vectors of the other units.
The fewer outliers there are, the larger $\maxeigxG$ will be relative to $\maxnormX^2$, and the more balance can be achieved.

We remark that no design can improve upon Proposition~\ref{prop:xM-cov-bound} without imposing structural restrictions on the covariates.
In particular, the scaling term $\maxnormX^2$ cannot be improved for general covariate vectors, as discussed in Section~\mainref{sec:new-perspective}.
In the example presented there, the orthogonal and large the covariate vector $\xv{k}$ may be considered an outlier.
Generally speaking, better covariate balancing guarantees will not be possible in the presence of outliers.

\begin{proof}[Proof of Proposition~\ref{prop:xM-cov-bound}]
	The proof follows a similar structure as the proof of Theorem~\mainref{thm:ht-mse-bound}, in that we also here extract the principal submatrices from the matrix inequality in Theorem~\mainref{thm:cov-bound}.
	The lower right $d$-by-$d$ block of $\Cov{\gsM \zv}$ is $\maxnormX^{-2}\paren{1 - \vto} \Cov{\xM^\tran \zv}$.
	The corresponding $d$-by-$d$ block of the matrix bound $\PgsM = \gsM \paren{\gsM^\tran \gsM}^{-1} \gsM^\tran$ is
	\begin{equation}
		\maxnormX^{-2} \paren{1 - \vto} \xM^\tran \paren*{\vto \unitM + \paren{1 - \vto} \maxnormX^{-2} \xM \xM^\tran }^{-1} \xM.
	\end{equation}
	After rearranging terms, this yields the inequality
	\begin{equation}
		\Cov{\xM^\tran \zv}
		\preceq
		\xM^\tran \paren*{\vto \unitM + \paren{1-\vto} \maxnormX^{-2} \xM \xM^\tran}^{-1} \xM .
	\end{equation}
	To prove the current proposition, we will show that we may re-write this matrix upper bound as
	\begin{equation}
		\xM^\tran \paren*{\vto \unitM + \paren{1-\vto} \maxnormX^{-2} \xM \xM^\tran}^{-1} \xM
		=
		\paren[\Big]{ \vto \paren{\xM^\tran \xM}^\pinv + \paren{1 - \vto} \paren{ \maxnormX^{2} \projX}^\pinv }^\pinv
	\end{equation}
	We do so by reasoning about the singular value decomposition of the covariate matrix $\xM$.
	To this end, let $\xM = \mat{U} \mat{\Sigma} \mat{V}^\tran$ be the singular value decomposition.
	We only consider the case where $d \leq n$, as the case where $d > n$ follows in a similar manner.
	If $d \leq n$, then  $\mat{U}$ is a $n$-by-$n$ orthogonal matrix, $\mat{\Sigma}$ is an $n$-by-$n$ diagonal matrix with non-negative diagonal entries, and $\mat{V}$ is a $d$-by-$n$ matrix with orthogonal rows.
	Using the singular value decomposition and orthogonality properties of $\mat{U}$, we have that
	\begin{align}
		&\xM^\tran \paren*{\vto \unitM + \paren{1-\vto} \maxnormX^{-2} \xM \xM^\tran}^{-1} \xM \\
		&= \mat{V} \mat{\Sigma} \mat{U}^\tran \paren*{\vto \unitM + \paren{1 - \vto} \maxnormX^{-2} \mat{U} \mat{\Sigma} \mat{V}^\tran \mat{V} \mat{\Sigma} \mat{U}^\tran}^{-1} \mat{U} \mat{\Sigma} \mat{V}^\tran
		&\text{(SVD)}\\
		&= \mat{V} \mat{\Sigma} \mat{U}^\tran \paren*{\vto \mat{U} \mat{U}^\tran + \paren{1 - \vto} \maxnormX^{-2} \mat{U} \mat{\Sigma} \mat{V}^\tran \mat{V} \mat{\Sigma} \mat{U}^\tran}^{-1} \mat{U} \mat{\Sigma} \mat{V}^\tran
		&\text{($\mat{U} \mat{U}^\tran = \unitM$)}\\
		&=\mat{V} \mat{\Sigma} \mat{U}^\tran \paren*{ \mat{U} \paren*{\vto \unitM + \paren{1 - \vto} \maxnormX^{-2} \mat{\Sigma} \mat{V}^\tran \mat{V} \mat{\Sigma} } \mat{U}^\tran}^{-1}\mat{\Sigma} \mat{V}^\tran
		&\text{(distributing $\mat{U}$)} \\
		&= \mat{V} \mat{\Sigma} \mat{U}^\tran \mat{U} \paren*{\vto \unitM + \paren{1 - \vto} \maxnormX^{-2} \mat{\Sigma} \mat{V}^\tran \mat{V} \mat{\Sigma} }^{-1} \mat{U}^\tran \mat{U} \mat{\Sigma} \mat{V}^\tran
		&\text{(inverse and $\mat{U}^{-1} = \mat{U}^\tran$)} \\
		&= \mat{V} \mat{\Sigma} \paren*{\vto \unitM + \paren{1 - \vto} \maxnormX^{-2} \mat{\Sigma} \mat{V}^\tran \mat{V} \mat{\Sigma} }^{-1} \mat{\Sigma} \mat{V}^\tran
		&\text{($\mat{U} \mat{U}^\tran = \unitM$)} \\
	\end{align}
	We can compute the pseudo-inverse of this matrix as
	\begin{align}
		\paren*{\xM^\tran \paren*{\vto \unitM + \paren{1-\vto} \maxnormX^{-2} \xM \xM^\tran}^{-1} \xM}^\pinv
		&= \paren*{\mat{V} \mat{\Sigma} \paren*{\vto \unitM + \paren{1 - \vto} \maxnormX^{-2} \mat{\Sigma} \mat{V}^\tran \mat{V} \mat{\Sigma} }^{-1} \mat{\Sigma} \mat{V}^\tran}^\pinv \\
		&= \mat{V} \mat{\Sigma}^\pinv \paren*{\vto \unitM + \paren{1 - \vto} \maxnormX^{-2} \mat{\Sigma} \mat{V}^\tran \mat{V} \mat{\Sigma} } \mat{\Sigma}^\pinv \mat{V}^\tran \\
		&= \vto \mat{V} \paren{\mat{\Sigma}^\pinv}^2 \mat{V}^\tran + \paren{1-\vto} \maxnormX^{-2} \mat{V} \mat{\Sigma}^\pinv \mat{\Sigma} \mat{V}^\tran \mat{V} \mat{\Sigma} \mat{\Sigma}^\pinv \mat{V}^\tran \\
		&= \vto \mat{V} \paren{\mat{\Sigma}^\pinv}^2 \mat{V}^\tran + \paren{1-\vto} \maxnormX^{-2} (\mat{V} \mat{\Sigma}^\pinv \mat{\Sigma} \mat{V}^\tran)^2  ,
	\end{align}
	where the third equality follows from distributing the outer matrices.
	We analyze each term separately, beginning with the left term.
	Note that
	\begin{equation}
		\xM^\tran \xM
		= \mat{V} \mat{\Sigma} \mat{U}^\tran \mat{U} \mat{\Sigma} \mat{V}^\tran
		= \mat{V} \mat{\Sigma}^2 \mat{V}^\tran
	\end{equation}
	and so by the orthogonality of rows of $\mat{V}$, one can check that
	\begin{equation}
		\paren{\xM^\tran \xM}^\pinv
		= \mat{V} \paren{\mat{\Sigma}^2}^\pinv \mat{V}^\tran
		= \mat{V} \paren{\mat{\Sigma}^\pinv}^2 \mat{V}^\tran .
	\end{equation}
	The matrix in the second term is equal to the orthogonal projection matrix onto the row span of $\xM$.
	To see this, observe that $\mat{V} \mat{\Sigma}^\pinv \mat{\Sigma} \mat{V}^\tran$ is the sum of the outer products of the right singular vectors corresponding to positive singular values. Because these vectors form an orthonormal basis for the row span of $\xM$, the sum of their outer products is the projection matrix $\projX$.
	As $\projX^2 = \projX = \projX^\pinv$,
	\begin{equation}
		\paren{1-\vto} \maxnormX^{-2} (\mat{V} \mat{\Sigma}^\pinv \mat{\Sigma} \mat{V}^\tran)^2
		= \paren{1-\vto} \maxnormX^{-2} \projX^2
		= \paren{1-\vto} \maxnormX^{-2} \projX^\pinv
		= \paren{1-\vto} \paren{\maxnormX^2 \projX }^\pinv.
	\end{equation}
	Putting these two terms together, we arrive at
	\begin{equation}
		\paren*{\xM^\tran \paren*{\vto \unitM + \paren{1-\vto} \maxnormX^{-2} \xM \xM^\tran}^{-1} \xM}^\pinv
		= \vto \paren{\xM^\tran \xM}^\pinv + \paren{1-\vto} \paren{\maxnormX^2 \projX }^\pinv.
	\end{equation}
	The proof is completed by taking the pseudoinverse of both sides.
\end{proof}

\subsection{Computational barriers to improved covariate balance}\label{sec:computational-hardness}

In this section, we demonstrate that achieving more covariate balance than that which is guaranteed by Gram--Schmidt Walk design with $\vto = 0$ is computationally intractable.
Unless one imposes additional restrictions or assumptions, $\norm{\Cov{\xM^\tran \zv}}$ cannot be made smaller than $\maxnormX^2 = \max_{i \in [n]} \norm{\xvi}^2$.

First, let us provide a lower bound which holds for all designs, irrespective of computational considerations.
To see this, consider when unit $1$ has the covariate vector with largest norm that also is orthogonal to all other covariate vectors: $\iprod{\xv{1}, \xvi} = 0$ for all $i \geq 2$ and $\maxnormX^2 = \norm{\xv{1}}^2$.
In this case, choosing $\lfx = \xv{1} / \norm{\xv{1}}$ yields
\begin{equation}
	\norm{\Cov{\xM^\tran \zv}}
	= \max_{\norm{\lfx} = 1} \E[\bigg]{ \paren[\Big]{\sum_{i=1}^n \zi \iprod{\xvi, \lfx} }^2}
	\geq \E[\bigg]{ \paren[\Big]{\sum_{i=1}^n \zi \frac{\iprod{\xvi, \xv{1}} }{\norm{\xv{1}}}}^2 }
	= \norm{\xv{1}}^2.
\end{equation}
However, there exist covariate matrices $\xM$ for which $\norm{\Cov{\xM^\tran \zv}} < \maxnormX^2$ is possible.

Now let us return to computational tractability for general covariates. 
\cite{Charikar2011Tight} prove that, given an $n$-by-$n$ matrix $\xM$ with $\pm 1$ entries, it is NP-hard to determine whether
\begin{equation}
	\min_{\zv \in \setb{\pm 1}^n} \norm{\xM^\tran \zv}^2 \geq c\,n^2
	\quad \text{or} \quad
	\min_{\zv \in \setb{\pm 1}^n} \norm{\xM^\tran \zv}^2 = 0,
\end{equation}
where $c > 0$ is universal, but presently unspecified, constant.
We compare this hardness result to the covariate balance guarantees we prove for the Gram--Schmidt Walk design with $\vto = 0$.
The covariate balance guarantees of Proposition~\ref{prop:xM-cov-bound} imply that in this case,
\begin{equation}
	\E[\big]{ \norm{\xM^\tran \zv}^2 }
	= \tr{ \Cov{\xM^\tran z} }
	\leq \maxnormX^2 \tr{ \projX }
	\leq n^2 ,
\end{equation}
where the third inequality follows by properties of projection matrices and that $\xM$ has $\pm 1$ entries, so $\maxnormX^2 = n$.
Thus, improving the covariate balance by even a constant factor pushes up against the boundary of computational tractability.
This demonstrates that no computationally feasible design can provide a significantly better guarantee on expected covariate balance without assumptions on the structure of the covariates.

There has been additional progress made to demonstrate that covariate balance cannot be improved in general.
In particular, \citet{Zhang2022Hardness} proves the following, more directly applicable hardness result: suppose that you are given a set of covariates $\xM$ and you are guaranteed that either
\begin{enumerate}
	\item There exists a symmetric experimental design satisfying $\norm{\Cov{\xM^\tran \zv}} = 0$
	\item For every symmetric  experimental design, $\norm{\Cov{\xM^\tran \zv}} \geq c$ for some constant $c$.
\end{enumerate}
The recent result of \citet{Zhang2022Hardness} proves that for some sufficiently small constant $c \in (0,1)$, it is NP-Hard to distinguish whether (1) or (2) is true.
The result is also extended to arbitrary experimental designs for which individual treatment assignments are nearly uniform, i.e. $\Pr{\zi = 1} \approx p$ for all $i \in [n]$.
This hardness result demonstrates that without further assumptions on the covariates, no computationally feasible design can provide a significantly better $\xsc$ parameter than what is guaranteed by the Gram--Schmidt Walk design when $\vto = 0$.

%% file: tex/supp-implementation.tex
\section{Fast Implementation of the Design}\label{sec:gsw-implementation}

The most computationally intensive aspect of the Gram--Schmidt Walk is the computation of the step
direction $\gsuv{t}$.
Although it is defined as the solution to an optimization problem, it may be obtained efficiently by
solving a system of linear equations.
Computational speed ups may be obtained by pre-computing and maintaining a
certain matrix factorization, decreasing the cost of repeated linear system solves at each iteration.
In this section, we provide details of such an efficient implementation.

\subsection{Derivation of the step direction}
Recall that at each iteration $t$, the step direction $\gsuv{t}$ is defined as the vector which has
coordinates $\gsuvi{t}{i} = 0$ for $i \notin \alive_{t}$, coordinate $\gsuvi{t}{p_{t}} = 1$ for the pivot unit
$p_{t}$, and the remaining coordinates are the solution to
\[
\gsuvi{t}{\alive_{t} \setminus p_{t}}
= \argmin_{\gsuv{}} \norm{\gscv{p_{t}}
	+ \sum_{i \notin \alive_{t} \setminus p_{t}} \gsuvi{}{i} \gscv{i}}^2
	\enspace .
\]
The minimization above is a least squares problem and the solution may be obtained by solving a
system of linear equations.
Let $k$ be the number of units which are alive and not the pivot,
i.e., $k = \abs{\alive_{t} \setminus p_{t}}$, and let $\gsM_{t}$ be the $(n+d)$-by-$k$ matrix with
columns $\gscv{i}$ for $i \in \alive_{t} \setminus p_{t}$.
As the augmented covariate vectors are linearly independent, the coordinates $\gsuvi{t}{\alive_{t} \setminus p_{t}} $
that minimize the quantity $\norm*{ \gscv{p_{t}} + \gsM_{t} \gsuvi{t}{\alive_{t} \setminus p_{t}}  }^2$
are given by the normal equations
\[
\gsuvi{t}{\alive_{t} \setminus p_{t}}
= - \paren*{\gsM_{t}^{\tran} \gsM_{t}}^{-1} \gsM_{t}^{\tran} \gscv{p_{t}}
\enspace .
\]
Let $\xM_{t}$ denote the row-submatrix of $\xM$ with rows $\alive_{t} \setminus p_{t}$.
Using our specific form of $\gsM$, and by direct calculation and application of the Woodbury identity
lemma, we obtain that
\[
\paren*{\gsM_{t}^{\tran} \gsM_{t}}^{-1}
= \paren*{ \vto  \unitM_{k} + \maxnormX^{-2}(1 - \vto) \xM_{t} \xM_{t}^{\tran}}^{-1}
= {\vto}^{-1} \bracket*{ \unitM_{k} -
\xM_{t} \paren*{ \xM_{t}^{\tran} \xM_{t} + \frac{ \maxnormX^{2} \vto }{ 1- \vto } \unitM_{d}  }^{-1}
\xM_{t}^{\tran}
}
\enspace.
\]
By again using our specific form of input matrix $\gsM$, a direct calculation yields that
\[
\gsM_{t}^{\tran} \gscv{p_{t}} = \maxnormX^{-2} \paren{1 - \vto} \xM_{t} \xv{p_{t}}
\enspace.
\]
Thus, we obtain a form for the relevant coordinates in the update direction vector $\gsuv{t}$
\begin{equation}\label{eq:u_vec_special_case}
\gsuvi{t}{\alive_{t} \setminus p_{t}}
= - \paren*{\frac{1 - \vto}{ \maxnormX^{2} \vto}}
\underbrace{\xM_{t}}_{n \times d}
\bracket*{
	\xv{p_{t}}
	-
	\underbrace{\paren*{ \xM_{t}^{\tran} \xM_{t} + \frac{ \maxnormX^{2} \vto }{ 1- \vto } \unitM_{d}
	}^{-1}}_{d \times d}
	\underbrace{\xM_{t}^{\tran} \xM_{t}}_{d \times d}
	\xv{p_{t}}
}
\enspace,
\end{equation}
which involves smaller matrices of size $d \times d$, rather than $n \times n$.
In the next few paragraphs, we show how computing and maintaining factorizations of these smaller
matrices results in faster computations of the step direction $\gsuv{t}$. We are chiefly concerned with
computing and maintaining a factorization of the matrix
$\paren{ \xM_{t}^{\tran} \xM_{t} + \maxnormX^{2} \vto \paren{ 1- \vto }^{-1} \unitM_{d}}$.
We describe an implementation which uses the Cholesky factorization, although there are several
appropriate alternatives.

\subsection{Cholesky factorizations}
Here, we briefly review Cholesky factorizations and their computational properties.
The \emph{Cholesky factorizatio}n of an
$n$-by-$n$ symmetric positive definite matrix $\aM$ is the unique factorization $\aM = \lM
\lM^{\tran}$, where $\lM$ is lower triangular.
Given the matrix $\aM$, the matrix $\lM$ may be obtained using $\bigO{n^3}$ arithmetic operations.
Once the Cholesky factorization $\lM$ is obtained, solutions $\xv{}$ to the system of linear equations
$\aM \xv{} = \bv$ may be computed using $\bigO{n^2}$ arithmetic operations by using a forward-backward
algorithm which leverages the triangular structure of $\lM$.
In general, solving systems of linear equations takes $\bigO{n^{3}}$ arithmetic operations\footnote{
While there are algorithms based on fast matrix multiplication that are asymptotically faster, they do not meaningfully change this discussion for realistic values of $n$.}
and so if
many
linear system solves are required, then computing the factorization and using the faster
forward-backward algorithm yields computational speed-ups.
Suppose that $\aM$ is a positive definite matrix with Cholesky factorization $\aM = \lM \lM^{\tran}$ and
that the rank-1 updated matrix $\aM + \vv \vv^{\tran}$ has Cholesky factorization
$\aM + \vv \vv^{\tran} = \lM_{+} \lM_{+}^{\tran}$.
Given the original factorization $\lM$ and the vector $\vv$, the updated factorization $\lM_{+}$ may be
computed using $\bigO{n^2}$ arithmetic computations, without extra memory allocation.
Updating in this way is a much more efficient way to maintain the factorization than explicitly
computing $\aM + \vv \vv^{\tran}$ and its factorization directly.
The same technique may be used for rank-1 downdates $\aM - \vv \vv^{\tran}$ when the updated matrix
remains positive definite.
For more details, see \citet{Stewart1998Matrix, TrefethenBau}.

\subsection{Computing and maintaining factorizations}
Before the first pivot is chosen, we have that $\xM_t = \xM$, as no rows of $\xM$ have been decided.
Thus, we compute
$\paren{ \xM_{t}^{\tran} \xM_{t} + \maxnormX^{2} \vto \paren{ 1- \vto }^{-1} \unitM_{d}}$
directly and then compute a Cholesky
factorization. Computing the matrix directly requires $\bigO{nd^2}$ time and computing the
factorization
requires $\bigO{d^3}$ time.
Each time a variable $i \in [n]$ is frozen or chosen as the pivot, the set $\alive_{t} \setminus
p_{t}$ is updated and so we must update the factorization
$\paren{ \xM_{t}^{\tran} \xM_{t} + \maxnormX^{2} \vto \paren{ 1- \vto }^{-1} \unitM_{d}}$.
The update consists of removing the row vector $\xvi$ from $\xM_{t}$.
One can see that this corresponds to a rank-1 downdate to the entire matrix
$\paren{ \xM_{t}^{\tran} \xM_{t} + \maxnormX^{2} \vto \paren{ 1- \vto }^{-1} \unitM_{d}}$.
Rank-1 downdates to a Cholesky factorization may be computed in-place, using $\bigO{d^2}$
arithmetic
operations.
Because there will be at most $n$ rank-1 updates to this factorization, the total update cost is
$\bigO{nd^2}$ arithmetic operations.
Thus, the total computational cost of maintaining this Cholesky factorization is $\bigO{nd^2}$
arithmetic operations and $\bigO{d^2}$ memory.

\subsection{Computing step directions}
Assume that at each iteration, we have a Choleksy factorization of the matrix
$\paren{ \xM_{t}^{\tran} \xM_{t} + \maxnormX^{2} \vto \paren{ 1- \vto }^{-1} \unitM_{d}}$.
By \eqref{eq:u_vec_special_case}, we can solve for the relevant coordinates in
the step
direction $\gsuvi{t}{\alive_{t} \setminus p_{t}} $ using the following three computations:
\begin{enumerate}
	\item $\at{1} = \xM_{t}^{\tran} \xM_{t} \xv{p_{t}}$
	\item $\at{2} = \paren*{ \xM_{t}^{\tran} \xM_{t} + \maxnormX^{2} \vto \paren{ 1- \vto }^{-1}
	\unitM_{d}}^{-1} \at{1}$
	\item $\gsuvi{t}{\alive_{t} \setminus p_{t}}
	=  - \maxnormX^{-2} {\vto}^{-1} \paren{1 - \vto} \xM_{t} \paren*{\xv{p_{t}} - \at{2} }$
\end{enumerate}
If the matrix $\xM_{t}^{\tran} \xM_{t}$ is explicitly available at the beginning of each iteration, then
computing $\at{1}$ can be done in $\bigO{d^2}$ time by matrix-vector multiplication. While it is
possible to
maintain $\xM_{t}^{\tran} \xM_{t}$ explicitly, it requires an extra $\bigO{d^2}$ memory. On the other
hand,
if $\xM_{t}^{\tran} \xM_{t}$ is not explicitly available, then $\at{1}$ may be obtained from a factorization
of $\paren{ \xM_{t}^{\tran} \xM_{t} + \maxnormX^{2} \vto \paren{ 1- \vto }^{-1} \unitM_{d}}$, as
\[
\at{1}
= \paren*{ \xM_{t}^{\tran} \xM_{t} + \frac{\maxnormX^{2} \vto}{1- \vto } \unitM_{d}} \xv{p_{t}}
- \paren*{\frac{ \maxnormX^{2} \vto }{ 1- \vto }} \xv{p_{t}}
\enspace,
\]
which saves $\bigO{d^2}$ memory and incurs only a slightly larger arithmetic cost of $\bigO{d^2 + d}$.
Next, one may compute $\at{2}$ using $\bigO{d^2}$ arithmetic operations  via a forward-backward
solver
on the Cholesky factorization.
Finally, computing $\gsuvi{t}{\alive_{t} \setminus p_{t}} $ may be done in $\bigO{nd}$ operations via
matrix-vector multiplication.
Thus, the per iteration cost of computing $\gsuv{t}$ given a factorized
$\paren{ \xM_{t}^{\tran} \xM_{t} + \maxnormX^{2} \vto \paren{ 1- \vto }^{-1} \unitM_{d}}$
is $\bigO{nd + d^2}$ arithmetic
operations.
Because there are at most $n$ iterations, this leads to a total cost of $\bigO{n^2 d + nd^2}$ arithmetic
operations.
We remark that $\bigO{n}$ memory is required for storing vectors such as
$\gsuvi{t}{\alive_{t} \setminus p_{t}}$.

Thus, an assignment may be sampled from the Gram--Schmidt Walk design using
$\bigO{n^2 d}$ arithmetic computations and $\bigO{n + d^2}$ extra storage when
implemented
with these matrix factorizations.
There are several practical considerations when implementing this algorithm.
First, for what values of $n$ and $d$ is this practically feasible?
Of course, this depends on the computing infrastructure which is available to experimenters, but
roughly speaking, sampling from the Gram--Schmidt Walk is as computationally intensive as computing
all pairs of inner products of covariates $\xv{1}, \xv{2} \dots \xv{n} \in \Reals^{d}$.
Computing these inner products requires $\bigO{n^2 d}$ arithmetic operations and computing this
matrix
of inner products $\xM \xM^{\tran}$ is a pre-processing step of our implementation.
The analysis above shows that the remainder of the algorithm requires roughly the same number of
arithmetic operations.
Thus, sampling from the Gram--Schmidt Walk should be practically feasible in cases where computing
all inner products is practically feasible.
A second practical consideration are the computational speed-ups for sampling more than one
assignment from the design.
When sampling many assignments from the Gram--Schmidt Walk, we may greatly reduce the run time by computing the initial cholesky factorization of ${\paren{ \xM_{t}^{\tran} \xM_{t} + \maxnormX^{2} \vto \paren{ 1- \vto }^{-1} \unitM_{d}}}$ and re-using it for each sample.
Finally, we remark that although our focus is to speed up the Gram--Schmidt Walk when we use the augmented covariate vectors, similar matrix factorizations may also be used to decrease the asymptotic run time of the general Gram--Schmidt Walk.

\subsection{Proof of asymptotic runtime} \label{sec:asymptotic-runtime}

\begin{proposition}
	An assignment from the Gram--Schmidt Walk design can be sampled using $\bigO{n^2 d}$ arithmetic
	operations and $\bigO{n + d^2}$ additional storage.
\end{proposition}

\begin{proof}
As detailed in Section~\ref{sec:gsw-implementation}, these computational resource guarantees may be achieved by storing and maintaining a Cholesky factorization of the matrix
$ \paren{ \xM_{t}^{\tran} \xM_{t} + \maxnormX^{2} \vto \paren{ 1- \vto }^{-1} \unitM_{d}}$,
where $\xM_{t}$ denotes the row-submatrix of $\xM$ with rows $\alive_{t} \setminus p_{t}$.
Constructing the matrix $\xM^{\tran} \xM$ requires $\bigO{nd^2}$ arithmetic operations and
$\bigO{d^2}$ space.
Initially computing a Cholesky factorization of this matrix requires $\bigO{d^{3}}$ arithmetic operations
and may be done in place.
Updating the Cholesky factorization may be done using $\bigO{nd}$ arithmetic operations in place and
this is done at most $n$ times.
Thus, constructing and maintaining the Cholesky factorization requires at most $\bigO{n^2 d}$
arithmetic operations and $\bigO{d^2}$ space, assuming that $d \leq n$.

Finally, computing the step direction $\gsuvt$ at each iteration requires $\bigO{nd}$ arithmetic
operations and $\bigO{n}$ space given the above Cholesky factorization. This happens for at most $n$
iterations, yielding a total of $\bigO{n^2 d}$ arithmetic operations and $\bigO{n}$ space.
Thus, combining the computational requirements of maintaining the Cholesky factorizaiton and
computing the step directions $\gsuvt$ yields a total requirement of $\bigO{n^2 \xdim}$ arithmetic
operations and $\bigO{n + \xdim^2}$ additional storage to generate one assignment vector using the
Gram--Schmidt Walk.
\end{proof}

%% file: tex/supp-other-designs.tex
\section{Comparison with Other Designs}

\subsection{Rerandomization}

We compare Gram--Schmidt Walk design to rerandomization in two ways.
In Section~\ref{sec:rerand-compare-tradeoff} we compare how the two designs navigate the balance-robustness trade-off by way of an example.
In Section~\ref{sec:rerand-compare-limvar}, we compare the limiting variance between the two designs.

\subsubsection{Comparison of the Balance-Robustness Trade-off} \label{sec:rerand-compare-tradeoff}

Because rerandomization does not allow for imbalanced assignments by construction, it will improve covariate balance.
But, it is unclear how the acceptance threshold translates to covariate balance guarantees in terms of $\norm{\Cov{\xM^\tran \zv}}$, and how well rerandomization navigates the balance--robustness trade-off.

\begin{figure}[ht]
	\centering
	\includegraphics[width=0.85\textwidth]{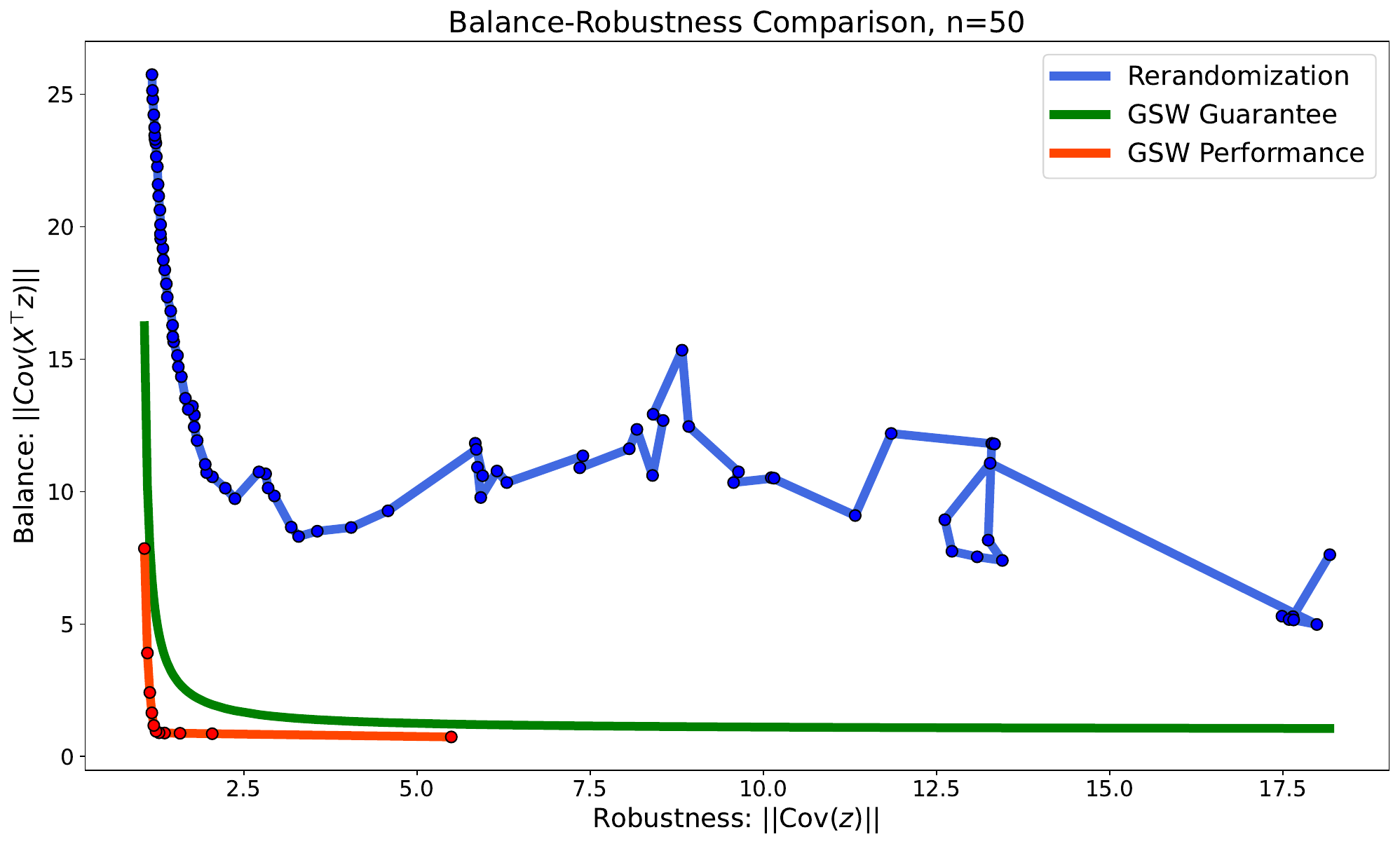}
	\caption{An instance of how different designs navigate the balance--robustness trade-off}
	\label{fig:investigate-rerand}
\end{figure}

It is beyond the scope of the current paper to analytically characterize the finite-sample behavior of rerandomization.
Instead, we consider a specific instance where rerandomization fails to adequately navigate the balance-robustness trade-off to illustrate the concern.
In the next section, we compare the limiting variance of rerandomization and the Gram--Schmidt Walk design.

Figure~\ref{fig:investigate-rerand} presents the results from a numerical simulation of the finite-sample properties of rerandomization.
There are $n=50$ units with $d=10$ dimensional covariate vectors, which are constructed in the following way:
the matrix of covariate vectors is given by
\[
\xM^\tran
=
\begin{bmatrix}
	\frac{1}{\sqrt{6}} \cdot \unitM_5 & \frac{2}{\sqrt{21}} \mat{J}_{5,25} & \frac{2}{\sqrt{21}} \mat{J}_{5,20} \\
	\frac{1}{\sqrt{6}} \cdot \mat{J}_5 & \frac{1}{\sqrt{21}} \onevec_5 \otimes \unitM_5 & - \frac{1}{\sqrt{21}} \onevec_4 \otimes \unitM_5
\end{bmatrix}
\mat{S} \mat{D}
\enspace,
\]
where $\mat{J}_{k,\ell}$ is the $k$-by-$\ell$ matrix whose entries are one, $\otimes$ is the Kronecker product, and $\mat{S}$ is a diagonal matrix whose entries alternate $\pm 1$,
and $\mat{D} = \mat{I} - \onevec \onevec^\tran / n$ is the operator that removes the mean from a vector.
The axes of the figure are the robustness and covariate balance terms.
The blue and red points are the trade-offs achieved by the rerandomization and the Gram--Schmidt Walk design, as their respective design parameters are varied.
The green line is the guarantee for the Gram--Schmidt Walk design, as described in Theorem~\mainref{theorem:gsw-satisfies-discrepancy}.

In this setting, rerandomization does not navigate the trade-off well.
For example, when $\norm{\Cov{\zv}} = 2$, corresponding to a worst-case variance that is twice as high as under the minimax design, rerandomization achieves covariate balance $\norm{\Cov{\xM^\tran \zv}} \approx 10$.
In contrast, the Gram--Schmidt Walk design almost perfectly balances all linear functions, in the sense $\norm{\Cov{\xM^\tran \zv}} \approx 0$, when $\norm{\Cov{\zv}} = 2$.
The reason that the curve for rerandomization is not monotonic is that the version of rerandomization described by \citet{Li2018Asymptotic} does not directly target $\norm{\Cov{\xM^\tran \zv}}$; improving covariate balance under the Mahalanobis distance does not necessarily improve $\norm{\Cov{\xM^\tran \zv}}$.

\begin{figure}[ht]
	\centering
	\begin{subfigure}{0.49\textwidth}
		\centering
		\includegraphics[width=\textwidth]{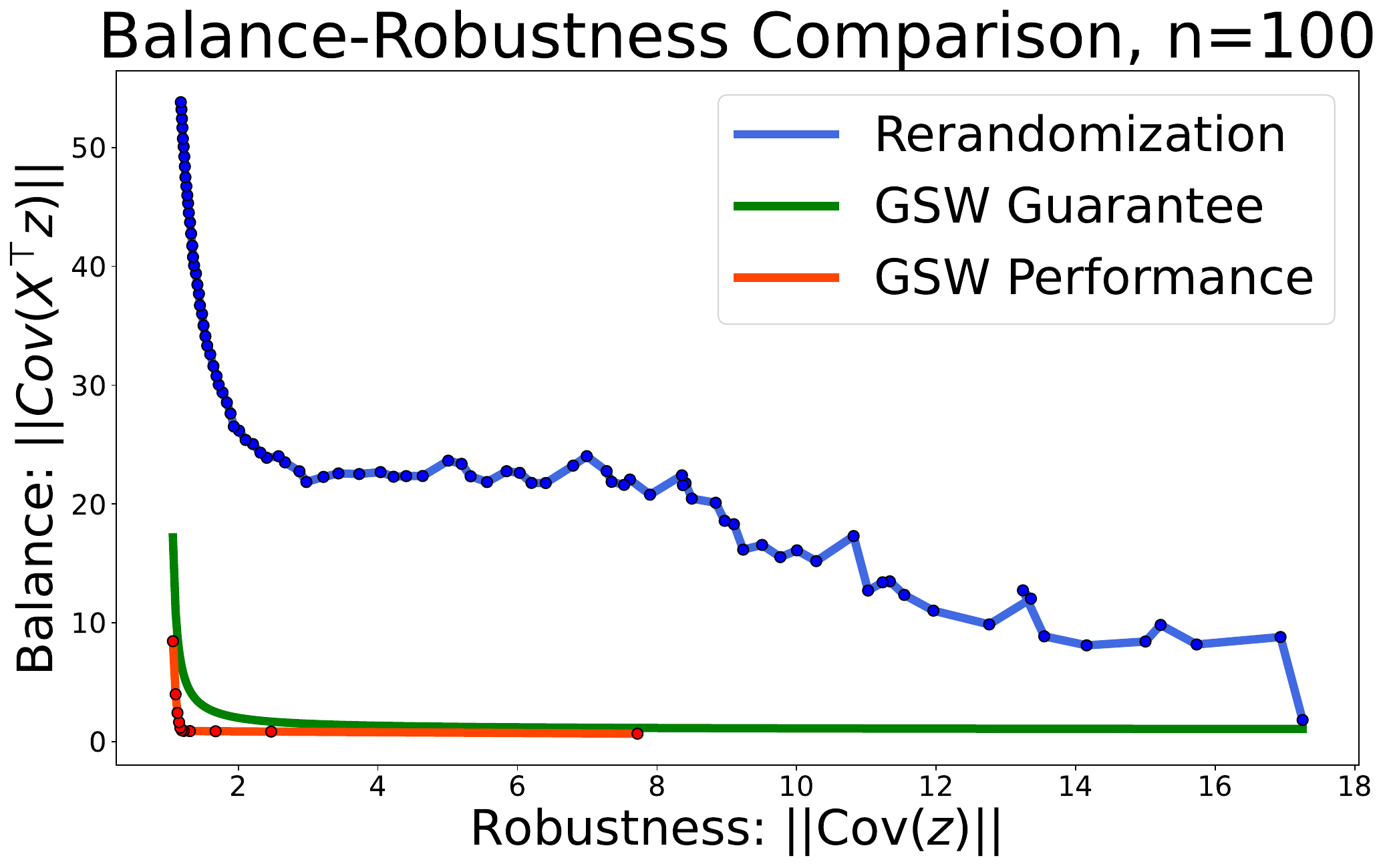}
		\caption{Comparison with $n=100$}
	\end{subfigure}
	\hfill
	\begin{subfigure}{0.49\textwidth}
		\centering
		\includegraphics[width=\textwidth]{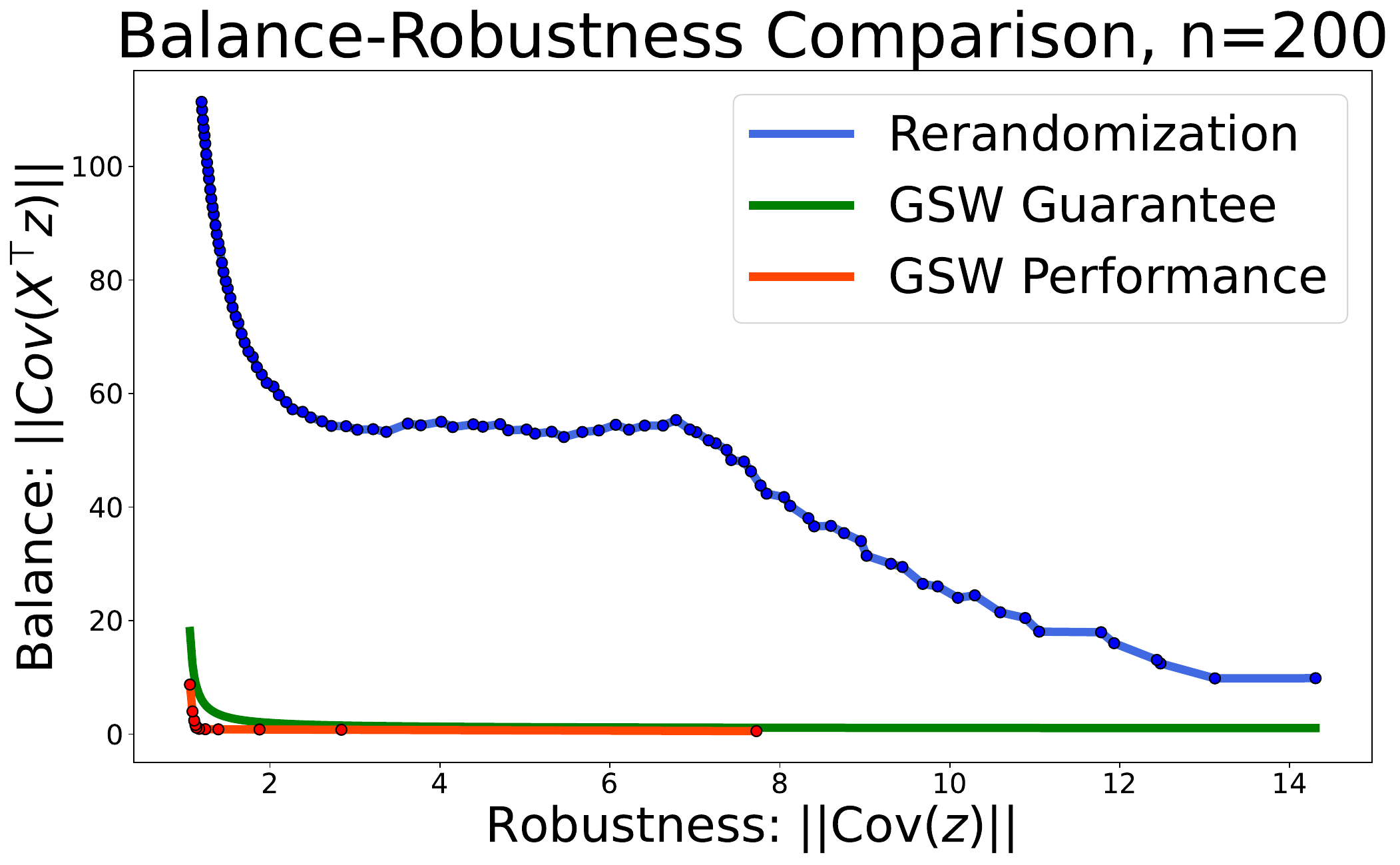}
		\caption{Comparison with $n=200$}
	\end{subfigure}
	\caption{Navigating the balance--robustness trade-off with larger instances}
	\label{fig:more-investigate-rerand}
\end{figure}

In Figure~\ref{fig:more-investigate-rerand} we extend the numerical simulations to $n=100$ and $n=200$ units.
The dimension and matrix structure remains the same.
For these larger sample sizes, re-randomization is even further away from the trade-off achieved by the Gram--Schmidt Walk design.
These simulations with larger sample sizes suggest that this phenomenon is not due to small sample size, but is rather an inherent aspect of the re-randomization design.

We present these examples to illustrate that there are situations where rerandomization navigates the balance--robustness trade-off poorly.
There are other situations where it matches or even exceeds the performance of the Gram--Schmidt Walk design.
The point here is that the rerandomization is not sufficiently understood in finite samples for experimenters to discern when rerandomization is expected to perform well.
In particular, rerandomization does not provide any guarantees with respect to the balance--robustness trade-off.

\subsubsection{Comparison of Limiting Variances} \label{sec:rerand-compare-limvar}

In this section, we prove Proposition~\mainref{prop:dominates-rerandomization} from the main body.
For completeness, we restate the proposition here.

\begin{refproposition}{\mainref{prop:dominates-rerandomization}}
\domrerand
\end{refproposition}

\begin{proof}
	Corollary~2 in \citet{Li2018Asymptotic} shows that the limiting variance of the difference-in-means estimator under rerandomization $\rerandlimvar$ is
	\begin{equation}
		\rerandlimvar = V_{\tau\tau} \bracket[\big]{ 1 - \paren{1 - v_{K, a}} R^2 },
	\end{equation}
	using the notation from their paper.
	Here, $V_{\tau\tau}$ is the limiting variance of the difference-in-means estimator under complete randomization, $v_{K, a}$ is the variance of a truncated random variable, depending on the number of covariates $K$ and the balance acceptance threshold $a$, and $R^2$ is a type of coefficient of determination, describing how predictive the covariates are of the potential outcomes.
	Lemma~\ref{lem:R2-derivation}, which is stated and proven later in this section, shows that we can write asymptotic variance of rerandomization as
	\begin{equation}
		\rerandlimvar = v_{K, a} \frac{\min_{\beta} \norm{ \pomv - \onevec \beta }^2}{n} + \paren{1 - v_{K, a}} \frac{\min_{\lfx} \norm{ \pomv - \xM \lfx }^2}{n},
	\end{equation}
	provided that Condition~1 in \citet{Li2018Asymptotic} holds and a constant is included among the covariates $\xM$.

	When $\xM$ includes a constant, we have $\min_{\beta} \norm{ \pomv - \onevec \beta }^2 \geq \min_{\lfx} \norm{ \pomv - \xM \lfx }^2$, and the inequality is strict whenever the covariates, excluding the constant, are at least somewhat predictive of $\pomv$.
	As noted by \citet{Li2018Asymptotic}, $v_{K, a} \in (0, 1)$, so
	\begin{equation}
		\rerandlimvar \geq v_{K, a} \frac{\min_{\lfx} \norm{ \pomv - \xM \lfx }^2}{n} + \paren{1 - v_{K, a}} \frac{\min_{\lfx} \norm{ \pomv - \xM \lfx }^2}{n} = \frac{\min_{\lfx} \norm{ \pomv - \xM \lfx }^2}{n} = \limvar.
	\end{equation}

	Lemma~\ref{lem:GSW-aVar-Li-assump} shows that the limiting variance of the Horvitz--Thompson estimator under the Gram--Schmidt Walk design is upper bounded by $\limvar = n^{-1} \min_{\lfx} \norm{ \pomv - \xM \lfx }^2$ provided that Condition~1 in \citet{Li2018Asymptotic} holds, that the second moment of the potential outcomes is asymptotically bounded, that a constant is included among the covariates, and that $\vto$ satisfies the rate condition in Theorem~\mainref{thm:asymp-variance}.
\end{proof}

\begin{lemma}\label{lem:R2-derivation}
	Suppose Condition~1 in \citet{Li2018Asymptotic} holds and a constant is included among the covariates, so that the first column of $\xM$ is $\onevec$.
	Then, the asymptotic variance $\rerandlimvar$ of the difference-in-means estimator under rerandomization is
	\begin{equation}
		\rerandlimvar = v_{K, a} \frac{\min_{\beta} \norm{ \pomv - \onevec \beta }^2}{n} + \paren{1 - v_{K, a}} \frac{\min_{\lfx} \norm{ \pomv - \xM \lfx }^2}{n},
	\end{equation}
	where $v_{K, a} \in (0, 1)$ is the ratio defined by \citet{Li2018Asymptotic}.
\end{lemma}

\begin{proof}
	As we consider symmetric designs, we have $r_1 = r_0 = 1/2$, using the notation of \citet{Li2018Asymptotic}.
	The proof should extend to arbitrary $r_1$, but we do not investigate that here.
	Furthermore, because the rerandomization procedure of \citet{Li2018Asymptotic} removes the means of its covariates, the inclusion of a constant column in $\xM$ does not change the design or $\rerandlimvar$.

	Under Condition~1 in \citet{Li2018Asymptotic}, their Corollary~2 applies, which shows that
	\begin{equation}
		\rerandlimvar = V_{\tau\tau} \bracket[\big]{ 1 - \paren{1 - v_{K, a}} R^2 },
	\end{equation}
	where
	\begin{multline}
		V_{\tau\tau} = 2 S^2_{a} + 2 S^2_{b} - S^2_{\tau},
		\qquad
		R^2 = \frac{2 S^2_{a|\mathbf{X}} + 2 S^2_{b|\mathbf{X}} - S^2_{\tau|\mathbf{X}}}{2 S^2_{a} + 2 S^2_{b} - S^2_{\tau}},
		\qquad
		S^2_{a|\mathbf{X}} = \mathbf{S}_{\mathbf{X},a}^\tran \paren{\mathbf{S}^2_{\mathbf{X}}}^{-1} \mathbf{S}_{\mathbf{X},a},
		\\
		\mathbf{S}_{\mathbf{X},a} = \frac{1}{n} \sum_{i=1}^n \paren{\tilde\xvi - \bar{\boldsymbol{x}}} \paren{\poai - \bar{a}},
		\qquad
		\mathbf{S}^2_{\mathbf{X}} = \frac{1}{n} \sum_{i=1}^n \paren{\tilde\xvi - \bar{\boldsymbol{x}}} \paren{\tilde\xvi - \bar{\boldsymbol{x}}}^\tran,
		\\
		S^2_{a} = \frac{1}{n} \sum_{i=1}^n \paren{\poai - \bar{a}}^2,
		\qquad
		\bar{a} = \frac{1}{n} \sum_{i=1}^n \poai,
		\qquad
		\bar{\boldsymbol{x}} = \frac{1}{n} \sum_{i=1}^n \tilde\xvi,
	\end{multline}
	and the definitions of $S^2_{b|\mathbf{X}}$, $S^2_{\tau|\mathbf{X}}$, $\mathbf{S}_{\mathbf{X},b}$, $\mathbf{S}_{\mathbf{X},\tau}$ $S^2_{b}$ and $S^2_{\tau}$ are the same as for $S^2_{a|\mathbf{X}}$, $\mathbf{S}_{\mathbf{X},a}$ and $S^2_{a}$, but with $\poai$ replaced either by $\pobi$ or by $\tau_i = \poai - \pobi$.
	The vector $\tilde\xvi$ is the covariate vector $\xvi$ for unit $i$ excluding the constant.
	See \citet{Li2018Asymptotic} for more details.

	Our definition of the asymptotic variance $\rerandlimvar$ is identical to the definition in \citet{Li2018Asymptotic}, but there is one superficial difference.
	We use the scaling $1/n$ in the definition of the population moments, while \citet{Li2018Asymptotic} use the scaling $1/(n-1)$.
	We do this to conform with the convention used in the current paper, which uses $1/n$ for the scaling of all population moments.%
	\footnote{\citet{Li2018Asymptotic} do not explain why they define their population moments using $1/(n-1)$ rather than $1/n$ as the scaling.
		Using $1/(n-1)$ suggests that they interpret these variables as estimators of some population quantities and apply a finite-sample adjustment, but the variables are the actual (finite) population moments, so there should be no reason to apply a sample adjustment.}
	Asymptotically, the difference is immaterial, because $(n - 1) / n \to 1$.
	That is, the asymptotic variance is the same no matter if we use the scaling $1/n$ or the scaling $1/(n-1)$.
	Nevertheless, $(n-1)/n < 1$ for all $n$, so the asymptotic variance is smaller with our scaling, meaning that the rescaling is to the advantage of rerandomization.

	Let $Q = 2 S^2_{a|\mathbf{X}} + 2 S^2_{b|\mathbf{X}} - S^2_{\tau|\mathbf{X}}$, and note that $R^2 = Q / V_{\tau\tau}$.
	Hence, we can write the asymptotic variance as
	\begin{equation}
		\rerandlimvar
		= V_{\tau\tau} - \paren{1 - v_{K, a}} Q
		= v_{K, a} V_{\tau\tau} + \paren{1 - v_{K, a}} \paren{V_{\tau\tau} - Q}.
	\end{equation}

	We will first consider $Q$.
	Let $\mat{D} = \unitM - \onevec \paren{\onevec^\tran \onevec}^{-1} \onevec^\tran$, where $\onevec$ is $n$-dimensional.
	Note that $\mat{D}$ is the projection onto the orthogonal complement of $\onevec$.
	Hence, the demeaned version of any vector $\vec{v} \in \Reals^n$ is given by $\mat{D} \vec{v}$.
	For example, the $i$th element of $\mat{D} \poav$ is equal to $\poai - \bar{a}$.
	This allows us to rewrite $\mathbf{S}_{\mathbf{X},a}$ and $\mathbf{S}^2_{\mathbf{X}}$ as
	\begin{equation}
		n \mathbf{S}_{\mathbf{X},a} = \paren{\mat{D} \tilde\xM}^\tran \mat{D} \poav = \tilde\xM^\tran \mat{D} \poav
		\qquad
		n \mathbf{S}^2_{\mathbf{X}} = \paren{\mat{D} \tilde\xM}^\tran \mat{D} \tilde\xM = \tilde\xM^\tran \mat{D} \tilde\xM,
	\end{equation}
	where $\tilde\xM$ is the covariate matrix excluding the constant.
	The second equalities follow from $\mat{D}^\tran \mat{D} = \mat{D}$.
	We can therefore write
	\begin{equation}
		n S^2_{a|\mathbf{X}} =
		\poav^\tran \mat{D}^\tran \tilde\xM \paren{\tilde\xM^\tran \mat{D} \tilde\xM}^{-1} \tilde\xM^\tran \mat{D} \poav
	\end{equation}
	Note that $\tilde\xM^\tran \mat{D}^\tran \mat{D} \tilde\xM\paren{\tilde\xM^\tran \mat{D} \tilde\xM}^{-1} = \unitM$, so
	\begin{equation}
		n S^2_{a|\mathbf{X}} =
		\poav^\tran \mat{D}^\tran \tilde\xM \paren{\tilde\xM^\tran \mat{D} \tilde\xM}^{-1} \tilde\xM^\tran \mat{D}^\tran \mat{D} \tilde\xM\paren{\tilde\xM^\tran \mat{D} \tilde\xM}^{-1} \tilde\xM^\tran \mat{D} \poav
		= \norm{\mat{P} \poav}^2.
	\end{equation}
	where $\mat{P} = \mat{D}\tilde\xM\paren{\tilde\xM^\tran \mat{D} \tilde\xM}^{-1} \tilde\xM^\tran \mat{D}$.
	Following the same argument, we have
	\begin{equation}
		n S^2_{b|\mathbf{X}} = \norm{\mat{P} \pobv}^2,
		\qquad
		n S^2_{\tau|\mathbf{X}} = \norm{\mat{P} \boldsymbol{\tau}}^2,
	\end{equation}
	where $\boldsymbol{\tau} = \poav - \pobv$.

	Note that
	\begin{equation}
		\norm{\mat{P} \boldsymbol{\tau}}^2
		= \iprod{\mat{P} \boldsymbol{\tau}, \mat{P} \boldsymbol{\tau}}
		= \norm{\mat{P} \poav}^2 - 2 \iprod{\mat{P} \poav, \mat{P} \pobv} + \norm{\mat{P} \pobv}^2,
	\end{equation}
	so
	\begin{equation}
		n Q
		= 2 n S^2_{a|\mathbf{X}} + 2 n S^2_{b|\mathbf{X}} -  n S^2_{\tau|\mathbf{X}}
		= \norm{\mat{P} \poav}^2 + 2 \iprod{\mat{P} \poav, \mat{P} \pobv} + \norm{\mat{P} \pobv}^2.
	\end{equation}
	Recall that $\pomv = \poav + \pobv$, so
	\begin{equation}
		\norm{\mat{P} \pomv}^2
		= \norm{\mat{P} \poav}^2 + 2 \iprod{\mat{P} \poav, \mat{P} \pobv} + \norm{\mat{P} \pobv}^2
		= n Q.
	\end{equation}
	That is, $Q = n^{-1} \norm{\mat{P} \pomv}^2$.

	Next, we consider $V_{\tau\tau} = 2 S^2_{a} + 2 S^2_{b} - S^2_{\tau}$.
	Following the same argument as above, we have
	\begin{equation}
		n S^2_{a} = \norm{\mat{D} \poav}^2,
		\qquad
		n S^2_{b} = \norm{\mat{D} \pobv}^2
		\qquadand
		n S^2_{\tau} = \norm{\mat{D} \boldsymbol{\tau}}^2.
	\end{equation}
	We also have
	\begin{equation}
		\norm{\mat{D} \pomv}^2
		=
		2 \norm{\mat{D} \poav}^2 + 2\norm{\mat{D} \pobv}^2 - \norm{\mat{D} \boldsymbol{\tau}}^2,
	\end{equation}
	so $V_{\tau\tau} = n^{-1} \norm{\mat{D} \pomv}^2$.

	Recall that $V_{\tau\tau} - Q$ appeared in the expression of the asymptotic variance.
	We can write this difference as
	\begin{equation}
		V_{\tau\tau} - Q
		= \frac{\norm{\mat{D} \pomv}^2 - \norm{\mat{P} \pomv}^2}{n}.
	\end{equation}
	Note that $\mat{D}\mat{P} = \mat{P}\mat{D} = \mat{P}\mat{P} = \mat{P}$, so
	\begin{equation}
		\norm{\mat{D} \pomv}^2 - \norm{\mat{P} \pomv}^2
		=
		\norm{\paren{\mat{D} - \mat{P}} \pomv}^2.
	\end{equation}
	Let $\mat{M} = \unitM - \xM\paren{\xM^\tran \xM}^{-1} \xM^\tran$, where $\xM$ is the covariate matrix including the constant.
	That is, $\xM$ is $\tilde\xM$ with $\onevec$ added as a column.
	The Frisch--Waugh--Lovell theorem implies that $\mat{M} = \mat{D} - \mat{P}$ \citep{Lovell2008Simple}.
	Therefore,
	\begin{equation}
		V_{\tau\tau} - Q
		= \frac{\norm{\paren{\mat{D} - \mat{P}} \pomv}^2}{n}
		= \frac{\norm{\mat{M} \pomv}^2}{n}.
	\end{equation}

	Taken together, we can write
	\begin{equation}
		V_{Re}
		= v_{K, a} \frac{\norm{\mat{D} \pomv}^2}{n} + \paren{1 - v_{K, a}} \frac{\norm{\mat{M} \pomv}^2}{n}.
	\end{equation}
	The proof is completed by noting that
	\begin{equation}
		\norm{\mat{D} \pomv}^2 = \min_{\beta} \norm{ \pomv - \onevec \beta }^2
		\qquadand
		\norm{\mat{M} \pomv}^2 = \min_{\lfx} \norm{ \pomv - \xM \lfx }^2.
		\tag*{\qedhere}
	\end{equation}
\end{proof}

\begin{lemma}\label{lem:GSW-aVar-Li-assump}
	Suppose that Condition~1 in \citet{Li2018Asymptotic} holds, that the second moment of the potential outcomes is asymptotically bounded, $n^{-1} \norm{\pomv}^2 = \bigO{1}$, and that a constant is included among the covariates, so that the first column of $\xM$ is $\onevec$, and that the design parameter is selected to satisfy $1 - \vto = \littleO{1}$ and $1 - \vto = \littleOmega[\big]{\maxnormX^2 / n}$.
	Then, the asymptotic variance of the Horvitz--Thompson estimator under the Gram--Schmidt Walk design is upper bounded by $\min_{\lfx} \norm{ \pomv - \xM \lfx }^2$.
\end{lemma}

\begin{proof}
	The proof largely mirrors the proof of Theorem~\mainref{thm:asymp-variance}.
	As in that proof, we can write
	\begin{equation}
		n \Var[\big]{\htest} - \frac{1}{n} \norm{ \rpomv }^2
		\leq \frac{1 - \vto}{\vto n} \norm{ \rpomv }^2 + \frac{\maxnormX^2}{\paren{1 - \vto} n} \norm[\big]{\lsfx}^2.
	\end{equation}
	The first term is addressed in the same way as in the proof of Theorem~\mainref{thm:asymp-variance}, yielding
	\begin{equation}
		\frac{1 - \vto}{\vto n} \norm{ \rpomv }^2 \leq \frac{1 - \vto}{\vto n} \norm{ \pomv }^2 = \littleO{1}.
	\end{equation}

	As in the proof of Lemma~\ref{lem:R2-derivation}, let $\mat{D} = \unitM - \onevec \paren{\onevec^\tran \onevec}^{-1} \onevec^\tran$ and let $\tilde\xM$ denote the covariate matrix $\xM$ excluding the constant.
	Define $\tilde\lfx = \paren{n^{-1} \tilde\xM^\tran \mat{D} \tilde\xM}^{-1} n^{-1} \tilde\xM^\tran \mat{D} \pomv$.
	Condition~1 in \citet{Li2018Asymptotic} stipulates that $n^{-1} \tilde\xM^\tran \mat{D} \tilde\xM$ and $n^{-1} \tilde\xM^\tran \mat{D} \pomv = n^{-1} \tilde\xM^\tran \mat{D} \paren{\poav + \pobv}$ have finite limits, and that $n^{-1} \tilde\xM^\tran \mat{D} \tilde\xM$ is nonsingular.
	This implies that $\norm{\tilde\lfx}^2 = \bigO{1}$.
	Note that the Frisch--Waugh--Lovell theorem implies that $\lsfx$ is the concatenation of the average potential outcome $\bar{\mu} = \paren{\onevec^\tran \onevec}^{-1} \onevec^\tran \pomv$ and $\tilde\lfx$.
	Bounded second moment of the potential outcomes implies that their average is bounded: $\bar{\mu}^2 = \bigO{1}$.
	Therefore, $\norm{\lsfx}^2 = \bar{\mu}^2 + \norm{\tilde\lfx}^2 = \bigO{1}$.
	\citet{Li2018Asymptotic} consider $d$ as fixed, so with $1 - \vto = \littleOmega[\big]{\maxnormX^2 / n}$, we have
	\begin{equation}
		\frac{\maxnormX^2}{\paren{1 - \vto} n} \norm[\big]{\lsfx}^2 = \littleO{1}.
		\tag*{\qedhere}
	\end{equation}
\end{proof}

\subsection{Matched pair design}

In this section, we prove Proposition~\mainref{prop:compare-w-matching}, which establishes that a match pair design must incur a large amount of covariate imbalance when the covariate vectors are uniformly drawn from the sphere.
We restate the proposition here for completeness.

\begin{refproposition}{\mainref{prop:compare-w-matching}}
	\comparematching
\end{refproposition}
\begin{proof}
	Let $M$ be the set of pairs of units in the matching.
	The covariance matrix of the discrepancy vector is given as
	\[
	\Cov{\xM^\tran \zv} = \sum_{(i,j) \in M} (\xvi - \xvj)(\xvi - \xvj)^\tran
	\enspace.
	\]
	Recall that the operator norm of a positive semidefinite matrix is its largest eigenvalue and the trace is the sum of its eigenvalues.
	Using this and linearity of trace, we may obtain the lower bound:
	\[
	\norm[\big]{\Cov{\xM^\tran \zv}}
	\geq \frac{1}{d} \cdot \tr[\big]{\Cov{\xM^\tran \zv}}
	= \frac{1}{d} \sum_{(i,j) \in M} \tr[\big]{(\xvi - \xvj)(\xvi - \xvj)^\tran }
	= \frac{1}{d} \sum_{(i,j) \in M} \norm{\xvi - \xvj}^2 \enspace.
	\]

	In the remainder of the proof, we use $B(r, \vec{x}) \triangleq \setb{\vec{y} \in \Reals^d : \norm{\vec{x}-\vec{y}} \leq r}$ to denote the ball of radius $r$ centered at vector $\vec{x}$.
	For each unit $i \in [n]$, let $m_i \in [n]$ be the unit whose covariate vector $\xv{m_i}$ is closest to $\xv{i}$ and let $\delta_i$ denote this distance, i.e. $\norm{\xvi - \xv{m_i}} = \min_{k \neq i} \norm{\xvi - \xv{k}} \triangleq \delta_i$.
	Given some radius $r > 0$, the probability that $\delta_i \geq r$ is the probability that all of the other covariate vectors lie outside the ball of radius $r$ centered at $\xvi$, which is at least
	\[
	\Pr{\delta_i \geq r}
	= \paren[\Bigg]{1 - \frac{\Vol{B(r, \xvi) \cap B(1,\zerovec)} }{\Vol{B(1,\zerovec)}}}^{n-1}
	\geq \paren[\Bigg]{1 - \frac{\Vol{B(r, \xvi) }}{\Vol{B(1,\zerovec)}} }^{n-1}
	\geq (1 - r^d)^{n-1}
	\enspace.
	\]
	We apply this with the radius $r = \paren{\frac{2}{dn}}^{1/d}$, and calculate
	\begin{align*}
		\Pr{\delta_i \geq r}
		&\geq (1 - r^d)^{n-1}
		&\text{(from above)}\\
		&= (1 - 2/(dn))^{n-1}
		&\text{(choice of $r$)}\\
		&\geq (1-1/n)^{n-1}
		&\text{$d \geq 2$}\\
		&\geq 1 / e
		\enspace.
	\end{align*}
	Thus, the expected squared distance between the random vector $\xvi$ and its nearest neighbor $\xv{m_i}$ is at least
	\[
	\E{\norm{\xvi - \xv{m_i}}^2 }
	\geq r^2 \cdot \Pr{\delta_i \geq r}
	\geq \paren[\Big]{\frac{2}{dn}}^{2/d} \cdot \frac{1}{e}
	\enspace.
	\]
	And, we obtain the following bound on the expected norm of the covariance matrix:
	\begin{align*}
		\E[\Big]{\norm[\big]{\Cov{\xM^\tran \zv}}}
		&\geq \E[\Big]{\frac{1}{d} \sum_{(i,j) \in M} \norm{\xvi - \xvj}^2} \\
		& \geq \E[\Big]{\frac{1}{d} \cdot \frac{1}{2} \sum_{i \in [n]} \norm{\xvi - \xv{m_i}}^2} \\
		&= \frac{1}{2d}  \sum_{i \in [n]} \E[\Big]{\norm{\xvi - \xv{m_i}}^2} \\
		&\geq \frac{1}{2e} \cdot (2/d)^{2/d}  \cdot \frac{n^{1 -2/d}}{d} \enspace.
	\end{align*}
	The result follows by observing that $(2/d)^{2/d} \geq 1/e^{1/e}$ for all $d \geq 1$,
	$2 e^{1+1/e} \leq 8$, and $\xi \leq 1$ because the vectors were chosen from the unit ball.
\end{proof}

%% file: tex/supp-extensions.tex
\section{Extensions of Main Results}\label{sec:supp-extensions}

In this section, we present several extensions of the results in the main paper.
In Section~\ref{sec:non-uniform-probs-supp}, we discuss how to extend the Gram--Schmidt Walk design and its analysis to incorporate non-uniform treatment probabilities.
In Section~\ref{sec:ext-other-matrix-funs}, we discuss how our results can be extended to bounds on other matrix functions of the covariance matrices.
In Section~\ref{sec:different-moment-defs}, we discuss a generalized notion of robustness under different moment definitions on the potential outcomes and how a modified Gram--Schmidt Walk design may address this.
In Section~\ref{sec:supp-balanced-gsw}, we discuss a modified version of the Gram--Schmidt Walk design which assigns treatment groups of equal sizes.

\input{\texpath/supp-nonuniform-probs}

\input{\texpath/supp-ext-other-matrix-funs}

\input{\texpath/supp-different-moment-defs}

\input{\texpath/supp-balanced-gsw}

\input{\texpath/supp-alt-cis}

%% file: tex/supp-nonuniform-probs.tex
\subsection{Non-uniform treatment probabilities} \label{sec:non-uniform-probs-supp}

The Gram--Schmidt Walk design can be extended to allow arbitrary assignment probabilities.
We achieve this by changing the initial fractional assignments of the algorithm.
The experimenter provides a parameter vector $\pzv = \paren{\pz{1}, \dotsc, \pz{n}} \in \paren{0, 1}^n$ specifying the desired first-order assignment probability for each unit.
The first step of the algorithm in Section~\mainref{sec:gsw-design} is then modified so that $\gsv{1} \gets 2 \pzv - \onevec$.
The following corollary is a direct consequence of the martingale property of the fractional updates.

\begin{appcorollary}\label{cor:arbitrary-marginal-probs}
	Under the non-uniform Gram--Schmidt Walk design,
	\begin{equation}
		\Pr{\zi = 1} = \pzi
		\qquadtext{for all}
		i \in [n].
	\end{equation}
\end{appcorollary}

The properties of the original version of the design can be extended to the non-uniform version.
To do so, we redefine the vector $\pomv$ as
\begin{equation}
	\gpomv = \paren[\bigg]{\frac{\poae{1}}{2 \pz{1}} + \frac{\pobe{1}}{2 \paren{1 - \pz{1}}}, \dotsc, \frac{\poae{n}}{2 \pz{n}} + \frac{\pobe{n}}{2 \paren{1 - \pz{n}}}}.
\end{equation}
In this vector, each potential outcome is weighted by the probability that it is observed.
If $\pzv = 0.5 \times \onevec$, then $\gpomv = \pomv$, which replicates the uniform version of the design.
The mean squared error of the Horvitz--Thompson estimator can now be expressed as
\begin{equation}
	\E[\big]{\paren{\htest - \ate}^2}	= \frac{1}{n^2} \gpomv^\tran \Cov{\zv} \gpomv.
\end{equation}
This extends Lemma~\mainref{lemma:mse-expression} to any experimental design with non-deterministic assignments.
In particular, Theorems~\mainref{thm:cov-bound} and~\mainref{thm:sub-gaussian} hold for the non-uniform version of the design, so all properties that follow from these theorems also apply to the extended version when $\gpomv$ is substituted for $\pomv$.

%% file: tex/supp-ext-other-matrix-funs.tex
\subsection{Extension to Other Matrix Functions}\label{sec:ext-other-matrix-funs}

In the main paper, we were interested in bounding the operator norms of the matrices $\norm{\Cov{\zv}}$ and $\norm{\Cov{\xM^\tran \zv}}$.
However, experimenters may be interested in other measures of these matrices $\Cov{\zv}$ and $\Cov{\xM^\tran \zv}$ apart from the operator norm, including the Frobenius norm, trace, and determinant.
In this section, we show how the matrix inequality in Theorem~\mainref{thm:cov-bound} can give bounds on such matrix functions.

It is possible to use functions of $\Cov{\xM^\tran \zv}$ other than the operator norm as measures of covariate balance.
Examples include the trace norm and the Frobenius norm of $\Cov{\xM^\tran \zv}$.
But none of these alternative norms provide a guarantee on the mean square error.
That is, the mean square error might be large even if the covariates are perfectly predictive and we make these alternative norms relatively small.
For this reason, we believe the operator norm is a more useful measure of covariate balance; if the operator norm is small, we are guaranteed that all linear functions of the covariates are balanced.
However, if an experimenter has more detailed knowledge about the characteristics of the function $\lfx$, they might prefer to use one of these alternative norms.
For example, the trace norm will be more appropriate if the experimenter adopts a Bayesian perspective, specifying a mean-zero, isotropic prior on the linear function $\lfx$.
In this section, we will show how the existing guarantees of the Gram--Schmidt Walk design yield guarantees on a class of alternative matrix functions.

Before continuing, we remark that although bounds on these additional matrix functions can be derived under the Gram--Schmidt Walk design, the design itself may not be the experimental design which minimizes these additional matrix functions.
Indeed, the design is most well-suited to minimizing the operator norm, up to the limit of computational feasibility.

In what follows, we let $\mathcal{S}^n$ be the set of $n$-by-$n$ real symmetric matrices.

\begin{definition}\label{def:app-operator-monotone}
	A real-valued matrix function $g: \mathcal{S}^n \rightarrow \Reals$ is said to be \emph{operator monotone} if $\mat{A} \preceq \mat{B}$ implies that $g(\mat{A}) \leq g(\mat{B})$.
\end{definition}

\begin{definition}\label{def:app-operator-scaling}
	A real-valued matrix function $g: \mathcal{S}^n \rightarrow \Reals$ is said to be \emph{$h$-scaling} for a function $h: \Reals \rightarrow \Reals$ if $g( \alpha \cdot \mat{A}) = h(\alpha) \cdot g(\mat{A}) $ for all matrices $\mat{A} \in \mathcal{S}^n$ and scalars $\alpha \in \Reals$.
\end{definition}

We now list several examples of operator monotone $h$-scaling functions.
In the following examples, we focus on the monotone and scaling properties of real-valued matrix functions when restricted to positive semidefinite matrices and non-negative scalings.

\begin{itemize}
	\item \textbf{Operator Norm}: The operator norm $g(\mat{A}) = \max_{\vec{x} \neq 0} \norm{\mat{A} \vec{x}} / \norm{\vec{x}}$ is operator monotone over positive semidefinite matrices and $h$-scaling with the identity $h(t) = t$.
	\item \textbf{Frobenius Norm}: The Frobenius norm $g( \mat{A} ) = \sqrt{ \tr{ \mat{A}^\tran \mat{A} } }$ is operator monotone over positive semidefinite matrices and $h$-scaling with identity $h(t) = t$.
	\item \textbf{Trace}: The trace is $g(\mat{A}) = \tr{\mat{A}}$ is operator monotone over positive semidefinite matrices and $h$-scaling with the identity $h(t) = t$.
	\item \textbf{Schatten $p$-Norm}: Let the eigenvalues of an $n$-by-$n$ matrix $\mat{A}$ be denoted $\eigvali (\mat{A})$ for $i \in [n]$. The Schatten $p$-norm $g(\mat{A}) = \paren{\sum_{i=1}^n \eigvali (\mat{A})^p  }^{1/p}$ is operator monotone over positive semdefinite matrices and $h$-scaling with the identity $h(t) = t$.
	\item \textbf{Determinant}: The determinant $g(\mat{A}) = \det( \mat{A})$ is operator monotone over positive semidefinite matrices and $h$-scaling with $h(t) = t^n$, where $n$ is the dimension of the matrix $\mat{A}$.
\end{itemize}

The following proposition uses the matrix inequality in Theorem~\mainref{thm:cov-bound} to derive bounds on operator monotone and $h$-scaling matrix functions.
In a slight abuse of notation, we use the same real-valued matrix function $g$ to refer to matrices of different dimensions.

\begin{proposition} \label{claim:extension-other-matrix-funs}
	Let $g: \mathcal{S}^n \rightarrow \Reals$ be an operator monotone over positive semidefinite matrices and $h$-scaling real valued matrix function.
	Then, the assignments under the \gswdesign satisfy the following bounds:
	\[
	g( \Cov{\zv} ) \leq h\paren[\Big]{\frac{1}{\vto}} g(\unitM_n) 
	\quadand
	g(\Cov{\xM^\tran \zv}) \leq h \paren[\Big]{ \frac{\maxnormX^2}{1 - \vto} } g(\unitM_d)
	\enspace.
	\]
\end{proposition}
\begin{proof}
	As established in the proof of Theorem~\mainref{theorem:gsw-satisfies-discrepancy}, the random treatment assignment under \gswdesign satisfies the following two matrix inequalities: $ \Cov{\zv} \preceq 1/ \vto \cdot \unitM$ and $ \Cov{\xM^\tran \zv} \preceq \maxnormX^2 / (1-\vto) \cdot \unitM $.
	The result is obtained by applying the definition of operator monotonicity and scaling.
\end{proof}

%% file: tex/supp-different-moment-defs.tex
\subsection{Different moment definitions} \label{sec:different-moment-defs}

In this section, we reexamine the notion of robustness{\textemdash}and the resulting balance robustness trade-off{\textemdash}when different moment conditions are used to specify the relevant set of potential outcomes.

Previously, we used the moment condition $(1/n) \cdot \sum_{i=1}^n (\poai + \pobi)^2$ to measure what we referred to as the ``average magnitude'' of the potential outcomes.
For a given design, Lemma~\mainref{lemma:robustness} showed that the worst-case mean squared error of the Horvitz--Thompson estimator over the set of potential outcomes with bounded average magnitude is proportional to $\norm{\Cov{\zv}}$, the operator norm of the covariance matrix of the assignment vector.
This motivated $\norm{\Cov{\zv}}$ as a measure of robustness of the design, in a worst-case sense.
Changing the moment definition on the potential outcomes will change the notion of robustness.

We consider moment definitions which may be expressed as norms of the $2n$-dimensional vector of potential outcomes, denoted $\vec{\nu} = \paren{\poae{1}, \dots \poae{n}, \pobe{1}, \dots \pobe{n}} \in \Reals^{2n}$.
Let $\norm{\cdot}_V$ be an arbitrary norm on $\Reals^{2n}$ which will act as our moment definition.
We refer to this as the $V$-moment of the potential outcome vectors.
Examples of a few reasonable $V$-moments include:

\begin{enumerate}
	\item Sum of Average Magnitudes: $\norm{\vec{\nu}} = (1/n) \cdot \sum_{i=1}^n \poai^2 + (1/n) \cdot \sum_{i=1}^n \pobi^2$.
	\item Maximum Absolute Outcome: $\norm{\vec{\nu}} = \max_{i \in [n]} \max \paren[\big]{ \abs{\poai}, \abs{\pobi} }$.
\end{enumerate}

Next, we formally define the general concept of an operator norm.
Given an $n$-by-$m$ matrix $\mat{A}$ and norms $\norm{\cdot}_U$ and $\norm{\cdot}_V$ defined on $\Reals^n$ and $\Reals^m$, respectively, the \emph{$V \rightarrow U$ operator norm} of $\mat{A}$ is defined as
\[
\norm{\mat{A}}_{V \rightarrow U} = \max_{ \norm{\vec{x}}_V \leq 1} \norm{\mat{A} \vec{x}}_U \enspace.
\] 
The $V \rightarrow U$ operator norm defines a norm on $n$-by-$m$ matrices.
Throughout the main body of the paper, we have been using the operator norm on symmetric matrices where the norms $\norm{\cdot}_U$ and $\norm{\cdot}_V$ are both the $\ell_2$-norm.
This $\ell_2 \rightarrow \ell_2$ operator norm is also known as the spectral norm, as it is equal to the largest magnitude of the eigenvalues of $\mat{A}$.
Generally speaking, arbitrary $V \rightarrow U$ operator norms will not have such clean characterizations and are sometimes NP-Hard to compute.
The \emph{operator norm bound} is the following inequality, which holds for all vectors $\vec{x}$: $\norm{\mat{A} \vec{x} }_U \leq \norm{\vec{x}}_V \cdot \norm{\mat{A}}_{V \rightarrow U}$.

The following lemma demonstrates the general relevance of operator norms in the study of robust experimental designs.

\begin{applemma}\label{lemma:general-robustness}
	Consider a design satisfying $\Pr{\zi = 1} = 1/2$ for all units $i \in [n]$.
	Let $\norm{\cdot}_V$ be an arbitrary norm on $\Reals^n$.
	The worst case mean squared error of the Horvitz--Thompson estimator over all potential outcomes with bounded $V$-moment is
	\[
	\max_{ \norm{\vec{\nu}}_V \leq M } \E[\big]{ \paren{\ate - \htest}^2} = \frac{M^2}{n^2} \cdot \norm{\Cov{\mat{R^\tran} \zv}^{1/2}}_{V \rightarrow \ell_2}^2  
	\enspace,
	\]
	where $\mat{R} = [ \unitM_n  \ \unitM_n ]$ is the $n$-by-$2n$ matrix of horizontally concatenated $n$-dimensional identity matrices.
\end{applemma}
\begin{proof}
	Observe that the sum potential outcome vector $\pomv = \paren{\poae{1} + \pobe{1}, \dots \poae{n} + \pobe{n}}$ may be expressed as $\pomv = \mat{R} \vec{\nu}$.
	Lifting from the proof of Lemma~\mainref{lemma:mse-expression} (proved in Supplement~\ref{sec:supp-mse-expression}), the error of the Horvitz--Thompson estimator is equal to 
	\[
	\ate - \htest 
	= \frac{1}{n} \iprod{\zv, \pomv}
	= \frac{1}{n} \iprod{\zv, \mat{R} \vec{\nu} }
	= \frac{1}{n} \iprod{\mat{R}^\tran \zv, \vec{\nu}} 
	\enspace.
	\]
	Thus, by taking squares and expectations, we have that the mean squared error may be expressed as
	\[
	\E[\big]{\paren{\ate - \htest}^2}
	= \frac{1}{n^2} \E[\big]{\iprod{\mat{R}^\tran \zv, \vec{\nu}}^2}
	= \frac{1}{n^2} \vec{\nu}^\tran \Cov{\mat{R}^\tran \zv} \vec{\nu}
	= \frac{1}{n^2} \cdot  \norm{ \Cov{\mat{R}^\tran \zv}^{1/2} \vec{\nu} }^2_{\ell_2} \enspace.
	\]
	Using the operator norm bound, we have that the mean squared error may be bounded as
	\[
	n^2 \cdot \E[\big]{ \paren{\ate - \htest}^2} 
	= \norm{ \Cov{\mat{R}^\tran \zv}^{1/2} \vec{\nu} }^2_{\ell_2} 
	\leq \norm{\vec{\nu}}_V^2 \cdot \norm{\Cov{\mat{R}^\tran \zv}^{1/2}}_{V \rightarrow \ell_2}^2 
	\leq M^2 \cdot \norm{\Cov{\mat{R}^\tran \zv}^{1/2}}_{V \rightarrow \ell_2}^2 
	\enspace,
	\]
	where the third inequality follows by the restriction that $\norm{\vec{\nu}}_V \leq M$.
	The desired result follows by observing that the inequalities above are tight over the set of all potential outcome vectors satisfying $\norm{\vec{\nu}}_V \leq M$.
\end{proof}

Lemma~\ref{lemma:general-robustness} above motivates $\norm{\Cov{\mat{R^\tran} \zv}^{1/2}}_{V \rightarrow \ell_2}^2$ as a more general definition of robustness, which depends on the underlying norm $\norm{\cdot}_V$ being used to define the moment conditions on the vector of potential outcomes.
Note that when the underlying norm is $\norm{\vec{\nu}}_V = \norm{\pomv}_{\ell_2} = (1/n) \cdot \sum_{i=1}^n (\poai + \pobi)^2$, then $\norm{\Cov{\mat{R^\tran} \zv}^{1/2}}_{V \rightarrow \ell_2}^2$ reduces to the spectral norm of $\Cov{\zv}$ used throughout the main body of the paper.

This more general notion of robustness suggests different balance-robustness trade-offs.
In particular, experimenters might wish to make both operator norms $\norm{\Cov{\mat{R^\tran} \zv}^{1/2}}_{V \rightarrow \ell_2}^2$ and $\norm{\Cov{\xM^\tran \zv}}_{\ell_2 \rightarrow \ell_2}$ small, where the first captures robustness and the second captures covariate balance.
The following proposition shows that for a broad class of moment definitions, a modified version of the \gswdesign may still be used to navigate this trade-off.

\begin{appproposition}\label{prop:gsw-for-generalized-tradeoffs}
	Consider a norm defined by a positive definite quadratic form, $\norm{\vec{\nu}}_V = \vec{\nu}^\tran \mat{\Lambda} \vec{\nu}$.
	Let $\zv$ be a random assignment vector obtained by running the Gram--Schmidt Walk algorithm with input augmented covariate vectors defined as
	\[
	\vec{b}_i = 
	\begin{bmatrix}
		\frac{\sqrt{\vto}}{\eta} \vec{q}_i \\
		\frac{\sqrt{1-\vto}}{\maxnormX} \xvi
	\end{bmatrix}
	\enspace,
	\]
	where $\vto \in [0,1]$, $\vec{q}_i$ is the $i$th column of the matrix $\mat{\Lambda}^{-1/2} \mat{R}^\tran$, and $\eta = \max_{i \in [n]} \norm{\vec{q}_i}_{\ell_2}$.
	Then, the random assignment vector satisfies the two following bounds:
	\[
	\norm{\Cov{\mat{R^\tran} \zv}^{1/2}}_{V \rightarrow \ell_2}^2 \leq \frac{1}{\vto} \cdot \max_{i \in [n]} \norm{\vec{q}_i}_{\ell_2}^2
	\quad \text{and} \quad
	\norm{\Cov{\xM^\tran \zv}}_{\ell_2 \rightarrow \ell_2} \leq \frac{1}{1 - \vto} \cdot \max_{i \in [n]} \norm{\xvi}_{\ell_2}^2
	\enspace.
	\]
\end{appproposition}

The proof of Proposition~\ref{prop:gsw-for-generalized-tradeoffs} follows from Theorem~\mainref{thm:cov-bound} in a similar manner as Theorem~\mainref{theorem:gsw-satisfies-discrepancy}.
The central insight is that for norms defined by a positive definite quadratic form, (i.e. $\norm{\vec{\nu}}_V = \vec{\nu}^\tran \mat{\Lambda} \vec{\nu}$), the $V \rightarrow \ell_2$ operator norm is equal to the usual spectral norm of a conjugated matrix, i.e. $\norm{\mat{A}}_{V \rightarrow \ell_2} = \norm{\mat{\Lambda}^{-1/2} \mat{A} \mat{\Lambda}^{-1/2}}_{\ell_2 \rightarrow \ell_2}$

It may be possible to extend notions of robustness even further by considering moment conditions defined by \emph{semi-norms}, which are functions $\norm{\cdot}_V : \Reals^{2n} \rightarrow \Reals_{\geq 0}$ that satisfy all properties of norms except that it is possible for $\norm{\vec{\nu}}_V = 0$ when $\vec{\nu} \neq 0$.
One relevant example is the \emph{population variance} of the potential outcomes, defined as
\[
\norm{\vec{\nu}}_V =
\frac{1}{n} \cdot \sum_{i=1}^n \paren[\Big]{ \poai - \frac{1}{n} \sum_{j=1}^n \poaj }^2
+ \frac{1}{n} \cdot \sum_{i=1}^n \paren[\Big]{ \pobi - \frac{1}{n} \sum_{j=1}^n \pobj }^2
\enspace.
\]
Note that the seminorm ball $\setb{ \vec{\nu} \in \Reals^{2n} \mid \norm{\vec{\nu}}_V \leq M }$ is generally unbounded, as it contains the entire span of the vectors for which $\norm{\vec{\nu}}_V = 0$.
In this case, any design for which $\max_{ \norm{\vec{\nu}}_V \leq M} \E{\paren{\ate - \htest}^2}$ is finite must satisfy the strong property that $\norm{\Cov{\mat{R}^\tran \zv}^{1/2} \vec{\nu}}_{\ell_2} = 0$ for all $\vec{\nu}$ satisfying $\norm{\vec{\nu}} = 0$.
In other words, the nullspace of $\Cov{\mat{R}^\tran \zv}$ must contain the span of vectors $\vec{\nu}$ satisfying $\norm{\vec{\nu}}_V = 0$.
In the context of the population variance moment condition, then this means that a design which has finite robustness parameter must assign equal number of units to both treatment groups with probability 1.
This essentially recovers the results of \citet{Wu1981Robustness,Kallus2018Optimal,Bai2023Why,Nordin2022Properties}, who arrive at similar conclusions via worst-case analyses.

Investigating more general notions of robustness and understanding the relevance of these trade-offs are interesting future directions which lie beyond the scope of the current paper.

%% file: tex/supp-balanced-gsw.tex
\subsection{Treatment Group Sizes and the Gram--Schmidt Walk design} \label{sec:supp-balanced-gsw}

\newcommand{\ntreated}{n_+}
\newcommand{\ncontrol}{n_-}

The distribution of treatment group sizes produced by the Gram--Schmidt Walk design 
can never be much worse than that produced by the Bernoulli design.
We can make it better by adding a constant covariate to $\xM$.

In this section, we investigate the size of the treatment groups under the Gram--Schmidt Walk design.
We show that the concentration of the distribution of treatment group sizes is always almost as tight as that of the Bernoulli design, and that we can make the treatment group sizes more similar by including a constant among the covariates.
However, the treatment groups will generally not have the same size with probability 1.
In the second part of this section, we introduce a modification of the Gram--Schmidt Walk design which ensures that the two treatment groups have equal sizes with probability 1.

\subsubsection{Treatment Group Sizes Under Gram--Schmidt Walk design} \label{sec:treatment-group-size-analysis}

Let $\ntreated = \sumin \indicator{\zi = 1}$ and $\ncontrol = \sumin \indicator{\zi = -1 }$ be the (random) sizes of treatment and control groups, respectively.
The Gram--Schmidt Walk design is symmetric (e.g. $\Pr{\zi = 1} = 1/2$) and so the difference between the group sizes is zero in expectation, $\E{ \ntreated - \ncontrol } = 0$.
The following proposition provides a bound on the variance of the difference between these group sizes
  when we add a covariate that takes the value $\alpha$ for every unit.

\begin{appproposition}\label{prop:treatment-groups-var-const-vec}
    If we run the Gram--Schmidt Walk design on the matrix $\xM$ with a constant column $\alpha \onevec$ appended, then
	the variance of the difference of the treatment group sizes may be bounded as
	\[
	\Var{ \ntreated - \ncontrol }
	\leq \paren[\Big]{ \frac{\vto}{ n } + \frac{(1 - \vto) \alpha^2}{\xi^2 + \alpha^2} }^{-1}
	\enspace,
	\]
	where $\xi$ denotes the maximum row norm of $\xM$ excluding the constant column.
In particular, if no column is added or if $\alpha = 0$, the variance is at most $n / \vto$.
\end{appproposition}

Note that when $\vto = 1$, the (unnormalized) difference between the treatment groups $\ntreated - \ncontrol$ has variance equal to $n$.
This is the group size balance achieved by the Bernoulli design.
On the other hand, when $\vto = 0$, then the variance is bounded by $1 + (\xi / \alpha)^2$.
If $\alpha = \xi$, we have $\Pr{\abs{\ntreated - \ncontrol} \leq 7} \geq 15/16$, which is close to the balance achieved by complete randomization if $n$ is not very small.
For intermediate values of the design parameter $\vto \in [0,1]$, the variance is bounded by the weighted harmonic mean of these two values.
Hence, whenever $\vto < 1$ and $\alpha^2 > \xi^2 / (n - 1)$, the Gram--Schmidt Walk design will achieve more group size balance than the Bernoulli design.
Furthermore, more balance is achieved the larger $\alpha$ is.
Note that $\maxnormX^2 / \paren{n - 1} \to 0$ at a fast rate under our assumptions, so any $\alpha > 0$ satisfies the condition for sufficiently large $n$.

We now provide a proof of Proposition~\suppref{prop:treatment-groups-var-const-vec}, which follows from the matrix inequality of Theorem~\mainref{thm:cov-bound}.
We remark that subgaussian tail bounds on the difference of the treatment group sizes $\ntreated - \ncontrol$ may be obtained in a similar manner from Theorem~\mainref{thm:sub-gaussian}.

\begin{proof}[Proof of Proposition~\suppref{prop:treatment-groups-var-const-vec}]
	Let $\onevec$ be the $n$-dimensional vector of ones.
	As $\xi^2$ is an upper bound on the squared norm of the original covariate vectors, $\xi^2 + \alpha^2$ is an upper bound on the squared norm after we append the column $\alpha \onevec$.

	Let $\xMx{\alpha}$ be the matrix $\xM$ with a constant column $\alpha \onevec$ appended.
	The gram matrix $\xMx{\alpha} \xMx{\alpha}^\tran$ satisfies 
	\[
		\xMx{\alpha} 	\xMx{\alpha}^\tran
		= \xM \xM^\tran + \alpha^2 \onevec \onevec^\tran 
		\succeq
		\alpha^2 \onevec \onevec^\tran .
	\]
	Using this together with the matrix inequality in Theorem~\mainref{thm:cov-bound}, we have that
	\[
	\Cov{\zv}
	\preceq \paren[\Big]{\vto \unitM + 
	 \frac{1 - \vto}{\xi^2 + \alpha^2} \xMx{\alpha} \xMx{\alpha}^\tran }^{-1}
	\preceq \paren[\Big]{\vto \unitM +  \frac{(1 - \vto) \alpha^2}{\xi^2 + \alpha^2}
	\onevec \onevec^\tran }^{-1} .
	\]

	The variance of the difference in treatment group sizes $\ntreated - \ncontrol$ is a quadratic form in the covariance matrix $\Cov{\zv}$ and so we can use the upper bound above:
	\[
	\Var{ \ntreated - \ncontrol }
	= \onevec^\tran \Cov{\zv} \onevec
	\leq \onevec^\tran \paren[\Big]{
		\vto \unitM +  \frac{(1 - \vto) \alpha^2}{\xi^2 + \alpha^2}
		\onevec \onevec^\tran }^{-1}
	    \onevec
	= \paren[\Big]{ \frac{\vto}{n} + \frac{(1 - \vto) \alpha^2}{\xi^2 + \alpha^2 } }^{-1}
	\enspace.
	\qedhere
	\]
\end{proof}

\subsubsection{Gram--Schmidt Walk design with Fixed-Size Treatment Groups}\label{sec:gsw-fixed-group-sizes}

Although covariate balancing under the Gram--Schmidt Walk design will typically yield treatment groups of similar size, the design does not guarantee fixed sizes of the treatment groups.
In this section, we show that the design can be extended to strictly enforce a desired number of treated units.

The group-balanced Gram--Schmidt Walk design is obtained by changing the construction of the step direction.
In particular, to fix the size of the treatment groups, one may choose the step direction according to the following constrained optimization:
\[
\begin{aligned}
	\gsuvt \gets \
	& \underset{\gsuv{}}{\text{argmin}}
	& & \norm{ \gsM \gsunv }^2 \\
	& \text{subject to}
	& & \gsuni = 0 \text{ for all } i \notin \alive \\
	& & & \gsun{p} = 1 \\
	& & & \sum_{i=1}^n \gsuni = 0
\end{aligned}
\]
The only difference here is that we have added an additional constraint that the sum of the coordinates of the step direction is zero.
The only exception to using this rule is that when only one unit remaining with a fractional assignment, in which case the step direction is the corresponding standard basis vector.

The modification ensures that the number of treated units $\ntreated = \sumin \indicator{\zi = 1}$ is as close as possible the expected number of treated units $\E{\ntreated} = \sumin \pzi$.
This is demonstrated by the following proposition.

\begin{appproposition} \label{prop:balanced_gsw}
	With probability one under the modified group-balanced Gram--Schmidt Walk design,
	\begin{equation}
		\abs[\big]{ \ntreated - \E{\ntreated}} < 1.
	\end{equation}
	If $\E{\ntreated}$ is an integer, then $\ntreated = \E{\ntreated}$ with probability one.
\end{appproposition}

\begin{proof}
	Note that for any assignment vector $\zv \in \setb{ \pm 1}^n$, the difference between the sizes of the two treatment groups is given by $\ntreated - \ncontrol = \iprod{ \onevec , \zv}$.
	Taking expectations, we arrive at
	\begin{equation}
		\E{ \ntreated - \ncontrol}
		= \E{ \iprod{ \onevec , \zv} }
		= \iprod{ \onevec , \E{\zv}}
		= \iprod{ \onevec, \zv_1},
	\end{equation}
	where $\zv_1$ is the initial fractional assignment and the last equality follows by the martingale property.
	We can express the difference between group sizes in terms of the iterative updates made by the group-balanced Gram--Schmidt Walk as
	\begin{equation}
		\ntreated - \ncontrol
		= \iprod{ \onevec , \zv}
		= \iprod{ \onevec, \sum_{t=1}^T \gsdt \gsuvt + \zv_1 }
		= \sum_{t=1}^T \gsdt \iprod{\onevec , \gsuvt} + \iprod{\onevec , \zv_1}
		= \sum_{t=1}^T \gsdt \iprod{\onevec , \gsuvt} + \E{\ntreated - \ncontrol}
		.
	\end{equation}
	For all but the final iteration, there is at least one alive unit which is not the pivot.
	Thus, by the additional constraint in the group-balanced Gram--Schmidt Walk, we have that $\iprod{\onevec , \gsuvt} = 0$ for $t=1, 2, \dots T - 1$.
	This means that in the above sum, all terms are zero except possibly the last term corresponding to the final iteration $T$.
	Applying this and rearranging the expressions above yields
	\begin{equation}
		\paren{\ntreated - \ncontrol} - \E{\ntreated - \ncontrol} = \gsd{T} \iprod{\onevec , \gsuv{T}} .
	\end{equation}
	The remainder of the proof considers two cases of the final iteration.
	The first case is that there is more than one alive unit at the final iteration.
	In this case, the additional balancing constraint ensures that $\iprod{ \onevec, \gsuv{T}} = 0$.
	Thus, we have that $\ntreated - \ncontrol = \E{\ntreated - \ncontrol}$ when there is more than one alive unit at the last iteration.

	The second case to consider is that the pivot is the only alive unit at the last iteration.
	In this case, we have that the update vector $\gsuv{T}$ has $1$ in the entry corresponding to the pivot and $0$ in the remaining entries.
	Thus, we have that $\iprod{ \onevec, \gsuv{T}} = 1$ in this case.
	The two possible values of the step size $\gsd{T}$ are $1 - \gsve{T}{p}$ and $1 + \gsve{T}{p}$.
	Because $\gsve{T}{p} \in \paren{-1,1}$, we have that $\abs{\gsd{T}} < 2$, regardless of which possible value is chosen.
	Thus, we obtain the upper bound
	\begin{equation}
		\abs{\paren{\ntreated - \ncontrol} - \E{\ntreated - \ncontrol}}
		= \abs{\gsd{T} \iprod{\onevec , \gsuv{T}}}
		= \abs{\gsd{T}} \cdot \abs{\iprod{\onevec , \gsuv{T}}}
		= \abs{\gsd{T}}
		< 2.
	\end{equation}
	The desired result follows from simple manipulation of terms.
	Because $\ntreated + \ncontrol = n$, we have that $\ncontrol = n - \ntreated$.
	Substituting this into the term on the left hand side, we obtain
	\begin{equation}
		\paren{\ntreated - \ncontrol} - \E{\ntreated - \ncontrol}
		= \paren{2 \ntreated - n} - \paren{2 \E{\ntreated} - n}
		= 2 \paren{ \ntreated - \E{\ntreated}}
	\end{equation}
	and now the upper bound above yields that $\abs{\ntreated - \E{\ntreated}} < 1$, as desired.

	It directly follows that if $\E{\ntreated}$ is an integer, then $\ntreated = \E{\ntreated}$ with probability one.
	To see this, observe that if $\E{\ntreated}$ is an integer, then $\ntreated - \E{\ntreated}$ is also an integer.
	Thus, the condition $\abs{\ntreated - \E{\ntreated}} < 1$ implies that $\ntreated - \E{\ntreated} = 0$ so that these two quantities are equal.
\end{proof}

The main reason for fixing the group sizes is that the estimator becomes invariant to constant shifts in the potential outcomes.
Experimenters often find this a desirable property because the average treatment effect is itself invariant to such shifts.
While we find that this strict balance modification does not typically significantly affect the behavior of the design, none of our analysis of the Gram--Schmidt Walk design described in the main paper applies to this modified version.
The reason is a technical one: namely, that the orthogonality of the updates between pivot phases no longer holds once this additional constraint is added.

We remark that adding this strict balance constraint does not increase the overall runtime of sampling an assignment vector.
In particular, the modified step size is the solution to a system of linear equations and so the same techniques as described in Section~\suppref{sec:gsw-implementation} may be used here.

%% file: tex/supp-alt-cis.tex
\subsection{Alternative Confidence Intervals} \label{sec:supp-alt-cis}

In this section, we discuss modified confidence intervals which may be more appropriate in finite samples.

The first possible modification is to replace $\evb$ with $\evb / \vto$ in the confidence intervals.
The limiting variance presumes that $\vto$ is close to one, but that might not be the case in small samples.
As shown by Proposition~\ref{coro:asymp-mse-fixed-vto}, the ratio $\vb / \vto$ is an upper bound on the limiting variance when $\vto$ does not approach one, making it more appropriate to use when $\vto$ is far from one in finite samples.
This modification also produces confidence intervals that are valid when the design parameter $\vto$ does not approach one.

The second modification is to replace $\evb$ with upper bound $\ridgeloss$ from Theorem~\mainref{thm:ht-mse-bound}.
This is a finite-sample bound on the variance, removing all asymptotic approximations of the variance.
In particular, unlike the limiting variance in Theorem~\mainref{thm:asymp-variance}, $\ridgeloss$ does not presume that the sample is sufficiently large so we are able to perfectly balance all linear functions of the covariates.
We provide a consistent, conservative estimator of $\ridgeloss$ in Section~\ref{sec:supp-ridge-estimator}.
These modifications are advisable when the sample is small, as the limiting variance can be anti-conservative in such setting.
However, these alternative characterizations of the variance will often be loose, also asymptotically,  meaning that the confidence intervals tend to be overly conservative.

Finally, the confidence interval in Theorem~\mainref{prop:confidence-intervals} uses an asymptotic approximation of the sampling distribution of the estimator.
An alternative is to use the finite-sample tail bounds described in Corollary~\mainref{coro:ht-tail-bound}.
Even if the tail bounds are valid in finite samples, the confidence intervals based on them will not be finite-sample valid, because $\ridgeloss$, which is used in the tail bounds, is typically unknown.
The estimator of $\ridgeloss$ mentioned above can be used also here, in which case the confidence interval would be asymptotically valid, but with fewer asymptotic approximations than the other intervals.
This third modification will increase conservativeness both in finite samples and asymptotically.

When the treatment effect is constant between units, an exact, finite-sample valid confidence interval can be constructed by inverting a Fisher-type randomization test.

%% file: tex/supp-additional-proofs.tex
\section{Additional Proofs} \label{sec:additional-proofs}

In this section, we provide additional proofs of various results contained in the main paper.

\input{\texpath/supp-mse-of-ht}

\input{\texpath/supp-var-bound}

\input{\texpath/supp-asymptotics}

\input{\texpath/supp-confidence-intervals}

%% file: tex/supp-mse-of-ht.tex
\subsection{MSE of the Horvitz--Thompson estimator (Lemma~\mainref{lemma:mse-expression})} \label{sec:supp-mse-expression}

In this section, we prove Lemma~\mainref{lemma:mse-expression}, which presents an expression for the mean squared error of the Horvitz Thompson estimator under an arbitrary design satisfying $\Pr{\zi = 1} = 1/2$ for all units $i \in [n]$.
We begin by restating Lemma~\mainref{lemma:mse-expression} below.

\begin{reflemma}{\mainref{lemma:mse-expression}}
	\mseexpression
\end{reflemma}
\begin{proof}
	Recall that the average treatment effect and Horvitz--Thompson estimator can be written as
	\begin{equation}
		\ate = \frac{1}{n} \iprod{\onevec, \poav - \pobv}
		\qquadand
		\htest = \frac{2}{n} \iprod{\zv, \oov}.
	\end{equation}
	By expressing the observed outcome as $\ooi = \poai \paren{ \frac{1 + \zi}{2}} + \pobi \paren{ \frac{1 - \zi}{2}}$, we see that
	\begin{equation}
		n \htest
		= 2 \iprod{\zv, \oov}
		= \iprod{\zv, \poav + \pobv} + \iprod{\onevec, \poav - \pobv}
		= \iprod{\zv, \pomv} + n \ate 
		\enspace,
	\end{equation}
	where we have used that $\zi^2 = 1$.
	Thus, we have that the error of the Horvitz--Thompson estimator is $\htest - \ate = \iprod{\zv, \pomv}/ n$.
	The expectation of the square of this expression is
	\begin{equation}
		\E[\big]{\paren{\htest - \ate}^2}
		= \frac{1}{n^2} \E[\big]{ \iprod{\zv, \pomv}^2 }
		= \frac{1}{n^2} \pomv^\tran \E[\big]{\zv \zv^\tran} \pomv,
	\end{equation}
	because $\pomv$ is not random.
	The proof is completed by noting that $\E{\zv \zv^\tran} = \Cov{\zv}$ because $\E{\zv} = \zerovec$ when $\Pr{\zi = 1} = 1 / 2$ for all $i \in [n]$.
	The desired result is obtained by rearranging terms.
\end{proof}

\subsection{Worst-case MSE (Lemma~\mainref{lemma:robustness})}

\begin{reflemma}{\mainref{lemma:robustness}}
	\lemmarobustness
\end{reflemma}

\begin{proof}
	By Lemma~\mainref{lemma:mse-expression} and using the operator norm inequality, we have that for all $\pomv \in \POspace{M}$,
	\[
	\E[\big]{\paren{\htest - \ate}^2}
	= \frac{1}{n^2} \pomv^\tran \Cov{\zv} \pomv
	\leq \frac{1}{n} \cdot \frac{1}{n} \norm{\pomv}^2 \cdot \norm{ \Cov{\zv} } 
	\leq \frac{M}{n}  \norm{ \Cov{\zv} } 
	\enspace.
	\]
	Moreover, this inequality is tight in the sense that there exists $\pomv \in \POspace{M}$ where equality holds.
	This follows from the definition of the operator norm.
\end{proof}

\subsection{Bernoulli design is min-max (Proposition~\mainref{coro:bernoulli-min-max})}

\begin{refproposition}{\mainref{coro:bernoulli-min-max}}
	\corobernoulliminmax
\end{refproposition}

\begin{proof}
	If a design is symmetric, it means that $\Pr{\zi = 1} = 1/2$ for all $i \in [n]$, which implies that $\Var{\zi} = 1$ for all $i \in [n]$.
	This means that $\tr{\Cov{\zv}} = n$.
	Using this, we can obtain the following lower bound on the operator norm of $\Cov{\zv}$:
	\[
	1 = \frac{1}{n} \cdot \tr{ \Cov{\zv} } = \frac{1}{n} \sumin \eigvali \leq \eigvalmax = \norm{\Cov{\zv}} \enspace.
	\]
	To see that equality holds for the Bernoulli design, observe that $\Cov{\zv} = \unitM$, so that $\norm{\Cov{\zv}} = 1$.
\end{proof}

\subsection{Balance-Robustness Trade-off Exists (Proposition~\mainref{prop:trade-off})}

\begin{refproposition}{\mainref{prop:trade-off}}
	\proptradeoff
\end{refproposition}
\begin{proof}
	By Proposition~\mainref{coro:bernoulli-min-max}, any design that achieves maximal robustness, in the sense of minimizing $\norm{ \Cov{\zv} }$, must have the covariance matrix $\Cov{\zv} = \unitM$.
	Therefore, the covariate balance operator norm for any such design is
	\[
	\norm{ \Cov{\xM^\tran \zv} }
	= \norm{ \xM^\tran \Cov{\zv} \xM }
	= \norm{ \xM^\tran \xM }
	= \sigma_{\max}(\xM)^2
	\enspace.
	\]
	Thus, by assumption,
	\[
	\norm{ \Cov{\xM^\tran \zv} }
	= \sigma_{\max}(\xM)^2
	> \maxnormX^2.
	\]
	As shown in Theorem~\mainref{theorem:gsw-satisfies-discrepancy}, the Gram--Schmidt Walk design with $\vto = 0$ gives the guarantee $\norm{\Cov{\xM^\tran \zv} } \leq \maxnormX^2$.
	Hence, a design that minimizes $\norm{ \Cov{\zv} }$ does not yield the minimal covariate balance operator norm.
\end{proof}

\subsection{Balance-Robustness Trade-off and MSE (Theorem~\mainref{thm:utility-of-new-discrepancy-formulation})}

In this section, we prove Theorem~\mainref{thm:utility-of-new-discrepancy-formulation}, which provides a bound on the mean squared error of the Horvitz--Thompson estimator that depends on the robustness and covariate balance guarantees.
We restate the theorem here for completeness.

\begin{reftheorem}{\mainref{thm:utility-of-new-discrepancy-formulation}}
	\mseimplications
\end{reftheorem}
\begin{proof}
	\begin{proof}
		Let $\lfx \in \Reals^d$ be an arbitrary vector and let $\fpomv = \xM^\tran \lfx$ and $\rpomv = \pomv - \fpomv$ so that $\pomv = \rpomv + \fpomv$.
		First, we use Lemma~\mainref{lemma:mse-expression} together with a generalized arithmetic-geometric (AM-GM) inequality to separate the mean squared error into two parts: one which depends on the linear prediction $\fpomv$ and the other which depends on the residual $\rpomv$.
		For all $\eta > 0$,
		\begin{align*}
			n^2 \E{\paren{\ate - \htest}^2} 
			&= \pomv^\tran \Cov{\zv} \pomv 
			\\
			&= \paren{\fpomv + \rpomv}^\tran \Cov{\zv} \paren{\fpomv + \rpomv} 
			\\
			&= \fpomv^\tran \Cov{\zv} \fpomv + \rpomv \Cov{\zv} \rpomv + 2 \fpomv^\tran \Cov{\zv} \rpomv
			\\
			&\leq \paren{1 + \eta^2} \ \fpomv^\tran \Cov{\zv} \fpomv 
			+ \paren{1 + \eta^{-2}} \ \rpomv^\tran \Cov{\zv} \rpomv 
			\\
			&= \paren{1 + \eta^2} \ \lfx^\tran \Cov{\xM^\tran \zv} \lfx 
			+ \paren{1 + \eta^{-2}} \ \rpomv^\tran \Cov{\zv} \rpomv
			\enspace.
			\intertext{By applying the operator norm bound to the quadratic forms and using the bounds on the two operator norms guaranteed by Problem~\ref{problem:distributional-discrepancy}, we have that }
			&\leq \paren{1 + \eta^2} \ \norm{ \Cov{\xM^\tran \zv} } \norm{\lfx}^2
			+ \paren{1 + \eta^{-2}} \ \norm{\Cov{\zv}} \norm{\rpomv}^2 \\
			& \leq \paren{1 + \eta^2} \xsc \norm{\lfx}^2
			+ \paren{1 + \eta^{-2}} \zsc \norm{\rpomv}^2 \enspace.
			\intertext{Choosing the value of $\eta^2 = \sqrt{\zsc \norm{\rpomv}^2 / \xsc \norm{\lfx}^2}$ to minimize this upper bound and recalling that the residuals are defined by $\rpomv = \pomv - \xM \lfx$, we have that}
			& \leq \zsc \norm{\pomv - \xM \lfx}^2
			+ \xsc \norm{\lfx}^2
			+ 2 \sqrt{\zsc \xsc} \norm{\pomv - \xM \lfx} \cdot \norm{\lfx}
			\enspace.
		\end{align*}
		Note that this upper bound holds for an arbitrary vector $\lfx$.
		The theorem follows by minimizing over all such $\lfx$ and dividing both sides by $n$.
	\end{proof}
\end{proof}

%% file: tex/supp-var-bound.tex
\subsection{GSW-Design Navigation of Trade-off (Theorem~\mainref{theorem:gsw-satisfies-discrepancy})}

\begin{reftheorem}{\mainref{theorem:gsw-satisfies-discrepancy}}
	\theoremgswsatisfiesdiscrepancy
\end{reftheorem}
\begin{proof}
	All projection matrices are less than the identity matrix in the Loewner order.
	Thus, Theorem~\mainref{thm:cov-bound} implies that $\Cov{\gsM \zv} \preceq \PgsM \preceq \unitM$.
	Observe that the covariance matrix of $\Cov{\gsM \zv}$ can be written in block form as
	\[
	\Cov{\gsM \zv}
	= \begin{bmatrix}
		\vto \Cov{\zv}
		&\maxnormX^{-1}\sqrt{\vto \paren{1 - \vto}} \Cov{\xM^\tran \zv, \zv}^\tran\;
		\\[0.25em]
		\;\maxnormX^{-1}\sqrt{\vto \paren{1 - \vto}} \Cov{\xM^\tran \zv, \zv} \quad
		&\maxnormX^{-2} \paren{1 - \vto} \Cov{\xM^\tran \zv}
	\end{bmatrix}.
	\]
	By extracting the upper left and lower right blocks in the matrix inequality $\Cov{\gsM \zv} \preceq \unitM$ and rearranging terms, we have that $ \Cov{\zv} \preceq \vto^{-1} \unitM$ and $ \maxnormX^{-2} \Cov{\xM^\tran \zv} \preceq \paren{1-\vto}^{-1} \unitM $.
	The proof is completed by taking the operator norm of both sides of these inequalities.
\end{proof}

\subsection{Analysis of the mean squared error (Theorem~\mainref{thm:ht-mse-bound})} \label{sec:app-mse-analysis}

We begin by analyzing the mean squared error of the Horvitz--Thompson estimator under the Gram--Schmidt Walk design.
We start by presenting the relationship between the quadratic form in matrix $\rlMat$ and the loss of ridge regression.

\begin{applemma}\label{lem:ridgeloss-expression}
	Let $\xM$ be an arbitrary $n$-by-$\xdim$ matrix with maximum row norm $\maxnormX = \max_{i \in [n]} \norm{\xvi}$.
	For all $\vto \in (0,1)$ and $\pomv \in \Reals^n$,
	\begin{equation}
	n \ridgeloss
	=
	\pomv^\tran \rlMat \pomv
	=
	\pomv^\tran \paren[\big]{ \vto \unitM + \paren{1 - \vto} \maxnormX^{-2} \xM^{\tran} \xM  }^{-1} \pomv
	=
	\min_{\lfx \in \Reals^\xdim} \bracket[\bigg]{
		\frac{1}{\vto}\norm*{ \pomv - \xM \lfx}^2
		+ \frac{\maxnormX^2}{1 - \vto} \norm*{\lfx}^2
	}.
	\end{equation}
\end{applemma}

\begin{proof}
	Let $\lfxr$ be the optimal linear function in the minimization term above.
	Note that multiplying the objective function by $\vto > 0$ does not change the minimizer $\lfxr$, and so
	\begin{equation}
	\lfxr
	=
	\argmin_{\lfx \in \Reals^\xdim} \bracket[\bigg]{
		\frac{1}{\vto} \norm*{ \pomv - \xM \lfx}^2
		+ \frac{\maxnormX^2 }{1 - \vto} \norm*{\lfx}^2}
	=
	\argmin_{\lfx \in \Reals^\xdim} \bracket[\bigg]{
		\norm*{ \pomv - \xM \lfx}^2
		+ \frac{\maxnormX^2 \vto}{1 - \vto} \norm*{\lfx}^2}
	\enspace,
	\end{equation}
	which has closed-form solution \citep[see, e.g.,][p. 64]{Hastie2009Elements}:
	\begin{equation}
	\lfxr
	=
	\paren[\bigg]{\xM^\tran \xM + \frac{\maxnormX^2 \vto}{1 - \vto} \unitM }^{-1} \xM^\tran \pomv
	=
	\rM^{-1} \xM^{\tran} \pomv
	\enspace,
	\end{equation}
	where we have defined
	$\rM =  \xM^{\tran} \xM + \frac{\maxnormX^2 \vto}{1 - \vto} \unitM $.
	We next consider each of the terms in the objective function when we substitute the optimal $\lfxr$.
	The second term becomes
	\[
	\frac{\maxnormX^2 }{1 - \vto} \norm*{\lfxr}^{2}
	= \frac{\maxnormX^2 }{1 - \vto} \norm*{ \rM^{-1} \xM^\tran \pomv }^{2}
	= \frac{\maxnormX^2 }{1 - \vto} \pomv^{\tran} \xM \rM^{-2} \xM^{\tran} \pomv.
	\]
	The first term becomes
	\begin{align*}
	\frac{1}{\vto} \norm*{ \pomv - \xM \lfxr }^2
	&= \frac{1}{\vto} \norm*{ \pomv - \xM \rM^{-1} \xM^{\tran} \pomv }^2
	= \frac{1}{\vto} \norm*{ \paren*{\unitM - \xM \rM^{-1} \xM^{\tran}} \pomv }^2 \\
	&= \frac{1}{\vto} \pomv^{\tran} \paren*{\unitM - \xM \rM^{-1} \xM^{\tran}}^{2} \pomv \\
	&= \frac{1}{\vto} \pomv^{\tran}
	\paren*{\unitM - 2 \xM \rM^{-1} \xM^{\tran} + \xM \rM^{-1} \xM^{\tran} \xM \rM^{-1} \xM^\tran }
	\pomv \\
	&= \frac{1}{\vto} \pomv^{\tran}
	\paren*{\unitM - \xM \bracket*{ 2 \rM^{-1} - \rM^{-1} \xM^{\tran} \xM \rM^{-1}} \xM^{\tran}}
	\pomv \\
	&= \frac{1}{\vto} \pomv^{\tran}
	\paren*{\unitM - \xM \bracket*{ 2 \rM^{-1} - \rM^{-2} \xM^{\tran} \xM} \xM^{\tran}}
	\pomv
	\enspace,
	\end{align*}
	where the last line follows from the fact that $\rM^{-1}$ and $\xM^{\tran} \xM$ commute.
	To see that the matrices $\rM^{-1}$ and $\xM \xM^{\tran}$ commute, first observe that
	$\rM = \frac{\maxnormX^2 \vto}{1 - \vto} \unitM +  \xM^{\tran} \xM$ has the same eigenvectors as
	$\xM^{\tran} \xM$.
	It follows that $\rM^{-1}$ also has the same eigenvectors as $\xM^{\tran} \xM$.
	Thus, the two matrices $\rM^{-1}$ and $\xM^{\tran} \xM$ are simultaneously diagonalizable and
	therefore commute.

	Substituting these separate calculations into the objective function, we obtain the optimal value
	\begin{align}
	\frac{1}{\vto} & \norm*{ \pomv - \xM \lfxr}^2
	+ \frac{\maxnormX^2}{1 - \vto} \norm*{\lfxr}^2 \\
	&= \frac{1}{\vto} \pomv^{\tran}
	\paren*{\unitM - \xM \bracket*{ 2 \rM^{-1} - \rM^{-2} \xM^{\tran} \xM} \xM^{\tran}}
	\pomv
	+ \frac{\maxnormX^2 }{1 - \vto} \pomv^{\tran} \xM \rM^{-2} \xM^{\tran} \pomv \\
	&= \frac{1}{\vto} \pomv^{\tran}
	\paren*{\unitM - \xM
		\bracket*{ 2 \rM^{-1} - \rM^{-2} \xM^{\tran} \xM - \frac{\vto \maxnormX^{2}}{1 - \vto} \rM^{-2} }
		\xM^{\tran}}
	\pomv \\
	&= \frac{1}{\vto} \pomv^{\tran}
	\paren*{\unitM - \xM
		\bracket*{ 2 \rM^{-1} - \rM^{-2}
			\paren*{ \xM^{\tran} \xM + \frac{\vto \maxnormX^{2}}{1 - \vto} \unitM }
		}\xM^{\tran}}
	\pomv \\
	&= \frac{1}{\vto} \pomv^{\tran}
	\paren*{\unitM - \xM \bracket*{ 2 \rM^{-1} - \rM^{-2} \rM}\xM^{\tran}}
	\pomv \\
	&= \frac{1}{\vto} \pomv^{\tran} \paren*{\unitM - \xM \rM^{-1} \xM^{\tran}} \pomv
	\end{align}

	To complete the proof, we apply the Woodbury identity which asserts that for appropriately sized
	matrices $\boldsymbol{U}$, $\boldsymbol{V}$, and $\boldsymbol{C}$,
	${
		\paren*{\unitM + \boldsymbol{U} \boldsymbol{C} \boldsymbol{V}}^{-1}
		=
		\unitM - \boldsymbol{U} \paren*{\boldsymbol{C}^{-1} + \boldsymbol{V}\boldsymbol{U} }^{-1}
		\boldsymbol{V}
	}$, given that the inverses exist.
	Applying the Woodbury identity with $\boldsymbol{U} = \xM$, $\boldsymbol{V} = \xM^{\tran}$, and
	$\boldsymbol{C} = \frac{1-\vto}{ \maxnormX^{2} \vto} \unitM$, we obtain
	\begin{align*}
	\frac{1}{\vto} \paren*{\unitM - \xM \rM^{-1} \xM^{\tran}}
	&= \frac{1}{\vto} \paren*{\unitM - \xM
		\paren*{\frac{\maxnormX^{2} \vto}{1 - \vto} \unitM + \xM^{\tran} \xM }^{-1}
		\xM^{\tran}
	} \\
	&= \frac{1}{\vto} \paren*{ \unitM +  \frac{ \maxnormX^{-2}(1-\vto)}{ \vto} \xM^{\tran} \xM }^{-1}
	= \paren*{ \vto \unitM +   \maxnormX^{-2}(1-\vto) \xM^{\tran} \xM }^{-1}. \qedhere
	\end{align*}
\end{proof}

Using this lemma, we are now ready to establish the improved mean squared error analysis of the Horvitz--Thompson estimator under the \gswdesign.

\begin{reftheorem}{\mainref{thm:ht-mse-bound}}
	\htmsebound
\end{reftheorem}
\begin{proof}
	In Lemma~\mainref{lemma:mse-expression}, we established that the mean squared error of the Horvitz--Thompson estimator is a quadratic form in the covariance matrix of assignments, $\Cov{\zv}$.
	We can obtain a bound on this matrix using the inequality in Theorem~\mainref{thm:cov-bound}.
	The upper left $n$-by-$n$ block of $\Cov{\gsM \zv}$ is $\vto \Cov{\zv}$.
	The corresponding block of the projection matrix $\PgsM$ in Theorem~\mainref{thm:cov-bound} is $\vto \rlMat$ where
	\begin{equation}
		\rlMat = \paren[\big]{ \vto \unitM + \paren{1-\vto} \maxnormX^{-2} \xM \xM^\tran}^{-1}.
	\end{equation}

	If $\mat{A} \preceq \mat{B}$, then any two principal submatrices corresponding to the same row and column set $S$ satisfy the inequality $\mat{A}_S \preceq \mat{B}_S$.
	It follows that $\Cov{\zv} \preceq \rlMat$.
	Using the definition of the Loewner partial order together with Lemma\ref{lem:ridgeloss-expression}, we obtain
	\[
	\E{\paren{\ate - \htest}^2} = \frac{1}{n^2} \pomv^\tran \Cov{\zv} \pomv
	\leq \frac{1}{n^2} \pomv^\tran \rlMat \pomv
	= L / n \enspace. \qedhere
	\]
\end{proof}

%% file: tex/supp-asymptotics.tex
\subsection{When balancing improves precision} \label{sec:when-balancing-improves}

The following corollary tells us that it is almost always beneficial to seek at least some covariate balance when using the Gram--Schmidt Walk design.

\begin{corollary} \label{corr:opt-vto-lt-one}
	If the scaled sum of cross-moments between covariates and potential outcomes is greater than the second moment of potential outcomes, $\maxnormX^{-2} \norm{ \xM^\tran \pomv }^2 > \norm{ \pomv }^2$, then the design parameter $\vto$ that minimizes the mean squared error is less than one.
\end{corollary}

The cross-moments capture the predictiveness of the covariates, so $\norm{ \xM^\tran \pomv }^2$ becomes larger as the covariates become more predictive.
Typically, we expect $\norm{ \xM^\tran \pomv }^2$ to be much larger than $\norm{ \pomv }^2$ unless the sample is very small and the covariates are close to completely unpredictive.
To see this, note that $\norm{ \xM^\tran \pomv }^2$ tends to grow at an $n^2$-rate if the covariates are at least somewhat predictive, while $\norm{ \pomv }^2$ tends to grow at an $n$-rate.
The factor $\maxnormX^{2}$ captures the scaling of the covariates and the presence of outliers.
As we discuss in Section~\mainref{sec:asymptotic-analysis}, a reasonable growth rate of $\maxnormX^{2}$ is $\xdim \logf{n}$, meaning that the scaling will generally not be consequential.
This tells us that it is almost always beneficial to at least partially balance the covariates, so we should set $\vto < 1$.
One exception is small experiments with nearly unpredictive covariates, where $\vto = 1$ may be optimal.

\begin{proof}[Proof of Corollary~\ref{corr:opt-vto-lt-one}]
	We begin by letting
	\begin{equation}
		\rlMatv{\vto}
		= \paren[\Big]{\vto \unitM + \paren{1 - \vto} \maxnormX^{-2} \xM \xM^\tran }^{-1}.
	\end{equation}
	We can write $\minloss{\vto} = n^{-1} \pomv^\tran \rlMatv{\vto} \pomv$, and
	\begin{equation}
		\frac{d \minloss{\vto}}{d \vto}
		= \frac{1}{n} \pomv^\tran \rlMatv{\vto} \paren[\Big]{\maxnormX^{-2} \xM \xM^\tran - \unitM} \rlMatv{\vto} \pomv.
	\end{equation}
	Note that $\rlMatv{1} = \unitM$, implying that
	\begin{equation}
		\frac{d \minloss{\vto}}{d \vto}\Bigr|_{\vto=1} > 0
		\iff
		\pomv^\tran \paren[\Big]{\maxnormX^{-2} \xM \xM^\tran - \unitM} \pomv > 0
		\iff
		\maxnormX^{-2} \norm{ \xM^\tran \pomv }^2 > \norm{ \pomv }^2.
	\end{equation}
	Note that $\minloss{1} = \norm{ \pomv }^2$, meaning that the inequality in Theorem~\mainref{thm:ht-mse-bound} is an equality when $\vto = 1$.
	Thus, the derivative of the mean squared error coincide of the derivative of the bound at $\vto = 1$.
\end{proof}

%% file: tex/supp-confidence-intervals.tex
\subsection{Tail bound on Horvitz--Thompson estimator (Corollary~\mainref{coro:ht-tail-bound})}

\newcommand{\tscale}{t}

\begin{refcorollary}{\mainref{coro:ht-tail-bound}}
	\httailbound
\end{refcorollary}
\begin{proof}
	We prove the bound for the upper tail.
	The proof for the lower tail is identical.
	For any $\tscale > 0$, we have
	\[
	\Pr[\big]{\htest - \ate \geq \teerr}
	\leq
	\expf{- \tscale \teerr} \E[\big]{\expf[\big]{\tscale \paren{\htest - \ate}}}.
	\]
	This can be shown either as a consequence of Markov's inequality or from the exponential inequality $\indicator{x \geq 0} \leq \expf{\tscale x}$.
	The proof of Lemma~\mainref{lemma:mse-expression} (proved in Supplement~\ref{sec:supp-mse-expression}) shows that $\htest - \ate = \iprod{\zv, \pomv} / n$.
	The columns of $\gsM$ are linearly independent by construction, so we can define a vector $\arbvec = \tscale n^{-1} \gsM \paren[\big]{ \gsM^\tran \gsM }^{-1} \pomv$.
	This~allows us to write
	\[
	\E[\big]{\expf[\big]{\tscale \paren{\htest - \ate}}}
	=
	\E[\big]{ \expf[\big]{\tscale n^{-1} \iprod{\zv, \pomv} } }
	=
	\E[\big]{ \expf[\big]{ \iprod{\gsM \zv, \arbvec} } }.
	\]
	Theorem~\mainref{thm:sub-gaussian} upper bounds the right-hand side by $\expf{ \norm{\arbvec}^2 / 2}$.
	For the current choice of $\arbvec$, the squared norm simplifies to
	\begin{equation}
		\norm{\arbvec}^2
		=
		\frac{\tscale^2}{n^2} \pomv^\tran \paren[\big]{ \gsM^\tran \gsM }^{-1} \pomv
		=
		\frac{\tscale^2 \ridgeloss}{n},
	\end{equation}
	where the final equality follows from Lemma~\suppref{lem:ridgeloss-expression}.
	Taken together, we obtain
	\begin{equation}
		\Pr[\big]{\htest - \ate \geq \teerr}
		\leq
		\expf[\bigg]{ \frac{\tscale^2 \ridgeloss}{2n} - \tscale \teerr }.
	\end{equation}
	The proof is completed by setting $\tscale = \teerr n / \ridgeloss$.
\end{proof}

%% file: tex/supp-simulations.tex
\newcommand{\mw}{\rule{1.5cm}{0pt}}

\newcommand{\statstab}[1]{%
\begin{table}
\centering
\caption{Performance of various designs relative to the Bernoulli design when $n = #1$}\label{tab:stats-#1}
\begin{tabular}{l rr rrrr}\toprule
\mw & \mw & \mw & \mw & \mw & \mw & \mw \\[-\arraystretch\normalbaselineskip]
 & & & \multicolumn{4}{c}{Mean square error} \\ \cmidrule(lr){4-7}
 & Robustness & Balance & A & B & C & D \\\midrule
\input{tabs/stats-#1.tex}\bottomrule
\end{tabular}
\end{table}
}

\newcommand{\citab}[1]{%
\begin{table}
\centering
\caption{Coverage and width of confidence intervals $n = #1$}\label{tab:ci-#1}
\begin{tabular}{l rr rr}\toprule
%\mw & \mw & \mw & \mw & \mw & \mw & \mw \\[-\arraystretch\normalbaselineskip]
 & \multicolumn{2}{c}{Main CI} & \multicolumn{2}{c}{Alternative CI} \\ \cmidrule(lr){2-3} \cmidrule(lr){4-5}
 & Coverage & Width & Coverage & Width\\\midrule
\input{tabs/ci-#1.tex}\bottomrule
\end{tabular}
\end{table}
}

\section{Numerical Illustrations}\label{sec:full-simulations}

\subsection{Setting}

The simulations are based on data from an experiment by \citet{Groh2016Macroinsurance}.
The experiment investigates how insurance against macroeconomic shocks affects microenterprises in Egypt.
The sample consisted of $2961$ enterprises that were clients of Egypt's largest microfinance institution.
The authors offered the insurance to a randomly selected subset of the enterprises, using a combination of stratification and the matched pair design.
After three to seven months, they measured various outcomes and estimated the treatment effects by comparing the two treatment groups.
The estimates indicate that the insurance had little impact on the enterprises.

Our aim here is not to recreate the exact empirical setting in the original experiment.
This is generally not possible because we never observe all potential outcomes.
The purpose is instead to use the data from \citet{Groh2016Macroinsurance} to create a plausible empirical setting.

\paragraph{Experimental Units}
We run the simulations with $n \in \braces{30, 296, 2960}$ units from the original data set.
For each sample size $n$, we use the first $n$ units in the data set according to the original order of the data.
For the largest sample size, we omit one unit to make the sample size even to accommodate the matched pair designs, which requires an even number of units.

\paragraph{Covariates}
The covariates we will seek to balance are the $14$ covariates that \citet{Groh2016Macroinsurance} use in their matched paired design.
These covariates include the owner's gender, risk aversion, sales after revolution, and profits in previous months, to name a few.
However, two of these covariates are almost collinear.
They are indicator variables of missingness of two other covariates, and we collapse them into a single covariate using disjunction.
The covariates are demeaned, normalized and decorrelated before treatment assignment so that they are in scaled isotropic position: $\xM^\tran \xM = n \unitM$.

\paragraph{Outcomes}
We investigate four outcomes, some of which are based on real outcomes from the study of \citet{Groh2016Macroinsurance} and some which are synthetic.
In all cases, the two potential outcomes of each unit are the same: $\poai = \pobi$.
As showed in the paper, the mean square error depends on the potential outcomes only through their sum $\poai + \pobi$.
The outcomes are:
\begin{enumerate}[label=\Alph*.]
	\item
	\textbf{New Workers}:
	This is a binary outcome from the original study of whether the enterprises have hired a new worker after treatment assignment.
	This is the actual outcome in the experiment, which potentially is affected by treatment.
	However, we do not attempt to remove this treatment effect, only demeaning and normalizing it.
	The purpose here is not to recreate the true potential outcomes in the original experiment, which would be impossible without strong assumptions.
	The purpose is instead to create potential outcomes that are empirically plausible.
	The covariates are not particularly informative of this outcome.
	The adjusted coefficient of determination ($R^2$) for sample $n = 2960$ is $0.039$.

	\item
	\textbf{Worst-Case Outcomes}:
	This is an artificially generated outcome to represent a worst-case scenario.
	We generate the outcome based on the largest eigenvectors of three of the designs in the study: the matched pair design, rerandomization and the Gram--Schmidt Walk design with $\vto = 0.01$.
	In particular, the outcome vector is the largest eigenvector of the sum of the outer products of the eigenvectors from the three designs.
	These designs all seek covariate balance, and they are subsequently some of the least robust designs in the study.
	By construction, the covariates are almost completely uninformative of this potential outcome.

	\item
	\textbf{Profits}:
	This is another outcome from the original study, namely the profits of the enterprises after treatment assignment.
	Like above, we do not attempt to remove any potential treatment effect, but we demean and normalize the outcome.
	Two of the covariates are the enterprises' profits at baseline before treatment assignment, so the covariates are highly predictive of this outcome.
	The coefficient of determination is $0.314$.

	\item
	\textbf{Linearly Predictive Outcomes}:
	This is another artificially generated outcome to represent a setting where the covariates are perfectly predictive of the potential outcomes.
	The potential outcomes for each unit is the sum of its first five covariates: $\poai = \pobi = \iprod{\vec{v}, \xvi}$, where $\vec{v}$ is vector whose first five elements are one and the remaining elements are zero.
\end{enumerate}

\paragraph{Designs}
We consider five different experimental designs:
\begin{enumerate}
	\item
	\textbf{Bernoulli}:
		The fully randomized design, in which the treatments are assigned independently.
	\item
	\textbf{Complete}:
		The group-balanced randomization design, in which each treatment group is ensured to contain exactly half of the units.
	\item
	\textbf{Matched Pair}:
		The matched pair design using the network flow algorithm by \citet{Greevy2004Optimal} to construct optimal pairs.
	\item
	\textbf{Rerandomization}:
		The rerandomization procedure described by \citet{Li2018Asymptotic}, with the acceptance threshold $a$ is selected so $\Pr{\chi^2_{K} \leq a} = 0.001$, as recommended by \citet{Li2018Asymptotic}.

	\item
	\textbf{GSW}:
		The Gram--Schmidt Walk design as described in the current paper.
		We set the parameter $\vto$ to five values ranging from focusing mostly on balance to focusing mostly on robustness: $0.01$, $0.1$, $0.5$, $0.9$ and $0.99$.
\end{enumerate}

\subsection{Mean Square Error}

Tables~\ref{tab:stats-30},~\ref{tab:stats-296}~and~\ref{tab:stats-2960} present the operator norms and mean square errors under the different designs.
The other two tables have the same structure, but present the results for sample sizes $n = 30$ and $n = 2960$.

We begin by focusing on Table~\ref{tab:stats-296}, with $n=296$ units.
The simulation results corroborate the theoretical results.
Designs that limit the amount of randomness, such as the matched pair design, rerandomization and the Gram--Schmidt Walk design with small $\vto$, are less robust, as indicated by a larger $\norm{\Cov{\zv}}$.
However, they also achieve better covariate balance, as indicated by a smaller $\norm{\Cov{\xM^\tran \zv}}$.
The most covariate balance is achieved by the Gram–Schmidt Walk design with $\vto = 0.01$, where the covariate imbalance is $3\%$ of the imbalance under the Bernoulli design.
Still, the Gram–Schmidt Walk design with $\vto = 0.01$ is more robust than the matched pair design and almost as robust as rerandomization.

The mean square errors largely reflect the operator norms; designs that achieve more balance perform better when covariates are predictive, and perform worse when they are not.
All designs performs virtually identically for outcome ``A,'' which means that this particular outcome is not explained well by the covariates, nor is it the worst-case outcome for these designs.
The outcome in column ``B'' is created to illustrate a setting where covariate balance is not helpful, and we see that the mean square error follows the robustness measure closely, with the exception of the matched pair design.
The outcome in this setting is constructed to be closer to the worst-cases for rerandomization and the Gram–Schmidt Walk design.
If we had generated the outcomes to be closer to the worst-case for the matched pair design, the mean square error for the matched pair design would be closer to two, mirroring its robustness guarantee.
The final two columns show that the mean square error is smaller when the covariates are predictive for designs that seek covariate balance.
The covariates are perfectly predictive in column ``D,'' capturing what could be seen as a best-case scenario.

The results from the smaller and larger samples largely mirror the results when $n = 296$.
One difference is that rerandomization achieves more balance, at the cost of less robustness, when $n = 30$ compared to the Gram--Schmidt Walk design with $\vto = 0.01$.
When $n = 296$, the Gram--Schmidt Walk design with $\vto = 0.01$ gave more balance, at the cost of less robustness, compared to rerandomization.

When $n = 2960$, all designs except the matched pair design have similar robustness.
There are, however, large difference in balance.
The most balance is achieved by Gram--Schmidt Walk design with $\vto = 0.01$, yielding a balance operator norm that is approximately 41 times smaller than under the matched pair design and approximately 57 times smaller than under rerandomization.
This is reflected in the mean square errors when the covariates are predictive of the outcomes.
The one exception is outcome C, for which the matched pair design yields slightly lower mean square error than the Gram--Schmidt Walk design.
However, for outcome D, the mean square error is more than 31 times smaller under the Gram--Schmidt Walk design than under the matched pair design.

For all simulations, but especially for $n = 2960$, the Monte Carlo estimates of the robustness operator norm is upwards biased.
For example, we know that the operator norm is $1$ under the Bernoulli design no matter the sample size, but the simulations gives $1.01$, $1.03$, and $1.11$, respectively for the different sample sizes.
Readers should keep this in mind when interpreting the operator norms.
The bias will be slightly larger when the true operator norm is small, but the bias will not change the ranking.

\begin{table}
\centering
\caption{Performance of various designs relative to the Bernoulli design when $n = 30$}\label{tab:stats-30}
\begin{tabular}{l rr rrrr}\toprule
\mw & \mw & \mw & \mw & \mw & \mw & \mw \\[-\arraystretch\normalbaselineskip]
 & & & \multicolumn{4}{c}{Mean square error} \\ \cmidrule(lr){4-7}
 & Robustness & Balance & A & B & C & D \\\midrule
\input{tabs/stats-30.tex}\bottomrule
\end{tabular}
\end{table}

\begin{table}
\centering
\caption{Performance of various designs relative to the Bernoulli design when $n = 296$}\label{tab:stats-296}
\begin{tabular}{l rr rrrr}\toprule
\mw & \mw & \mw & \mw & \mw & \mw & \mw \\[-\arraystretch\normalbaselineskip]
 & & & \multicolumn{4}{c}{Mean square error} \\ \cmidrule(lr){4-7}
 & Robustness & Balance & A & B & C & D \\\midrule
\input{tabs/stats-296.tex}\bottomrule
\end{tabular}
\end{table}

\begin{table}
\centering
\caption{Performance of various designs relative to the Bernoulli design when $n = 2960$}\label{tab:stats-2960}
\begin{tabular}{l rr rrrr}\toprule
\mw & \mw & \mw & \mw & \mw & \mw & \mw \\[-\arraystretch\normalbaselineskip]
 & & & \multicolumn{4}{c}{Mean square error} \\ \cmidrule(lr){4-7}
 & Robustness & Balance & A & B & C & D \\\midrule
\input{tabs/stats-2960.tex}\bottomrule
\end{tabular}
\end{table}

\clearpage
\subsection{Confidence Intervals}

Tables~\ref{tab:ci-30},~\ref{tab:ci-296}~and~\ref{tab:ci-2960} present the coverage and width of the confidence intervals described in Section~\mainref{sec:confidence-intervals} in the main paper for the three different sample sizes.
The table also present the alternative confidence interval based on the finite-sample variance bound $\ridgeloss$, as described in Section~\ref{sec:supp-alt-cis}.
Finally, for comparison, the tables present the coverage and width of conventional Neyman-style confidence intervals under the Bernoulli and complete randomization designs.
The simulation setting is identical to above, with the exception that the sharp null no longer holds.
Instead, the potential outcomes are the two outcomes from the original study.
That is, potential outcome $\poai$ is outcome A from above, and potential outcome $\pobi$ is outcome C from above.
This is because, unlike the mean square error, the behavior of the confidence intervals is not invariant to whether the sharp null holds.
The nominal confidence level for all confidence intervals is $95\%$.

Table~\ref{tab:ci-30} shows that the main confidence intervals undercover when $\vto \geq 0.5$ and $n = 30$.
The coverage rate is around $93\%$, so it is not dramatic undercoverage, but still notable.
The reason for this is that the asymptotic approximation used in the main confidence intervals presumes that all linear functions have been balance, and this is not true for $\vto$ close to one when $n$ is small.
When $\vto = 0.1$ and $\vto = 0.01$, coverage exceeds the nominal level when $n = 30$, and the width of the intervals is approximately $25\%$ narrower than under the Bernoulli and complete randomization designs.
Note that the variance bound introduces some conservativeness in this setting, so the confidence intervals when $\vto = 0.1$ and $\vto = 0.01$ may undercover for some potential outcomes.
The alternative confidence intervals have coverage that meet or exceed the nominal level, although the width is sometimes dramatically larger than the width of the confidence intervals under the Bernoulli design.

When $n = 296$ and $n = 2960$, as presented in Tables~\ref{tab:ci-296}~and~\ref{tab:ci-2960}, the coverage of the main confidence intervals exceed the nominal level for all values of $\vto$.
The widths of the intervals are approximately $10\%$ to $14\%$ narrower than under the Bernoulli and complete randomization designs.
The alternative confidence intervals have coverage exceeding the nominal levels, but are wider than the main intervals.

\begin{table}
\centering
\caption{Coverage and width of confidence intervals $n = 30$}\label{tab:ci-30}
\begin{tabular}{l rr rr}\toprule
%\mw & \mw & \mw & \mw & \mw & \mw & \mw \\[-\arraystretch\normalbaselineskip]
 & \multicolumn{2}{c}{Main CI} & \multicolumn{2}{c}{Alternative CI} \\ \cmidrule(lr){2-3} \cmidrule(lr){4-5}
 & Coverage & Width & Coverage & Width\\\midrule
\input{tabs/ci-30.tex}\bottomrule
\end{tabular}
\end{table}

\begin{table}
\centering
\caption{Coverage and width of confidence intervals $n = 296$}\label{tab:ci-296}
\begin{tabular}{l rr rr}\toprule
%\mw & \mw & \mw & \mw & \mw & \mw & \mw \\[-\arraystretch\normalbaselineskip]
 & \multicolumn{2}{c}{Main CI} & \multicolumn{2}{c}{Alternative CI} \\ \cmidrule(lr){2-3} \cmidrule(lr){4-5}
 & Coverage & Width & Coverage & Width\\\midrule
\input{tabs/ci-296.tex}\bottomrule
\end{tabular}
\end{table}

\begin{table}
\centering
\caption{Coverage and width of confidence intervals $n = 2960$}\label{tab:ci-2960}
\begin{tabular}{l rr rr}\toprule
%\mw & \mw & \mw & \mw & \mw & \mw & \mw \\[-\arraystretch\normalbaselineskip]
 & \multicolumn{2}{c}{Main CI} & \multicolumn{2}{c}{Alternative CI} \\ \cmidrule(lr){2-3} \cmidrule(lr){4-5}
 & Coverage & Width & Coverage & Width\\\midrule
\input{tabs/ci-2960.tex}\bottomrule
\end{tabular}
\end{table}

\clearpage

%% file: tabs/stats-30.tex
Bernoulli       & 1.01 & 1.00 & 1.00 & 1.00 & 1.00 & 1.00 \\
Complete        & 1.04 & 1.03 & 1.03 & 0.94 & 1.03 & 1.03 \\
Matched Pair    & 2.01 & 1.97 & 1.15 & 1.58 & 0.47 & 1.04 \\
Rerandomization & 4.57 & 0.35 & 0.76 & 4.11 & 1.24 & 0.18 \\
GSW 0.99        & 1.01 & 1.00 & 1.00 & 1.00 & 1.00 & 0.99 \\
GSW 0.90        & 1.07 & 0.98 & 0.99 & 1.06 & 0.99 & 0.94 \\
GSW 0.50        & 1.48 & 0.87 & 0.93 & 1.45 & 0.97 & 0.65 \\
GSW 0.10        & 2.61 & 0.57 & 0.81 & 2.55 & 0.95 & 0.26 \\
GSW 0.01        & 3.17 & 0.52 & 0.80 & 3.01 & 0.91 & 0.21 \\

%% file: tabs/stats-296.tex
Bernoulli       & 1.03 & 1.00 & 1.00 & 1.00 & 1.00 & 1.00 \\
Complete        & 1.04 & 1.00 & 1.00 & 1.00 & 1.00 & 1.00 \\
Matched Pair    & 2.05 & 0.42 & 1.02 & 1.19 & 0.58 & 0.24 \\
Rerandomization & 1.45 & 0.18 & 0.99 & 1.44 & 0.68 & 0.17 \\
GSW 0.99        & 1.04 & 0.98 & 1.00 & 1.01 & 0.99 & 0.97 \\
GSW 0.90        & 1.08 & 0.81 & 0.99 & 1.07 & 0.91 & 0.78 \\
GSW 0.50        & 1.30 & 0.33 & 0.99 & 1.29 & 0.72 & 0.28 \\
GSW 0.10        & 1.50 & 0.07 & 0.99 & 1.50 & 0.63 & 0.05 \\
GSW 0.01        & 1.56 & 0.03 & 0.98 & 1.56 & 0.62 & 0.02 \\

%% file: tabs/stats-2960.tex
Bernoulli       & 1.11 & 1.000 & 1.00 & 1.00 & 1.00 & 1.000 \\
Complete        & 1.11 & 1.000 & 1.00 & 1.00 & 1.00 & 1.001 \\
Matched Pair    & 2.16 & 0.123 & 1.00 & 1.46 & 0.67 & 0.063 \\
Rerandomization & 1.12 & 0.171 & 0.97 & 1.04 & 0.74 & 0.171 \\
GSW 0.99        & 1.11 & 0.842 & 0.99 & 1.00 & 0.95 & 0.841 \\
GSW 0.90        & 1.11 & 0.327 & 0.97 & 1.00 & 0.79 & 0.325 \\
GSW 0.50        & 1.12 & 0.052 & 0.96 & 1.00 & 0.70 & 0.051 \\
GSW 0.10        & 1.12 & 0.008 & 0.96 & 1.01 & 0.69 & 0.006 \\
GSW 0.01        & 1.12 & 0.003 & 0.96 & 1.05 & 0.69 & 0.002 \\

%% file: tabs/ci-30.tex
Bernoulli & 98.7 & 1.00 &  &  \\
Complete  & 97.6 & 0.98 &  &  \\
GSW 0.99  & 92.7 & 0.75 &  98.0 & 0.94 \\
GSW 0.90  & 92.8 & 0.75 &  95.0 & 0.79 \\
GSW 0.50  & 93.7 & 0.76 &  98.8 & 1.06 \\
GSW 0.10  & 95.3 & 0.77 & 100.0 & 2.43 \\
GSW 0.01  & 95.6 & 0.77 & 100.0 & 7.73 \\

%% file: tabs/ci-296.tex
Bernoulli & 98.4 & 1.00 &  &  \\
Complete  & 98.4 & 1.00 &  &  \\
GSW 0.99  & 96.3 & 0.86 &  98.1 & 0.97 \\
GSW 0.90  & 96.7 & 0.86 &  98.0 & 0.94 \\
GSW 0.50  & 97.6 & 0.86 &  99.8 & 1.22 \\
GSW 0.10  & 98.1 & 0.86 & 100.0 & 2.71 \\
GSW 0.01  & 98.2 & 0.86 & 100.0 & 8.56 \\

%% file: tabs/ci-2960.tex
Bernoulli & 98.8 & 1.00 &  &  \\
Complete  & 98.8 & 1.00 &  &  \\
GSW 0.99  & 97.9 & 0.90 &  98.4 & 0.94 \\
GSW 0.90  & 98.7 & 0.90 &  99.1 & 0.96 \\
GSW 0.50  & 99.0 & 0.90 & 100.0 & 1.28 \\
GSW 0.10  & 99.1 & 0.90 & 100.0 & 2.86 \\
GSW 0.01  & 99.1 & 0.90 & 100.0 & 9.04 \\